%% file: main.tex
\documentclass[conference]{IEEEtran}
\IEEEoverridecommandlockouts
\usepackage{cite}
\usepackage{amsmath,amssymb,amsfonts}
\usepackage{algorithmic}
\usepackage{graphicx}
\usepackage{textcomp}
\usepackage{xcolor}
\def\BibTeX{{\rm B\kern-.05em{\sc i\kern-.025em b}\kern-.08em
    T\kern-.1667em\lower.7ex\hbox{E}\kern-.125emX}}

\usepackage{booktabs}  
\usepackage{gnuplot-lua-tikz}

\usepackage{dsfont}
\DeclareMathOperator{\Tr}{Tr}
\newcommand{\Ell}{\mathcal{L}}

\newcommand{\R}{\mathds{R}}
\newcommand{\C}{\mathds{C}}

\newcommand{\bfa}{\boldsymbol{\alpha}}
\newcommand{\bs}[1]{{\boldsymbol{#1}}}

\usepackage{xcolor}

\begin{document}

\title{Quandary: An open-source C++ package for high-performance optimal control of open quantum systems}

\author{
\IEEEauthorblockN{Stefanie G{\"u}nther and N. Anders Petersson}
\IEEEauthorblockA{\textit{Center for Applied Scientific Computing} \\
\textit{Lawrence Livermore National Laboratory}\\
Livermore, CA, USA \\
\{guenther5, petersson1\}@llnl.gov}
 \and
 \IEEEauthorblockN{Jonathan L. DuBois}
 \IEEEauthorblockA{\textit{Quantum Coherent Device Physics} \\
 \textit{Lawrence Livermore National Laboratory}\\
 Livermore, CA, USA\\
 dubois9@llnl.gov}
}

\maketitle

\begin{abstract}
Quantum optimal control can be used to shape the control pulses for realizing unitary and non-unitary transformations of quantum states. These control pulses provide the fundamental interface between the quantum compiler and the quantum hardware. Most current software for quantum optimal control (e.g. Qutip\cite{johansson2012qutip} or Krotov~\cite{goerz2019krotov}) is restricted to run on shared memory platforms, limiting their applicability to smaller quantum systems, in particular if interactions with the environment are taken into account. This paper gives an overview of the open-source code Quandary, which is designed to solve quantum control problems in larger open quantum systems modelled by Lindblad's master equation. Implemented in C++, Quandary uses the message passing paradigm for distributed memory computers that enables scalability to large numbers of compute cores. Accompanied by numerical examples, this paper presents an overview on existing theoretical developments for open optimal quantum control realizing state-to-state transfer, unitary gate optimization as well as state-preparation, and presents the numerical tools and implementation aspect as realized in Quandary, for deployment on modern high-performance computing platforms. 
\end{abstract}


\section{Introduction}

Designing optimal controls for driving a quantum system to a desired final state is a crucial task in several application areas, for example NMR spectroscopy \cite{PhysRevA.63.032308, nielsen2010, TOSNER2009120} and molecular physics \cite{shapiro2012quantum}. More recently, the rapidly evolving field of quantum information processing and quantum computing has attracted much interest in quantum optimal control, e.g. for quantum state preparation \cite{Rojan2014arbitrary}, realization of logical gate operations \cite{doi:10.1080/09500340802344933, PhysRevLett.89.188301}, 
and quantum error correction \cite{waldherr2014quantum, gaitan2008quantum}. See also the excellent review of quantum control theory, tools and applications in \cite{koch2016controlling} and references therein.

\begin{figure}
    \centering
    \input{figures/statetostate/fidelities.tex}
   \caption{Fidelity at $t=10$ns for a zero-to-one state transfer for various $T_1$ decay times. "Closed system" optimization is performed without $T_1$ decay terms, only enabling those terms when evaluating the resulting fidelity. In the "open system" case the decay terms are included during the optimization.}
    \label{fig:0to1_openvsclosed}
\end{figure}
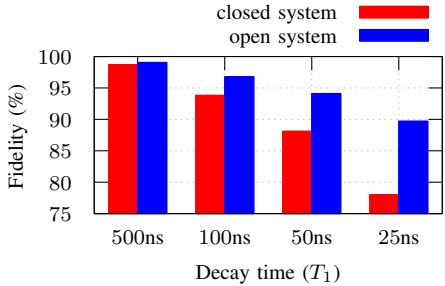

Numerical optimal control of quantum systems is a challenging problem. First of all, the numerical methods must be chosen carefully to properly reflect the physical properties of a quantum system.
Secondly, the computational complexity grows quickly because the dimension of the underlying Hilbert space $N$ scales exponentially with the number of qubits or qudits in the system -- for example, $N=2^Q$ for $Q$ qubits. 
When a closed quantum system is considered, the state vector is in $\C^N$ and its time evolution is governed by Schrodinger's equation. However, quantum systems are never fully separated from their environment. The importance of taking system-environment interaction into account is exemplified in Figure \ref{fig:0to1_openvsclosed}, where control pulses were optimized for a state-to-state transfer from {the} $|0\rangle$ to the $|1\rangle$ state. In the first case (red bars), the optimization is performed in a closed system setting and the target is reached with $99.9999\%$ fidelity. However, when those controls are evaluated in an open system setting, taking decay and dephasing operators into account, the fidelity drops quickly as the decay times get shorter. In the second case (blue bars), the control pulses were directly optimized in the open system setting, leading to greatly improved target state fidelities.
When taking system-environment interactions into account throughout the optimization, the system state is described by it's density matrix $\rho$ with dimension $\C^{N\times N}$, e.g. $N^2=2^{2Q}$ for $Q$ qubits. As the number of qubits to be taken into account increases, this quickly becomes computationally demanding, both in terms of memory requirement as well as compute time, indicating the need for 
high-performance compute (HPC) platforms. 

Open-source software packages for optimal control of open quantum systems include the popular Python-based quantum toolbox QuTiP~\cite{johansson2012qutip}, and the Krotov package~\cite{goerz2019krotov} in which the optimization algorithm of the same name is implemented. These Python packages provide an easy-to-use and flexible interface. However, they are restricted to run on shared memory platforms. Parallel processing is restricted to multi-threading and concurrent execution of independent tasks. For this reason, their applicability is restricted to model smaller quantum systems. Similar scalability restrictions apply to Matlab packages for optimal quantum control, such as Dynamo~\cite{machnes2011comparing} or Spinach \cite{HOGBEN2011179}. For an extensive list of open-source quantum software packages, see \cite{openquantumprojects, quantikilist}. 
Some HPC capabilities for open quantum dynamics are provided by the open-source package QuaC \cite{QuaCGithub}, which is written in C++ and based on the PETSc library \cite{petsc-web-page}. However, Quac only simulates open system dynamics and lacks optimal control capabilities.

In this paper we present an overview of the open-source code Quandary~\cite{quandaryGithub}, which is designed for scalable simulations and optimal control of open quantum systems, aiming at large {HPC} platforms. Quandary is written in C++, which provides a portable implementation for distributed memory computing based on the message passing paradigm. 

{In Quandary, the underlying quantum dynamics is modeled by Lindblad's master equation and a flexible user interface (examples provided on GitHub~\cite{quandaryGithub}) allows for a variety of composite systems to be considered, see Section \ref{sec:mastereq}. {It uses a compact control pulse parameterization based on B-spline wavelets that modulate the amplitude and phase of carrier waves.}
As described in Section \ref{sec:optimproblem}, Quandary uses gradient-based optimal control to shape {the} control pulses for driving a quantum system to perform {state and gate transformations}. Gradient computations are carried out using the discrete adjoint method to yield exact gradients of the discretized objective function. Three common optimal control scenarios are considered: state-to-state transfer, unconditional state preparation, and unitary gate transformations. For the latter two cases, Quandary implements several recent theoretical results that significantly reduce the computational complexity of the quantum control problem. Detailed numerical examples are provided to compare features of the implementation and illustrate convergence properties.
}

{{Targeting HPC platforms, Quandary provides three} levels of computational parallelism. First, independent solves of Lindblad's master equation are distributed over different compute cores. Secondly, the quantum state in each solve is distributed over multiple cores to perform linear algebra operations in parallel, based on the PETSc library~\cite{petsc-web-page}. A third, more experimental, level of parallelism applies a multigrid-in-time algorithm for solving Lindblad's master equation, using the XBraid library~\cite{xbraid}. Detailed weak and strong {scaling studies} are reported in Section \ref{sec:scaling}, focusing on parallel performance of the linear algebra operations.
Conclusions are given in Section~\ref{sec:conclusions}. 
}

\section{Simulation of open quantum dynamics}\label{sec:mastereq}

Quandary models the dynamics of composite open quantum systems that are driven by external control pulses. The composite system is composed of $Q\geq 1$ subsystems, where each subsystem $q$ can have an arbitrary (finite) number of energy levels, $n_q\geq 1$. Each subsystem can, for example, model a qubit ($n_q=2$), a qudit ($n_q>2$), or a resonator cavity ($n_q \gg 2$). The state of the composite system is described by its density matrix $\rho(t) \in \C^{N\times N}$, where $N = \prod_{q=1}^Q n_q$.

The quantum dynamics is modeled by Lindblad's master equation~\cite{manzano2020lindblad, lidar2019lecture}:
\begin{align}\label{eq:mastereq}
 \dot \rho(t) = &-i\left[H(t),\rho(t)\right] + \Ell\left(\rho(t)\right), \quad 0\leq t \leq T,
\end{align}
subject to an initial condition. To slow down the oscillatory time variation in the density matrix, Quandary solves Lindblad's equation in a rotating frame of reference.
The Lindbladian $\Ell(\rho)$ holds collapse operators for modeling Markovian interactions between the quantum system and its environment. Both decay and dephasing can be modeled in each of the subsystems:
\begin{align}
    \Ell\left(\rho\right) =  \sum_{q=1}^Q \sum_{l=1}^2  \Ell_{lq} \rho \Ell_{lq}^{\dagger} - \frac 1 2 \left( \Ell_{lq}^{\dagger}\Ell_{lq} \rho + \rho\Ell_{lq}^{\dagger} \Ell_{lq}\right).
\end{align}
Here, the collapse operators acting on subsystem $q$ are $\Ell_{1q} := \frac{1}{\sqrt{T_1^q}} a_q$ (decay) and $\Ell_{2q} :=  \frac{1}{\sqrt{T_2^q}} a_q^{\dagger}a_q$ (dephasing) with decay and dephasing times $T_1^q$ and  $T_2^q$. The lowering and raising operators acting on subsystem $q$ are denoted $a_q$ and $a_q^\dagger$, respectively.
The Hamiltonian is decomposed into a time-independent system part and a time-varying control part, which models the action of external control fields. 
In the system Hamiltonian, Quandary implements two common coupling models for circuit QED devices, characterized by the cross-Kerr coefficients $\xi_{pq}$ and/or the Jaynes-Cummings coupling coefficients $g_{pq}$:
\begin{align}
  H_{sys} := \sum_{q=1}^Q &\left({\left(\omega_q - \omega^{rot}_q\right)} a_q^\dagger a_q - \frac{\xi_q}{2} a_q^{\dagger}a_q^{\dagger}a_q a_q\right. +\nonumber \\
  & \ \  \left.\sum_{p>q} g_{pq} \left(a_p^\dagger a_q + a_pa_q^\dagger\right) - \xi_{pq} a_p^{\dagger}a_p a_q^{\dagger} a_q  \right),
\end{align}
{where} $\omega_q$ is the 0-1 transition frequency, $\xi_q$ is the self-Kerr coefficient{, and $\omega_q^{rot}$ denotes the frequency of rotation in subsystem $q$}. We note that the above form of the Jaynes-Cummings coupling term assumes that all rotational frequencies are equal; the general form implemented in Quandary includes the time-dependent factors {$\exp(\pm i (\omega^{rot}_p - \omega^{rot}_q) t)$ that pre-multiply the terms {$a_p^\dagger a_q$ and $a_pa_q^\dagger$, respectively}.}

\begin{figure}
    \centering
    \includegraphics[width=.23\textwidth]{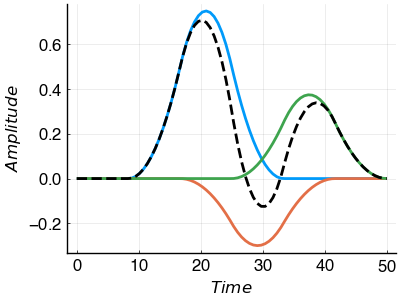}
    \includegraphics[width=.23\textwidth]{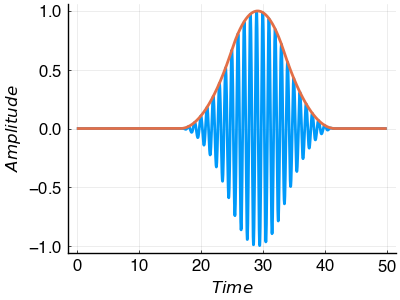}
    \caption{Left: A B-spline function (black dashed line) is a weighted sum of B-spline basis functions with compact support (colored lines). Right: The envelope and phase of the carrier wave is controlled by the B-spline.}
    \label{fig:bspline}
\end{figure}

The action of external fields on subsystem $q$ is modeled by the lab-frame control Hamiltonian {$f_q(t)(a_q + a_q^\dagger)$}, where the control pulse is the form $f_q(t) = 2\, \mbox{Re}\{d_q(t) e^{i t \omega^{rot}_q} \}$. After applying the rotating frame approximation, this leads to the control Hamiltonian 
\begin{align}
   H_c(t) = \sum_{q=1}^Q d_q(t) a_q + d_q^*(t) a_q^\dagger.
\end{align} 
In Quandary, the control pulses in the rotating frame are parameterized by B-spline basis functions that modulate the amplitude and phase of a number of carrier waves: 
\begin{align}\label{eq:rotctrl}
  d_q(\alpha_q,t) &= \sum_{s=1}^{N_s}S_s(t) \sum_{f=1}^{N_f} \alpha_{q}^{s,f}\,e^{i t \Omega_q^f}.
\end{align}
The control pulse for subsystem $q$ consists of $N_s$ basis functions $S_s(t)$ and $N_f$ carrier frequencies $\Omega_q^f$, with control parameters $\alpha^{q}_{s,f} \in \C$, giving a total of $2N_s N_f$ real-valued control parameters per subsystem. 
The basis functions $S_s(t)$ are chosen to be piece-wise quadratic B-spline wavelets\cite{Unser97}, which have local support in time and are continuously differentiable. The wavelets are centered on a uniform grid in time 
where the knot spacing is $\Delta \tau = T/(N_s-2)$. Figure \ref{fig:bspline} exemplifies the control parameterization with three basis functions (left) serving as envelopes for a carrier wave (right). 
Parameterizing the control functions using B-splines with carrier waves provides a compact alternative to discretizing the control functions on the same time step as Lindblad's master equation. Transitions between energy levels in a quantum system are caused by resonance, where the resonant frequencies can be determined from the coefficients in the system Hamiltonian~\cite{petersson2021optimal}.
By choosing the frequencies of the carrier waves such that they coincide with the resonant frequencies, the amplitude and phase of the carrier waves can vary on a much slower time scale than the carrier waves. Thus, the number of control parameters can be independent of, and significantly smaller than, the number of time steps for solving Lindblad's equation.
In contrast to other control parameterizations using basis functions (e.g. Fourier modes or Legendre polynomials, see e.g. \cite{caneva2011chopped}), the B-spline wavelets are local in time. Hence, each control parameter only influences the envelope function locally.

In its forward integration mode, Quandary evolves Lindblad's equation in time using a given control vector $\bs{\alpha}$ and a given initial state.
To enable accurate long time-integration and avoid numerical dissipation, Quandary employs the symplectic implicit midpoint rule (IMR), which is a second-order accurate Runge-Kutta scheme. Further details on the time-integration scheme as it is implemented in Quandary are provided in Appendix \ref{app:IMR}.

\section{Gradient-based optimal control}\label{sec:optimproblem}

In addition to simulating the time-evolution of an open quantum system, Quandary can optimize the control vector $\bs{\alpha}$ to drive the system towards a desired target state at a final time $T>0$. 
{To that end, Quandary minimizes an objective function $\cal G(\bs{\alpha})$ of the following form:}
\begin{align}
  {\cal G}(\boldsymbol{\alpha}) &=\frac{1}{M} \sum_{i=1}^{M} \beta_i J(\rho^{tar}_i, \rho_i(T)) \nonumber \\
     &\ + \gamma_1 \int_0^T \beta(t) J(\rho^{tar}_i, \rho_i(t)) \, \mathrm{d} t + \frac{\gamma_2}{2} \| \bfa \|^2_2. \label{eq:optimproblem}
\end{align}
The realized state, $\rho_i(t)$, is the solution of Lindblad's master equation \eqref{eq:mastereq}, driven by external control fields corresponding to the control vector $\boldsymbol{\alpha}$. The realized state is subject to the initial condition $\rho_i(0)$, as will be discussed below.
The first term in ${\cal G}(\bs{\alpha})$ is (a weighted average of) a {merit functional} $J$ that quantifies the discrepancy between the desired target state, $\rho^{tar}_i$, and the realized state, $\rho_i(T)$, evaluated at the final time $T$.
The second term in ${\cal G}(\bs{\alpha})$ serves as a penalty that can be added {with parameter $\gamma_1 \geq 0$}. This term drives the system towards the target state before the final time is reached. Here, the weight function is chosen to be $\beta(t) =  \frac{1}{a}\exp\left(-\left(\frac{t-T}{a}\right)^2\right)$, where $a>0$ is a tunable parameter.
The third term in ${\cal G}(\bs{\alpha})$ is a Tikhonov regularization that can {be added with parameter $\gamma_2 \geq 0$}. This term regularizes the optimization problem by favoring control vectors with smaller norm.

The following choices for the {merit functional} $J$ are available in Quandary:
\begin{align}
 J_{Frob}(\rho^{tar},\rho) &= \frac 12 \| \rho^{tar} - \rho||^2_F \\ 
 J_{Tr}(\rho^{tar},\rho)  &= 1 - \mbox{Tr}\left((\rho^{tar})^\dagger\rho\right) \\
 J_{N_m}(\rho)  &= \mbox{Tr} \left( N_m \rho \right). \label{eq:Jmeasure}
\end{align}
As explained in more detail below, the first two functionals, $J_{Frob}$ and $J_{Tr}$, are most commonly used for realizing state-to-state transfer and unitary gate transformations (Section \ref{sec:statetostate} and Section \ref{sec:gateoptim}, respectively). 
The {merit functional} $J_{N_m}$ is primarily intended to reduce the computational complexity for the unconditional preparation of pure states (Section \ref{sec:purestateprep}).

The optimization problem is solved numerically by employing preconditioned gradient-based (L-BFGS) updates to the control vector $\bs{\alpha}$.
The gradient of the objective function with respect to the control vector is computed by solving the adjoint Lindblad equation, which propagates sensitivities backwards in time. All components of the gradient can then be calculated at a computational cost that is independent of the number of control parameters.
Quandary follows the first-discretize-then-optimize approach and derives the discrete adjoint equation for gradient computation from the discretized Lindblad equation, see Appendix~\ref{app:adjoint}.
This choice ensures that the gradient of the objective function is consistent with the numerical evaluation of the objective function, and hence is exact on the discretized (implementation) level -- not only in the limit as the step size goes to zero, but also for any finite step size in the time discretization. Further details on the discretization of the adjoint Lindblad equation and the gradient computation are given in Appendix~\ref{app:adjoint}. 

\subsection{State-to-state transfer}\label{sec:statetostate}
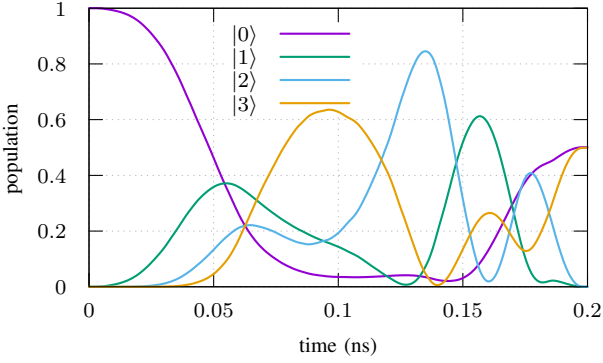
\begin{figure}
    \centering
    \input{figures/FockSuperp03/population.tex}
    \caption{State-to-State transfer: Optimized evolution of energy level occupation for the preparation of the Fock superposition state $\frac{1}{\sqrt{2}}\left(|0\rangle + |3\rangle\right)$. At $T=0.2$ns, the target state is reached with $99.999\%$ fidelity.
    }
    \label{fig:Fock_population}
\end{figure}

\begin{figure}
    \centering
    \input{figures/FockSuperp03/optim_history.tex}
    \caption{Optimization history for the state-to-state transfer problem towards the Fock-superposition state.}
    \label{fig:Fock_optimhistory}
\end{figure}
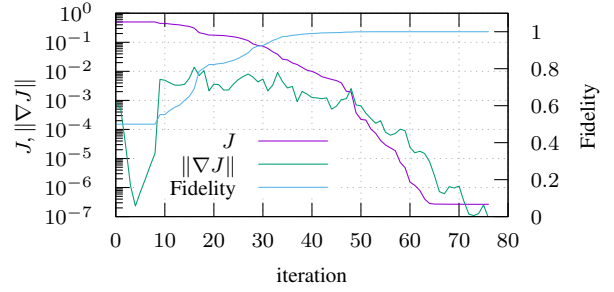

In the simplest setting, the {objective function} \eqref{eq:optimproblem} represents a state-to-state transfer problem that aims to drive one specific initial state $\rho(0)$ towards a given target state $\rho^{tar}$. In this case, the number of initial states in ${\cal G}(\bs{\alpha})$ is $M=1$. 
As an example, Figure \ref{fig:Fock_population} demonstrates the optimized evolution of a $4$-level qudit\footnote{System parameters for one qudit modelled with $n_q=4$ energy levels: $\omega_q/2\pi = 4.416$GHz, $\xi_q/2\pi=230$MHz, $T_1 = 93.79\mu$s, $T_2 = 102.52\mu$s. Control pulses are parameterized with $N_s=10$ B-spline basis functions with carrier wave frequencies of $\Omega_q/2\pi = \{0.0, -\xi_{pq}\}$.}
driven from the ground state $\rho(0) = |0\rangle\langle 0|$ to the Fock-superposition state $\rho^{tar} = \frac 12 \left(|0\rangle + |3\rangle\right)\left(\langle 0| + \langle 3|\right)$. To assess the quality of the optimized solution, for pure target states, Quandary measures the fidelity using 
\begin{align}\label{eq:fidelity}
    F(\rho^{tar},\rho) = \Tr \left(\left(\rho^{tar}\right)^\dagger \rho\right),\quad\rho^{tar} = |\psi\rangle\langle \psi|.
\end{align}
At $T=0.2$ns, the Fock superposition state is reached with $99.999\%$ fidelity. 
The convergence of the L-BFGS iteration is shown in Figure \ref{fig:Fock_optimhistory}. The {merit functional}, here $J_{Frob}$, decreases monotonically to below $10^{-6}$, and the reduction of the gradient norm by 4 orders of magnitude indicates that the optimization problem was successfully solved.

\subsection{Unconditional state preparation}\label{sec:purestateprep} 
In the unconditional state preparation problem, the optimization task is to find control pulses that drive \textit{any} initial state to a given target state $\rho^{tar}$, i.e.~the target is the same for all initial states. In contrast to a state-to-state transfer, this unconditional preparation of a target state typically requires the quantum system of interest to be coupled to a dissipative channel\cite{weiss2012quantum}, allowing for entropy flow from the system of interest to a bath. With Quandary, the unconditional state preparation capability can, for example, be used to optimize for a ground-state reset of multiple 
qudits, coupled to a dissipative readout cavity.

To account for any initial state, the optimization problem
generally needs to consider a basis of states that span all possible initial states $\rho(0)$, resulting in $M=N^2$ terms in the objective function \eqref{eq:optimproblem}. 
Quandary utilizes the following $N^2$ pure basis states to span the space of any initial state:
\begin{align} \label{eq:basismats}
B^{kj} := \frac 12 \left( |k\rangle \langle k| + |j\rangle \langle j| \right) +  \begin{cases} 
          0 & \text{if } \, k=j \\ 
        \frac 12 \left( |k\rangle \langle j| + |j\rangle \langle k| \right) & \text{if } \, k<j \\
        \frac i2 \left( |j\rangle \langle k| - |k\rangle  |j\rangle \right) & \text{if } \, k>j
      \end{cases}
\end{align}
for $k,j=0,\dots,N-1$. 
In contrast to other density matrix parameterizations, the basis matrices $B^{kj}$ are constructed in such a way that each element is itself a density matrix that represents a (pure) quantum state~\cite{groundstatepaper}. This ensures a physically meaningful propagation when solving Lindblad's master equation.

The benefit of these basis matrices becomes apparent when the target state is pure. In this case, it is suggested in~\cite{groundstatepaper} to utilize the {merit functional} $J_{N_m}$. In $J_{N_m}$ (see \eqref{eq:Jmeasure}), the observational operator $N_m$ is a diagonal matrix, with diagonal elements $\lambda_l = |l-m|\, \forall\, l=0,\dots, N-1$. Here, the integer $m\geq 0$ corresponds to the pure target $\rho^{tar}=|m\rangle\langle m|$ in the computational basis.
It is straightforward to verify that $J_{N_m}(\rho)\geq 0$ and that $J_{N_m}(\rho)= 0$, if and only if $\rho = |m\rangle\langle m|$.
Since the solution of Lindblad's master equation is linear with respect to the initial state, and the {merit functional} $J_{N_m}(\rho(T))$ is linear in $\rho(T)$, it is sufficient to consider one ensemble state 
\begin{align}
    \rho_s(0) := \frac {1}{N^2} \sum_{kj} B^{kj},
\end{align}
as the only initial state for evaluating the objective function ${\cal G}(\bs{\alpha})$, see~\cite{groundstatepaper} for further details. This approach reduces the computational complexity from solving $M = N^2$ to $M=1$ Lindblad equations per objective function evaluation. 

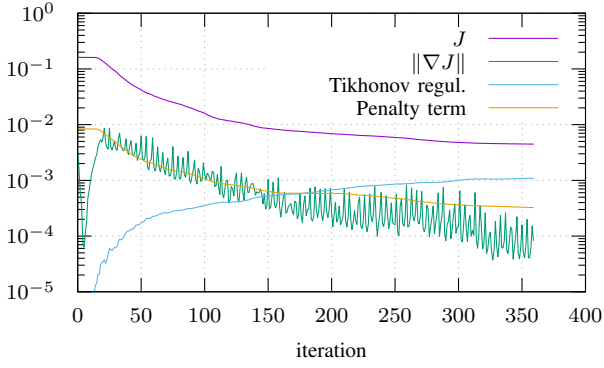
\begin{figure}
    \centering
    \input{figures/2x2x20_reset/optim_history.tex}
    \caption{Optimization history for the optimal reset of a qubit-qubit-cavity system~\cite{groundstatepaper}.}
    \label{fig:2x2x20_optimhistory}
\end{figure}

\begin{figure}
    \centering
    \input{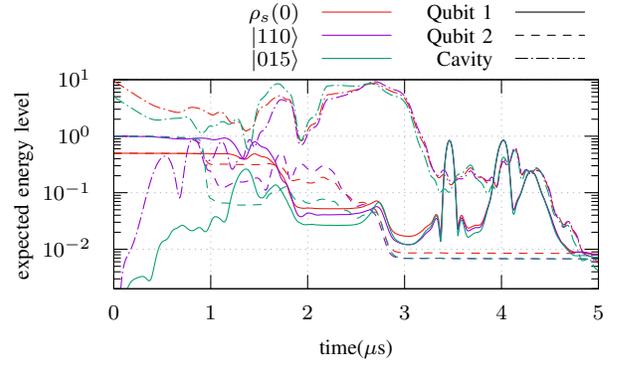}
    \caption{Optimal reset of a qubit-qubit-cavity system: Expected energy level over time for various initial states~\cite{groundstatepaper}.}
    \label{fig:2x2x20_expected}
\end{figure}

\begin{figure}
    \centering
    \input{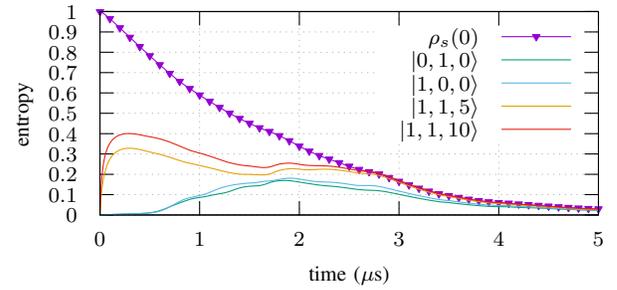}
    \caption{Evolution of the von Neumann entropy for various initial states, driven by the control pulses from the optimal reset of a qubit-qubit-cavity system.}
    \label{fig:2x2x20_neumann}
\end{figure}

An important application for unconditional state preparation is ground state reset, which aims to drive a system from any initial state to the ground state of the zero's energy level, hence $\rho^{tar} = |0\rangle\langle 0|\in\C^{N\times N}$. Fast reset is required in many application scenarios, for example in schemes for error correction~\cite{divincenzo2000physical, barrett2013simulating, reed2012realization}. 
In that case, the merit functional $J_{N_0}$ measures the expected energy level of the final state $\rho(T)$. As an example, Figure \ref{fig:2x2x20_optimhistory}, demonstrates the optimization history for resetting two qubits with $n_{1}=n_{2}=2$, coupled to a readout cavity with $n_{3}=20$ energy levels\footnote{System parameters for the optimal reset of two qubits in a cavity: $\omega_q / 2\pi = \{4.41666, 4.510, 6.84081\}$GHz, $\xi_q/2\pi = \{230.56, 251.0, 0\}$MHz, $\xi_{pq} = \{0.001, 1.176, 1.2\}$MHz, $T_1 = \{80,90,0.3892\}\mu$s, $T_2 = \{26,30,0\}\mu$s. Control pulses are parameterized with $N_s=100$ B-spline wavelets with carrier wave frequencies $\Omega_q/2\pi = 0$}, from any initial qudit-qudit-cavity state to the ground state $\rho^{tar} = |000\rangle\langle 000|$. The optimization problem is solved with $M=1$ in the objective function ${\cal G}(\bs{\alpha})$, utilizing the ensemble state $\rho_s(0)$ to span all initial states in the Hilbert space of dimension $N=80$. 
Utilizing the optimized control pulses, Figure \ref{fig:2x2x20_expected} plots the evolution of expected energy levels for different (arbitrarily selected) initial states. It demonstrates how the different states coalesce to the ensemble state before they simultaneously decay towards the ground state.
At $T=5\mu$s the system reaches the ground state with an average fidelity of 99.18\% for the first qubit, 99.16\% for the second qubit, and 99.4\% for the cavity. Here the fidelity is averaged by evaluating \eqref{eq:fidelity} over all (pure) basis states $B^{kl}$.
Figure \ref{fig:2x2x20_neumann} shows the evolution of the normalized von Neumann entropy
\begin{align}
    S(\rho(t)) = -\frac{1}{\log(N)}\Tr\left(\rho(t) \log \left( \rho(t) \right)\right),
\end{align}
for various initial states. It illustrates that the ensemble state $\rho_s$ is maximally mixed and represents a 'worst-case' initial state for ground state reset.

\subsection{Logical gate optimization}\label{sec:gateoptim}

Most prominently, quantum optimal control is utilized for realizing logical gate operations, represented by a unitary matrix $V\in \C^{N\times N}$. Here the goal is to determine control pulses that drive any initial state $\rho(0)$ to the unitarily transformed target state $\rho^{tar} = V\rho(0)V^\dagger$. In contrast to state-to-state transfer and state preparation, in this case the target state depends on the initial state itself. In general, a basis of initial states would need to be taken into account in the objective function, leading to $M=N^2$ evaluations of Lindblad's equation. However, it was shown in \cite{Reich-Gualdi-Koch-2013, goerz2014optimal} that it is sufficient to consider only three specific initial states throughout the optimization process, independent of the Hilbert space dimension $N$. Those initial states are designed such that they distinguish between any two unitaries in that space and are given by
\begin{align}
    \rho_1(0) &:= \sum_{j=0}^{N-1} \lambda_j {|j\rangle \langle j| }  \quad \text{for} \quad \lambda_j \geq 0, \, \lambda_i \neq \lambda_j,\\
    \rho_2(0) &:= \bs{\psi}_1 \bs{\psi}_1^\dagger, \quad
    \bs{\psi}_1 = \frac{1}{\sqrt{N}}
    \begin{bmatrix}
        1, 1, \cdots, 1
    \end{bmatrix}^T, \\
    \rho_3(0) &= \frac 1N I_N.
\end{align}
Note also that the three initial states are not sufficient to deduce a bound on the realized gate error. Instead this can be computed by propagating at least $N+1$ states with the optimized control vector, see \cite{goerz2014optimal}.
When the objective function \eqref{eq:optimproblem} is based on the three initial states, it is suggested in \cite{goerz2014optimal} to weigh the contributions from each of those states to emphasize the first state, e.g. by choosing $\beta_1 = 20.0, \beta_2=\beta_3=1.0$. Further, when using $J_{Tr}$ as {merit functional}, it is advised to scale each contribution by the purity of $\rho_i(0)$.

\begin{figure}
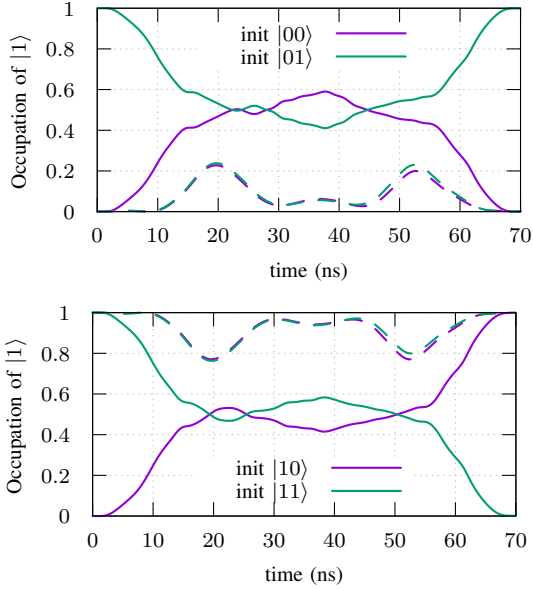

    \centering
    \input{figures/cnot_gate/population_init00_01.tex}
    \input{figures/cnot_gate/population_init10_11.tex}
    \caption{CNOT gate optimization: Occupation of the $|1\rangle$-state for qubit 1 (dashed lines) and qubit 2 (solid lines), for various initial states where qubit 1 is initially in the $|0\rangle$-state (top panel) or in the $|1\rangle$-state (bottom panel). A gate fidelity of $99.89\%$ is reached at $T=70$ns.}
    \label{fig:cnot_population}
\end{figure}

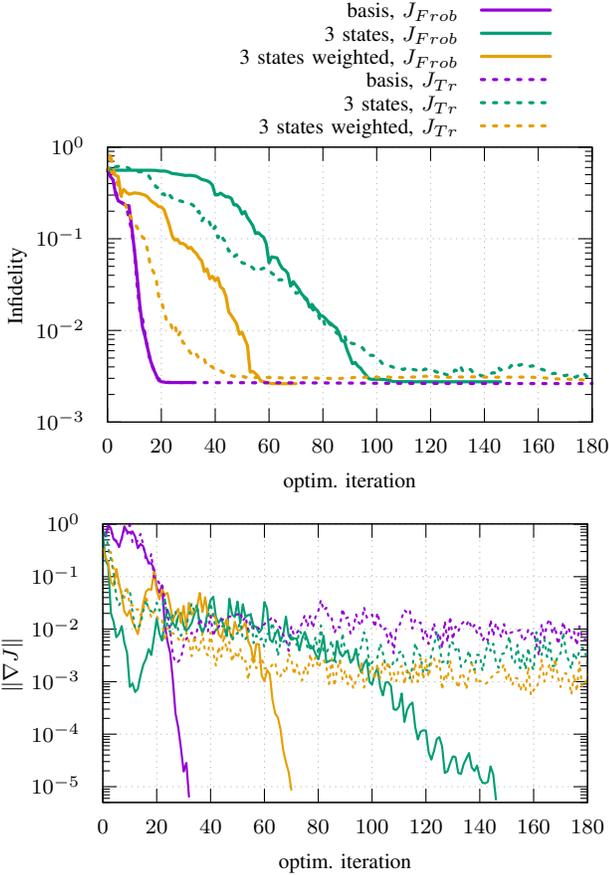
\begin{figure}
    \centering
    \input{figures/cnot_gate/optim_objective.tex}
    \input{figures/cnot_gate/optim_gradient.tex}
    \caption{Convergence history of CNOT gate optimization: Average infidelity, $1-F$ (top), and norm of the gradient (bottom) for the {merit functionals} $J_{Frob}$ and $J_{Tr}$, considering either three (weighted) initial states, or the full basis of initial states during the optimization.}
    \label{fig:cnot_optim_history}
\end{figure}

Both {merit functionals} $J_{Tr}$ and $J_{Frob}$ are commonly used for unitary gate optimization. Note that they differ by the average impurity of $\rho^{tar}=V\rho(0)V^\dagger$ and $\rho(T)$:
\begin{align}
    J_{Frob} &= \frac{1}{2} \mbox{Tr}\left((\rho^{tar})^2\right) - \mbox{Tr}\left((\rho^{tar})^\dagger \rho(T)\right) + \frac{1}{2} \mbox{Tr}\left(\rho(T)^2\right) \nonumber \\
     &= J_{Tr} - \frac{1}{2} \left(1-\mbox{Tr}((\rho^{tar})^2) \right) - \frac{1}{2} \left(1 - \mbox{Tr}(\rho(T)^2\right).\nonumber 
\end{align}
While the purity of $\rho^{tar}$ is constant throughout the optimization, the purity of $\rho(T)$ may vary, potentially hampering optimization convergence when $J_{Tr}$ is used in the objective function.
Below, we analyze the performance of the two {merit functionals} when applied to a CNOT gate optimization problem, spanning two qubits.\footnote{The system parameters for the CNOT gate optimization example are: $\omega_q/2\pi =  \{4.105, 4.812\}$GHz $\xi_q/2\pi =  \{219.8, 225.2\}$MHz, $\xi_{12}/2\pi = 10$MHz, $T_1 = \{{56, 58}\}\mu$s, $T_2 = \{28,28\}\mu$s. We assume a gate duration of $T=70$ ns. The controls are parameterized with $N_s=50$ basis functions with carrier wave frequencies $\Omega_q^1 = 0$ and $\Omega_q^2 = -\xi_{12}$.}
The CNOT gate performs a NOT operation on the second qubit, controlled by the state of the first qubit, as exemplified in Figure \ref{fig:cnot_population} through the population of the $|1\rangle$-state in each qubit for four different initial states. 
The performance of the two {merit functionals} $J_{Frob}$ and $J_{Tr}$ is demonstrated in Figure \ref{fig:cnot_optim_history}. It shows the optimization history in terms of the infidelity $1-F$ (gate error) for either {merit functional}, using either a full basis of initial states ($M=N^2$), or only the three initial states ($M=3$), in ${\cal G}(\bs{\alpha})$. We also evaluate the influence of the weights $\beta_i = \{20,1,1\}/ \mbox{Tr}\left(\rho_i(0)^2\right)$ in the objective function. While all combinations of {merit functionals}, initial conditions and weights, eventually reach about the same level of average fidelity (top panel of Figure \ref{fig:cnot_optim_history}), only the {merit functional} $J_{Frob}$ results in an optimization history that terminates with a small gradient norm (bottom panel), indicating its superiority over $J_{Tr}$. Further, the optimization converges in fewer iterations when using a full basis of initial conditions than when only three initial states are considered, and the importance of choosing appropriate weights $\beta_i$ is visible. 
In terms of iteration counts and gradient reduction, the best performance is achieved for $J_{Frob}$ with a full basis of initial conditions. 
For modest problem sizes, one may account for the increase in computational complexity by distributing the initial conditions over $N^2$ compute units. However, for larger problem sizes, the computational complexity will be more favorable if the objective function is evaluated on the weighted sum over three initial states only.

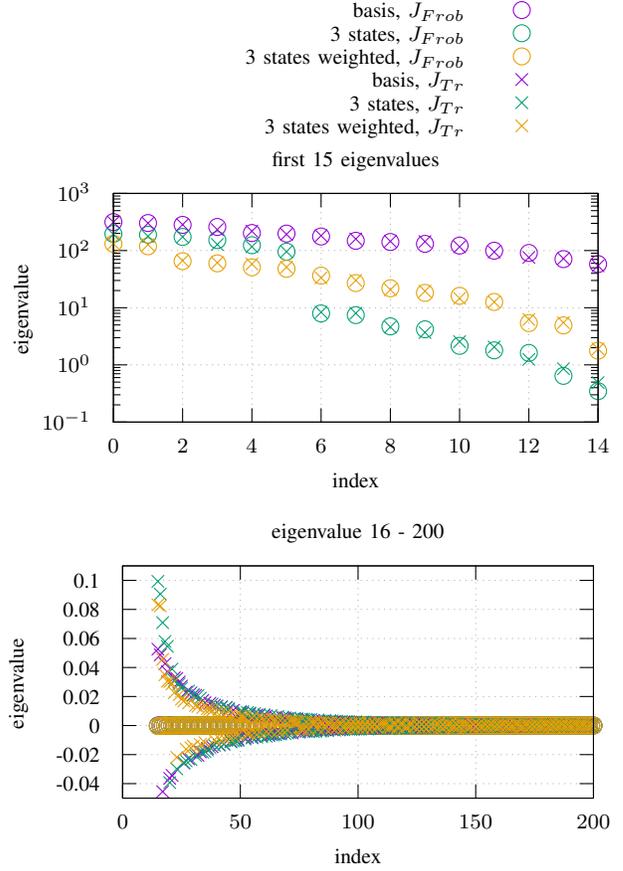
\begin{figure}
    \centering
    \input{figures/cnot_gate/Hessian_eig_1to15.tex}
    \input{figures/cnot_gate/Hessian_eig_16to200.tex}
    \caption{Eigenvalues of the Hessian of $J_{Frob}$ and $J_{Tr}$ evaluated at the optimized CNOT control parameters. The dominant 15 eigenvalues are shown on a log-scale on the top. The bottom panel shows the remaining (positive and negative) eigenvalues on a linear scale.}
    \label{fig:cnot_hessian}
\end{figure}

To further investigate the properties of the two {merit functionals}, in Figure~\ref{fig:cnot_hessian} we numerically evaluate the eigenvalues of the Hessian of the objective function ${\cal G}(\bs{\alpha})$ at the numerical optima. For all combinations of {merit functionals} and initial conditions, the first 15 eigenvalues are orders of magnitudes larger than the rest. This observation appears to be closely related to theoretical results in \cite{larocca2020exploiting}, suggesting that a minimum of $15$ basis functions are required to parameterize the control functions for an arbitrary unitary two-qubit gate transformation. 
The bottom panel of Figure~\ref{fig:cnot_hessian} plots the remaining 185 eigenvalues. Here, negative eigenvalues are visible, indicating that the objective function is non-convex. For the {merit functional} $J_{Tr}$, the Hessian of the objective function has a multitude of negative eigenvalues of the order $O(10^{-2})$. In contrast, the {merit functional} $J_{Frob}$ only leads to a Hessian with a few negative eigenvalues of the order $O(10^{-4})$. This observation provides insight into why $J_{Frob}$ leads to superior optimization convergence compared to $J_{Tr}$.

Quandary's object-oriented implementation allows developers to easily extend the predefined gate set to suit their particular simulation and optimization requirements.
For example, Appendix \ref{app:swap-14} demonstrates optimal control towards a customized SWAP gate spanning four fully coupled qubits. 

\section{Parallel Scalability}\label{sec:scaling}

Written in C++, Quandary is portable and is designed for modern {HPC} architectures. 
Three levels of parallelism are implemented, each based on the message-passing paradigm.
In the first and most trivial level, Quandary distributes independent solves of Lindblads master equation onto multiple compute units, such that different initial conditions in the objective function are propagated simultaneously. Parallel speedup for distributing initial conditions is perfect because such simulations are independent of one another, eliminating the need for communication other than a final gather operation to sum up the contributions to the total objective function (data omitted to conserve space). 

As a second level {of parallelism}, Quandary distributes the quantum state (in vectorized form, see Appendix \ref{app:IMR}) onto multiple compute cores. Here, Quandary relies on PETSc\cite{petsc-web-page} to perform parallel linear algebra across multiple distributed-memory nodes. System matrices are distributed accordingly and stored in parallel using the sparse matrix AIJ format in PETSc. In the following, we present strong and weak scaling results 
for solving Lindblad's master equation and the corresponding adjoint equation, for one initial condition. Note that the achieved speedups from distributed linear algebra multiplies with the speedup for distributing independent runs over different initial conditions from the first level of parallelism. All computations reported below were performed on the Quartz cluster at Lawrence Livermore National Laboratory\footnote{Quartz has 3,018 nodes based on Intel Broadwell E5-2695 v4 processors, with 36 cores per node, 128 GB node memory, and an Intel Omni-Path 100 Gb/s interconnect.}.

\begin{figure}
    \centering
    \input{figures/petsc_scaling/strong_scaling.tex}
    \caption{Strong-scaling study for distributed linear algebra. The different colors correspond to Hilbert space dimensions of $N=\{256, 1024, 4096\}$.}
    \label{fig:petsc_scaling_strongscaling}
\end{figure}
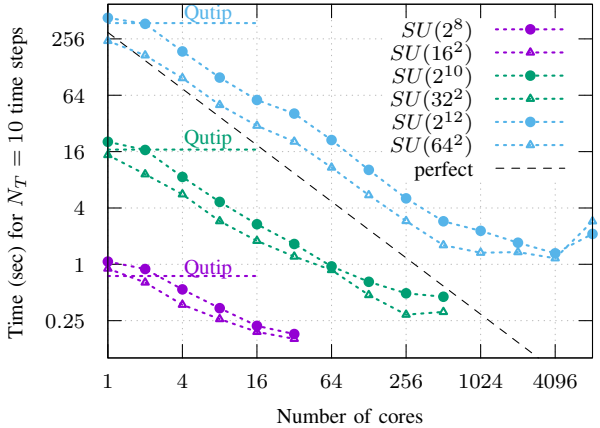

Figure \ref{fig:petsc_scaling_strongscaling} presents strong-scaling results for integrating the discretized master equation for $N_T=10$ time steps, for an increasing number of compute cores. The different colors correspond to different problem sizes. We compare fully coupled qubit systems with Hilbert space dimensions $N=2^Q$, as well as systems of the same dimension that couple two qudits with $n_q>1$ energy levels, resulting in $N=n_q^2$. Note that the cases $Q=\{8, 10, 12\}$ and $n_q=\{16, 32, 64\}$ both yield $N=\{ 256, 1024, 4096 \}$, corresponding to a density matrix with $N^2 = \{65\,536, 1\,048\,576, 16\,777\,216\}$ elements -- distributed over multiple cores. For a given problem size, the runtimes reported in Figure \ref{fig:petsc_scaling_strongscaling} demonstrate significant strong scalability when increasing the number of compute nodes. For comparison, we also show timings for solving Lindblad's equation with the same Hamiltonian and collapse operators, using the Python package Qutip\cite{johansson2012qutip}. Here, Qutip's \texttt{mesolve()} routine was used to integrate Lindblad's equation for $10$ time steps with the second-order Adam's time-integration scheme. While Qutip was configured to use multi-threading, it's scalability is limited and most beneficial for performing independent runs (e.g., the first level of Quandary parallelism). No speedup was obtained by running \texttt{mesolve()} in multi-threaded mode for the test cases considered here, and the timings reported were obtained for runs using a single thread. In contrast, distributing the quantum state vector onto multiple compute cores in Quandary scales reasonably well to large numbers of compute cores, with drastic reductions in run times. 
From Figure \ref{fig:petsc_scaling_strongscaling}, we can also observe weak scalability when both the problem size as well as the compute resources are increased simultaneously (horizontal slices of the same figure). 

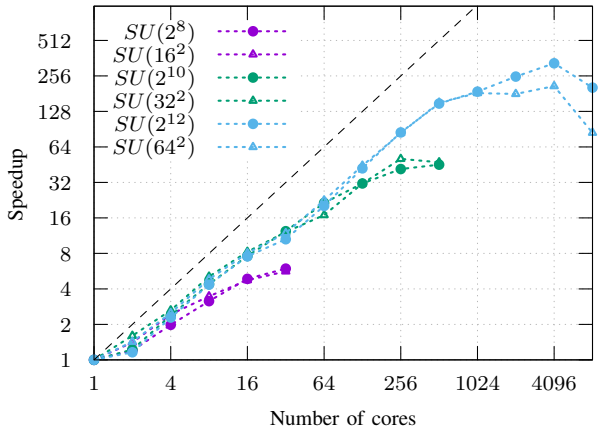
\begin{figure}
    \centering
    \input{figures/petsc_scaling/speedup.tex}
    \caption{Speedup achieved from distributed linear algebra over serial computation. The different colors correspond to Hilbert space dimensions of $N=\{256, 1024, 4096\}$.}
    \label{fig:petsc_scaling_speedup}
\end{figure}

The above results are presented in a different way in Figure \ref{fig:petsc_scaling_speedup}. Here, we illustrate the attainable speedup from parallel computations, compared to serial computations on a single core.
For the biggest test case considered here, with $N = 2^{12}$, a maximum speedup of up to 328x is achieved, reducing the run time for performing $10$ time-step from about 8 minutes to about 1.5 seconds.
Since typically many (thousands) of time-steps are needed to evolve the density matrix to the final time, the speedups achieved from spatial distribution are essential for solving quantum control problems in larger quantum system. 

\begin{figure}
    \centering
    \input{figures/petsc_scaling/weakscaling.tex}
    \caption{Time (minutes) to evaluate $\nabla {\cal G}$ for a 0-1 transfer problem with duration $T=1$ns.}
    \label{fig:01-transfer:timegradient}
\end{figure}
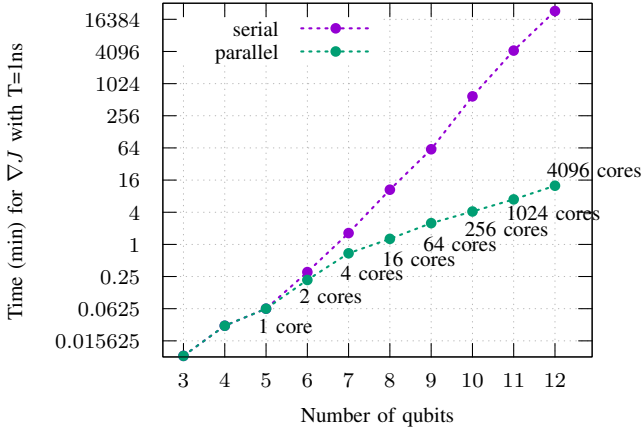

We next investigate the computational costs for one gradient evaluation in a state-to-state transfer problem, as function of the problem size. We consider $Q$ coupled qubits and hence system dimensions of $N=2^Q$, and evaluate the gradient of the objective function, $\nabla_\alpha {\cal G}$ based on the merit functional $J_{Frob}${, for a} state-to-state problem that transforms the ground state $|0\dots 0\rangle$ to the target $|1\dots 1\rangle$-state. To make the timing study tractable, we assume a duration of $T=1$ns. Figure \ref{fig:01-transfer:timegradient} compares serial and parallel run times when the state is distributed onto multiple compute cores. Note that adding one qubit to the system quadruples the total number of elements in the density matrix. To keep the number of elements per core fixed, the number of cores is increased by a factor of four each time a qubit is added. For eight and more qubits, this number of cores is calibrated to agree with the minimum run time achieved from the strong-scaling study in Figure \ref{fig:petsc_scaling_strongscaling}. 

From Figure \ref{fig:01-transfer:timegradient}, it is apparent that a serial computation of the gradient quickly becomes intractable as the number of qubits increases (the serial execution times shown for $Q\geq 10$ are extrapolated from running $10$ time-steps only, and multiplying with the corresponding number of required steps to reach $T=1$ns). In contrast, parallel execution puts optimal control of up to 12 qubits in the realm of the possible. 
The moderate increase in parallel compute time, when simultaneously increasing the number of qubits and the number of cores, can be explained in two ways. On the one hand, the number of time-steps needed to accurately solve Lindbad's master equation to a fixed time increases {quadratically} with the number of qubits, i.e.~{$N_T = {\cal O}(Q^2)$}. For {example}, the 12-qubit {case} was using 4.4 times more time steps than the 6-qubit case.\footnote{The time-step size required to resolve the quantum dynamics up to a fixed accuracy tolerance can be estimated based on the eigenvalues of the system Hamiltonian. We considered a fixed coupling term $\xi_{pq}=const$ between all qubits for this scaling study, which simplifies the eigenvalue estimation and leads to the time-step estimate for integrating the system with $Q$ qubits, $\Delta t^{(Q)}/\Delta t^{(2)} = 2/(Q(Q-1))$. The number of time-steps to integrate up to a fixed time is inversely proportional to the time step. Thus, 12-qubits require about $12\cdot 11 / (6\cdot 5)=4.4$ times more time steps than the 6-qubit case.}
On the other hand, the most significant computational tasks during the implicit time stepping consists of solving a linear system of equations. During each iteration of GMRES (see Appendix \ref{app:IMR}), the super-operator matrix is applied to a vector of size $N^2$. While the super-operator matrix is very sparse, the number of non-zero entries per row scales {quadratically} with the number of qubits. 
The computational complexity of each {GMRES} iteration  therefore scales as ${\cal O}(N^2 Q^2)$, {where $Q=\log_2 N$}. {Taking both of these effects into account, the calculation of the gradient of the objective function scales as ${\cal O}(N^2 (\log_2 N)^4)$}, whereas the increase in the number of computational cores "only" increases as ${\cal O}(N^2)$.

A third, more experimental, level of parallelism is also available in Quandary. This approach targets concurrency across the time dimension by interfacing with the parallel-in-time library XBraid \cite{xbraid}. XBraid distributes the discrete time steps onto multiple compute cores, and performs a parallel non-intrusive multigrid-in-time (MGRIT) algorithm, which Quandary can use to solve the forward and adjoint Lindblad equations. We remark that the design of efficient MGRIT methods for wave propagation problems is still under active research. The speedup from the MGRIT algorithm hence depends on the specific test case and an in-depth study will be published elsewhere.

\section{Conclusions}\label{sec:conclusions}
The Quandary code can be used for simulating the evolution of an open quantum system for a given control vector. It can also be used to solve the optimal control problem for driving the quantum system towards a desired state, aiming at three common tasks. First, for realizing a state-to-state transformation. Secondly, for approximating a unitary state transformation, and finally for unconditional pure state preparation. In the latter two cases, Quandary implements recent theoretical results that significantly reduce the computational burden during the optimization.

Because the computational burden grows rapidly with the size of the quantum system, Quandary implements a message passing paradigm through the PETSc library, based on the MPI library. This approach allows Quandary to run on modern HPC platforms, where the computational tasks can be distributed over many nodes in a cluster, where each node has many cores. We have demonstrated weak and strong scaling properties for solving Lindblad's equation forwards in time, and for evaluating an objective function and its gradient.

While the evaluation of the objective function and its gradient have been shown to scale well on a parallel machine, the L-BFGS algorithm sometimes suffers from poor convergence for larger quantum systems. These convergence problems will be the subject of further investigations.

\appendix

\subsection{Numerical time-integration of Lindblad's master equation}\label{app:IMR}

Quandary solves Lindblad's master equation in vectorized form for the vector $q(t) := \mbox{vec}(\rho(t)) \in \C^{N^2}$. The vectorized Lindblad master equation is given by 
\begin{align}\label{eq:mastereq_vectorized}
    \dot q(t) = &M(t) q(t), \quad \forall \, t\in[0,T] \\
    M(t) := & -i\left(I_N\otimes H(t) - H^T(t) \otimes I_N\right) \nonumber \\
    &\quad +\,\sum_{q=1}^{Q}\sum_{l=1}^2 
    \Ell_{lq}\otimes \Ell_{lq} \notag
    \\
    & \quad \quad - \frac 1 2 \left( I_N\otimes
    \Ell^T_{lq}\Ell_{lq} + \Ell^T_{lq}\Ell_{lq} \otimes I_N \right) \notag.
\end{align}
Note that, in the absence of Lindbladian collapse operators, the system matrix $M(t)$ of this linear ordinary differential equation (ODE) becomes skew-symmetric, resulting in a pure wave propagation problem. For this reason we choose the symplectic Implicit Midpoint Rule (IMR)~\cite{hairer2006geometric} scheme to evolve $q(t)$ forward in time, on a uniform grid in time, $t_j = j\Delta t, j=0,\dots, N_T$, with constant time step $\Delta t = T/N_T$. The IMR scheme is a Runge-Kutta method with accuracy ${\cal O}(\Delta t)^2$. 
The approximations $q^j \approx q(t_j)$ are given by 
\begin{align} \label{eq:IMR}
    &q^{j+1} = q^j + \Delta t \, M^{j+\frac 12} z^j, \quad j=0,\dots,N_T -1, 
\end{align}
where the stage variable $z^j$ solves the linear system
\begin{align} \label{eq:linsystem_primal}
\quad \left( I-\frac{\Delta t}{2} M^{j+\frac 1 2} \right)z^j = q^j,
\end{align}
and where $M^{j+\frac 1 2} := M\left(t_j + \frac{\Delta t}{2}\right)$. Quandary solves the linear system using the iterative GMRES method~\cite{saad1986gmres}.

\subsection{Gradient computation using the discrete adjoint approach}\label{app:adjoint}
Quandary solves the optimization problem by employing a gradient-based L-BFGS~\cite{nocedal2006numerical} optimization scheme to iteratively update the control vector $\boldsymbol{\alpha}\in \R^{2QN_sN_f}$ based on the gradient of the objective function ${\cal G}(\bs{\alpha})$. 
The gradient is computed by using the adjoint approach, which allows all components of the gradient to be evaluated at computational costs that is independent of the number of control parameters. We follow the first-discretize-then-optimize approach and derive the adjoint equations from the discretized optimization problem in order to get a gradient that is exact on the discretized (implementation) level and consistent with the forward evaluation. 

Let the adjoint variables be denoted $\hat{q}^j$. Since the IMR is symmetric, the consistent adjoint time-integration scheme is also the IMR, and reads
\begin{align}
    \hat q^j = \hat q^{j+1} + \Delta t \left(M^{j+\frac 12}\right)^\dag \hat z^j + \gamma_1 \beta(t_j)\partial_{q^j} J(q^j)^\dag,
\end{align}
for all $j = N_T - 1,\dots 0$, where the last term results from the discretized integral penalty term in the objective function, acting as a source term for the discrete adjoint equations. 
The terminal condition is given by $\hat q^{N_T} = \frac 1 M \beta \partial_{\partial q^{N_T}} J$ which represents the local sensitivity of the objective function evaluated at final time $T$, and the adjoint equations can then be solved backwards in time from $j=N_T-1, \dots, 0$. Here, the adjoint stage variables $\hat z^j$ in each time-step solve the adjoint linear system 
\begin{align}
\left(I-\frac{\Delta t}{2}\left(M^{j+\frac 1 2}\right)^\dag\right) \hat z^j =  \hat q^{j+1} \label{eq:linsystem_adjoint}
\end{align}
using GMRES iterations.

Using the state and the adjoint variables, the gradient of the objective function with respect to the control parameters $\boldsymbol{\alpha}$ is given then by 
\begin{align}\label{eq:reducedgradient}
    \nabla {\cal G}(\bs{\alpha}) =  \sum_{j=0}^{N_T-1} \Delta t\left(\nabla_{\bfa}{M^{j+\frac 12}}\,z^j\right)^\dag \hat z^j + \gamma_2 \bfa.
\end{align}
where $z^j$ and $\hat z^j$ solve the linear systems \eqref{eq:linsystem_primal} and \eqref{eq:linsystem_adjoint}, respectively.

\underline{Remark:} The above expressions can be derived either by traversing the flow graph to evaluate the objective function ${\cal G}(\bs{\alpha})$ in reverse order and applying the chain rule to propagate local sensitivities backwards through the flow graph {as is done, for example, in the reverse mode of Algorithmic/Automatic Differentiation~\cite{griewank1989automatic}}. Alternatively (and equivalently), they follow from saddle point conditions for the Lagrange functional associated with the discretized optimization problem.

\subsection{Customized SWAP gate spanning 4 qubits}\label{app:swap-14}
Quandary's predefined gate set for logical gate optimization can be easily extended to suit specific needs. This feature is of particular importance as it can serve as a test bed for developing specialized compilers to perform specific tasks. Here, we demonstrate optimization towards a custom logical gate that spans $4$ qubits, performing a SWAP-14 operation on the first and the last qubit, while leaving the other two qubits in their respective state\footnote{System parameters for the SWAP-14 optimization on 4 qubits: $\omega_q/2\pi = \{5.1771,4.9639,4.91526,4.8118\}$GHz, $\xi_q/2\pi = \{334.9, 321.8, 341.0, 350.1\}$MHz, $\xi_{pq}/2\pi = 100$MHz, $T_1 = \{93.79, 91.67, 91.87, 95.67\}\mu$s, $T_2 = \{102.52, 101.2, 112.34, 105.43\}\mu$s.}. This SWAP-14 gate performs the following logical operation:
\\[1ex] 
\begin{centering}
\begin{tabular}{@ { } lcr@ { }}
   \toprule
      $\rho(0)$ &$\leftrightarrow$ &$\rho(T)$  \\
   \midrule
      $|0bc0\rangle$ &$\leftrightarrow$ &$|0bc0\rangle$ \\
      $|0bc1\rangle$ &$\leftrightarrow$ &$|1bc0\rangle$ \\
      $|1bc0\rangle$ &$\leftrightarrow$ &$|0bc1\rangle$ \\
      $|1bc1\rangle$ &$\leftrightarrow$ &$|1bc1\rangle$ \\
   \bottomrule
  \end{tabular} 
\end{centering}
\\[1ex]
for $b,c\in\{0,1\}$, representing the second and third qubit. The resulting unitary gate matrix $V\in\R^{N\times N}$, $N=2^4$, has 1's on the diagonal at positions where no operation is performed (first and last row of the table). It has symmetric pairs of off-diagonal 1's at positions corresponding to the swaps in the two middle rows of the table.

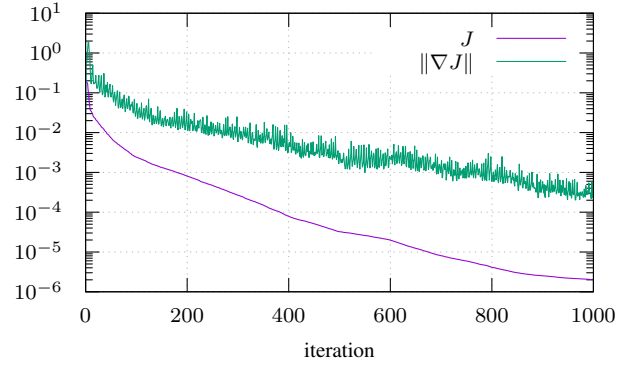
\begin{figure}
    \centering
    \input{figures/swap03/optim_history.tex}
    \caption{SWAP gate to swap the state of qubit 1 with qubit 4: Optimization history using the merit functional $J_{Frob}$.}
    \label{fig:swap03_optimhistory}
\end{figure}

We consider a gate duration of $T=10$ns, and parameterize the controls with $N_s=50$ B-spline basis functions with carrier wave frequencies $\Omega_q/2\pi = \{0.0, -\xi_{pq}, -2\xi_{pq}, -3\xi_{pq}\}$. 
Figure \ref{fig:swap03_optimhistory} shows the optimization convergence history in terms of objective function and gradient norm reduction. It is apparent that this test case takes significantly more iterations than other test cases presented here, which we attribute to the more complex optimization landscape when four qubits are to be controlled. Nevertheless, the optimized control pulses realize the desired SWAP gate on 4 qubits with an average gate fidelity of $99.68\%$. 

\section*{Acknowledgment}

This work was made possible by financial support from DoE ASCR, award SCW-1683, and from LLNL through LDRD project 20-ERD-028.
This work was performed under the auspices of the U.S. Department of Energy by Lawrence Livermore National Laboratory under Contract DE-AC52-07NA27344. This is contribution LLNL-PROC-827486.



\bibliographystyle{IEEEtran}
\bibliography{IEEEabrv,mybib}
%



\end{document}

%% file: figures/statetostate/fidelities.tex
\begin{tikzpicture}[gnuplot]
\tikzset{every node/.append style={font={\footnotesize}}}
\path (0.000,0.000) rectangle (6.500,4.000);
\gpcolor{color=gp lt color axes}
\gpsetlinetype{gp lt axes}
\gpsetdashtype{gp dt axes}
\gpsetlinewidth{0.50}
\draw[gp path] (1.320,0.985)--(5.947,0.985);
\gpcolor{color=gp lt color border}
\gpsetlinetype{gp lt border}
\gpsetdashtype{gp dt solid}
\gpsetlinewidth{1.00}
\draw[gp path] (1.320,0.985)--(1.500,0.985);
\draw[gp path] (5.947,0.985)--(5.767,0.985);
\node[gp node right] at (1.136,0.985) {$75$};
\gpcolor{color=gp lt color axes}
\gpsetlinetype{gp lt axes}
\gpsetdashtype{gp dt axes}
\gpsetlinewidth{0.50}
\draw[gp path] (1.320,1.403)--(5.947,1.403);
\gpcolor{color=gp lt color border}
\gpsetlinetype{gp lt border}
\gpsetdashtype{gp dt solid}
\gpsetlinewidth{1.00}
\draw[gp path] (1.320,1.403)--(1.500,1.403);
\draw[gp path] (5.947,1.403)--(5.767,1.403);
\node[gp node right] at (1.136,1.403) {$80$};
\gpcolor{color=gp lt color axes}
\gpsetlinetype{gp lt axes}
\gpsetdashtype{gp dt axes}
\gpsetlinewidth{0.50}
\draw[gp path] (1.320,1.821)--(5.947,1.821);
\gpcolor{color=gp lt color border}
\gpsetlinetype{gp lt border}
\gpsetdashtype{gp dt solid}
\gpsetlinewidth{1.00}
\draw[gp path] (1.320,1.821)--(1.500,1.821);
\draw[gp path] (5.947,1.821)--(5.767,1.821);
\node[gp node right] at (1.136,1.821) {$85$};
\gpcolor{color=gp lt color axes}
\gpsetlinetype{gp lt axes}
\gpsetdashtype{gp dt axes}
\gpsetlinewidth{0.50}
\draw[gp path] (1.320,2.239)--(5.947,2.239);
\gpcolor{color=gp lt color border}
\gpsetlinetype{gp lt border}
\gpsetdashtype{gp dt solid}
\gpsetlinewidth{1.00}
\draw[gp path] (1.320,2.239)--(1.500,2.239);
\draw[gp path] (5.947,2.239)--(5.767,2.239);
\node[gp node right] at (1.136,2.239) {$90$};
\gpcolor{color=gp lt color axes}
\gpsetlinetype{gp lt axes}
\gpsetdashtype{gp dt axes}
\gpsetlinewidth{0.50}
\draw[gp path] (1.320,2.657)--(5.947,2.657);
\gpcolor{color=gp lt color border}
\gpsetlinetype{gp lt border}
\gpsetdashtype{gp dt solid}
\gpsetlinewidth{1.00}
\draw[gp path] (1.320,2.657)--(1.500,2.657);
\draw[gp path] (5.947,2.657)--(5.767,2.657);
\node[gp node right] at (1.136,2.657) {$95$};
\gpcolor{color=gp lt color axes}
\gpsetlinetype{gp lt axes}
\gpsetdashtype{gp dt axes}
\gpsetlinewidth{0.50}
\draw[gp path] (1.320,3.075)--(5.947,3.075);
\gpcolor{color=gp lt color border}
\gpsetlinetype{gp lt border}
\gpsetdashtype{gp dt solid}
\gpsetlinewidth{1.00}
\draw[gp path] (1.320,3.075)--(1.500,3.075);
\draw[gp path] (5.947,3.075)--(5.767,3.075);
\node[gp node right] at (1.136,3.075) {$100$};
\gpcolor{color=gp lt color axes}
\gpsetlinetype{gp lt axes}
\gpsetdashtype{gp dt axes}
\gpsetlinewidth{0.50}
\draw[gp path] (1.898,0.985)--(1.898,3.075);
\gpcolor{color=gp lt color border}
\gpsetlinetype{gp lt border}
\gpsetdashtype{gp dt solid}
\gpsetlinewidth{1.00}
\draw[gp path] (1.898,0.985)--(1.898,1.165);
\draw[gp path] (1.898,3.075)--(1.898,2.895);
\node[gp node center] at (1.898,0.677) {$500$ns};
\gpcolor{color=gp lt color axes}
\gpsetlinetype{gp lt axes}
\gpsetdashtype{gp dt axes}
\gpsetlinewidth{0.50}
\draw[gp path] (3.055,0.985)--(3.055,3.075);
\gpcolor{color=gp lt color border}
\gpsetlinetype{gp lt border}
\gpsetdashtype{gp dt solid}
\gpsetlinewidth{1.00}
\draw[gp path] (3.055,0.985)--(3.055,1.165);
\draw[gp path] (3.055,3.075)--(3.055,2.895);
\node[gp node center] at (3.055,0.677) {$100$ns};
\gpcolor{color=gp lt color axes}
\gpsetlinetype{gp lt axes}
\gpsetdashtype{gp dt axes}
\gpsetlinewidth{0.50}
\draw[gp path] (4.212,0.985)--(4.212,3.075);
\gpcolor{color=gp lt color border}
\gpsetlinetype{gp lt border}
\gpsetdashtype{gp dt solid}
\gpsetlinewidth{1.00}
\draw[gp path] (4.212,0.985)--(4.212,1.165);
\draw[gp path] (4.212,3.075)--(4.212,2.895);
\node[gp node center] at (4.212,0.677) {$50$ns};
\gpcolor{color=gp lt color axes}
\gpsetlinetype{gp lt axes}
\gpsetdashtype{gp dt axes}
\gpsetlinewidth{0.50}
\draw[gp path] (5.369,0.985)--(5.369,3.075);
\gpcolor{color=gp lt color border}
\gpsetlinetype{gp lt border}
\gpsetdashtype{gp dt solid}
\gpsetlinewidth{1.00}
\draw[gp path] (5.369,0.985)--(5.369,1.165);
\draw[gp path] (5.369,3.075)--(5.369,2.895);
\node[gp node center] at (5.369,0.677) {$25$ns};
\draw[gp path] (1.320,3.075)--(1.320,0.985)--(5.947,0.985)--(5.947,3.075)--cycle;
\node[gp node center,rotate=-270] at (0.292,2.030) {Fidelity (\%)};
\node[gp node center] at (3.633,0.215) {Decay time ($T_1$)};
\node[gp node right] at (4.663,3.666) {closed system};
\gpfill{rgb color={1.000,0.000,0.000}} (4.847,3.589)--(5.763,3.589)--(5.763,3.743)--(4.847,3.743)--cycle;
\gpcolor{rgb color={1.000,0.000,0.000}}
\draw[gp path] (4.847,3.589)--(5.763,3.589)--(5.763,3.743)--(4.847,3.743)--cycle;
\gpfill{rgb color={1.000,0.000,0.000}} (1.513,0.985)--(1.899,0.985)--(1.899,2.969)--(1.513,2.969)--cycle;
\draw[gp path] (1.513,0.985)--(1.513,2.968)--(1.898,2.968)--(1.898,0.985)--cycle;
\gpfill{rgb color={1.000,0.000,0.000}} (2.670,0.985)--(3.056,0.985)--(3.056,2.559)--(2.670,2.559)--cycle;
\draw[gp path] (2.670,0.985)--(2.670,2.558)--(3.055,2.558)--(3.055,0.985)--cycle;
\gpfill{rgb color={1.000,0.000,0.000}} (3.826,0.985)--(4.213,0.985)--(4.213,2.081)--(3.826,2.081)--cycle;
\draw[gp path] (3.826,0.985)--(3.826,2.080)--(4.212,2.080)--(4.212,0.985)--cycle;
\gpfill{rgb color={1.000,0.000,0.000}} (4.983,0.985)--(5.370,0.985)--(5.370,1.237)--(4.983,1.237)--cycle;
\draw[gp path] (4.983,0.985)--(4.983,1.236)--(5.369,1.236)--(5.369,0.985)--cycle;
\gpcolor{color=gp lt color border}
\node[gp node right] at (4.663,3.358) {open system};
\gpfill{rgb color={0.000,0.000,1.000}} (4.847,3.281)--(5.763,3.281)--(5.763,3.435)--(4.847,3.435)--cycle;
\gpcolor{rgb color={0.000,0.000,1.000}}
\draw[gp path] (4.847,3.281)--(5.763,3.281)--(5.763,3.435)--(4.847,3.435)--cycle;
\gpfill{rgb color={0.000,0.000,1.000}} (1.898,0.985)--(2.285,0.985)--(2.285,2.997)--(1.898,2.997)--cycle;
\draw[gp path] (1.898,0.985)--(1.898,2.996)--(2.284,2.996)--(2.284,0.985)--cycle;
\gpfill{rgb color={0.000,0.000,1.000}} (3.055,0.985)--(3.442,0.985)--(3.442,2.808)--(3.055,2.808)--cycle;
\draw[gp path] (3.055,0.985)--(3.055,2.807)--(3.441,2.807)--(3.441,0.985)--cycle;
\gpfill{rgb color={0.000,0.000,1.000}} (4.212,0.985)--(4.598,0.985)--(4.598,2.584)--(4.212,2.584)--cycle;
\draw[gp path] (4.212,0.985)--(4.212,2.583)--(4.597,2.583)--(4.597,0.985)--cycle;
\gpfill{rgb color={0.000,0.000,1.000}} (5.369,0.985)--(5.755,0.985)--(5.755,2.215)--(5.369,2.215)--cycle;
\draw[gp path] (5.369,0.985)--(5.369,2.214)--(5.754,2.214)--(5.754,0.985)--cycle;
\gpcolor{color=gp lt color border}
\draw[gp path] (1.320,3.075)--(1.320,0.985)--(5.947,0.985)--(5.947,3.075)--cycle;
\gpdefrectangularnode{gp plot 1}{\pgfpoint{1.320cm}{0.985cm}}{\pgfpoint{5.947cm}{3.075cm}}
\end{tikzpicture}

%% file: figures/FockSuperp03/population.tex
\begin{tikzpicture}[gnuplot]
\tikzset{every node/.append style={font={\footnotesize}}}
\path (0.000,0.000) rectangle (8.500,5.000);
\gpcolor{color=gp lt color axes}
\gpsetlinetype{gp lt axes}
\gpsetdashtype{gp dt axes}
\gpsetlinewidth{0.50}
\draw[gp path] (1.320,0.985)--(7.947,0.985);
\gpcolor{color=gp lt color border}
\gpsetlinetype{gp lt border}
\gpsetdashtype{gp dt solid}
\gpsetlinewidth{1.00}
\draw[gp path] (1.320,0.985)--(1.500,0.985);
\draw[gp path] (7.947,0.985)--(7.767,0.985);
\node[gp node right] at (1.136,0.985) {$0$};
\gpcolor{color=gp lt color axes}
\gpsetlinetype{gp lt axes}
\gpsetdashtype{gp dt axes}
\gpsetlinewidth{0.50}
\draw[gp path] (1.320,1.726)--(7.947,1.726);
\gpcolor{color=gp lt color border}
\gpsetlinetype{gp lt border}
\gpsetdashtype{gp dt solid}
\gpsetlinewidth{1.00}
\draw[gp path] (1.320,1.726)--(1.500,1.726);
\draw[gp path] (7.947,1.726)--(7.767,1.726);
\node[gp node right] at (1.136,1.726) {$0.2$};
\gpcolor{color=gp lt color axes}
\gpsetlinetype{gp lt axes}
\gpsetdashtype{gp dt axes}
\gpsetlinewidth{0.50}
\draw[gp path] (1.320,2.467)--(7.947,2.467);
\gpcolor{color=gp lt color border}
\gpsetlinetype{gp lt border}
\gpsetdashtype{gp dt solid}
\gpsetlinewidth{1.00}
\draw[gp path] (1.320,2.467)--(1.500,2.467);
\draw[gp path] (7.947,2.467)--(7.767,2.467);
\node[gp node right] at (1.136,2.467) {$0.4$};
\gpcolor{color=gp lt color axes}
\gpsetlinetype{gp lt axes}
\gpsetdashtype{gp dt axes}
\gpsetlinewidth{0.50}
\draw[gp path] (1.320,3.209)--(7.947,3.209);
\gpcolor{color=gp lt color border}
\gpsetlinetype{gp lt border}
\gpsetdashtype{gp dt solid}
\gpsetlinewidth{1.00}
\draw[gp path] (1.320,3.209)--(1.500,3.209);
\draw[gp path] (7.947,3.209)--(7.767,3.209);
\node[gp node right] at (1.136,3.209) {$0.6$};
\gpcolor{color=gp lt color axes}
\gpsetlinetype{gp lt axes}
\gpsetdashtype{gp dt axes}
\gpsetlinewidth{0.50}
\draw[gp path] (1.320,3.950)--(3.129,3.950);
\draw[gp path] (4.965,3.950)--(7.947,3.950);
\gpcolor{color=gp lt color border}
\gpsetlinetype{gp lt border}
\gpsetdashtype{gp dt solid}
\gpsetlinewidth{1.00}
\draw[gp path] (1.320,3.950)--(1.500,3.950);
\draw[gp path] (7.947,3.950)--(7.767,3.950);
\node[gp node right] at (1.136,3.950) {$0.8$};
\gpcolor{color=gp lt color axes}
\gpsetlinetype{gp lt axes}
\gpsetdashtype{gp dt axes}
\gpsetlinewidth{0.50}
\draw[gp path] (1.320,4.691)--(7.947,4.691);
\gpcolor{color=gp lt color border}
\gpsetlinetype{gp lt border}
\gpsetdashtype{gp dt solid}
\gpsetlinewidth{1.00}
\draw[gp path] (1.320,4.691)--(1.500,4.691);
\draw[gp path] (7.947,4.691)--(7.767,4.691);
\node[gp node right] at (1.136,4.691) {$1$};
\gpcolor{color=gp lt color axes}
\gpsetlinetype{gp lt axes}
\gpsetdashtype{gp dt axes}
\gpsetlinewidth{0.50}
\draw[gp path] (1.320,0.985)--(1.320,4.691);
\gpcolor{color=gp lt color border}
\gpsetlinetype{gp lt border}
\gpsetdashtype{gp dt solid}
\gpsetlinewidth{1.00}
\draw[gp path] (1.320,0.985)--(1.320,1.165);
\draw[gp path] (1.320,4.691)--(1.320,4.511);
\node[gp node center] at (1.320,0.677) {$0$};
\gpcolor{color=gp lt color axes}
\gpsetlinetype{gp lt axes}
\gpsetdashtype{gp dt axes}
\gpsetlinewidth{0.50}
\draw[gp path] (2.977,0.985)--(2.977,4.691);
\gpcolor{color=gp lt color border}
\gpsetlinetype{gp lt border}
\gpsetdashtype{gp dt solid}
\gpsetlinewidth{1.00}
\draw[gp path] (2.977,0.985)--(2.977,1.165);
\draw[gp path] (2.977,4.691)--(2.977,4.511);
\node[gp node center] at (2.977,0.677) {$0.05$};
\gpcolor{color=gp lt color axes}
\gpsetlinetype{gp lt axes}
\gpsetdashtype{gp dt axes}
\gpsetlinewidth{0.50}
\draw[gp path] (4.634,0.985)--(4.634,3.274);
\draw[gp path] (4.634,4.506)--(4.634,4.691);
\gpcolor{color=gp lt color border}
\gpsetlinetype{gp lt border}
\gpsetdashtype{gp dt solid}
\gpsetlinewidth{1.00}
\draw[gp path] (4.634,0.985)--(4.634,1.165);
\draw[gp path] (4.634,4.691)--(4.634,4.511);
\node[gp node center] at (4.634,0.677) {$0.1$};
\gpcolor{color=gp lt color axes}
\gpsetlinetype{gp lt axes}
\gpsetdashtype{gp dt axes}
\gpsetlinewidth{0.50}
\draw[gp path] (6.290,0.985)--(6.290,4.691);
\gpcolor{color=gp lt color border}
\gpsetlinetype{gp lt border}
\gpsetdashtype{gp dt solid}
\gpsetlinewidth{1.00}
\draw[gp path] (6.290,0.985)--(6.290,1.165);
\draw[gp path] (6.290,4.691)--(6.290,4.511);
\node[gp node center] at (6.290,0.677) {$0.15$};
\gpcolor{color=gp lt color axes}
\gpsetlinetype{gp lt axes}
\gpsetdashtype{gp dt axes}
\gpsetlinewidth{0.50}
\draw[gp path] (7.947,0.985)--(7.947,4.691);
\gpcolor{color=gp lt color border}
\gpsetlinetype{gp lt border}
\gpsetdashtype{gp dt solid}
\gpsetlinewidth{1.00}
\draw[gp path] (7.947,0.985)--(7.947,1.165);
\draw[gp path] (7.947,4.691)--(7.947,4.511);
\node[gp node center] at (7.947,0.677) {$0.2$};
\draw[gp path] (1.320,4.691)--(1.320,0.985)--(7.947,0.985)--(7.947,4.691)--cycle;
\node[gp node center,rotate=-270] at (0.292,2.838) {population};
\node[gp node center] at (4.633,0.215) {time (ns)};
\node[gp node right] at (3.681,4.352) {$|0\rangle$};
\gpcolor{rgb color={0.580,0.000,0.827}}
\gpsetlinewidth{2.00}
\draw[gp path] (3.865,4.352)--(4.781,4.352);
\draw[gp path] (1.320,4.691)--(1.337,4.691)--(1.353,4.691)--(1.370,4.691)--(1.386,4.691)%
  --(1.403,4.691)--(1.419,4.691)--(1.436,4.690)--(1.453,4.689)--(1.469,4.688)--(1.486,4.687)%
  --(1.502,4.686)--(1.519,4.684)--(1.535,4.682)--(1.552,4.679)--(1.569,4.677)--(1.585,4.674)%
  --(1.602,4.671)--(1.618,4.668)--(1.635,4.665)--(1.651,4.661)--(1.668,4.658)--(1.684,4.654)%
  --(1.701,4.650)--(1.718,4.645)--(1.734,4.641)--(1.751,4.636)--(1.767,4.630)--(1.784,4.625)%
  --(1.800,4.618)--(1.817,4.612)--(1.834,4.605)--(1.850,4.597)--(1.867,4.589)--(1.883,4.580)%
  --(1.900,4.571)--(1.916,4.561)--(1.933,4.551)--(1.950,4.539)--(1.966,4.528)--(1.983,4.515)%
  --(1.999,4.502)--(2.016,4.488)--(2.032,4.473)--(2.049,4.458)--(2.066,4.442)--(2.082,4.425)%
  --(2.099,4.409)--(2.115,4.391)--(2.132,4.373)--(2.148,4.354)--(2.165,4.334)--(2.182,4.314)%
  --(2.198,4.294)--(2.215,4.273)--(2.231,4.252)--(2.248,4.231)--(2.264,4.208)--(2.281,4.185)%
  --(2.297,4.161)--(2.314,4.136)--(2.331,4.109)--(2.347,4.081)--(2.364,4.052)--(2.380,4.022)%
  --(2.397,3.991)--(2.413,3.961)--(2.430,3.930)--(2.447,3.898)--(2.463,3.866)--(2.480,3.834)%
  --(2.496,3.800)--(2.513,3.766)--(2.529,3.731)--(2.546,3.694)--(2.563,3.656)--(2.579,3.618)%
  --(2.596,3.580)--(2.612,3.542)--(2.629,3.505)--(2.645,3.468)--(2.662,3.431)--(2.679,3.393)%
  --(2.695,3.354)--(2.712,3.315)--(2.728,3.275)--(2.745,3.236)--(2.761,3.198)--(2.778,3.160)%
  --(2.795,3.123)--(2.811,3.085)--(2.828,3.048)--(2.844,3.010)--(2.861,2.972)--(2.877,2.934)%
  --(2.894,2.896)--(2.910,2.857)--(2.927,2.818)--(2.944,2.779)--(2.960,2.739)--(2.977,2.699)%
  --(2.993,2.660)--(3.010,2.619)--(3.026,2.579)--(3.043,2.539)--(3.060,2.500)--(3.076,2.461)%
  --(3.093,2.423)--(3.109,2.386)--(3.126,2.349)--(3.142,2.313)--(3.159,2.277)--(3.176,2.241)%
  --(3.192,2.205)--(3.209,2.168)--(3.225,2.132)--(3.242,2.096)--(3.258,2.060)--(3.275,2.024)%
  --(3.292,1.989)--(3.308,1.955)--(3.325,1.922)--(3.341,1.890)--(3.358,1.860)--(3.374,1.831)%
  --(3.391,1.802)--(3.408,1.775)--(3.424,1.749)--(3.441,1.723)--(3.457,1.699)--(3.474,1.675)%
  --(3.490,1.652)--(3.507,1.630)--(3.523,1.609)--(3.540,1.588)--(3.557,1.569)--(3.573,1.550)%
  --(3.590,1.532)--(3.606,1.514)--(3.623,1.498)--(3.639,1.482)--(3.656,1.467)--(3.673,1.452)%
  --(3.689,1.438)--(3.706,1.425)--(3.722,1.412)--(3.739,1.399)--(3.755,1.387)--(3.772,1.375)%
  --(3.789,1.363)--(3.805,1.352)--(3.822,1.341)--(3.838,1.330)--(3.855,1.319)--(3.871,1.309)%
  --(3.888,1.299)--(3.905,1.290)--(3.921,1.281)--(3.938,1.272)--(3.954,1.264)--(3.971,1.256)%
  --(3.987,1.248)--(4.004,1.241)--(4.021,1.233)--(4.037,1.226)--(4.054,1.219)--(4.070,1.213)%
  --(4.087,1.207)--(4.103,1.201)--(4.120,1.196)--(4.136,1.191)--(4.153,1.186)--(4.170,1.181)%
  --(4.186,1.176)--(4.203,1.171)--(4.219,1.167)--(4.236,1.162)--(4.252,1.158)--(4.269,1.155)%
  --(4.286,1.151)--(4.302,1.148)--(4.319,1.146)--(4.335,1.143)--(4.352,1.141)--(4.368,1.138)%
  --(4.385,1.136)--(4.402,1.134)--(4.418,1.132)--(4.435,1.130)--(4.451,1.128)--(4.468,1.126)%
  --(4.484,1.124)--(4.501,1.123)--(4.518,1.121)--(4.534,1.120)--(4.551,1.119)--(4.567,1.118)%
  --(4.584,1.117)--(4.600,1.117)--(4.617,1.116)--(4.634,1.116)--(4.650,1.115)--(4.667,1.115)%
  --(4.683,1.115)--(4.700,1.115)--(4.716,1.115)--(4.733,1.114)--(4.749,1.114)--(4.766,1.114)%
  --(4.783,1.113)--(4.799,1.113)--(4.816,1.112)--(4.832,1.112)--(4.849,1.112)--(4.865,1.112)%
  --(4.882,1.112)--(4.899,1.112)--(4.915,1.112)--(4.932,1.113)--(4.948,1.113)--(4.965,1.113)%
  --(4.981,1.114)--(4.998,1.114)--(5.015,1.115)--(5.031,1.115)--(5.048,1.116)--(5.064,1.117)%
  --(5.081,1.117)--(5.097,1.118)--(5.114,1.119)--(5.131,1.120)--(5.147,1.121)--(5.164,1.122)%
  --(5.180,1.122)--(5.197,1.123)--(5.213,1.124)--(5.230,1.125)--(5.246,1.126)--(5.263,1.127)%
  --(5.280,1.128)--(5.296,1.129)--(5.313,1.130)--(5.329,1.131)--(5.346,1.132)--(5.362,1.133)%
  --(5.379,1.134)--(5.396,1.135)--(5.412,1.136)--(5.429,1.136)--(5.445,1.137)--(5.462,1.137)%
  --(5.478,1.138)--(5.495,1.138)--(5.512,1.138)--(5.528,1.138)--(5.545,1.138)--(5.561,1.138)%
  --(5.578,1.138)--(5.594,1.137)--(5.611,1.136)--(5.628,1.136)--(5.644,1.134)--(5.661,1.133)%
  --(5.677,1.131)--(5.694,1.129)--(5.710,1.127)--(5.727,1.125)--(5.744,1.122)--(5.760,1.119)%
  --(5.777,1.116)--(5.793,1.113)--(5.810,1.110)--(5.826,1.106)--(5.843,1.103)--(5.859,1.099)%
  --(5.876,1.095)--(5.893,1.090)--(5.909,1.086)--(5.926,1.082)--(5.942,1.078)--(5.959,1.075)%
  --(5.975,1.072)--(5.992,1.069)--(6.009,1.067)--(6.025,1.065)--(6.042,1.063)--(6.058,1.063)%
  --(6.075,1.062)--(6.091,1.062)--(6.108,1.062)--(6.125,1.063)--(6.141,1.064)--(6.158,1.066)%
  --(6.174,1.068)--(6.191,1.070)--(6.207,1.073)--(6.224,1.077)--(6.241,1.082)--(6.257,1.088)%
  --(6.274,1.094)--(6.290,1.102)--(6.307,1.110)--(6.323,1.119)--(6.340,1.129)--(6.357,1.140)%
  --(6.373,1.152)--(6.390,1.165)--(6.406,1.179)--(6.423,1.195)--(6.439,1.211)--(6.456,1.228)%
  --(6.472,1.247)--(6.489,1.266)--(6.506,1.287)--(6.522,1.309)--(6.539,1.331)--(6.555,1.355)%
  --(6.572,1.380)--(6.588,1.406)--(6.605,1.432)--(6.622,1.459)--(6.638,1.487)--(6.655,1.515)%
  --(6.671,1.545)--(6.688,1.574)--(6.704,1.604)--(6.721,1.635)--(6.738,1.667)--(6.754,1.699)%
  --(6.771,1.731)--(6.787,1.763)--(6.804,1.796)--(6.820,1.829)--(6.837,1.862)--(6.854,1.895)%
  --(6.870,1.928)--(6.887,1.961)--(6.903,1.994)--(6.920,2.027)--(6.936,2.059)--(6.953,2.091)%
  --(6.970,2.123)--(6.986,2.154)--(7.003,2.185)--(7.019,2.215)--(7.036,2.245)--(7.052,2.274)%
  --(7.069,2.302)--(7.085,2.330)--(7.102,2.356)--(7.119,2.380)--(7.135,2.404)--(7.152,2.427)%
  --(7.168,2.448)--(7.185,2.468)--(7.201,2.486)--(7.218,2.504)--(7.235,2.520)--(7.251,2.536)%
  --(7.268,2.550)--(7.284,2.562)--(7.301,2.574)--(7.317,2.585)--(7.334,2.595)--(7.351,2.604)%
  --(7.367,2.612)--(7.384,2.619)--(7.400,2.627)--(7.417,2.634)--(7.433,2.641)--(7.450,2.649)%
  --(7.467,2.657)--(7.483,2.666)--(7.500,2.676)--(7.516,2.687)--(7.533,2.698)--(7.549,2.709)%
  --(7.566,2.720)--(7.583,2.731)--(7.599,2.742)--(7.616,2.752)--(7.632,2.762)--(7.649,2.772)%
  --(7.665,2.781)--(7.682,2.789)--(7.698,2.797)--(7.715,2.805)--(7.732,2.811)--(7.748,2.817)%
  --(7.765,2.823)--(7.781,2.828)--(7.798,2.832)--(7.814,2.835)--(7.831,2.838)--(7.848,2.839)%
  --(7.864,2.841)--(7.881,2.841)--(7.897,2.841)--(7.914,2.841)--(7.930,2.841)--(7.947,2.841);
\gpcolor{color=gp lt color border}
\node[gp node right] at (3.681,4.044) {$|1\rangle$};
\gpcolor{rgb color={0.000,0.620,0.451}}
\draw[gp path] (3.865,4.044)--(4.781,4.044);
\draw[gp path] (1.320,0.985)--(1.337,0.985)--(1.353,0.985)--(1.370,0.985)--(1.386,0.985)%
  --(1.403,0.985)--(1.419,0.985)--(1.436,0.986)--(1.453,0.987)--(1.469,0.988)--(1.486,0.989)%
  --(1.502,0.990)--(1.519,0.992)--(1.535,0.994)--(1.552,0.997)--(1.569,0.999)--(1.585,1.002)%
  --(1.602,1.005)--(1.618,1.008)--(1.635,1.011)--(1.651,1.014)--(1.668,1.018)--(1.684,1.022)%
  --(1.701,1.026)--(1.718,1.030)--(1.734,1.035)--(1.751,1.040)--(1.767,1.045)--(1.784,1.051)%
  --(1.800,1.057)--(1.817,1.063)--(1.834,1.070)--(1.850,1.078)--(1.867,1.086)--(1.883,1.094)%
  --(1.900,1.103)--(1.916,1.113)--(1.933,1.123)--(1.950,1.133)--(1.966,1.145)--(1.983,1.157)%
  --(1.999,1.169)--(2.016,1.183)--(2.032,1.196)--(2.049,1.211)--(2.066,1.226)--(2.082,1.241)%
  --(2.099,1.256)--(2.115,1.272)--(2.132,1.289)--(2.148,1.306)--(2.165,1.324)--(2.182,1.342)%
  --(2.198,1.360)--(2.215,1.378)--(2.231,1.397)--(2.248,1.416)--(2.264,1.435)--(2.281,1.455)%
  --(2.297,1.475)--(2.314,1.497)--(2.331,1.519)--(2.347,1.542)--(2.364,1.566)--(2.380,1.590)%
  --(2.397,1.614)--(2.413,1.639)--(2.430,1.663)--(2.447,1.687)--(2.463,1.711)--(2.480,1.735)%
  --(2.496,1.759)--(2.513,1.784)--(2.529,1.809)--(2.546,1.835)--(2.563,1.861)--(2.579,1.886)%
  --(2.596,1.911)--(2.612,1.935)--(2.629,1.958)--(2.645,1.981)--(2.662,2.003)--(2.679,2.025)%
  --(2.695,2.047)--(2.712,2.068)--(2.728,2.089)--(2.745,2.109)--(2.761,2.128)--(2.778,2.147)%
  --(2.795,2.164)--(2.811,2.181)--(2.828,2.197)--(2.844,2.212)--(2.861,2.227)--(2.877,2.241)%
  --(2.894,2.255)--(2.910,2.268)--(2.927,2.280)--(2.944,2.292)--(2.960,2.303)--(2.977,2.313)%
  --(2.993,2.322)--(3.010,2.330)--(3.026,2.338)--(3.043,2.344)--(3.060,2.349)--(3.076,2.354)%
  --(3.093,2.357)--(3.109,2.359)--(3.126,2.361)--(3.142,2.361)--(3.159,2.361)--(3.176,2.359)%
  --(3.192,2.357)--(3.209,2.354)--(3.225,2.350)--(3.242,2.344)--(3.258,2.338)--(3.275,2.331)%
  --(3.292,2.322)--(3.308,2.313)--(3.325,2.304)--(3.341,2.294)--(3.358,2.283)--(3.374,2.272)%
  --(3.391,2.260)--(3.408,2.248)--(3.424,2.235)--(3.441,2.223)--(3.457,2.210)--(3.474,2.197)%
  --(3.490,2.183)--(3.507,2.170)--(3.523,2.157)--(3.540,2.143)--(3.557,2.130)--(3.573,2.116)%
  --(3.590,2.103)--(3.606,2.089)--(3.623,2.076)--(3.639,2.063)--(3.656,2.049)--(3.673,2.037)%
  --(3.689,2.024)--(3.706,2.011)--(3.722,1.999)--(3.739,1.986)--(3.755,1.974)--(3.772,1.962)%
  --(3.789,1.950)--(3.805,1.938)--(3.822,1.926)--(3.838,1.915)--(3.855,1.903)--(3.871,1.891)%
  --(3.888,1.880)--(3.905,1.869)--(3.921,1.859)--(3.938,1.848)--(3.954,1.838)--(3.971,1.828)%
  --(3.987,1.817)--(4.004,1.807)--(4.021,1.797)--(4.037,1.787)--(4.054,1.777)--(4.070,1.768)%
  --(4.087,1.759)--(4.103,1.750)--(4.120,1.742)--(4.136,1.734)--(4.153,1.726)--(4.170,1.717)%
  --(4.186,1.709)--(4.203,1.700)--(4.219,1.692)--(4.236,1.683)--(4.252,1.675)--(4.269,1.667)%
  --(4.286,1.658)--(4.302,1.651)--(4.319,1.643)--(4.335,1.636)--(4.352,1.629)--(4.368,1.622)%
  --(4.385,1.615)--(4.402,1.608)--(4.418,1.601)--(4.435,1.595)--(4.451,1.589)--(4.468,1.582)%
  --(4.484,1.576)--(4.501,1.570)--(4.518,1.564)--(4.534,1.557)--(4.551,1.551)--(4.567,1.544)%
  --(4.584,1.537)--(4.600,1.531)--(4.617,1.524)--(4.634,1.517)--(4.650,1.510)--(4.667,1.503)%
  --(4.683,1.495)--(4.700,1.487)--(4.716,1.479)--(4.733,1.470)--(4.749,1.462)--(4.766,1.453)%
  --(4.783,1.445)--(4.799,1.437)--(4.816,1.429)--(4.832,1.421)--(4.849,1.413)--(4.865,1.404)%
  --(4.882,1.395)--(4.899,1.385)--(4.915,1.375)--(4.932,1.364)--(4.948,1.353)--(4.965,1.342)%
  --(4.981,1.330)--(4.998,1.318)--(5.015,1.307)--(5.031,1.295)--(5.048,1.283)--(5.064,1.271)%
  --(5.081,1.260)--(5.097,1.249)--(5.114,1.237)--(5.131,1.226)--(5.147,1.215)--(5.164,1.204)%
  --(5.180,1.193)--(5.197,1.181)--(5.213,1.170)--(5.230,1.158)--(5.246,1.147)--(5.263,1.137)%
  --(5.280,1.126)--(5.296,1.115)--(5.313,1.105)--(5.329,1.095)--(5.346,1.085)--(5.362,1.075)%
  --(5.379,1.065)--(5.396,1.055)--(5.412,1.046)--(5.429,1.038)--(5.445,1.031)--(5.462,1.025)%
  --(5.478,1.020)--(5.495,1.016)--(5.512,1.014)--(5.528,1.012)--(5.545,1.013)--(5.561,1.014)%
  --(5.578,1.018)--(5.594,1.023)--(5.611,1.029)--(5.628,1.038)--(5.644,1.049)--(5.661,1.062)%
  --(5.677,1.078)--(5.694,1.096)--(5.710,1.117)--(5.727,1.140)--(5.744,1.166)--(5.760,1.193)%
  --(5.777,1.223)--(5.793,1.254)--(5.810,1.288)--(5.826,1.323)--(5.843,1.362)--(5.859,1.405)%
  --(5.876,1.453)--(5.893,1.504)--(5.909,1.559)--(5.926,1.616)--(5.942,1.675)--(5.959,1.735)%
  --(5.975,1.794)--(5.992,1.853)--(6.009,1.911)--(6.025,1.969)--(6.042,2.027)--(6.058,2.087)%
  --(6.075,2.150)--(6.091,2.213)--(6.108,2.277)--(6.125,2.342)--(6.141,2.406)--(6.158,2.469)%
  --(6.174,2.530)--(6.191,2.591)--(6.207,2.650)--(6.224,2.707)--(6.241,2.761)--(6.257,2.813)%
  --(6.274,2.862)--(6.290,2.908)--(6.307,2.952)--(6.323,2.993)--(6.340,3.032)--(6.357,3.068)%
  --(6.373,3.102)--(6.390,3.133)--(6.406,3.161)--(6.423,3.186)--(6.439,3.207)--(6.456,3.225)%
  --(6.472,3.239)--(6.489,3.248)--(6.506,3.254)--(6.522,3.254)--(6.539,3.250)--(6.555,3.242)%
  --(6.572,3.229)--(6.588,3.212)--(6.605,3.190)--(6.622,3.165)--(6.638,3.136)--(6.655,3.103)%
  --(6.671,3.067)--(6.688,3.027)--(6.704,2.983)--(6.721,2.936)--(6.738,2.885)--(6.754,2.832)%
  --(6.771,2.775)--(6.787,2.716)--(6.804,2.654)--(6.820,2.590)--(6.837,2.524)--(6.854,2.457)%
  --(6.870,2.388)--(6.887,2.317)--(6.903,2.246)--(6.920,2.174)--(6.936,2.101)--(6.953,2.029)%
  --(6.970,1.956)--(6.986,1.884)--(7.003,1.813)--(7.019,1.743)--(7.036,1.674)--(7.052,1.608)%
  --(7.069,1.544)--(7.085,1.483)--(7.102,1.426)--(7.119,1.372)--(7.135,1.322)--(7.152,1.277)%
  --(7.168,1.236)--(7.185,1.199)--(7.201,1.166)--(7.218,1.138)--(7.235,1.113)--(7.251,1.093)%
  --(7.268,1.077)--(7.284,1.064)--(7.301,1.055)--(7.317,1.048)--(7.334,1.044)--(7.351,1.043)%
  --(7.367,1.044)--(7.384,1.046)--(7.400,1.050)--(7.417,1.055)--(7.433,1.059)--(7.450,1.063)%
  --(7.467,1.066)--(7.483,1.067)--(7.500,1.067)--(7.516,1.065)--(7.533,1.062)--(7.549,1.059)%
  --(7.566,1.055)--(7.583,1.050)--(7.599,1.046)--(7.616,1.041)--(7.632,1.036)--(7.649,1.031)%
  --(7.665,1.026)--(7.682,1.021)--(7.698,1.017)--(7.715,1.012)--(7.732,1.008)--(7.748,1.004)%
  --(7.765,1.000)--(7.781,0.997)--(7.798,0.994)--(7.814,0.991)--(7.831,0.989)--(7.848,0.987)%
  --(7.864,0.986)--(7.881,0.985)--(7.897,0.985)--(7.914,0.985)--(7.930,0.985)--(7.947,0.985);
\gpcolor{color=gp lt color border}
\node[gp node right] at (3.681,3.736) {$|2\rangle$};
\gpcolor{rgb color={0.337,0.706,0.914}}
\draw[gp path] (3.865,3.736)--(4.781,3.736);
\draw[gp path] (1.320,0.985)--(1.337,0.985)--(1.353,0.985)--(1.370,0.985)--(1.386,0.985)%
  --(1.403,0.985)--(1.419,0.985)--(1.436,0.985)--(1.453,0.985)--(1.469,0.985)--(1.486,0.985)%
  --(1.502,0.985)--(1.519,0.985)--(1.535,0.985)--(1.552,0.985)--(1.569,0.985)--(1.585,0.985)%
  --(1.602,0.985)--(1.618,0.985)--(1.635,0.985)--(1.651,0.985)--(1.668,0.985)--(1.684,0.985)%
  --(1.701,0.985)--(1.718,0.985)--(1.734,0.985)--(1.751,0.985)--(1.767,0.985)--(1.784,0.986)%
  --(1.800,0.986)--(1.817,0.986)--(1.834,0.986)--(1.850,0.986)--(1.867,0.986)--(1.883,0.987)%
  --(1.900,0.987)--(1.916,0.987)--(1.933,0.988)--(1.950,0.988)--(1.966,0.989)--(1.983,0.989)%
  --(1.999,0.990)--(2.016,0.991)--(2.032,0.991)--(2.049,0.992)--(2.066,0.993)--(2.082,0.995)%
  --(2.099,0.996)--(2.115,0.997)--(2.132,0.999)--(2.148,1.000)--(2.165,1.002)--(2.182,1.004)%
  --(2.198,1.006)--(2.215,1.008)--(2.231,1.011)--(2.248,1.013)--(2.264,1.016)--(2.281,1.019)%
  --(2.297,1.023)--(2.314,1.026)--(2.331,1.030)--(2.347,1.035)--(2.364,1.040)--(2.380,1.045)%
  --(2.397,1.050)--(2.413,1.056)--(2.430,1.062)--(2.447,1.069)--(2.463,1.076)--(2.480,1.083)%
  --(2.496,1.091)--(2.513,1.099)--(2.529,1.108)--(2.546,1.117)--(2.563,1.127)--(2.579,1.138)%
  --(2.596,1.148)--(2.612,1.159)--(2.629,1.171)--(2.645,1.182)--(2.662,1.194)--(2.679,1.206)%
  --(2.695,1.219)--(2.712,1.233)--(2.728,1.247)--(2.745,1.261)--(2.761,1.275)--(2.778,1.289)%
  --(2.795,1.303)--(2.811,1.317)--(2.828,1.332)--(2.844,1.347)--(2.861,1.363)--(2.877,1.378)%
  --(2.894,1.394)--(2.910,1.410)--(2.927,1.427)--(2.944,1.444)--(2.960,1.461)--(2.977,1.478)%
  --(2.993,1.495)--(3.010,1.513)--(3.026,1.530)--(3.043,1.548)--(3.060,1.565)--(3.076,1.582)%
  --(3.093,1.598)--(3.109,1.614)--(3.126,1.629)--(3.142,1.644)--(3.159,1.659)--(3.176,1.673)%
  --(3.192,1.687)--(3.209,1.700)--(3.225,1.713)--(3.242,1.726)--(3.258,1.738)--(3.275,1.749)%
  --(3.292,1.759)--(3.308,1.768)--(3.325,1.776)--(3.341,1.782)--(3.358,1.788)--(3.374,1.793)%
  --(3.391,1.797)--(3.408,1.800)--(3.424,1.802)--(3.441,1.804)--(3.457,1.805)--(3.474,1.805)%
  --(3.490,1.804)--(3.507,1.802)--(3.523,1.799)--(3.540,1.796)--(3.557,1.792)--(3.573,1.788)%
  --(3.590,1.784)--(3.606,1.780)--(3.623,1.775)--(3.639,1.770)--(3.656,1.765)--(3.673,1.759)%
  --(3.689,1.754)--(3.706,1.748)--(3.722,1.742)--(3.739,1.735)--(3.755,1.729)--(3.772,1.721)%
  --(3.789,1.714)--(3.805,1.706)--(3.822,1.697)--(3.838,1.689)--(3.855,1.680)--(3.871,1.672)%
  --(3.888,1.663)--(3.905,1.655)--(3.921,1.647)--(3.938,1.639)--(3.954,1.632)--(3.971,1.625)%
  --(3.987,1.618)--(4.004,1.611)--(4.021,1.604)--(4.037,1.597)--(4.054,1.591)--(4.070,1.584)%
  --(4.087,1.579)--(4.103,1.574)--(4.120,1.570)--(4.136,1.567)--(4.153,1.564)--(4.170,1.560)%
  --(4.186,1.556)--(4.203,1.553)--(4.219,1.551)--(4.236,1.549)--(4.252,1.550)--(4.269,1.552)%
  --(4.286,1.554)--(4.302,1.557)--(4.319,1.560)--(4.335,1.564)--(4.352,1.567)--(4.368,1.572)%
  --(4.385,1.578)--(4.402,1.583)--(4.418,1.589)--(4.435,1.594)--(4.451,1.598)--(4.468,1.602)%
  --(4.484,1.607)--(4.501,1.612)--(4.518,1.620)--(4.534,1.629)--(4.551,1.639)--(4.567,1.650)%
  --(4.584,1.661)--(4.600,1.673)--(4.617,1.687)--(4.634,1.701)--(4.650,1.716)--(4.667,1.733)%
  --(4.683,1.750)--(4.700,1.768)--(4.716,1.787)--(4.733,1.806)--(4.749,1.825)--(4.766,1.843)%
  --(4.783,1.861)--(4.799,1.878)--(4.816,1.895)--(4.832,1.913)--(4.849,1.932)--(4.865,1.954)%
  --(4.882,1.978)--(4.899,2.004)--(4.915,2.032)--(4.932,2.062)--(4.948,2.093)--(4.965,2.126)%
  --(4.981,2.159)--(4.998,2.193)--(5.015,2.227)--(5.031,2.263)--(5.048,2.299)--(5.064,2.335)%
  --(5.081,2.373)--(5.097,2.410)--(5.114,2.448)--(5.131,2.486)--(5.147,2.525)--(5.164,2.564)%
  --(5.180,2.605)--(5.197,2.648)--(5.213,2.691)--(5.230,2.735)--(5.246,2.779)--(5.263,2.822)%
  --(5.280,2.866)--(5.296,2.911)--(5.313,2.958)--(5.329,3.007)--(5.346,3.060)--(5.362,3.114)%
  --(5.379,3.169)--(5.396,3.226)--(5.412,3.283)--(5.429,3.339)--(5.445,3.394)--(5.462,3.446)%
  --(5.478,3.497)--(5.495,3.547)--(5.512,3.596)--(5.528,3.644)--(5.545,3.692)--(5.561,3.739)%
  --(5.578,3.784)--(5.594,3.828)--(5.611,3.869)--(5.628,3.908)--(5.644,3.945)--(5.661,3.980)%
  --(5.677,4.012)--(5.694,4.040)--(5.710,4.064)--(5.727,4.084)--(5.744,4.100)--(5.760,4.110)%
  --(5.777,4.117)--(5.793,4.118)--(5.810,4.115)--(5.826,4.108)--(5.843,4.096)--(5.859,4.077)%
  --(5.876,4.053)--(5.893,4.021)--(5.909,3.983)--(5.926,3.937)--(5.942,3.886)--(5.959,3.830)%
  --(5.975,3.770)--(5.992,3.707)--(6.009,3.642)--(6.025,3.574)--(6.042,3.503)--(6.058,3.426)%
  --(6.075,3.345)--(6.091,3.258)--(6.108,3.168)--(6.125,3.076)--(6.141,2.983)--(6.158,2.889)%
  --(6.174,2.794)--(6.191,2.700)--(6.207,2.606)--(6.224,2.512)--(6.241,2.419)--(6.257,2.328)%
  --(6.274,2.238)--(6.290,2.151)--(6.307,2.065)--(6.323,1.980)--(6.340,1.898)--(6.357,1.818)%
  --(6.373,1.739)--(6.390,1.663)--(6.406,1.590)--(6.423,1.519)--(6.439,1.452)--(6.456,1.389)%
  --(6.472,1.331)--(6.489,1.278)--(6.506,1.230)--(6.522,1.188)--(6.539,1.152)--(6.555,1.121)%
  --(6.572,1.096)--(6.588,1.078)--(6.605,1.064)--(6.622,1.057)--(6.638,1.056)--(6.655,1.060)%
  --(6.671,1.070)--(6.688,1.085)--(6.704,1.106)--(6.721,1.131)--(6.738,1.162)--(6.754,1.198)%
  --(6.771,1.239)--(6.787,1.283)--(6.804,1.332)--(6.820,1.385)--(6.837,1.440)--(6.854,1.499)%
  --(6.870,1.561)--(6.887,1.625)--(6.903,1.690)--(6.920,1.757)--(6.936,1.824)--(6.953,1.891)%
  --(6.970,1.958)--(6.986,2.024)--(7.003,2.088)--(7.019,2.150)--(7.036,2.208)--(7.052,2.263)%
  --(7.069,2.313)--(7.085,2.358)--(7.102,2.398)--(7.119,2.431)--(7.135,2.458)--(7.152,2.478)%
  --(7.168,2.491)--(7.185,2.498)--(7.201,2.497)--(7.218,2.490)--(7.235,2.476)--(7.251,2.456)%
  --(7.268,2.431)--(7.284,2.400)--(7.301,2.365)--(7.317,2.327)--(7.334,2.284)--(7.351,2.239)%
  --(7.367,2.191)--(7.384,2.141)--(7.400,2.088)--(7.417,2.034)--(7.433,1.977)--(7.450,1.920)%
  --(7.467,1.862)--(7.483,1.803)--(7.500,1.743)--(7.516,1.684)--(7.533,1.626)--(7.549,1.570)%
  --(7.566,1.515)--(7.583,1.462)--(7.599,1.412)--(7.616,1.363)--(7.632,1.317)--(7.649,1.274)%
  --(7.665,1.233)--(7.682,1.195)--(7.698,1.161)--(7.715,1.129)--(7.732,1.101)--(7.748,1.076)%
  --(7.765,1.054)--(7.781,1.035)--(7.798,1.020)--(7.814,1.007)--(7.831,0.998)--(7.848,0.992)%
  --(7.864,0.988)--(7.881,0.986)--(7.897,0.985)--(7.914,0.985)--(7.930,0.985)--(7.947,0.985);
\gpcolor{color=gp lt color border}
\node[gp node right] at (3.681,3.428) {$|3\rangle$};
\gpcolor{rgb color={0.902,0.624,0.000}}
\draw[gp path] (3.865,3.428)--(4.781,3.428);
\draw[gp path] (1.320,0.985)--(1.337,0.985)--(1.353,0.985)--(1.370,0.985)--(1.386,0.985)%
  --(1.403,0.985)--(1.419,0.985)--(1.436,0.985)--(1.453,0.985)--(1.469,0.985)--(1.486,0.985)%
  --(1.502,0.985)--(1.519,0.985)--(1.535,0.985)--(1.552,0.985)--(1.569,0.985)--(1.585,0.985)%
  --(1.602,0.985)--(1.618,0.985)--(1.635,0.985)--(1.651,0.985)--(1.668,0.985)--(1.684,0.985)%
  --(1.701,0.985)--(1.718,0.985)--(1.734,0.985)--(1.751,0.985)--(1.767,0.985)--(1.784,0.985)%
  --(1.800,0.985)--(1.817,0.985)--(1.834,0.985)--(1.850,0.985)--(1.867,0.985)--(1.883,0.985)%
  --(1.900,0.985)--(1.916,0.985)--(1.933,0.985)--(1.950,0.985)--(1.966,0.985)--(1.983,0.985)%
  --(1.999,0.985)--(2.016,0.985)--(2.032,0.985)--(2.049,0.985)--(2.066,0.985)--(2.082,0.985)%
  --(2.099,0.985)--(2.115,0.985)--(2.132,0.985)--(2.148,0.986)--(2.165,0.986)--(2.182,0.986)%
  --(2.198,0.986)--(2.215,0.986)--(2.231,0.986)--(2.248,0.986)--(2.264,0.987)--(2.281,0.987)%
  --(2.297,0.987)--(2.314,0.987)--(2.331,0.988)--(2.347,0.988)--(2.364,0.989)--(2.380,0.989)%
  --(2.397,0.990)--(2.413,0.991)--(2.430,0.991)--(2.447,0.992)--(2.463,0.993)--(2.480,0.994)%
  --(2.496,0.996)--(2.513,0.997)--(2.529,0.999)--(2.546,1.000)--(2.563,1.002)--(2.579,1.005)%
  --(2.596,1.007)--(2.612,1.010)--(2.629,1.012)--(2.645,1.015)--(2.662,1.018)--(2.679,1.022)%
  --(2.695,1.026)--(2.712,1.030)--(2.728,1.035)--(2.745,1.040)--(2.761,1.045)--(2.778,1.051)%
  --(2.795,1.056)--(2.811,1.063)--(2.828,1.069)--(2.844,1.076)--(2.861,1.084)--(2.877,1.092)%
  --(2.894,1.101)--(2.910,1.111)--(2.927,1.121)--(2.944,1.132)--(2.960,1.143)--(2.977,1.156)%
  --(2.993,1.169)--(3.010,1.183)--(3.026,1.199)--(3.043,1.215)--(3.060,1.232)--(3.076,1.249)%
  --(3.093,1.268)--(3.109,1.287)--(3.126,1.306)--(3.142,1.327)--(3.159,1.349)--(3.176,1.373)%
  --(3.192,1.397)--(3.209,1.423)--(3.225,1.451)--(3.242,1.480)--(3.258,1.511)--(3.275,1.542)%
  --(3.292,1.575)--(3.308,1.609)--(3.325,1.644)--(3.341,1.680)--(3.358,1.715)--(3.374,1.751)%
  --(3.391,1.787)--(3.408,1.823)--(3.424,1.859)--(3.441,1.896)--(3.457,1.933)--(3.474,1.970)%
  --(3.490,2.007)--(3.507,2.044)--(3.523,2.082)--(3.540,2.119)--(3.557,2.156)--(3.573,2.192)%
  --(3.590,2.228)--(3.606,2.263)--(3.623,2.297)--(3.639,2.331)--(3.656,2.365)--(3.673,2.398)%
  --(3.689,2.430)--(3.706,2.462)--(3.722,2.494)--(3.739,2.525)--(3.755,2.557)--(3.772,2.588)%
  --(3.789,2.619)--(3.805,2.650)--(3.822,2.682)--(3.838,2.713)--(3.855,2.744)--(3.871,2.774)%
  --(3.888,2.804)--(3.905,2.832)--(3.921,2.860)--(3.938,2.886)--(3.954,2.912)--(3.971,2.938)%
  --(3.987,2.963)--(4.004,2.988)--(4.021,3.012)--(4.037,3.036)--(4.054,3.059)--(4.070,3.081)%
  --(4.087,3.102)--(4.103,3.121)--(4.120,3.138)--(4.136,3.154)--(4.153,3.171)--(4.170,3.188)%
  --(4.186,3.205)--(4.203,3.221)--(4.219,3.237)--(4.236,3.251)--(4.252,3.263)--(4.269,3.273)%
  --(4.286,3.282)--(4.302,3.290)--(4.319,3.297)--(4.335,3.303)--(4.352,3.309)--(4.368,3.313)%
  --(4.385,3.317)--(4.402,3.320)--(4.418,3.323)--(4.435,3.327)--(4.451,3.331)--(4.468,3.335)%
  --(4.484,3.339)--(4.501,3.341)--(4.518,3.341)--(4.534,3.340)--(4.551,3.338)--(4.567,3.335)%
  --(4.584,3.330)--(4.600,3.325)--(4.617,3.319)--(4.634,3.312)--(4.650,3.304)--(4.667,3.295)%
  --(4.683,3.286)--(4.700,3.276)--(4.716,3.265)--(4.733,3.255)--(4.749,3.245)--(4.766,3.236)%
  --(4.783,3.227)--(4.799,3.219)--(4.816,3.210)--(4.832,3.200)--(4.849,3.189)--(4.865,3.176)%
  --(4.882,3.161)--(4.899,3.145)--(4.915,3.127)--(4.932,3.107)--(4.948,3.087)--(4.965,3.065)%
  --(4.981,3.043)--(4.998,3.021)--(5.015,2.997)--(5.031,2.973)--(5.048,2.948)--(5.064,2.922)%
  --(5.081,2.896)--(5.097,2.869)--(5.114,2.842)--(5.131,2.814)--(5.147,2.785)--(5.164,2.756)%
  --(5.180,2.725)--(5.197,2.693)--(5.213,2.660)--(5.230,2.627)--(5.246,2.594)--(5.263,2.561)%
  --(5.280,2.527)--(5.296,2.491)--(5.313,2.454)--(5.329,2.413)--(5.346,2.370)--(5.362,2.325)%
  --(5.379,2.278)--(5.396,2.230)--(5.412,2.181)--(5.429,2.132)--(5.445,2.084)--(5.462,2.037)%
  --(5.478,1.991)--(5.495,1.945)--(5.512,1.898)--(5.528,1.851)--(5.545,1.803)--(5.561,1.755)%
  --(5.578,1.707)--(5.594,1.659)--(5.611,1.611)--(5.628,1.564)--(5.644,1.517)--(5.661,1.470)%
  --(5.677,1.425)--(5.694,1.380)--(5.710,1.337)--(5.727,1.297)--(5.744,1.258)--(5.760,1.223)%
  --(5.777,1.190)--(5.793,1.160)--(5.810,1.133)--(5.826,1.108)--(5.843,1.086)--(5.859,1.065)%
  --(5.876,1.046)--(5.893,1.031)--(5.909,1.019)--(5.926,1.011)--(5.942,1.007)--(5.959,1.007)%
  --(5.975,1.010)--(5.992,1.017)--(6.009,1.026)--(6.025,1.038)--(6.042,1.052)--(6.058,1.070)%
  --(6.075,1.090)--(6.091,1.113)--(6.108,1.138)--(6.125,1.165)--(6.141,1.193)--(6.158,1.223)%
  --(6.174,1.254)--(6.191,1.285)--(6.207,1.317)--(6.224,1.350)--(6.241,1.384)--(6.257,1.417)%
  --(6.274,1.451)--(6.290,1.485)--(6.307,1.519)--(6.323,1.553)--(6.340,1.587)--(6.357,1.620)%
  --(6.373,1.653)--(6.390,1.685)--(6.406,1.716)--(6.423,1.747)--(6.439,1.776)--(6.456,1.804)%
  --(6.472,1.830)--(6.489,1.853)--(6.506,1.875)--(6.522,1.895)--(6.539,1.913)--(6.555,1.928)%
  --(6.572,1.941)--(6.588,1.951)--(6.605,1.959)--(6.622,1.964)--(6.638,1.967)--(6.655,1.967)%
  --(6.671,1.965)--(6.688,1.960)--(6.704,1.953)--(6.721,1.943)--(6.738,1.931)--(6.754,1.917)%
  --(6.771,1.901)--(6.787,1.884)--(6.804,1.864)--(6.820,1.842)--(6.837,1.819)--(6.854,1.795)%
  --(6.870,1.769)--(6.887,1.743)--(6.903,1.716)--(6.920,1.689)--(6.936,1.662)--(6.953,1.635)%
  --(6.970,1.609)--(6.986,1.584)--(7.003,1.560)--(7.019,1.538)--(7.036,1.518)--(7.052,1.501)%
  --(7.069,1.486)--(7.085,1.475)--(7.102,1.467)--(7.119,1.462)--(7.135,1.461)--(7.152,1.464)%
  --(7.168,1.471)--(7.185,1.482)--(7.201,1.496)--(7.218,1.514)--(7.235,1.536)--(7.251,1.561)%
  --(7.268,1.589)--(7.284,1.619)--(7.301,1.652)--(7.317,1.686)--(7.334,1.723)--(7.351,1.760)%
  --(7.367,1.799)--(7.384,1.839)--(7.400,1.881)--(7.417,1.924)--(7.433,1.968)--(7.450,2.014)%
  --(7.467,2.061)--(7.483,2.110)--(7.500,2.160)--(7.516,2.210)--(7.533,2.259)--(7.549,2.308)%
  --(7.566,2.356)--(7.583,2.402)--(7.599,2.447)--(7.616,2.490)--(7.632,2.531)--(7.649,2.570)%
  --(7.665,2.606)--(7.682,2.640)--(7.698,2.671)--(7.715,2.700)--(7.732,2.726)--(7.748,2.749)%
  --(7.765,2.769)--(7.781,2.787)--(7.798,2.801)--(7.814,2.813)--(7.831,2.822)--(7.848,2.828)%
  --(7.864,2.832)--(7.881,2.833)--(7.897,2.834)--(7.914,2.834)--(7.930,2.834)--(7.947,2.834);
\gpcolor{color=gp lt color border}
\gpsetlinewidth{1.00}
\draw[gp path] (1.320,4.691)--(1.320,0.985)--(7.947,0.985)--(7.947,4.691)--cycle;
\gpdefrectangularnode{gp plot 1}{\pgfpoint{1.320cm}{0.985cm}}{\pgfpoint{7.947cm}{4.691cm}}
\end{tikzpicture}

%% file: figures/FockSuperp03/optim_history.tex
\begin{tikzpicture}[gnuplot]
\tikzset{every node/.append style={font={\footnotesize}}}
\path (0.000,0.000) rectangle (8.500,4.000);
\gpcolor{color=gp lt color axes}
\gpsetlinetype{gp lt axes}
\gpsetdashtype{gp dt axes}
\gpsetlinewidth{0.50}
\draw[gp path] (1.504,0.985)--(6.719,0.985);
\gpcolor{color=gp lt color border}
\gpsetlinetype{gp lt border}
\gpsetdashtype{gp dt solid}
\gpsetlinewidth{1.00}
\draw[gp path] (1.504,0.985)--(1.684,0.985);
\node[gp node right] at (1.320,0.985) {$10^{-7}$};
\draw[gp path] (1.504,1.101)--(1.594,1.101);
\draw[gp path] (1.504,1.169)--(1.594,1.169);
\draw[gp path] (1.504,1.218)--(1.594,1.218);
\draw[gp path] (1.504,1.255)--(1.594,1.255);
\draw[gp path] (1.504,1.286)--(1.594,1.286);
\draw[gp path] (1.504,1.312)--(1.594,1.312);
\draw[gp path] (1.504,1.334)--(1.594,1.334);
\draw[gp path] (1.504,1.354)--(1.594,1.354);
\gpcolor{color=gp lt color axes}
\gpsetlinetype{gp lt axes}
\gpsetdashtype{gp dt axes}
\gpsetlinewidth{0.50}
\draw[gp path] (1.504,1.372)--(1.747,1.372);
\draw[gp path] (4.503,1.372)--(6.719,1.372);
\gpcolor{color=gp lt color border}
\gpsetlinetype{gp lt border}
\gpsetdashtype{gp dt solid}
\gpsetlinewidth{1.00}
\draw[gp path] (1.504,1.372)--(1.684,1.372);
\node[gp node right] at (1.320,1.372) {$10^{-6}$};
\draw[gp path] (1.504,1.488)--(1.594,1.488);
\draw[gp path] (1.504,1.556)--(1.594,1.556);
\draw[gp path] (1.504,1.604)--(1.594,1.604);
\draw[gp path] (1.504,1.642)--(1.594,1.642);
\draw[gp path] (1.504,1.672)--(1.594,1.672);
\draw[gp path] (1.504,1.698)--(1.594,1.698);
\draw[gp path] (1.504,1.721)--(1.594,1.721);
\draw[gp path] (1.504,1.740)--(1.594,1.740);
\gpcolor{color=gp lt color axes}
\gpsetlinetype{gp lt axes}
\gpsetdashtype{gp dt axes}
\gpsetlinewidth{0.50}
\draw[gp path] (1.504,1.758)--(1.747,1.758);
\draw[gp path] (4.503,1.758)--(6.719,1.758);
\gpcolor{color=gp lt color border}
\gpsetlinetype{gp lt border}
\gpsetdashtype{gp dt solid}
\gpsetlinewidth{1.00}
\draw[gp path] (1.504,1.758)--(1.684,1.758);
\node[gp node right] at (1.320,1.758) {$10^{-5}$};
\draw[gp path] (1.504,1.875)--(1.594,1.875);
\draw[gp path] (1.504,1.943)--(1.594,1.943);
\draw[gp path] (1.504,1.991)--(1.594,1.991);
\draw[gp path] (1.504,2.028)--(1.594,2.028);
\draw[gp path] (1.504,2.059)--(1.594,2.059);
\draw[gp path] (1.504,2.085)--(1.594,2.085);
\draw[gp path] (1.504,2.107)--(1.594,2.107);
\draw[gp path] (1.504,2.127)--(1.594,2.127);
\gpcolor{color=gp lt color axes}
\gpsetlinetype{gp lt axes}
\gpsetdashtype{gp dt axes}
\gpsetlinewidth{0.50}
\draw[gp path] (1.504,2.145)--(6.719,2.145);
\gpcolor{color=gp lt color border}
\gpsetlinetype{gp lt border}
\gpsetdashtype{gp dt solid}
\gpsetlinewidth{1.00}
\draw[gp path] (1.504,2.145)--(1.684,2.145);
\node[gp node right] at (1.320,2.145) {$10^{-4}$};
\draw[gp path] (1.504,2.261)--(1.594,2.261);
\draw[gp path] (1.504,2.329)--(1.594,2.329);
\draw[gp path] (1.504,2.377)--(1.594,2.377);
\draw[gp path] (1.504,2.415)--(1.594,2.415);
\draw[gp path] (1.504,2.446)--(1.594,2.446);
\draw[gp path] (1.504,2.471)--(1.594,2.471);
\draw[gp path] (1.504,2.494)--(1.594,2.494);
\draw[gp path] (1.504,2.514)--(1.594,2.514);
\gpcolor{color=gp lt color axes}
\gpsetlinetype{gp lt axes}
\gpsetdashtype{gp dt axes}
\gpsetlinewidth{0.50}
\draw[gp path] (1.504,2.531)--(6.719,2.531);
\gpcolor{color=gp lt color border}
\gpsetlinetype{gp lt border}
\gpsetdashtype{gp dt solid}
\gpsetlinewidth{1.00}
\draw[gp path] (1.504,2.531)--(1.684,2.531);
\node[gp node right] at (1.320,2.531) {$10^{-3}$};
\draw[gp path] (1.504,2.648)--(1.594,2.648);
\draw[gp path] (1.504,2.716)--(1.594,2.716);
\draw[gp path] (1.504,2.764)--(1.594,2.764);
\draw[gp path] (1.504,2.801)--(1.594,2.801);
\draw[gp path] (1.504,2.832)--(1.594,2.832);
\draw[gp path] (1.504,2.858)--(1.594,2.858);
\draw[gp path] (1.504,2.880)--(1.594,2.880);
\draw[gp path] (1.504,2.900)--(1.594,2.900);
\gpcolor{color=gp lt color axes}
\gpsetlinetype{gp lt axes}
\gpsetdashtype{gp dt axes}
\gpsetlinewidth{0.50}
\draw[gp path] (1.504,2.918)--(6.719,2.918);
\gpcolor{color=gp lt color border}
\gpsetlinetype{gp lt border}
\gpsetdashtype{gp dt solid}
\gpsetlinewidth{1.00}
\draw[gp path] (1.504,2.918)--(1.684,2.918);
\node[gp node right] at (1.320,2.918) {$10^{-2}$};
\draw[gp path] (1.504,3.034)--(1.594,3.034);
\draw[gp path] (1.504,3.102)--(1.594,3.102);
\draw[gp path] (1.504,3.151)--(1.594,3.151);
\draw[gp path] (1.504,3.188)--(1.594,3.188);
\draw[gp path] (1.504,3.219)--(1.594,3.219);
\draw[gp path] (1.504,3.245)--(1.594,3.245);
\draw[gp path] (1.504,3.267)--(1.594,3.267);
\draw[gp path] (1.504,3.287)--(1.594,3.287);
\gpcolor{color=gp lt color axes}
\gpsetlinetype{gp lt axes}
\gpsetdashtype{gp dt axes}
\gpsetlinewidth{0.50}
\draw[gp path] (1.504,3.304)--(6.719,3.304);
\gpcolor{color=gp lt color border}
\gpsetlinetype{gp lt border}
\gpsetdashtype{gp dt solid}
\gpsetlinewidth{1.00}
\draw[gp path] (1.504,3.304)--(1.684,3.304);
\node[gp node right] at (1.320,3.304) {$10^{-1}$};
\draw[gp path] (1.504,3.421)--(1.594,3.421);
\draw[gp path] (1.504,3.489)--(1.594,3.489);
\draw[gp path] (1.504,3.537)--(1.594,3.537);
\draw[gp path] (1.504,3.575)--(1.594,3.575);
\draw[gp path] (1.504,3.605)--(1.594,3.605);
\draw[gp path] (1.504,3.631)--(1.594,3.631);
\draw[gp path] (1.504,3.654)--(1.594,3.654);
\draw[gp path] (1.504,3.673)--(1.594,3.673);
\gpcolor{color=gp lt color axes}
\gpsetlinetype{gp lt axes}
\gpsetdashtype{gp dt axes}
\gpsetlinewidth{0.50}
\draw[gp path] (1.504,3.691)--(6.719,3.691);
\gpcolor{color=gp lt color border}
\gpsetlinetype{gp lt border}
\gpsetdashtype{gp dt solid}
\gpsetlinewidth{1.00}
\draw[gp path] (1.504,3.691)--(1.684,3.691);
\node[gp node right] at (1.320,3.691) {$10^{0}$};
\gpcolor{color=gp lt color axes}
\gpsetlinetype{gp lt axes}
\gpsetdashtype{gp dt axes}
\gpsetlinewidth{0.50}
\draw[gp path] (1.504,0.985)--(1.504,3.691);
\gpcolor{color=gp lt color border}
\gpsetlinetype{gp lt border}
\gpsetdashtype{gp dt solid}
\gpsetlinewidth{1.00}
\draw[gp path] (1.504,0.985)--(1.504,1.165);
\draw[gp path] (1.504,3.691)--(1.504,3.511);
\node[gp node center] at (1.504,0.677) {$0$};
\gpcolor{color=gp lt color axes}
\gpsetlinetype{gp lt axes}
\gpsetdashtype{gp dt axes}
\gpsetlinewidth{0.50}
\draw[gp path] (2.156,0.985)--(2.156,1.221);
\draw[gp path] (2.156,2.145)--(2.156,3.691);
\gpcolor{color=gp lt color border}
\gpsetlinetype{gp lt border}
\gpsetdashtype{gp dt solid}
\gpsetlinewidth{1.00}
\draw[gp path] (2.156,0.985)--(2.156,1.165);
\draw[gp path] (2.156,3.691)--(2.156,3.511);
\node[gp node center] at (2.156,0.677) {$10$};
\gpcolor{color=gp lt color axes}
\gpsetlinetype{gp lt axes}
\gpsetdashtype{gp dt axes}
\gpsetlinewidth{0.50}
\draw[gp path] (2.808,0.985)--(2.808,1.221);
\draw[gp path] (2.808,2.145)--(2.808,3.691);
\gpcolor{color=gp lt color border}
\gpsetlinetype{gp lt border}
\gpsetdashtype{gp dt solid}
\gpsetlinewidth{1.00}
\draw[gp path] (2.808,0.985)--(2.808,1.165);
\draw[gp path] (2.808,3.691)--(2.808,3.511);
\node[gp node center] at (2.808,0.677) {$20$};
\gpcolor{color=gp lt color axes}
\gpsetlinetype{gp lt axes}
\gpsetdashtype{gp dt axes}
\gpsetlinewidth{0.50}
\draw[gp path] (3.460,0.985)--(3.460,1.221);
\draw[gp path] (3.460,2.145)--(3.460,3.691);
\gpcolor{color=gp lt color border}
\gpsetlinetype{gp lt border}
\gpsetdashtype{gp dt solid}
\gpsetlinewidth{1.00}
\draw[gp path] (3.460,0.985)--(3.460,1.165);
\draw[gp path] (3.460,3.691)--(3.460,3.511);
\node[gp node center] at (3.460,0.677) {$30$};
\gpcolor{color=gp lt color axes}
\gpsetlinetype{gp lt axes}
\gpsetdashtype{gp dt axes}
\gpsetlinewidth{0.50}
\draw[gp path] (4.112,0.985)--(4.112,1.221);
\draw[gp path] (4.112,2.145)--(4.112,3.691);
\gpcolor{color=gp lt color border}
\gpsetlinetype{gp lt border}
\gpsetdashtype{gp dt solid}
\gpsetlinewidth{1.00}
\draw[gp path] (4.112,0.985)--(4.112,1.165);
\draw[gp path] (4.112,3.691)--(4.112,3.511);
\node[gp node center] at (4.112,0.677) {$40$};
\gpcolor{color=gp lt color axes}
\gpsetlinetype{gp lt axes}
\gpsetdashtype{gp dt axes}
\gpsetlinewidth{0.50}
\draw[gp path] (4.763,0.985)--(4.763,3.691);
\gpcolor{color=gp lt color border}
\gpsetlinetype{gp lt border}
\gpsetdashtype{gp dt solid}
\gpsetlinewidth{1.00}
\draw[gp path] (4.763,0.985)--(4.763,1.165);
\draw[gp path] (4.763,3.691)--(4.763,3.511);
\node[gp node center] at (4.763,0.677) {$50$};
\gpcolor{color=gp lt color axes}
\gpsetlinetype{gp lt axes}
\gpsetdashtype{gp dt axes}
\gpsetlinewidth{0.50}
\draw[gp path] (5.415,0.985)--(5.415,3.691);
\gpcolor{color=gp lt color border}
\gpsetlinetype{gp lt border}
\gpsetdashtype{gp dt solid}
\gpsetlinewidth{1.00}
\draw[gp path] (5.415,0.985)--(5.415,1.165);
\draw[gp path] (5.415,3.691)--(5.415,3.511);
\node[gp node center] at (5.415,0.677) {$60$};
\gpcolor{color=gp lt color axes}
\gpsetlinetype{gp lt axes}
\gpsetdashtype{gp dt axes}
\gpsetlinewidth{0.50}
\draw[gp path] (6.067,0.985)--(6.067,3.691);
\gpcolor{color=gp lt color border}
\gpsetlinetype{gp lt border}
\gpsetdashtype{gp dt solid}
\gpsetlinewidth{1.00}
\draw[gp path] (6.067,0.985)--(6.067,1.165);
\draw[gp path] (6.067,3.691)--(6.067,3.511);
\node[gp node center] at (6.067,0.677) {$70$};
\gpcolor{color=gp lt color axes}
\gpsetlinetype{gp lt axes}
\gpsetdashtype{gp dt axes}
\gpsetlinewidth{0.50}
\draw[gp path] (6.719,0.985)--(6.719,3.691);
\gpcolor{color=gp lt color border}
\gpsetlinetype{gp lt border}
\gpsetdashtype{gp dt solid}
\gpsetlinewidth{1.00}
\draw[gp path] (6.719,0.985)--(6.719,1.165);
\draw[gp path] (6.719,3.691)--(6.719,3.511);
\node[gp node center] at (6.719,0.677) {$80$};
\draw[gp path] (6.719,0.985)--(6.539,0.985);
\node[gp node left] at (6.903,0.985) {$0$};
\draw[gp path] (6.719,1.477)--(6.539,1.477);
\node[gp node left] at (6.903,1.477) {$0.2$};
\draw[gp path] (6.719,1.969)--(6.539,1.969);
\node[gp node left] at (6.903,1.969) {$0.4$};
\draw[gp path] (6.719,2.461)--(6.539,2.461);
\node[gp node left] at (6.903,2.461) {$0.6$};
\draw[gp path] (6.719,2.953)--(6.539,2.953);
\node[gp node left] at (6.903,2.953) {$0.8$};
\draw[gp path] (6.719,3.445)--(6.539,3.445);
\node[gp node left] at (6.903,3.445) {$1$};
\draw[gp path] (1.504,3.691)--(1.504,0.985)--(6.719,0.985)--(6.719,3.691)--cycle;
\node[gp node center,rotate=-270] at (0.292,2.338) {$J, \|\nabla J\|$};
\node[gp node center,rotate=-270] at (7.793,2.338) {Fidelity};
\node[gp node center] at (4.111,0.215) {iteration};
\node[gp node right] at (3.219,1.991) {$J$};
\gpcolor{rgb color={0.580,0.000,0.827}}
\draw[gp path] (3.403,1.991)--(4.319,1.991);
\draw[gp path] (1.504,3.575)--(1.569,3.575)--(1.634,3.575)--(1.700,3.575)--(1.765,3.575)%
  --(1.830,3.575)--(1.895,3.575)--(1.960,3.575)--(2.026,3.575)--(2.091,3.556)--(2.156,3.556)%
  --(2.221,3.548)--(2.286,3.541)--(2.351,3.536)--(2.417,3.525)--(2.482,3.515)--(2.547,3.493)%
  --(2.612,3.432)--(2.677,3.418)--(2.743,3.401)--(2.808,3.400)--(2.873,3.396)--(2.938,3.394)%
  --(3.003,3.386)--(3.069,3.378)--(3.134,3.366)--(3.199,3.347)--(3.264,3.327)--(3.329,3.283)%
  --(3.394,3.262)--(3.460,3.254)--(3.525,3.223)--(3.590,3.178)--(3.655,3.144)--(3.720,3.080)%
  --(3.786,3.070)--(3.851,3.030)--(3.916,3.008)--(3.981,2.970)--(4.046,2.933)--(4.112,2.917)%
  --(4.177,2.889)--(4.242,2.848)--(4.307,2.829)--(4.372,2.811)--(4.437,2.795)--(4.503,2.764)%
  --(4.568,2.653)--(4.633,2.644)--(4.698,2.367)--(4.763,2.278)--(4.829,2.270)--(4.894,2.152)%
  --(4.959,2.093)--(5.024,1.999)--(5.089,1.939)--(5.155,1.900)--(5.220,1.862)--(5.285,1.708)%
  --(5.350,1.636)--(5.415,1.444)--(5.480,1.400)--(5.546,1.333)--(5.611,1.218)--(5.676,1.160)%
  --(5.741,1.152)--(5.806,1.151)--(5.872,1.151)--(5.937,1.150)--(6.002,1.150)--(6.067,1.150)%
  --(6.132,1.149)--(6.198,1.149)--(6.263,1.149)--(6.328,1.149)--(6.393,1.149)--(6.458,1.149);
\gpcolor{color=gp lt color border}
\node[gp node right] at (3.219,1.683) {$\|\nabla J\|$};
\gpcolor{rgb color={0.000,0.620,0.451}}
\draw[gp path] (3.403,1.683)--(4.319,1.683);
\draw[gp path] (1.504,2.500)--(1.569,2.420)--(1.634,1.939)--(1.700,1.394)--(1.765,1.126)%
  --(1.830,1.311)--(1.895,1.480)--(1.960,1.658)--(2.026,1.828)--(2.091,2.810)--(2.156,2.800)%
  --(2.221,2.765)--(2.286,2.735)--(2.351,2.733)--(2.417,2.745)--(2.482,2.826)--(2.547,2.973)%
  --(2.612,2.862)--(2.677,2.928)--(2.743,2.659)--(2.808,2.747)--(2.873,2.741)--(2.938,2.669)%
  --(3.003,2.670)--(3.069,2.749)--(3.134,2.808)--(3.199,2.850)--(3.264,2.882)--(3.329,2.841)%
  --(3.394,2.777)--(3.460,2.815)--(3.525,2.658)--(3.590,2.733)--(3.655,2.902)--(3.720,2.782)%
  --(3.786,2.697)--(3.851,2.717)--(3.916,2.515)--(3.981,2.650)--(4.046,2.616)--(4.112,2.570)%
  --(4.177,2.564)--(4.242,2.554)--(4.307,2.565)--(4.372,2.416)--(4.437,2.454)--(4.503,2.542)%
  --(4.568,2.553)--(4.633,2.688)--(4.698,2.471)--(4.763,2.458)--(4.829,2.361)--(4.894,2.291)%
  --(4.959,2.226)--(5.024,2.261)--(5.089,2.108)--(5.155,2.069)--(5.220,2.091)--(5.285,2.148)%
  --(5.350,2.129)--(5.415,1.908)--(5.480,1.834)--(5.546,1.855)--(5.611,1.824)--(5.676,1.714)%
  --(5.741,1.530)--(5.806,1.317)--(5.872,1.290)--(5.937,1.384)--(6.002,1.366)--(6.067,1.391)%
  --(6.132,1.189)--(6.198,1.018)--(6.263,0.999)--(6.328,1.026)--(6.393,1.141)--(6.451,0.985);
\gpcolor{color=gp lt color border}
\node[gp node right] at (3.219,1.375) {Fidelity};
\gpcolor{rgb color={0.337,0.706,0.914}}
\draw[gp path] (3.403,1.375)--(4.319,1.375);
\draw[gp path] (1.504,2.214)--(1.569,2.215)--(1.634,2.215)--(1.700,2.215)--(1.765,2.215)%
  --(1.830,2.215)--(1.895,2.215)--(1.960,2.215)--(2.026,2.215)--(2.091,2.342)--(2.156,2.343)%
  --(2.221,2.397)--(2.286,2.438)--(2.351,2.467)--(2.417,2.530)--(2.482,2.585)--(2.547,2.687)%
  --(2.612,2.918)--(2.677,2.961)--(2.743,3.008)--(2.808,3.009)--(2.873,3.020)--(2.938,3.026)%
  --(3.003,3.046)--(3.069,3.064)--(3.134,3.091)--(3.199,3.128)--(3.264,3.164)--(3.329,3.228)%
  --(3.394,3.254)--(3.460,3.262)--(3.525,3.293)--(3.590,3.329)--(3.655,3.351)--(3.720,3.380)%
  --(3.786,3.384)--(3.851,3.397)--(3.916,3.403)--(3.981,3.411)--(4.046,3.418)--(4.112,3.420)%
  --(4.177,3.424)--(4.242,3.429)--(4.307,3.430)--(4.372,3.432)--(4.437,3.433)--(4.503,3.435)%
  --(4.568,3.440)--(4.633,3.440)--(4.698,3.444)--(4.763,3.444)--(4.829,3.444)--(4.894,3.445)%
  --(4.959,3.445)--(5.024,3.445)--(5.089,3.445)--(5.155,3.445)--(5.220,3.445)--(5.285,3.445)%
  --(5.350,3.445)--(5.415,3.445)--(5.480,3.445)--(5.546,3.445)--(5.611,3.445)--(5.676,3.445)%
  --(5.741,3.445)--(5.806,3.445)--(5.872,3.445)--(5.937,3.445)--(6.002,3.445)--(6.067,3.445)%
  --(6.132,3.445)--(6.198,3.445)--(6.263,3.445)--(6.328,3.445)--(6.393,3.445)--(6.458,3.445);
\gpcolor{color=gp lt color border}
\draw[gp path] (1.504,3.691)--(1.504,0.985)--(6.719,0.985)--(6.719,3.691)--cycle;
\gpdefrectangularnode{gp plot 1}{\pgfpoint{1.504cm}{0.985cm}}{\pgfpoint{6.719cm}{3.691cm}}
\end{tikzpicture}

%% file: figures/2x2x20_reset/optim_history.tex
\begin{tikzpicture}[gnuplot]
\tikzset{every node/.append style={font={\footnotesize}}}
\path (0.000,0.000) rectangle (8.500,5.000);
\gpcolor{color=gp lt color axes}
\gpsetlinetype{gp lt axes}
\gpsetdashtype{gp dt axes}
\gpsetlinewidth{0.50}
\draw[gp path] (1.196,0.985)--(7.947,0.985);
\gpcolor{color=gp lt color border}
\gpsetlinetype{gp lt border}
\gpsetdashtype{gp dt solid}
\gpsetlinewidth{1.00}
\draw[gp path] (1.196,0.985)--(1.376,0.985);
\draw[gp path] (7.947,0.985)--(7.767,0.985);
\node[gp node right] at (1.012,0.985) {$10^{-5}$};
\draw[gp path] (1.196,1.208)--(1.286,1.208);
\draw[gp path] (7.947,1.208)--(7.857,1.208);
\draw[gp path] (1.196,1.339)--(1.286,1.339);
\draw[gp path] (7.947,1.339)--(7.857,1.339);
\draw[gp path] (1.196,1.431)--(1.286,1.431);
\draw[gp path] (7.947,1.431)--(7.857,1.431);
\draw[gp path] (1.196,1.503)--(1.286,1.503);
\draw[gp path] (7.947,1.503)--(7.857,1.503);
\draw[gp path] (1.196,1.562)--(1.286,1.562);
\draw[gp path] (7.947,1.562)--(7.857,1.562);
\draw[gp path] (1.196,1.611)--(1.286,1.611);
\draw[gp path] (7.947,1.611)--(7.857,1.611);
\draw[gp path] (1.196,1.654)--(1.286,1.654);
\draw[gp path] (7.947,1.654)--(7.857,1.654);
\draw[gp path] (1.196,1.692)--(1.286,1.692);
\draw[gp path] (7.947,1.692)--(7.857,1.692);
\gpcolor{color=gp lt color axes}
\gpsetlinetype{gp lt axes}
\gpsetdashtype{gp dt axes}
\gpsetlinewidth{0.50}
\draw[gp path] (1.196,1.726)--(7.947,1.726);
\gpcolor{color=gp lt color border}
\gpsetlinetype{gp lt border}
\gpsetdashtype{gp dt solid}
\gpsetlinewidth{1.00}
\draw[gp path] (1.196,1.726)--(1.376,1.726);
\draw[gp path] (7.947,1.726)--(7.767,1.726);
\node[gp node right] at (1.012,1.726) {$10^{-4}$};
\draw[gp path] (1.196,1.949)--(1.286,1.949);
\draw[gp path] (7.947,1.949)--(7.857,1.949);
\draw[gp path] (1.196,2.080)--(1.286,2.080);
\draw[gp path] (7.947,2.080)--(7.857,2.080);
\draw[gp path] (1.196,2.172)--(1.286,2.172);
\draw[gp path] (7.947,2.172)--(7.857,2.172);
\draw[gp path] (1.196,2.244)--(1.286,2.244);
\draw[gp path] (7.947,2.244)--(7.857,2.244);
\draw[gp path] (1.196,2.303)--(1.286,2.303);
\draw[gp path] (7.947,2.303)--(7.857,2.303);
\draw[gp path] (1.196,2.353)--(1.286,2.353);
\draw[gp path] (7.947,2.353)--(7.857,2.353);
\draw[gp path] (1.196,2.396)--(1.286,2.396);
\draw[gp path] (7.947,2.396)--(7.857,2.396);
\draw[gp path] (1.196,2.433)--(1.286,2.433);
\draw[gp path] (7.947,2.433)--(7.857,2.433);
\gpcolor{color=gp lt color axes}
\gpsetlinetype{gp lt axes}
\gpsetdashtype{gp dt axes}
\gpsetlinewidth{0.50}
\draw[gp path] (1.196,2.467)--(7.947,2.467);
\gpcolor{color=gp lt color border}
\gpsetlinetype{gp lt border}
\gpsetdashtype{gp dt solid}
\gpsetlinewidth{1.00}
\draw[gp path] (1.196,2.467)--(1.376,2.467);
\draw[gp path] (7.947,2.467)--(7.767,2.467);
\node[gp node right] at (1.012,2.467) {$10^{-3}$};
\draw[gp path] (1.196,2.691)--(1.286,2.691);
\draw[gp path] (7.947,2.691)--(7.857,2.691);
\draw[gp path] (1.196,2.821)--(1.286,2.821);
\draw[gp path] (7.947,2.821)--(7.857,2.821);
\draw[gp path] (1.196,2.914)--(1.286,2.914);
\draw[gp path] (7.947,2.914)--(7.857,2.914);
\draw[gp path] (1.196,2.985)--(1.286,2.985);
\draw[gp path] (7.947,2.985)--(7.857,2.985);
\draw[gp path] (1.196,3.044)--(1.286,3.044);
\draw[gp path] (7.947,3.044)--(7.857,3.044);
\draw[gp path] (1.196,3.094)--(1.286,3.094);
\draw[gp path] (7.947,3.094)--(7.857,3.094);
\draw[gp path] (1.196,3.137)--(1.286,3.137);
\draw[gp path] (7.947,3.137)--(7.857,3.137);
\draw[gp path] (1.196,3.175)--(1.286,3.175);
\draw[gp path] (7.947,3.175)--(7.857,3.175);
\gpcolor{color=gp lt color axes}
\gpsetlinetype{gp lt axes}
\gpsetdashtype{gp dt axes}
\gpsetlinewidth{0.50}
\draw[gp path] (1.196,3.209)--(7.947,3.209);
\gpcolor{color=gp lt color border}
\gpsetlinetype{gp lt border}
\gpsetdashtype{gp dt solid}
\gpsetlinewidth{1.00}
\draw[gp path] (1.196,3.209)--(1.376,3.209);
\draw[gp path] (7.947,3.209)--(7.767,3.209);
\node[gp node right] at (1.012,3.209) {$10^{-2}$};
\draw[gp path] (1.196,3.432)--(1.286,3.432);
\draw[gp path] (7.947,3.432)--(7.857,3.432);
\draw[gp path] (1.196,3.562)--(1.286,3.562);
\draw[gp path] (7.947,3.562)--(7.857,3.562);
\draw[gp path] (1.196,3.655)--(1.286,3.655);
\draw[gp path] (7.947,3.655)--(7.857,3.655);
\draw[gp path] (1.196,3.727)--(1.286,3.727);
\draw[gp path] (7.947,3.727)--(7.857,3.727);
\draw[gp path] (1.196,3.785)--(1.286,3.785);
\draw[gp path] (7.947,3.785)--(7.857,3.785);
\draw[gp path] (1.196,3.835)--(1.286,3.835);
\draw[gp path] (7.947,3.835)--(7.857,3.835);
\draw[gp path] (1.196,3.878)--(1.286,3.878);
\draw[gp path] (7.947,3.878)--(7.857,3.878);
\draw[gp path] (1.196,3.916)--(1.286,3.916);
\draw[gp path] (7.947,3.916)--(7.857,3.916);
\gpcolor{color=gp lt color axes}
\gpsetlinetype{gp lt axes}
\gpsetdashtype{gp dt axes}
\gpsetlinewidth{0.50}
\draw[gp path] (1.196,3.950)--(3.719,3.950);
\draw[gp path] (7.763,3.950)--(7.947,3.950);
\gpcolor{color=gp lt color border}
\gpsetlinetype{gp lt border}
\gpsetdashtype{gp dt solid}
\gpsetlinewidth{1.00}
\draw[gp path] (1.196,3.950)--(1.376,3.950);
\draw[gp path] (7.947,3.950)--(7.767,3.950);
\node[gp node right] at (1.012,3.950) {$10^{-1}$};
\draw[gp path] (1.196,4.173)--(1.286,4.173);
\draw[gp path] (7.947,4.173)--(7.857,4.173);
\draw[gp path] (1.196,4.303)--(1.286,4.303);
\draw[gp path] (7.947,4.303)--(7.857,4.303);
\draw[gp path] (1.196,4.396)--(1.286,4.396);
\draw[gp path] (7.947,4.396)--(7.857,4.396);
\draw[gp path] (1.196,4.468)--(1.286,4.468);
\draw[gp path] (7.947,4.468)--(7.857,4.468);
\draw[gp path] (1.196,4.527)--(1.286,4.527);
\draw[gp path] (7.947,4.527)--(7.857,4.527);
\draw[gp path] (1.196,4.576)--(1.286,4.576);
\draw[gp path] (7.947,4.576)--(7.857,4.576);
\draw[gp path] (1.196,4.619)--(1.286,4.619);
\draw[gp path] (7.947,4.619)--(7.857,4.619);
\draw[gp path] (1.196,4.657)--(1.286,4.657);
\draw[gp path] (7.947,4.657)--(7.857,4.657);
\gpcolor{color=gp lt color axes}
\gpsetlinetype{gp lt axes}
\gpsetdashtype{gp dt axes}
\gpsetlinewidth{0.50}
\draw[gp path] (1.196,4.691)--(7.947,4.691);
\gpcolor{color=gp lt color border}
\gpsetlinetype{gp lt border}
\gpsetdashtype{gp dt solid}
\gpsetlinewidth{1.00}
\draw[gp path] (1.196,4.691)--(1.376,4.691);
\draw[gp path] (7.947,4.691)--(7.767,4.691);
\node[gp node right] at (1.012,4.691) {$10^{0}$};
\gpcolor{color=gp lt color axes}
\gpsetlinetype{gp lt axes}
\gpsetdashtype{gp dt axes}
\gpsetlinewidth{0.50}
\draw[gp path] (1.196,0.985)--(1.196,4.691);
\gpcolor{color=gp lt color border}
\gpsetlinetype{gp lt border}
\gpsetdashtype{gp dt solid}
\gpsetlinewidth{1.00}
\draw[gp path] (1.196,0.985)--(1.196,1.165);
\draw[gp path] (1.196,4.691)--(1.196,4.511);
\node[gp node center] at (1.196,0.677) {$0$};
\gpcolor{color=gp lt color axes}
\gpsetlinetype{gp lt axes}
\gpsetdashtype{gp dt axes}
\gpsetlinewidth{0.50}
\draw[gp path] (2.040,0.985)--(2.040,4.691);
\gpcolor{color=gp lt color border}
\gpsetlinetype{gp lt border}
\gpsetdashtype{gp dt solid}
\gpsetlinewidth{1.00}
\draw[gp path] (2.040,0.985)--(2.040,1.165);
\draw[gp path] (2.040,4.691)--(2.040,4.511);
\node[gp node center] at (2.040,0.677) {$50$};
\gpcolor{color=gp lt color axes}
\gpsetlinetype{gp lt axes}
\gpsetdashtype{gp dt axes}
\gpsetlinewidth{0.50}
\draw[gp path] (2.884,0.985)--(2.884,4.691);
\gpcolor{color=gp lt color border}
\gpsetlinetype{gp lt border}
\gpsetdashtype{gp dt solid}
\gpsetlinewidth{1.00}
\draw[gp path] (2.884,0.985)--(2.884,1.165);
\draw[gp path] (2.884,4.691)--(2.884,4.511);
\node[gp node center] at (2.884,0.677) {$100$};
\gpcolor{color=gp lt color axes}
\gpsetlinetype{gp lt axes}
\gpsetdashtype{gp dt axes}
\gpsetlinewidth{0.50}
\draw[gp path] (3.728,0.985)--(3.728,3.279);
\draw[gp path] (3.728,4.511)--(3.728,4.691);
\gpcolor{color=gp lt color border}
\gpsetlinetype{gp lt border}
\gpsetdashtype{gp dt solid}
\gpsetlinewidth{1.00}
\draw[gp path] (3.728,0.985)--(3.728,1.165);
\draw[gp path] (3.728,4.691)--(3.728,4.511);
\node[gp node center] at (3.728,0.677) {$150$};
\gpcolor{color=gp lt color axes}
\gpsetlinetype{gp lt axes}
\gpsetdashtype{gp dt axes}
\gpsetlinewidth{0.50}
\draw[gp path] (4.572,0.985)--(4.572,3.279);
\draw[gp path] (4.572,4.511)--(4.572,4.691);
\gpcolor{color=gp lt color border}
\gpsetlinetype{gp lt border}
\gpsetdashtype{gp dt solid}
\gpsetlinewidth{1.00}
\draw[gp path] (4.572,0.985)--(4.572,1.165);
\draw[gp path] (4.572,4.691)--(4.572,4.511);
\node[gp node center] at (4.572,0.677) {$200$};
\gpcolor{color=gp lt color axes}
\gpsetlinetype{gp lt axes}
\gpsetdashtype{gp dt axes}
\gpsetlinewidth{0.50}
\draw[gp path] (5.415,0.985)--(5.415,3.279);
\draw[gp path] (5.415,4.511)--(5.415,4.691);
\gpcolor{color=gp lt color border}
\gpsetlinetype{gp lt border}
\gpsetdashtype{gp dt solid}
\gpsetlinewidth{1.00}
\draw[gp path] (5.415,0.985)--(5.415,1.165);
\draw[gp path] (5.415,4.691)--(5.415,4.511);
\node[gp node center] at (5.415,0.677) {$250$};
\gpcolor{color=gp lt color axes}
\gpsetlinetype{gp lt axes}
\gpsetdashtype{gp dt axes}
\gpsetlinewidth{0.50}
\draw[gp path] (6.259,0.985)--(6.259,3.279);
\draw[gp path] (6.259,4.511)--(6.259,4.691);
\gpcolor{color=gp lt color border}
\gpsetlinetype{gp lt border}
\gpsetdashtype{gp dt solid}
\gpsetlinewidth{1.00}
\draw[gp path] (6.259,0.985)--(6.259,1.165);
\draw[gp path] (6.259,4.691)--(6.259,4.511);
\node[gp node center] at (6.259,0.677) {$300$};
\gpcolor{color=gp lt color axes}
\gpsetlinetype{gp lt axes}
\gpsetdashtype{gp dt axes}
\gpsetlinewidth{0.50}
\draw[gp path] (7.103,0.985)--(7.103,3.279);
\draw[gp path] (7.103,4.511)--(7.103,4.691);
\gpcolor{color=gp lt color border}
\gpsetlinetype{gp lt border}
\gpsetdashtype{gp dt solid}
\gpsetlinewidth{1.00}
\draw[gp path] (7.103,0.985)--(7.103,1.165);
\draw[gp path] (7.103,4.691)--(7.103,4.511);
\node[gp node center] at (7.103,0.677) {$350$};
\gpcolor{color=gp lt color axes}
\gpsetlinetype{gp lt axes}
\gpsetdashtype{gp dt axes}
\gpsetlinewidth{0.50}
\draw[gp path] (7.947,0.985)--(7.947,4.691);
\gpcolor{color=gp lt color border}
\gpsetlinetype{gp lt border}
\gpsetdashtype{gp dt solid}
\gpsetlinewidth{1.00}
\draw[gp path] (7.947,0.985)--(7.947,1.165);
\draw[gp path] (7.947,4.691)--(7.947,4.511);
\node[gp node center] at (7.947,0.677) {$400$};
\draw[gp path] (1.196,4.691)--(1.196,0.985)--(7.947,0.985)--(7.947,4.691)--cycle;
\node[gp node center] at (4.571,0.215) {iteration};
\node[gp node right] at (6.479,4.357) {$J$};
\gpcolor{rgb color={0.580,0.000,0.827}}
\draw[gp path] (6.663,4.357)--(7.579,4.357);
\draw[gp path] (1.196,4.108)--(1.213,4.105)--(1.230,4.104)--(1.247,4.104)--(1.264,4.103)%
  --(1.280,4.103)--(1.297,4.103)--(1.314,4.103)--(1.331,4.103)--(1.348,4.103)--(1.365,4.103)%
  --(1.382,4.103)--(1.399,4.103)--(1.415,4.102)--(1.432,4.102)--(1.449,4.100)--(1.466,4.093)%
  --(1.483,4.086)--(1.500,4.077)--(1.517,4.062)--(1.534,4.047)--(1.550,4.040)--(1.567,4.021)%
  --(1.584,4.012)--(1.601,3.992)--(1.618,3.983)--(1.635,3.964)--(1.652,3.953)--(1.669,3.936)%
  --(1.685,3.928)--(1.702,3.915)--(1.719,3.901)--(1.736,3.881)--(1.753,3.867)--(1.770,3.851)%
  --(1.787,3.838)--(1.804,3.828)--(1.820,3.812)--(1.837,3.800)--(1.854,3.786)--(1.871,3.774)%
  --(1.888,3.758)--(1.905,3.749)--(1.922,3.739)--(1.939,3.726)--(1.955,3.717)--(1.972,3.709)%
  --(1.989,3.699)--(2.006,3.691)--(2.023,3.680)--(2.040,3.672)--(2.057,3.658)--(2.074,3.652)%
  --(2.091,3.639)--(2.107,3.636)--(2.124,3.624)--(2.141,3.621)--(2.158,3.615)--(2.175,3.607)%
  --(2.192,3.603)--(2.209,3.590)--(2.226,3.586)--(2.242,3.580)--(2.259,3.573)--(2.276,3.566)%
  --(2.293,3.560)--(2.310,3.555)--(2.327,3.547)--(2.344,3.541)--(2.361,3.537)--(2.377,3.529)%
  --(2.394,3.525)--(2.411,3.521)--(2.428,3.516)--(2.445,3.511)--(2.462,3.508)--(2.479,3.500)%
  --(2.496,3.497)--(2.512,3.492)--(2.529,3.487)--(2.546,3.482)--(2.563,3.479)--(2.580,3.473)%
  --(2.597,3.467)--(2.614,3.459)--(2.631,3.454)--(2.647,3.446)--(2.664,3.443)--(2.681,3.435)%
  --(2.698,3.432)--(2.715,3.427)--(2.732,3.419)--(2.749,3.412)--(2.766,3.404)--(2.782,3.399)%
  --(2.799,3.391)--(2.816,3.385)--(2.833,3.380)--(2.850,3.373)--(2.867,3.367)--(2.884,3.360)%
  --(2.901,3.349)--(2.918,3.342)--(2.934,3.334)--(2.951,3.326)--(2.968,3.319)--(2.985,3.311)%
  --(3.002,3.305)--(3.019,3.300)--(3.036,3.294)--(3.053,3.290)--(3.069,3.285)--(3.086,3.279)%
  --(3.103,3.278)--(3.120,3.274)--(3.137,3.272)--(3.154,3.269)--(3.171,3.265)--(3.188,3.262)%
  --(3.204,3.259)--(3.221,3.258)--(3.238,3.255)--(3.255,3.253)--(3.272,3.250)--(3.289,3.245)%
  --(3.306,3.243)--(3.323,3.241)--(3.339,3.239)--(3.356,3.236)--(3.373,3.232)--(3.390,3.228)%
  --(3.407,3.225)--(3.424,3.220)--(3.441,3.217)--(3.458,3.214)--(3.474,3.209)--(3.491,3.205)%
  --(3.508,3.199)--(3.525,3.194)--(3.542,3.189)--(3.559,3.185)--(3.576,3.182)--(3.593,3.178)%
  --(3.609,3.174)--(3.626,3.170)--(3.643,3.167)--(3.660,3.165)--(3.677,3.163)--(3.694,3.162)%
  --(3.711,3.160)--(3.728,3.158)--(3.745,3.156)--(3.761,3.155)--(3.778,3.153)--(3.795,3.151)%
  --(3.812,3.149)--(3.829,3.147)--(3.846,3.145)--(3.863,3.143)--(3.880,3.141)--(3.896,3.140)%
  --(3.913,3.138)--(3.930,3.136)--(3.947,3.134)--(3.964,3.133)--(3.981,3.132)--(3.998,3.130)%
  --(4.015,3.129)--(4.031,3.128)--(4.048,3.126)--(4.065,3.125)--(4.082,3.124)--(4.099,3.123)%
  --(4.116,3.122)--(4.133,3.121)--(4.150,3.119)--(4.166,3.118)--(4.183,3.117)--(4.200,3.116)%
  --(4.217,3.115)--(4.234,3.113)--(4.251,3.112)--(4.268,3.110)--(4.285,3.110)--(4.301,3.108)%
  --(4.318,3.107)--(4.335,3.106)--(4.352,3.105)--(4.369,3.103)--(4.386,3.102)--(4.403,3.101)%
  --(4.420,3.100)--(4.436,3.098)--(4.453,3.097)--(4.470,3.096)--(4.487,3.095)--(4.504,3.094)%
  --(4.521,3.092)--(4.538,3.091)--(4.555,3.090)--(4.572,3.088)--(4.588,3.087)--(4.605,3.086)%
  --(4.622,3.085)--(4.639,3.084)--(4.656,3.083)--(4.673,3.083)--(4.690,3.082)--(4.707,3.081)%
  --(4.723,3.080)--(4.740,3.079)--(4.757,3.078)--(4.774,3.077)--(4.791,3.076)--(4.808,3.075)%
  --(4.825,3.073)--(4.842,3.072)--(4.858,3.072)--(4.875,3.071)--(4.892,3.069)--(4.909,3.068)%
  --(4.926,3.067)--(4.943,3.066)--(4.960,3.065)--(4.977,3.064)--(4.993,3.064)--(5.010,3.063)%
  --(5.027,3.062)--(5.044,3.061)--(5.061,3.061)--(5.078,3.060)--(5.095,3.059)--(5.112,3.058)%
  --(5.128,3.057)--(5.145,3.056)--(5.162,3.055)--(5.179,3.054)--(5.196,3.053)--(5.213,3.052)%
  --(5.230,3.051)--(5.247,3.050)--(5.263,3.049)--(5.280,3.048)--(5.297,3.046)--(5.314,3.045)%
  --(5.331,3.044)--(5.348,3.043)--(5.365,3.042)--(5.382,3.041)--(5.398,3.039)--(5.415,3.038)%
  --(5.432,3.038)--(5.449,3.036)--(5.466,3.035)--(5.483,3.034)--(5.500,3.033)--(5.517,3.032)%
  --(5.534,3.031)--(5.550,3.030)--(5.567,3.028)--(5.584,3.026)--(5.601,3.025)--(5.618,3.023)%
  --(5.635,3.021)--(5.652,3.018)--(5.669,3.017)--(5.685,3.015)--(5.702,3.013)--(5.719,3.011)%
  --(5.736,3.010)--(5.753,3.008)--(5.770,3.007)--(5.787,3.005)--(5.804,3.004)--(5.820,3.003)%
  --(5.837,3.001)--(5.854,2.999)--(5.871,2.998)--(5.888,2.997)--(5.905,2.995)--(5.922,2.994)%
  --(5.939,2.992)--(5.955,2.991)--(5.972,2.990)--(5.989,2.989)--(6.006,2.987)--(6.023,2.986)%
  --(6.040,2.985)--(6.057,2.984)--(6.074,2.982)--(6.090,2.982)--(6.107,2.981)--(6.124,2.979)%
  --(6.141,2.978)--(6.158,2.977)--(6.175,2.976)--(6.192,2.975)--(6.209,2.974)--(6.225,2.974)%
  --(6.242,2.973)--(6.259,2.972)--(6.276,2.972)--(6.293,2.971)--(6.310,2.970)--(6.327,2.969)%
  --(6.344,2.969)--(6.361,2.968)--(6.377,2.967)--(6.394,2.966)--(6.411,2.966)--(6.428,2.965)%
  --(6.445,2.964)--(6.462,2.964)--(6.479,2.963)--(6.496,2.962)--(6.512,2.962)--(6.529,2.961)%
  --(6.546,2.961)--(6.563,2.960)--(6.580,2.960)--(6.597,2.960)--(6.614,2.959)--(6.631,2.959)%
  --(6.647,2.959)--(6.664,2.958)--(6.681,2.958)--(6.698,2.957)--(6.715,2.957)--(6.732,2.957)%
  --(6.749,2.957)--(6.766,2.956)--(6.782,2.956)--(6.799,2.956)--(6.816,2.956)--(6.833,2.955)%
  --(6.850,2.955)--(6.867,2.955)--(6.884,2.955)--(6.901,2.954)--(6.917,2.954)--(6.934,2.954)%
  --(6.951,2.953)--(6.968,2.953)--(6.985,2.953)--(7.002,2.953)--(7.019,2.953)--(7.036,2.952)%
  --(7.052,2.952)--(7.069,2.952)--(7.086,2.952)--(7.103,2.951)--(7.120,2.951)--(7.137,2.951)%
  --(7.154,2.951)--(7.171,2.951)--(7.188,2.950)--(7.204,2.950)--(7.221,2.950)--(7.238,2.950)%
  --(7.255,2.950);
\gpcolor{color=gp lt color border}
\node[gp node right] at (6.479,4.049) {$\|\nabla J \|$};
\gpcolor{rgb color={0.000,0.620,0.451}}
\draw[gp path] (6.663,4.049)--(7.579,4.049);
\draw[gp path] (1.196,2.830)--(1.213,2.625)--(1.230,2.304)--(1.247,2.164)--(1.264,1.634)%
  --(1.280,1.578)--(1.297,1.744)--(1.314,1.922)--(1.331,2.226)--(1.348,2.333)--(1.365,2.457)%
  --(1.382,2.565)--(1.399,2.633)--(1.415,2.693)--(1.432,2.769)--(1.449,2.806)--(1.466,2.921)%
  --(1.483,2.971)--(1.500,3.008)--(1.517,2.976)--(1.534,3.041)--(1.550,3.159)--(1.567,2.972)%
  --(1.584,2.901)--(1.601,2.884)--(1.618,3.160)--(1.635,2.919)--(1.652,2.873)--(1.669,2.924)%
  --(1.685,2.992)--(1.702,2.893)--(1.719,2.866)--(1.736,3.015)--(1.753,3.097)--(1.770,2.934)%
  --(1.787,2.887)--(1.804,2.861)--(1.820,2.872)--(1.837,3.018)--(1.854,2.866)--(1.871,2.811)%
  --(1.888,2.927)--(1.905,2.994)--(1.922,2.823)--(1.939,2.791)--(1.955,2.827)--(1.972,2.959)%
  --(1.989,2.734)--(2.006,2.790)--(2.023,2.835)--(2.040,3.060)--(2.057,2.753)--(2.074,2.729)%
  --(2.091,2.783)--(2.107,3.050)--(2.124,2.648)--(2.141,2.606)--(2.158,2.724)--(2.175,2.747)%
  --(2.192,3.034)--(2.209,2.643)--(2.226,2.591)--(2.242,2.665)--(2.259,2.913)--(2.276,2.667)%
  --(2.293,2.584)--(2.310,2.644)--(2.327,2.837)--(2.344,2.610)--(2.361,2.544)--(2.377,2.686)%
  --(2.394,2.764)--(2.411,2.524)--(2.428,2.530)--(2.445,2.613)--(2.462,2.914)--(2.479,2.487)%
  --(2.496,2.484)--(2.512,2.573)--(2.529,2.788)--(2.546,2.494)--(2.563,2.484)--(2.580,2.606)%
  --(2.597,2.770)--(2.614,2.545)--(2.631,2.508)--(2.647,2.615)--(2.664,2.852)--(2.681,2.490)%
  --(2.698,2.431)--(2.715,2.519)--(2.732,2.601)--(2.749,2.738)--(2.766,2.474)--(2.782,2.419)%
  --(2.799,2.626)--(2.816,2.580)--(2.833,2.454)--(2.850,2.475)--(2.867,2.700)--(2.884,2.548)%
  --(2.901,2.474)--(2.918,2.561)--(2.934,2.542)--(2.951,2.490)--(2.968,2.469)--(2.985,2.589)%
  --(3.002,2.572)--(3.019,2.386)--(3.036,2.397)--(3.053,2.544)--(3.069,2.381)--(3.086,2.280)%
  --(3.103,2.650)--(3.120,2.288)--(3.137,2.233)--(3.154,2.364)--(3.171,2.465)--(3.188,2.467)%
  --(3.204,2.167)--(3.221,2.206)--(3.238,2.384)--(3.255,2.358)--(3.272,2.261)--(3.289,2.487)%
  --(3.306,2.565)--(3.323,2.231)--(3.339,2.181)--(3.356,2.369)--(3.373,2.564)--(3.390,2.324)%
  --(3.407,2.205)--(3.424,2.426)--(3.441,2.508)--(3.458,2.305)--(3.474,2.308)--(3.491,2.390)%
  --(3.508,2.420)--(3.525,2.428)--(3.542,2.367)--(3.559,2.271)--(3.576,2.316)--(3.593,2.298)%
  --(3.609,2.307)--(3.626,2.353)--(3.643,2.286)--(3.660,2.155)--(3.677,2.289)--(3.694,2.084)%
  --(3.711,2.125)--(3.728,2.481)--(3.745,2.090)--(3.761,2.105)--(3.778,2.244)--(3.795,2.444)%
  --(3.812,2.150)--(3.829,2.101)--(3.846,2.164)--(3.863,2.499)--(3.880,2.119)--(3.896,1.979)%
  --(3.913,2.117)--(3.930,2.242)--(3.947,2.305)--(3.964,2.031)--(3.981,2.051)--(3.998,2.123)%
  --(4.015,2.138)--(4.031,2.159)--(4.048,2.055)--(4.065,2.012)--(4.082,2.190)--(4.099,2.174)%
  --(4.116,1.955)--(4.133,2.016)--(4.150,2.298)--(4.166,2.009)--(4.183,1.905)--(4.200,2.152)%
  --(4.217,2.259)--(4.234,2.000)--(4.251,2.078)--(4.268,2.084)--(4.285,2.362)--(4.301,1.953)%
  --(4.318,1.926)--(4.335,1.992)--(4.352,2.380)--(4.369,2.070)--(4.386,1.970)--(4.403,2.084)%
  --(4.420,2.352)--(4.436,2.030)--(4.453,1.932)--(4.470,2.048)--(4.487,2.356)--(4.504,2.062)%
  --(4.521,1.974)--(4.538,2.058)--(4.555,2.359)--(4.572,2.041)--(4.588,1.852)--(4.605,2.013)%
  --(4.622,2.173)--(4.639,1.914)--(4.656,1.954)--(4.673,2.126)--(4.690,1.876)--(4.707,1.925)%
  --(4.723,1.982)--(4.740,2.347)--(4.757,1.945)--(4.774,1.845)--(4.791,1.969)--(4.808,2.311)%
  --(4.825,1.955)--(4.842,1.813)--(4.858,1.903)--(4.875,2.063)--(4.892,1.929)--(4.909,1.880)%
  --(4.926,2.167)--(4.943,2.244)--(4.960,1.856)--(4.977,1.828)--(4.993,1.934)--(5.010,2.279)%
  --(5.027,1.875)--(5.044,1.785)--(5.061,1.888)--(5.078,2.286)--(5.095,1.990)--(5.112,1.917)%
  --(5.128,2.030)--(5.145,2.384)--(5.162,1.980)--(5.179,1.725)--(5.196,1.894)--(5.213,2.016)%
  --(5.230,2.306)--(5.247,1.896)--(5.263,1.855)--(5.280,1.956)--(5.297,2.263)--(5.314,2.023)%
  --(5.331,1.826)--(5.348,1.930)--(5.365,2.167)--(5.382,1.909)--(5.398,1.885)--(5.415,1.878)%
  --(5.432,2.283)--(5.449,1.783)--(5.466,1.730)--(5.483,2.120)--(5.500,2.223)--(5.517,1.892)%
  --(5.534,1.858)--(5.550,1.984)--(5.567,2.406)--(5.584,1.848)--(5.601,1.857)--(5.618,1.984)%
  --(5.635,2.366)--(5.652,1.994)--(5.669,1.931)--(5.685,2.015)--(5.702,2.334)--(5.719,1.909)%
  --(5.736,1.890)--(5.753,2.000)--(5.770,2.319)--(5.787,1.837)--(5.804,1.783)--(5.820,1.850)%
  --(5.837,2.136)--(5.854,2.085)--(5.871,1.722)--(5.888,1.852)--(5.905,1.967)--(5.922,2.126)%
  --(5.939,1.758)--(5.955,1.891)--(5.972,2.031)--(5.989,2.299)--(6.006,1.724)--(6.023,1.891)%
  --(6.040,2.010)--(6.057,2.372)--(6.074,1.776)--(6.090,1.664)--(6.107,1.874)--(6.124,1.921)%
  --(6.141,2.073)--(6.158,1.692)--(6.175,1.817)--(6.192,2.003)--(6.209,2.154)--(6.225,1.765)%
  --(6.242,1.844)--(6.259,1.864)--(6.276,2.328)--(6.293,1.674)--(6.310,1.650)--(6.327,1.889)%
  --(6.344,2.141)--(6.361,1.839)--(6.377,1.655)--(6.394,1.914)--(6.411,1.999)--(6.428,1.724)%
  --(6.445,1.803)--(6.462,1.738)--(6.479,2.303)--(6.496,1.679)--(6.512,1.565)--(6.529,1.763)%
  --(6.546,2.130)--(6.563,1.845)--(6.580,1.554)--(6.597,1.634)--(6.614,2.032)--(6.631,1.491)%
  --(6.647,1.549)--(6.664,1.718)--(6.681,1.778)--(6.698,1.529)--(6.715,1.408)--(6.732,1.544)%
  --(6.749,1.919)--(6.766,1.597)--(6.782,1.451)--(6.799,1.633)--(6.816,2.003)--(6.833,1.570)%
  --(6.850,1.568)--(6.867,1.662)--(6.884,2.035)--(6.901,1.531)--(6.917,1.494)--(6.934,1.651)%
  --(6.951,1.848)--(6.968,1.809)--(6.985,1.419)--(7.002,1.578)--(7.019,1.721)--(7.036,1.982)%
  --(7.052,1.496)--(7.069,1.422)--(7.086,1.577)--(7.103,1.874)--(7.120,1.512)--(7.137,1.436)%
  --(7.154,1.595)--(7.171,1.948)--(7.188,1.442)--(7.204,1.435)--(7.221,1.568)--(7.238,1.859)%
  --(7.255,1.666);
\gpcolor{color=gp lt color border}
\node[gp node right] at (6.479,3.741) {Tikhonov regul.};
\gpcolor{rgb color={0.337,0.706,0.914}}
\draw[gp path] (6.663,3.741)--(7.579,3.741);
\draw[gp path] (1.417,0.985)--(1.432,1.065)--(1.449,1.108)--(1.466,1.213)--(1.483,1.216)%
  --(1.500,1.344)--(1.517,1.378)--(1.534,1.384)--(1.550,1.410)--(1.567,1.365)--(1.584,1.406)%
  --(1.601,1.477)--(1.618,1.563)--(1.635,1.549)--(1.652,1.534)--(1.669,1.557)--(1.685,1.573)%
  --(1.702,1.573)--(1.719,1.585)--(1.736,1.642)--(1.753,1.686)--(1.770,1.675)--(1.787,1.684)%
  --(1.804,1.707)--(1.820,1.746)--(1.837,1.766)--(1.854,1.788)--(1.871,1.790)--(1.888,1.802)%
  --(1.905,1.815)--(1.922,1.815)--(1.939,1.828)--(1.955,1.845)--(1.972,1.874)--(1.989,1.874)%
  --(2.006,1.875)--(2.023,1.884)--(2.040,1.901)--(2.057,1.914)--(2.074,1.919)--(2.091,1.936)%
  --(2.107,1.959)--(2.124,1.961)--(2.141,1.961)--(2.158,1.965)--(2.175,1.971)--(2.192,1.989)%
  --(2.209,1.989)--(2.226,1.991)--(2.242,1.998)--(2.259,2.008)--(2.276,2.014)--(2.293,2.019)%
  --(2.310,2.023)--(2.327,2.032)--(2.344,2.037)--(2.361,2.035)--(2.377,2.034)--(2.394,2.036)%
  --(2.411,2.038)--(2.428,2.043)--(2.445,2.049)--(2.462,2.060)--(2.479,2.062)--(2.496,2.061)%
  --(2.512,2.063)--(2.529,2.064)--(2.546,2.067)--(2.563,2.068)--(2.580,2.071)--(2.597,2.076)%
  --(2.614,2.080)--(2.631,2.083)--(2.647,2.087)--(2.664,2.094)--(2.681,2.092)--(2.698,2.092)%
  --(2.715,2.094)--(2.732,2.097)--(2.749,2.103)--(2.766,2.108)--(2.782,2.110)--(2.799,2.114)%
  --(2.816,2.118)--(2.833,2.119)--(2.850,2.122)--(2.867,2.127)--(2.884,2.130)--(2.901,2.136)%
  --(2.918,2.141)--(2.934,2.149)--(2.951,2.152)--(2.968,2.153)--(2.985,2.156)--(3.002,2.162)%
  --(3.019,2.162)--(3.036,2.164)--(3.053,2.167)--(3.069,2.170)--(3.086,2.173)--(3.103,2.175)%
  --(3.120,2.175)--(3.137,2.175)--(3.154,2.175)--(3.171,2.175)--(3.188,2.176)--(3.204,2.177)%
  --(3.221,2.177)--(3.238,2.178)--(3.255,2.179)--(3.272,2.180)--(3.289,2.182)--(3.306,2.183)%
  --(3.323,2.183)--(3.339,2.183)--(3.356,2.185)--(3.373,2.188)--(3.390,2.193)--(3.407,2.195)%
  --(3.424,2.199)--(3.441,2.204)--(3.458,2.205)--(3.474,2.209)--(3.491,2.214)--(3.508,2.226)%
  --(3.525,2.230)--(3.542,2.233)--(3.559,2.234)--(3.576,2.238)--(3.593,2.241)--(3.609,2.245)%
  --(3.626,2.250)--(3.643,2.252)--(3.660,2.253)--(3.677,2.255)--(3.694,2.256)--(3.711,2.257)%
  --(3.728,2.260)--(3.745,2.261)--(3.761,2.262)--(3.778,2.263)--(3.795,2.266)--(3.812,2.268)%
  --(3.829,2.269)--(3.846,2.271)--(3.863,2.274)--(3.880,2.275)--(3.896,2.275)--(3.913,2.276)%
  --(3.930,2.278)--(3.947,2.279)--(3.964,2.280)--(3.981,2.281)--(3.998,2.282)--(4.015,2.282)%
  --(4.031,2.283)--(4.048,2.283)--(4.065,2.284)--(4.082,2.285)--(4.099,2.286)--(4.116,2.286)%
  --(4.133,2.288)--(4.150,2.290)--(4.166,2.292)--(4.183,2.293)--(4.200,2.295)--(4.217,2.299)%
  --(4.234,2.299)--(4.251,2.301)--(4.268,2.303)--(4.285,2.308)--(4.301,2.308)--(4.318,2.308)%
  --(4.335,2.310)--(4.352,2.313)--(4.369,2.315)--(4.386,2.316)--(4.403,2.318)--(4.420,2.323)%
  --(4.436,2.325)--(4.453,2.326)--(4.470,2.326)--(4.487,2.329)--(4.504,2.330)--(4.521,2.332)%
  --(4.538,2.334)--(4.555,2.341)--(4.572,2.344)--(4.588,2.344)--(4.605,2.344)--(4.622,2.345)%
  --(4.639,2.348)--(4.656,2.350)--(4.673,2.353)--(4.690,2.354)--(4.707,2.355)--(4.723,2.357)%
  --(4.740,2.361)--(4.757,2.361)--(4.774,2.362)--(4.791,2.364)--(4.808,2.368)--(4.825,2.370)%
  --(4.842,2.372)--(4.858,2.373)--(4.875,2.377)--(4.892,2.379)--(4.909,2.381)--(4.926,2.385)%
  --(4.943,2.388)--(4.960,2.387)--(4.977,2.387)--(4.993,2.388)--(5.010,2.391)--(5.027,2.392)%
  --(5.044,2.392)--(5.061,2.393)--(5.078,2.396)--(5.095,2.396)--(5.112,2.397)--(5.128,2.399)%
  --(5.145,2.401)--(5.162,2.401)--(5.179,2.401)--(5.196,2.402)--(5.213,2.402)--(5.230,2.405)%
  --(5.247,2.405)--(5.263,2.405)--(5.280,2.406)--(5.297,2.409)--(5.314,2.412)--(5.331,2.412)%
  --(5.348,2.412)--(5.365,2.414)--(5.382,2.415)--(5.398,2.416)--(5.415,2.417)--(5.432,2.420)%
  --(5.449,2.419)--(5.466,2.419)--(5.483,2.418)--(5.500,2.419)--(5.517,2.420)--(5.534,2.421)%
  --(5.550,2.422)--(5.567,2.424)--(5.584,2.424)--(5.601,2.424)--(5.618,2.424)--(5.635,2.426)%
  --(5.652,2.429)--(5.669,2.430)--(5.685,2.432)--(5.702,2.435)--(5.719,2.435)--(5.736,2.435)%
  --(5.753,2.436)--(5.770,2.438)--(5.787,2.438)--(5.804,2.438)--(5.820,2.439)--(5.837,2.439)%
  --(5.854,2.440)--(5.871,2.439)--(5.888,2.439)--(5.905,2.439)--(5.922,2.440)--(5.939,2.440)%
  --(5.955,2.440)--(5.972,2.440)--(5.989,2.442)--(6.006,2.442)--(6.023,2.442)--(6.040,2.444)%
  --(6.057,2.447)--(6.074,2.448)--(6.090,2.448)--(6.107,2.449)--(6.124,2.452)--(6.141,2.457)%
  --(6.158,2.457)--(6.175,2.457)--(6.192,2.459)--(6.209,2.464)--(6.225,2.465)--(6.242,2.466)%
  --(6.259,2.468)--(6.276,2.473)--(6.293,2.473)--(6.310,2.473)--(6.327,2.474)--(6.344,2.476)%
  --(6.361,2.479)--(6.377,2.480)--(6.394,2.482)--(6.411,2.484)--(6.428,2.484)--(6.445,2.484)%
  --(6.462,2.485)--(6.479,2.486)--(6.496,2.486)--(6.512,2.486)--(6.529,2.485)--(6.546,2.485)%
  --(6.563,2.485)--(6.580,2.486)--(6.597,2.486)--(6.614,2.486)--(6.631,2.486)--(6.647,2.485)%
  --(6.664,2.485)--(6.681,2.484)--(6.698,2.484)--(6.715,2.484)--(6.732,2.484)--(6.749,2.484)%
  --(6.766,2.484)--(6.782,2.485)--(6.799,2.485)--(6.816,2.485)--(6.833,2.485)--(6.850,2.485)%
  --(6.867,2.486)--(6.884,2.486)--(6.901,2.487)--(6.917,2.487)--(6.934,2.487)--(6.951,2.488)%
  --(6.968,2.489)--(6.985,2.489)--(7.002,2.489)--(7.019,2.489)--(7.036,2.490)--(7.052,2.491)%
  --(7.069,2.491)--(7.086,2.492)--(7.103,2.493)--(7.120,2.494)--(7.137,2.494)--(7.154,2.494)%
  --(7.171,2.495)--(7.188,2.495)--(7.204,2.495)--(7.221,2.495)--(7.238,2.495)--(7.255,2.496);
\gpcolor{color=gp lt color border}
\node[gp node right] at (6.479,3.433) {Penalty term};
\gpcolor{rgb color={0.902,0.624,0.000}}
\draw[gp path] (6.663,3.433)--(7.579,3.433);
\draw[gp path] (1.196,3.157)--(1.213,3.154)--(1.230,3.152)--(1.247,3.152)--(1.264,3.152)%
  --(1.280,3.152)--(1.297,3.152)--(1.314,3.152)--(1.331,3.151)--(1.348,3.151)--(1.365,3.151)%
  --(1.382,3.151)--(1.399,3.151)--(1.415,3.150)--(1.432,3.150)--(1.449,3.149)--(1.466,3.144)%
  --(1.483,3.137)--(1.500,3.131)--(1.517,3.116)--(1.534,3.097)--(1.550,3.090)--(1.567,3.074)%
  --(1.584,3.066)--(1.601,3.048)--(1.618,3.049)--(1.635,3.025)--(1.652,3.013)--(1.669,2.994)%
  --(1.685,2.990)--(1.702,2.976)--(1.719,2.965)--(1.736,2.942)--(1.753,2.932)--(1.770,2.917)%
  --(1.787,2.904)--(1.804,2.894)--(1.820,2.879)--(1.837,2.870)--(1.854,2.857)--(1.871,2.843)%
  --(1.888,2.829)--(1.905,2.817)--(1.922,2.809)--(1.939,2.798)--(1.955,2.787)--(1.972,2.782)%
  --(1.989,2.771)--(2.006,2.763)--(2.023,2.752)--(2.040,2.748)--(2.057,2.738)--(2.074,2.731)%
  --(2.091,2.721)--(2.107,2.712)--(2.124,2.710)--(2.141,2.709)--(2.158,2.706)--(2.175,2.700)%
  --(2.192,2.696)--(2.209,2.685)--(2.226,2.681)--(2.242,2.676)--(2.259,2.669)--(2.276,2.663)%
  --(2.293,2.657)--(2.310,2.650)--(2.327,2.639)--(2.344,2.633)--(2.361,2.629)--(2.377,2.623)%
  --(2.394,2.611)--(2.411,2.610)--(2.428,2.607)--(2.445,2.603)--(2.462,2.601)--(2.479,2.595)%
  --(2.496,2.592)--(2.512,2.590)--(2.529,2.586)--(2.546,2.586)--(2.563,2.584)--(2.580,2.579)%
  --(2.597,2.579)--(2.614,2.574)--(2.631,2.569)--(2.647,2.563)--(2.664,2.558)--(2.681,2.554)%
  --(2.698,2.551)--(2.715,2.547)--(2.732,2.539)--(2.749,2.533)--(2.766,2.528)--(2.782,2.523)%
  --(2.799,2.519)--(2.816,2.512)--(2.833,2.508)--(2.850,2.503)--(2.867,2.496)--(2.884,2.491)%
  --(2.901,2.482)--(2.918,2.475)--(2.934,2.469)--(2.951,2.460)--(2.968,2.454)--(2.985,2.450)%
  --(3.002,2.447)--(3.019,2.443)--(3.036,2.438)--(3.053,2.435)--(3.069,2.431)--(3.086,2.427)%
  --(3.103,2.424)--(3.120,2.422)--(3.137,2.421)--(3.154,2.419)--(3.171,2.415)--(3.188,2.414)%
  --(3.204,2.411)--(3.221,2.410)--(3.238,2.408)--(3.255,2.407)--(3.272,2.404)--(3.289,2.400)%
  --(3.306,2.400)--(3.323,2.397)--(3.339,2.395)--(3.356,2.391)--(3.373,2.389)--(3.390,2.383)%
  --(3.407,2.381)--(3.424,2.379)--(3.441,2.375)--(3.458,2.374)--(3.474,2.372)--(3.491,2.369)%
  --(3.508,2.369)--(3.525,2.359)--(3.542,2.357)--(3.559,2.352)--(3.576,2.351)--(3.593,2.348)%
  --(3.609,2.343)--(3.626,2.335)--(3.643,2.335)--(3.660,2.334)--(3.677,2.328)--(3.694,2.327)%
  --(3.711,2.325)--(3.728,2.322)--(3.745,2.320)--(3.761,2.318)--(3.778,2.314)--(3.795,2.315)%
  --(3.812,2.309)--(3.829,2.306)--(3.846,2.304)--(3.863,2.301)--(3.880,2.300)--(3.896,2.300)%
  --(3.913,2.299)--(3.930,2.299)--(3.947,2.296)--(3.964,2.296)--(3.981,2.296)--(3.998,2.297)%
  --(4.015,2.297)--(4.031,2.296)--(4.048,2.296)--(4.065,2.295)--(4.082,2.296)--(4.099,2.296)%
  --(4.116,2.295)--(4.133,2.295)--(4.150,2.296)--(4.166,2.296)--(4.183,2.296)--(4.200,2.296)%
  --(4.217,2.297)--(4.234,2.296)--(4.251,2.294)--(4.268,2.295)--(4.285,2.288)--(4.301,2.293)%
  --(4.318,2.294)--(4.335,2.294)--(4.352,2.293)--(4.369,2.293)--(4.386,2.293)--(4.403,2.292)%
  --(4.420,2.293)--(4.436,2.293)--(4.453,2.293)--(4.470,2.292)--(4.487,2.293)--(4.504,2.292)%
  --(4.521,2.293)--(4.538,2.294)--(4.555,2.295)--(4.572,2.298)--(4.588,2.296)--(4.605,2.294)%
  --(4.622,2.293)--(4.639,2.293)--(4.656,2.293)--(4.673,2.294)--(4.690,2.293)--(4.707,2.291)%
  --(4.723,2.290)--(4.740,2.289)--(4.757,2.287)--(4.774,2.286)--(4.791,2.284)--(4.808,2.283)%
  --(4.825,2.280)--(4.842,2.280)--(4.858,2.278)--(4.875,2.278)--(4.892,2.276)--(4.909,2.274)%
  --(4.926,2.272)--(4.943,2.271)--(4.960,2.271)--(4.977,2.270)--(4.993,2.269)--(5.010,2.268)%
  --(5.027,2.266)--(5.044,2.265)--(5.061,2.264)--(5.078,2.262)--(5.095,2.261)--(5.112,2.260)%
  --(5.128,2.258)--(5.145,2.257)--(5.162,2.255)--(5.179,2.254)--(5.196,2.253)--(5.213,2.251)%
  --(5.230,2.248)--(5.247,2.248)--(5.263,2.247)--(5.280,2.245)--(5.297,2.242)--(5.314,2.239)%
  --(5.331,2.238)--(5.348,2.237)--(5.365,2.234)--(5.382,2.233)--(5.398,2.231)--(5.415,2.229)%
  --(5.432,2.228)--(5.449,2.225)--(5.466,2.225)--(5.483,2.223)--(5.500,2.219)--(5.517,2.219)%
  --(5.534,2.217)--(5.550,2.215)--(5.567,2.212)--(5.584,2.210)--(5.601,2.209)--(5.618,2.206)%
  --(5.635,2.203)--(5.652,2.198)--(5.669,2.196)--(5.685,2.194)--(5.702,2.191)--(5.719,2.189)%
  --(5.736,2.187)--(5.753,2.186)--(5.770,2.181)--(5.787,2.181)--(5.804,2.181)--(5.820,2.179)%
  --(5.837,2.177)--(5.854,2.174)--(5.871,2.173)--(5.888,2.171)--(5.905,2.170)--(5.922,2.166)%
  --(5.939,2.166)--(5.955,2.165)--(5.972,2.163)--(5.989,2.160)--(6.006,2.159)--(6.023,2.158)%
  --(6.040,2.157)--(6.057,2.155)--(6.074,2.154)--(6.090,2.153)--(6.107,2.152)--(6.124,2.150)%
  --(6.141,2.148)--(6.158,2.147)--(6.175,2.146)--(6.192,2.144)--(6.209,2.143)--(6.225,2.142)%
  --(6.242,2.141)--(6.259,2.141)--(6.276,2.140)--(6.293,2.138)--(6.310,2.138)--(6.327,2.137)%
  --(6.344,2.137)--(6.361,2.135)--(6.377,2.134)--(6.394,2.133)--(6.411,2.133)--(6.428,2.132)%
  --(6.445,2.131)--(6.462,2.130)--(6.479,2.130)--(6.496,2.129)--(6.512,2.129)--(6.529,2.128)%
  --(6.546,2.128)--(6.563,2.127)--(6.580,2.127)--(6.597,2.126)--(6.614,2.125)--(6.631,2.125)%
  --(6.647,2.125)--(6.664,2.125)--(6.681,2.123)--(6.698,2.123)--(6.715,2.123)--(6.732,2.123)%
  --(6.749,2.122)--(6.766,2.121)--(6.782,2.121)--(6.799,2.120)--(6.816,2.120)--(6.833,2.119)%
  --(6.850,2.119)--(6.867,2.119)--(6.884,2.118)--(6.901,2.117)--(6.917,2.117)--(6.934,2.116)%
  --(6.951,2.116)--(6.968,2.114)--(6.985,2.114)--(7.002,2.114)--(7.019,2.114)--(7.036,2.112)%
  --(7.052,2.112)--(7.069,2.112)--(7.086,2.111)--(7.103,2.110)--(7.120,2.109)--(7.137,2.109)%
  --(7.154,2.108)--(7.171,2.108)--(7.188,2.107)--(7.204,2.107)--(7.221,2.106)--(7.238,2.106)%
  --(7.255,2.105);
\gpcolor{color=gp lt color border}
\draw[gp path] (1.196,4.691)--(1.196,0.985)--(7.947,0.985)--(7.947,4.691)--cycle;
\gpdefrectangularnode{gp plot 1}{\pgfpoint{1.196cm}{0.985cm}}{\pgfpoint{7.947cm}{4.691cm}}
\end{tikzpicture}

%% file: figures/cnot_gate/optim_objective.tex
\begin{tikzpicture}[gnuplot]
\tikzset{every node/.append style={font={\footnotesize}}}
\path (0.000,0.000) rectangle (8.500,6.800);
\gpcolor{color=gp lt color axes}
\gpsetlinetype{gp lt axes}
\gpsetdashtype{gp dt axes}
\gpsetlinewidth{0.50}
\draw[gp path] (1.504,0.985)--(7.947,0.985);
\gpcolor{color=gp lt color border}
\gpsetlinetype{gp lt border}
\gpsetdashtype{gp dt solid}
\gpsetlinewidth{1.00}
\draw[gp path] (1.504,0.985)--(1.684,0.985);
\draw[gp path] (7.947,0.985)--(7.767,0.985);
\node[gp node right] at (1.320,0.985) {$10^{-3}$};
\draw[gp path] (1.504,1.352)--(1.594,1.352);
\draw[gp path] (7.947,1.352)--(7.857,1.352);
\draw[gp path] (1.504,1.567)--(1.594,1.567);
\draw[gp path] (7.947,1.567)--(7.857,1.567);
\draw[gp path] (1.504,1.719)--(1.594,1.719);
\draw[gp path] (7.947,1.719)--(7.857,1.719);
\draw[gp path] (1.504,1.837)--(1.594,1.837);
\draw[gp path] (7.947,1.837)--(7.857,1.837);
\draw[gp path] (1.504,1.934)--(1.594,1.934);
\draw[gp path] (7.947,1.934)--(7.857,1.934);
\draw[gp path] (1.504,2.015)--(1.594,2.015);
\draw[gp path] (7.947,2.015)--(7.857,2.015);
\draw[gp path] (1.504,2.086)--(1.594,2.086);
\draw[gp path] (7.947,2.086)--(7.857,2.086);
\draw[gp path] (1.504,2.149)--(1.594,2.149);
\draw[gp path] (7.947,2.149)--(7.857,2.149);
\gpcolor{color=gp lt color axes}
\gpsetlinetype{gp lt axes}
\gpsetdashtype{gp dt axes}
\gpsetlinewidth{0.50}
\draw[gp path] (1.504,2.204)--(7.947,2.204);
\gpcolor{color=gp lt color border}
\gpsetlinetype{gp lt border}
\gpsetdashtype{gp dt solid}
\gpsetlinewidth{1.00}
\draw[gp path] (1.504,2.204)--(1.684,2.204);
\draw[gp path] (7.947,2.204)--(7.767,2.204);
\node[gp node right] at (1.320,2.204) {$10^{-2}$};
\draw[gp path] (1.504,2.571)--(1.594,2.571);
\draw[gp path] (7.947,2.571)--(7.857,2.571);
\draw[gp path] (1.504,2.786)--(1.594,2.786);
\draw[gp path] (7.947,2.786)--(7.857,2.786);
\draw[gp path] (1.504,2.938)--(1.594,2.938);
\draw[gp path] (7.947,2.938)--(7.857,2.938);
\draw[gp path] (1.504,3.057)--(1.594,3.057);
\draw[gp path] (7.947,3.057)--(7.857,3.057);
\draw[gp path] (1.504,3.153)--(1.594,3.153);
\draw[gp path] (7.947,3.153)--(7.857,3.153);
\draw[gp path] (1.504,3.235)--(1.594,3.235);
\draw[gp path] (7.947,3.235)--(7.857,3.235);
\draw[gp path] (1.504,3.306)--(1.594,3.306);
\draw[gp path] (7.947,3.306)--(7.857,3.306);
\draw[gp path] (1.504,3.368)--(1.594,3.368);
\draw[gp path] (7.947,3.368)--(7.857,3.368);
\gpcolor{color=gp lt color axes}
\gpsetlinetype{gp lt axes}
\gpsetdashtype{gp dt axes}
\gpsetlinewidth{0.50}
\draw[gp path] (1.504,3.424)--(7.947,3.424);
\gpcolor{color=gp lt color border}
\gpsetlinetype{gp lt border}
\gpsetdashtype{gp dt solid}
\gpsetlinewidth{1.00}
\draw[gp path] (1.504,3.424)--(1.684,3.424);
\draw[gp path] (7.947,3.424)--(7.767,3.424);
\node[gp node right] at (1.320,3.424) {$10^{-1}$};
\draw[gp path] (1.504,3.791)--(1.594,3.791);
\draw[gp path] (7.947,3.791)--(7.857,3.791);
\draw[gp path] (1.504,4.005)--(1.594,4.005);
\draw[gp path] (7.947,4.005)--(7.857,4.005);
\draw[gp path] (1.504,4.158)--(1.594,4.158);
\draw[gp path] (7.947,4.158)--(7.857,4.158);
\draw[gp path] (1.504,4.276)--(1.594,4.276);
\draw[gp path] (7.947,4.276)--(7.857,4.276);
\draw[gp path] (1.504,4.372)--(1.594,4.372);
\draw[gp path] (7.947,4.372)--(7.857,4.372);
\draw[gp path] (1.504,4.454)--(1.594,4.454);
\draw[gp path] (7.947,4.454)--(7.857,4.454);
\draw[gp path] (1.504,4.525)--(1.594,4.525);
\draw[gp path] (7.947,4.525)--(7.857,4.525);
\draw[gp path] (1.504,4.587)--(1.594,4.587);
\draw[gp path] (7.947,4.587)--(7.857,4.587);
\gpcolor{color=gp lt color axes}
\gpsetlinetype{gp lt axes}
\gpsetdashtype{gp dt axes}
\gpsetlinewidth{0.50}
\draw[gp path] (1.504,4.643)--(7.947,4.643);
\gpcolor{color=gp lt color border}
\gpsetlinetype{gp lt border}
\gpsetdashtype{gp dt solid}
\gpsetlinewidth{1.00}
\draw[gp path] (1.504,4.643)--(1.684,4.643);
\draw[gp path] (7.947,4.643)--(7.767,4.643);
\node[gp node right] at (1.320,4.643) {$10^{0}$};
\gpcolor{color=gp lt color axes}
\gpsetlinetype{gp lt axes}
\gpsetdashtype{gp dt axes}
\gpsetlinewidth{0.50}
\draw[gp path] (1.504,0.985)--(1.504,4.643);
\gpcolor{color=gp lt color border}
\gpsetlinetype{gp lt border}
\gpsetdashtype{gp dt solid}
\gpsetlinewidth{1.00}
\draw[gp path] (1.504,0.985)--(1.504,1.165);
\draw[gp path] (1.504,4.643)--(1.504,4.463);
\node[gp node center] at (1.504,0.677) {$0$};
\gpcolor{color=gp lt color axes}
\gpsetlinetype{gp lt axes}
\gpsetdashtype{gp dt axes}
\gpsetlinewidth{0.50}
\draw[gp path] (2.220,0.985)--(2.220,4.643);
\gpcolor{color=gp lt color border}
\gpsetlinetype{gp lt border}
\gpsetdashtype{gp dt solid}
\gpsetlinewidth{1.00}
\draw[gp path] (2.220,0.985)--(2.220,1.165);
\draw[gp path] (2.220,4.643)--(2.220,4.463);
\node[gp node center] at (2.220,0.677) {$20$};
\gpcolor{color=gp lt color axes}
\gpsetlinetype{gp lt axes}
\gpsetdashtype{gp dt axes}
\gpsetlinewidth{0.50}
\draw[gp path] (2.936,0.985)--(2.936,4.643);
\gpcolor{color=gp lt color border}
\gpsetlinetype{gp lt border}
\gpsetdashtype{gp dt solid}
\gpsetlinewidth{1.00}
\draw[gp path] (2.936,0.985)--(2.936,1.165);
\draw[gp path] (2.936,4.643)--(2.936,4.463);
\node[gp node center] at (2.936,0.677) {$40$};
\gpcolor{color=gp lt color axes}
\gpsetlinetype{gp lt axes}
\gpsetdashtype{gp dt axes}
\gpsetlinewidth{0.50}
\draw[gp path] (3.652,0.985)--(3.652,4.643);
\gpcolor{color=gp lt color border}
\gpsetlinetype{gp lt border}
\gpsetdashtype{gp dt solid}
\gpsetlinewidth{1.00}
\draw[gp path] (3.652,0.985)--(3.652,1.165);
\draw[gp path] (3.652,4.643)--(3.652,4.463);
\node[gp node center] at (3.652,0.677) {$60$};
\gpcolor{color=gp lt color axes}
\gpsetlinetype{gp lt axes}
\gpsetdashtype{gp dt axes}
\gpsetlinewidth{0.50}
\draw[gp path] (4.368,0.985)--(4.368,4.643);
\gpcolor{color=gp lt color border}
\gpsetlinetype{gp lt border}
\gpsetdashtype{gp dt solid}
\gpsetlinewidth{1.00}
\draw[gp path] (4.368,0.985)--(4.368,1.165);
\draw[gp path] (4.368,4.643)--(4.368,4.463);
\node[gp node center] at (4.368,0.677) {$80$};
\gpcolor{color=gp lt color axes}
\gpsetlinetype{gp lt axes}
\gpsetdashtype{gp dt axes}
\gpsetlinewidth{0.50}
\draw[gp path] (5.083,0.985)--(5.083,4.643);
\gpcolor{color=gp lt color border}
\gpsetlinetype{gp lt border}
\gpsetdashtype{gp dt solid}
\gpsetlinewidth{1.00}
\draw[gp path] (5.083,0.985)--(5.083,1.165);
\draw[gp path] (5.083,4.643)--(5.083,4.463);
\node[gp node center] at (5.083,0.677) {$100$};
\gpcolor{color=gp lt color axes}
\gpsetlinetype{gp lt axes}
\gpsetdashtype{gp dt axes}
\gpsetlinewidth{0.50}
\draw[gp path] (5.799,0.985)--(5.799,4.643);
\gpcolor{color=gp lt color border}
\gpsetlinetype{gp lt border}
\gpsetdashtype{gp dt solid}
\gpsetlinewidth{1.00}
\draw[gp path] (5.799,0.985)--(5.799,1.165);
\draw[gp path] (5.799,4.643)--(5.799,4.463);
\node[gp node center] at (5.799,0.677) {$120$};
\gpcolor{color=gp lt color axes}
\gpsetlinetype{gp lt axes}
\gpsetdashtype{gp dt axes}
\gpsetlinewidth{0.50}
\draw[gp path] (6.515,0.985)--(6.515,4.643);
\gpcolor{color=gp lt color border}
\gpsetlinetype{gp lt border}
\gpsetdashtype{gp dt solid}
\gpsetlinewidth{1.00}
\draw[gp path] (6.515,0.985)--(6.515,1.165);
\draw[gp path] (6.515,4.643)--(6.515,4.463);
\node[gp node center] at (6.515,0.677) {$140$};
\gpcolor{color=gp lt color axes}
\gpsetlinetype{gp lt axes}
\gpsetdashtype{gp dt axes}
\gpsetlinewidth{0.50}
\draw[gp path] (7.231,0.985)--(7.231,4.643);
\gpcolor{color=gp lt color border}
\gpsetlinetype{gp lt border}
\gpsetdashtype{gp dt solid}
\gpsetlinewidth{1.00}
\draw[gp path] (7.231,0.985)--(7.231,1.165);
\draw[gp path] (7.231,4.643)--(7.231,4.463);
\node[gp node center] at (7.231,0.677) {$160$};
\gpcolor{color=gp lt color axes}
\gpsetlinetype{gp lt axes}
\gpsetdashtype{gp dt axes}
\gpsetlinewidth{0.50}
\draw[gp path] (7.947,0.985)--(7.947,4.643);
\gpcolor{color=gp lt color border}
\gpsetlinetype{gp lt border}
\gpsetdashtype{gp dt solid}
\gpsetlinewidth{1.00}
\draw[gp path] (7.947,0.985)--(7.947,1.165);
\draw[gp path] (7.947,4.643)--(7.947,4.463);
\node[gp node center] at (7.947,0.677) {$180$};
\draw[gp path] (1.504,4.643)--(1.504,0.985)--(7.947,0.985)--(7.947,4.643)--cycle;
\node[gp node center,rotate=-270] at (0.292,2.814) {Infidelity};
\node[gp node center] at (4.725,0.215) {optim. iteration};
\node[gp node right] at (6.291,6.466) {basis, $J_{Frob}$};
\gpcolor{rgb color={0.580,0.000,0.827}}
\gpsetlinewidth{3.00}
\draw[gp path] (6.475,6.466)--(7.391,6.466);
\draw[gp path] (1.504,4.334)--(1.540,4.254)--(1.576,4.211)--(1.611,4.027)--(1.647,3.927)%
  --(1.683,3.910)--(1.719,3.888)--(1.755,3.865)--(1.790,3.864)--(1.826,3.584)--(1.862,3.270)%
  --(1.898,2.923)--(1.934,2.536)--(1.969,2.253)--(2.005,2.118)--(2.041,1.952)--(2.077,1.802)%
  --(2.113,1.696)--(2.148,1.618)--(2.184,1.540)--(2.220,1.522)--(2.256,1.517)--(2.291,1.512)%
  --(2.327,1.511)--(2.363,1.511)--(2.399,1.511)--(2.435,1.511)--(2.470,1.511)--(2.506,1.511)%
  --(2.542,1.511)--(2.578,1.511)--(2.614,1.511)--(2.649,1.511);
\gpcolor{color=gp lt color border}
\node[gp node right] at (6.291,6.158) {3 states, $J_{Frob}$};
\gpcolor{rgb color={0.000,0.620,0.451}}
\draw[gp path] (6.475,6.158)--(7.391,6.158);
\draw[gp path] (1.504,4.334)--(1.540,4.338)--(1.576,4.337)--(1.611,4.336)--(1.647,4.336)%
  --(1.683,4.336)--(1.719,4.335)--(1.755,4.335)--(1.790,4.336)--(1.826,4.335)--(1.862,4.336)%
  --(1.898,4.335)--(1.934,4.335)--(1.969,4.335)--(2.005,4.335)--(2.041,4.335)--(2.077,4.335)%
  --(2.113,4.334)--(2.148,4.333)--(2.184,4.327)--(2.220,4.324)--(2.256,4.325)--(2.291,4.321)%
  --(2.327,4.308)--(2.363,4.292)--(2.399,4.288)--(2.435,4.284)--(2.470,4.278)--(2.506,4.272)%
  --(2.542,4.266)--(2.578,4.268)--(2.614,4.267)--(2.649,4.262)--(2.685,4.256)--(2.721,4.236)%
  --(2.757,4.213)--(2.793,4.205)--(2.828,4.195)--(2.864,4.175)--(2.900,4.138)--(2.936,4.007)%
  --(2.972,4.008)--(3.007,4.031)--(3.043,4.006)--(3.079,3.982)--(3.115,3.982)--(3.151,3.961)%
  --(3.186,3.874)--(3.222,3.876)--(3.258,3.833)--(3.294,3.753)--(3.330,3.786)--(3.365,3.706)%
  --(3.401,3.654)--(3.437,3.597)--(3.473,3.492)--(3.508,3.506)--(3.544,3.471)--(3.580,3.444)%
  --(3.616,3.353)--(3.652,3.103)--(3.687,3.184)--(3.723,3.175)--(3.759,3.170)--(3.795,3.085)%
  --(3.831,3.027)--(3.866,3.006)--(3.902,2.961)--(3.938,2.793)--(3.974,2.854)--(4.010,2.806)%
  --(4.045,2.759)--(4.081,2.695)--(4.117,2.654)--(4.153,2.624)--(4.189,2.515)--(4.224,2.551)%
  --(4.260,2.500)--(4.296,2.466)--(4.332,2.407)--(4.368,2.393)--(4.403,2.363)--(4.439,2.320)%
  --(4.475,2.295)--(4.511,2.262)--(4.547,2.237)--(4.582,2.172)--(4.618,2.068)--(4.654,2.060)%
  --(4.690,1.943)--(4.726,1.854)--(4.761,1.761)--(4.797,1.778)--(4.833,1.730)--(4.869,1.703)%
  --(4.904,1.652)--(4.940,1.612)--(4.976,1.564)--(5.012,1.554)--(5.048,1.554)--(5.083,1.554)%
  --(5.119,1.555)--(5.155,1.552)--(5.191,1.549)--(5.227,1.544)--(5.262,1.537)--(5.298,1.525)%
  --(5.334,1.524)--(5.370,1.524)--(5.406,1.523)--(5.441,1.523)--(5.477,1.523)--(5.513,1.522)%
  --(5.549,1.522)--(5.585,1.521)--(5.620,1.521)--(5.656,1.521)--(5.692,1.521)--(5.728,1.521)%
  --(5.764,1.521)--(5.799,1.521)--(5.835,1.521)--(5.871,1.521)--(5.907,1.521)--(5.943,1.521)%
  --(5.978,1.521)--(6.014,1.521)--(6.050,1.521)--(6.086,1.521)--(6.121,1.521)--(6.157,1.521)%
  --(6.193,1.521)--(6.229,1.521)--(6.265,1.521)--(6.300,1.521)--(6.336,1.521)--(6.372,1.521)%
  --(6.408,1.521)--(6.444,1.521)--(6.479,1.521)--(6.515,1.521)--(6.551,1.521)--(6.587,1.521)%
  --(6.623,1.521)--(6.658,1.521)--(6.694,1.521)--(6.730,1.521);
\gpcolor{color=gp lt color border}
\node[gp node right] at (6.291,5.850) {3 states weighted, $J_{Frob}$};
\gpcolor{rgb color={0.902,0.624,0.000}}
\draw[gp path] (6.475,5.850)--(7.391,5.850);
\draw[gp path] (1.504,4.334)--(1.540,4.319)--(1.576,4.296)--(1.611,4.267)--(1.647,4.252)%
  --(1.683,4.001)--(1.719,4.080)--(1.755,4.043)--(1.790,4.016)--(1.826,4.023)--(1.862,4.035)%
  --(1.898,4.030)--(1.934,4.025)--(1.969,4.013)--(2.005,4.004)--(2.041,3.977)--(2.077,3.933)%
  --(2.113,3.906)--(2.148,3.884)--(2.184,3.850)--(2.220,3.848)--(2.256,3.810)--(2.291,3.708)%
  --(2.327,3.601)--(2.363,3.463)--(2.399,3.407)--(2.435,3.416)--(2.470,3.363)--(2.506,3.362)%
  --(2.542,3.340)--(2.578,3.296)--(2.614,3.310)--(2.649,3.260)--(2.685,3.229)--(2.721,3.170)%
  --(2.757,3.130)--(2.793,2.988)--(2.828,3.020)--(2.864,2.892)--(2.900,2.925)--(2.936,2.867)%
  --(2.972,2.905)--(3.007,2.781)--(3.043,2.755)--(3.079,2.659)--(3.115,2.451)--(3.151,2.403)%
  --(3.186,2.349)--(3.222,2.290)--(3.258,2.153)--(3.294,2.180)--(3.330,2.088)--(3.365,2.002)%
  --(3.401,1.650)--(3.437,1.659)--(3.473,1.628)--(3.508,1.591)--(3.544,1.529)--(3.580,1.523)%
  --(3.616,1.511)--(3.652,1.504)--(3.687,1.497)--(3.723,1.497)--(3.759,1.497)--(3.795,1.496)%
  --(3.831,1.496)--(3.866,1.496)--(3.902,1.496)--(3.938,1.496)--(3.974,1.496)--(4.010,1.496);
\gpcolor{color=gp lt color border}
\node[gp node right] at (6.291,5.542) {basis, $J_{Tr}$};
\gpcolor{rgb color={0.580,0.000,0.827}}
\gpsetdashtype{gp dt 4}
\draw[gp path] (6.475,5.542)--(7.391,5.542);
\draw[gp path] (1.504,4.334)--(1.540,4.253)--(1.576,4.208)--(1.611,4.071)--(1.647,3.928)%
  --(1.683,3.911)--(1.719,3.887)--(1.755,3.863)--(1.790,3.680)--(1.826,3.546)--(1.862,3.268)%
  --(1.898,2.806)--(1.934,2.473)--(1.969,2.361)--(2.005,2.144)--(2.041,1.943)--(2.077,1.776)%
  --(2.113,1.703)--(2.148,1.593)--(2.184,1.560)--(2.220,1.544)--(2.256,1.528)--(2.291,1.517)%
  --(2.327,1.513)--(2.363,1.511)--(2.399,1.511)--(2.435,1.511)--(2.470,1.511)--(2.506,1.511)%
  --(2.542,1.511)--(2.578,1.511)--(2.614,1.511)--(2.649,1.510)--(2.685,1.510)--(2.721,1.510)%
  --(2.757,1.510)--(2.793,1.510)--(2.828,1.510)--(2.864,1.510)--(2.900,1.510)--(2.936,1.510)%
  --(2.972,1.509)--(3.007,1.509)--(3.043,1.509)--(3.079,1.509)--(3.115,1.509)--(3.151,1.509)%
  --(3.186,1.509)--(3.222,1.509)--(3.258,1.509)--(3.294,1.509)--(3.330,1.509)--(3.365,1.509)%
  --(3.401,1.508)--(3.437,1.508)--(3.473,1.508)--(3.508,1.508)--(3.544,1.508)--(3.580,1.508)%
  --(3.616,1.508)--(3.652,1.508)--(3.687,1.508)--(3.723,1.508)--(3.759,1.508)--(3.795,1.508)%
  --(3.831,1.507)--(3.866,1.507)--(3.902,1.507)--(3.938,1.507)--(3.974,1.507)--(4.010,1.507)%
  --(4.045,1.507)--(4.081,1.507)--(4.117,1.507)--(4.153,1.507)--(4.189,1.507)--(4.224,1.506)%
  --(4.260,1.506)--(4.296,1.506)--(4.332,1.506)--(4.368,1.506)--(4.403,1.506)--(4.439,1.505)%
  --(4.475,1.505)--(4.511,1.505)--(4.547,1.505)--(4.582,1.504)--(4.618,1.504)--(4.654,1.504)%
  --(4.690,1.504)--(4.726,1.504)--(4.761,1.504)--(4.797,1.504)--(4.833,1.503)--(4.869,1.503)%
  --(4.904,1.503)--(4.940,1.503)--(4.976,1.503)--(5.012,1.502)--(5.048,1.502)--(5.083,1.502)%
  --(5.119,1.502)--(5.155,1.502)--(5.191,1.501)--(5.227,1.501)--(5.262,1.501)--(5.298,1.501)%
  --(5.334,1.501)--(5.370,1.501)--(5.406,1.501)--(5.441,1.501)--(5.477,1.501)--(5.513,1.500)%
  --(5.549,1.500)--(5.585,1.500)--(5.620,1.500)--(5.656,1.500)--(5.692,1.499)--(5.728,1.499)%
  --(5.764,1.499)--(5.799,1.499)--(5.835,1.499)--(5.871,1.499)--(5.907,1.499)--(5.943,1.499)%
  --(5.978,1.499)--(6.014,1.499)--(6.050,1.499)--(6.086,1.499)--(6.121,1.499)--(6.157,1.499)%
  --(6.193,1.499)--(6.229,1.499)--(6.265,1.499)--(6.300,1.498)--(6.336,1.498)--(6.372,1.498)%
  --(6.408,1.498)--(6.444,1.498)--(6.479,1.498)--(6.515,1.498)--(6.551,1.498)--(6.587,1.498)%
  --(6.623,1.498)--(6.658,1.498)--(6.694,1.498)--(6.730,1.498)--(6.766,1.498)--(6.802,1.498)%
  --(6.837,1.498)--(6.873,1.498)--(6.909,1.498)--(6.945,1.498)--(6.981,1.498)--(7.016,1.498)%
  --(7.052,1.498)--(7.088,1.497)--(7.124,1.497)--(7.160,1.497)--(7.195,1.497)--(7.231,1.497)%
  --(7.267,1.497)--(7.303,1.497)--(7.338,1.497)--(7.374,1.497)--(7.410,1.497)--(7.446,1.497)%
  --(7.482,1.497)--(7.517,1.497)--(7.553,1.497)--(7.589,1.497)--(7.625,1.497)--(7.661,1.497)%
  --(7.696,1.497)--(7.732,1.497)--(7.768,1.497)--(7.804,1.497)--(7.840,1.497)--(7.875,1.497)%
  --(7.911,1.497)--(7.947,1.497);
\gpcolor{color=gp lt color border}
\node[gp node right] at (6.291,5.234) {3 states, $J_{Tr}$};
\gpcolor{rgb color={0.000,0.620,0.451}}
\draw[gp path] (6.475,5.234)--(7.391,5.234);
\draw[gp path] (1.504,4.334)--(1.540,4.360)--(1.576,4.418)--(1.611,4.382)--(1.647,4.387)%
  --(1.683,4.388)--(1.719,4.386)--(1.755,4.379)--(1.790,4.360)--(1.826,4.331)--(1.862,4.325)%
  --(1.898,4.318)--(1.934,4.315)--(1.969,4.313)--(2.005,4.309)--(2.041,4.280)--(2.077,4.218)%
  --(2.113,4.186)--(2.148,4.113)--(2.184,4.100)--(2.220,4.016)--(2.256,4.022)--(2.291,3.965)%
  --(2.327,4.010)--(2.363,3.967)--(2.399,3.959)--(2.435,3.923)--(2.470,3.929)--(2.506,3.917)%
  --(2.542,3.912)--(2.578,3.899)--(2.614,3.887)--(2.649,3.865)--(2.685,3.772)--(2.721,3.790)%
  --(2.757,3.657)--(2.793,3.718)--(2.828,3.695)--(2.864,3.634)--(2.900,3.530)--(2.936,3.587)%
  --(2.972,3.439)--(3.007,3.496)--(3.043,3.418)--(3.079,3.382)--(3.115,3.261)--(3.151,3.266)%
  --(3.186,3.186)--(3.222,3.212)--(3.258,3.153)--(3.294,3.104)--(3.330,3.094)--(3.365,3.054)%
  --(3.401,3.064)--(3.437,3.036)--(3.473,2.994)--(3.508,3.026)--(3.544,3.020)--(3.580,3.033)%
  --(3.616,3.023)--(3.652,3.016)--(3.687,2.965)--(3.723,2.943)--(3.759,2.869)--(3.795,2.880)%
  --(3.831,2.876)--(3.866,2.864)--(3.902,2.864)--(3.938,2.832)--(3.974,2.817)--(4.010,2.697)%
  --(4.045,2.711)--(4.081,2.695)--(4.117,2.642)--(4.153,2.605)--(4.189,2.541)--(4.224,2.498)%
  --(4.260,2.445)--(4.296,2.446)--(4.332,2.336)--(4.368,2.368)--(4.403,2.248)--(4.439,2.311)%
  --(4.475,2.274)--(4.511,2.212)--(4.547,2.173)--(4.582,2.149)--(4.618,2.162)--(4.654,2.150)%
  --(4.690,2.023)--(4.726,2.045)--(4.761,2.029)--(4.797,2.026)--(4.833,2.009)--(4.869,1.991)%
  --(4.904,1.963)--(4.940,1.902)--(4.976,1.869)--(5.012,1.875)--(5.048,1.868)--(5.083,1.844)%
  --(5.119,1.813)--(5.155,1.767)--(5.191,1.741)--(5.227,1.737)--(5.262,1.730)--(5.298,1.718)%
  --(5.334,1.698)--(5.370,1.698)--(5.406,1.694)--(5.441,1.683)--(5.477,1.708)--(5.513,1.686)%
  --(5.549,1.687)--(5.585,1.682)--(5.620,1.673)--(5.656,1.702)--(5.692,1.672)--(5.728,1.692)%
  --(5.764,1.679)--(5.799,1.683)--(5.835,1.678)--(5.871,1.653)--(5.907,1.645)--(5.943,1.603)%
  --(5.978,1.628)--(6.014,1.614)--(6.050,1.615)--(6.086,1.612)--(6.121,1.677)--(6.157,1.666)%
  --(6.193,1.732)--(6.229,1.662)--(6.265,1.683)--(6.300,1.701)--(6.336,1.679)--(6.372,1.634)%
  --(6.408,1.622)--(6.444,1.617)--(6.479,1.598)--(6.515,1.600)--(6.551,1.610)--(6.587,1.623)%
  --(6.623,1.641)--(6.658,1.639)--(6.694,1.719)--(6.730,1.677)--(6.766,1.686)--(6.802,1.701)%
  --(6.837,1.696)--(6.873,1.738)--(6.909,1.715)--(6.945,1.753)--(6.981,1.740)--(7.016,1.754)%
  --(7.052,1.736)--(7.088,1.740)--(7.124,1.728)--(7.160,1.733)--(7.195,1.713)--(7.231,1.706)%
  --(7.267,1.690)--(7.303,1.679)--(7.338,1.665)--(7.374,1.638)--(7.410,1.627)--(7.446,1.635)%
  --(7.482,1.622)--(7.517,1.618)--(7.553,1.611)--(7.589,1.601)--(7.625,1.615)--(7.661,1.580)%
  --(7.696,1.585)--(7.732,1.583)--(7.768,1.606)--(7.804,1.596)--(7.840,1.631)--(7.875,1.589)%
  --(7.911,1.651)--(7.947,1.620);
\gpcolor{color=gp lt color border}
\node[gp node right] at (6.291,4.926) {3 states weighted, $J_{Tr}$};
\gpcolor{rgb color={0.902,0.624,0.000}}
\draw[gp path] (6.475,4.926)--(7.391,4.926);
\draw[gp path] (1.504,4.334)--(1.540,4.572)--(1.576,4.371)--(1.611,4.256)--(1.647,4.104)%
  --(1.683,4.001)--(1.719,3.943)--(1.755,3.841)--(1.790,3.773)--(1.826,3.684)--(1.862,3.624)%
  --(1.898,3.575)--(1.934,3.508)--(1.969,3.445)--(2.005,3.416)--(2.041,3.253)--(2.077,3.047)%
  --(2.113,2.988)--(2.148,2.913)--(2.184,2.663)--(2.220,2.598)--(2.256,2.463)--(2.291,2.331)%
  --(2.327,2.337)--(2.363,2.235)--(2.399,2.289)--(2.435,2.116)--(2.470,2.180)--(2.506,2.043)%
  --(2.542,2.018)--(2.578,1.948)--(2.614,1.897)--(2.649,1.937)--(2.685,1.888)--(2.721,1.831)%
  --(2.757,1.852)--(2.793,1.818)--(2.828,1.717)--(2.864,1.735)--(2.900,1.720)--(2.936,1.668)%
  --(2.972,1.670)--(3.007,1.669)--(3.043,1.646)--(3.079,1.629)--(3.115,1.600)--(3.151,1.604)%
  --(3.186,1.592)--(3.222,1.596)--(3.258,1.588)--(3.294,1.598)--(3.330,1.582)--(3.365,1.583)%
  --(3.401,1.579)--(3.437,1.580)--(3.473,1.577)--(3.508,1.580)--(3.544,1.580)--(3.580,1.580)%
  --(3.616,1.580)--(3.652,1.577)--(3.687,1.577)--(3.723,1.576)--(3.759,1.576)--(3.795,1.575)%
  --(3.831,1.571)--(3.866,1.572)--(3.902,1.573)--(3.938,1.572)--(3.974,1.575)--(4.010,1.573)%
  --(4.045,1.576)--(4.081,1.578)--(4.117,1.574)--(4.153,1.577)--(4.189,1.575)--(4.224,1.574)%
  --(4.260,1.570)--(4.296,1.570)--(4.332,1.566)--(4.368,1.563)--(4.403,1.568)--(4.439,1.567)%
  --(4.475,1.565)--(4.511,1.564)--(4.547,1.565)--(4.582,1.565)--(4.618,1.567)--(4.654,1.571)%
  --(4.690,1.571)--(4.726,1.573)--(4.761,1.576)--(4.797,1.580)--(4.833,1.579)--(4.869,1.581)%
  --(4.904,1.578)--(4.940,1.585)--(4.976,1.580)--(5.012,1.582)--(5.048,1.582)--(5.083,1.582)%
  --(5.119,1.580)--(5.155,1.584)--(5.191,1.582)--(5.227,1.586)--(5.262,1.579)--(5.298,1.589)%
  --(5.334,1.582)--(5.370,1.582)--(5.406,1.578)--(5.441,1.580)--(5.477,1.581)--(5.513,1.578)%
  --(5.549,1.584)--(5.585,1.584)--(5.620,1.588)--(5.656,1.594)--(5.692,1.589)--(5.728,1.597)%
  --(5.764,1.590)--(5.799,1.600)--(5.835,1.591)--(5.871,1.591)--(5.907,1.583)--(5.943,1.581)%
  --(5.978,1.584)--(6.014,1.580)--(6.050,1.582)--(6.086,1.583)--(6.121,1.590)--(6.157,1.583)%
  --(6.193,1.585)--(6.229,1.584)--(6.265,1.587)--(6.300,1.587)--(6.336,1.586)--(6.372,1.587)%
  --(6.408,1.585)--(6.444,1.585)--(6.479,1.584)--(6.515,1.584)--(6.551,1.583)--(6.587,1.582)%
  --(6.623,1.582)--(6.658,1.582)--(6.694,1.582)--(6.730,1.582)--(6.766,1.582)--(6.802,1.582)%
  --(6.837,1.583)--(6.873,1.582)--(6.909,1.584)--(6.945,1.583)--(6.981,1.583)--(7.016,1.584)%
  --(7.052,1.580)--(7.088,1.582)--(7.124,1.577)--(7.160,1.583)--(7.195,1.577)--(7.231,1.577)%
  --(7.267,1.576)--(7.303,1.573)--(7.338,1.570)--(7.374,1.569)--(7.410,1.566)--(7.446,1.559)%
  --(7.482,1.560)--(7.517,1.558)--(7.553,1.557)--(7.589,1.557)--(7.625,1.556)--(7.661,1.554)%
  --(7.696,1.549)--(7.732,1.549)--(7.768,1.548)--(7.804,1.548)--(7.840,1.546)--(7.875,1.547)%
  --(7.911,1.548)--(7.947,1.549);
\gpcolor{color=gp lt color border}
\gpsetdashtype{gp dt solid}
\gpsetlinewidth{1.00}
\draw[gp path] (1.504,4.643)--(1.504,0.985)--(7.947,0.985)--(7.947,4.643)--cycle;
\gpdefrectangularnode{gp plot 1}{\pgfpoint{1.504cm}{0.985cm}}{\pgfpoint{7.947cm}{4.643cm}}
\end{tikzpicture}

%% file: figures/cnot_gate/optim_gradient.tex
\begin{tikzpicture}[gnuplot]
\tikzset{every node/.append style={font={\footnotesize}}}
\path (0.000,0.000) rectangle (8.500,5.000);
\gpcolor{color=gp lt color border}
\gpsetlinetype{gp lt border}
\gpsetdashtype{gp dt solid}
\gpsetlinewidth{1.00}
\draw[gp path] (1.504,0.985)--(1.594,0.985);
\draw[gp path] (7.947,0.985)--(7.857,0.985);
\draw[gp path] (1.504,1.040)--(1.594,1.040);
\draw[gp path] (7.947,1.040)--(7.857,1.040);
\draw[gp path] (1.504,1.087)--(1.594,1.087);
\draw[gp path] (7.947,1.087)--(7.857,1.087);
\draw[gp path] (1.504,1.128)--(1.594,1.128);
\draw[gp path] (7.947,1.128)--(7.857,1.128);
\draw[gp path] (1.504,1.163)--(1.594,1.163);
\draw[gp path] (7.947,1.163)--(7.857,1.163);
\gpcolor{color=gp lt color axes}
\gpsetlinetype{gp lt axes}
\gpsetdashtype{gp dt axes}
\gpsetlinewidth{0.50}
\draw[gp path] (1.504,1.195)--(7.947,1.195);
\gpcolor{color=gp lt color border}
\gpsetlinetype{gp lt border}
\gpsetdashtype{gp dt solid}
\gpsetlinewidth{1.00}
\draw[gp path] (1.504,1.195)--(1.684,1.195);
\draw[gp path] (7.947,1.195)--(7.767,1.195);
\node[gp node right] at (1.320,1.195) {$10^{-5}$};
\draw[gp path] (1.504,1.406)--(1.594,1.406);
\draw[gp path] (7.947,1.406)--(7.857,1.406);
\draw[gp path] (1.504,1.529)--(1.594,1.529);
\draw[gp path] (7.947,1.529)--(7.857,1.529);
\draw[gp path] (1.504,1.616)--(1.594,1.616);
\draw[gp path] (7.947,1.616)--(7.857,1.616);
\draw[gp path] (1.504,1.684)--(1.594,1.684);
\draw[gp path] (7.947,1.684)--(7.857,1.684);
\draw[gp path] (1.504,1.739)--(1.594,1.739);
\draw[gp path] (7.947,1.739)--(7.857,1.739);
\draw[gp path] (1.504,1.786)--(1.594,1.786);
\draw[gp path] (7.947,1.786)--(7.857,1.786);
\draw[gp path] (1.504,1.826)--(1.594,1.826);
\draw[gp path] (7.947,1.826)--(7.857,1.826);
\draw[gp path] (1.504,1.862)--(1.594,1.862);
\draw[gp path] (7.947,1.862)--(7.857,1.862);
\gpcolor{color=gp lt color axes}
\gpsetlinetype{gp lt axes}
\gpsetdashtype{gp dt axes}
\gpsetlinewidth{0.50}
\draw[gp path] (1.504,1.894)--(7.947,1.894);
\gpcolor{color=gp lt color border}
\gpsetlinetype{gp lt border}
\gpsetdashtype{gp dt solid}
\gpsetlinewidth{1.00}
\draw[gp path] (1.504,1.894)--(1.684,1.894);
\draw[gp path] (7.947,1.894)--(7.767,1.894);
\node[gp node right] at (1.320,1.894) {$10^{-4}$};
\draw[gp path] (1.504,2.104)--(1.594,2.104);
\draw[gp path] (7.947,2.104)--(7.857,2.104);
\draw[gp path] (1.504,2.227)--(1.594,2.227);
\draw[gp path] (7.947,2.227)--(7.857,2.227);
\draw[gp path] (1.504,2.314)--(1.594,2.314);
\draw[gp path] (7.947,2.314)--(7.857,2.314);
\draw[gp path] (1.504,2.382)--(1.594,2.382);
\draw[gp path] (7.947,2.382)--(7.857,2.382);
\draw[gp path] (1.504,2.437)--(1.594,2.437);
\draw[gp path] (7.947,2.437)--(7.857,2.437);
\draw[gp path] (1.504,2.484)--(1.594,2.484);
\draw[gp path] (7.947,2.484)--(7.857,2.484);
\draw[gp path] (1.504,2.525)--(1.594,2.525);
\draw[gp path] (7.947,2.525)--(7.857,2.525);
\draw[gp path] (1.504,2.560)--(1.594,2.560);
\draw[gp path] (7.947,2.560)--(7.857,2.560);
\gpcolor{color=gp lt color axes}
\gpsetlinetype{gp lt axes}
\gpsetdashtype{gp dt axes}
\gpsetlinewidth{0.50}
\draw[gp path] (1.504,2.592)--(7.947,2.592);
\gpcolor{color=gp lt color border}
\gpsetlinetype{gp lt border}
\gpsetdashtype{gp dt solid}
\gpsetlinewidth{1.00}
\draw[gp path] (1.504,2.592)--(1.684,2.592);
\draw[gp path] (7.947,2.592)--(7.767,2.592);
\node[gp node right] at (1.320,2.592) {$10^{-3}$};
\draw[gp path] (1.504,2.803)--(1.594,2.803);
\draw[gp path] (7.947,2.803)--(7.857,2.803);
\draw[gp path] (1.504,2.926)--(1.594,2.926);
\draw[gp path] (7.947,2.926)--(7.857,2.926);
\draw[gp path] (1.504,3.013)--(1.594,3.013);
\draw[gp path] (7.947,3.013)--(7.857,3.013);
\draw[gp path] (1.504,3.081)--(1.594,3.081);
\draw[gp path] (7.947,3.081)--(7.857,3.081);
\draw[gp path] (1.504,3.136)--(1.594,3.136);
\draw[gp path] (7.947,3.136)--(7.857,3.136);
\draw[gp path] (1.504,3.183)--(1.594,3.183);
\draw[gp path] (7.947,3.183)--(7.857,3.183);
\draw[gp path] (1.504,3.223)--(1.594,3.223);
\draw[gp path] (7.947,3.223)--(7.857,3.223);
\draw[gp path] (1.504,3.259)--(1.594,3.259);
\draw[gp path] (7.947,3.259)--(7.857,3.259);
\gpcolor{color=gp lt color axes}
\gpsetlinetype{gp lt axes}
\gpsetdashtype{gp dt axes}
\gpsetlinewidth{0.50}
\draw[gp path] (1.504,3.291)--(7.947,3.291);
\gpcolor{color=gp lt color border}
\gpsetlinetype{gp lt border}
\gpsetdashtype{gp dt solid}
\gpsetlinewidth{1.00}
\draw[gp path] (1.504,3.291)--(1.684,3.291);
\draw[gp path] (7.947,3.291)--(7.767,3.291);
\node[gp node right] at (1.320,3.291) {$10^{-2}$};
\draw[gp path] (1.504,3.501)--(1.594,3.501);
\draw[gp path] (7.947,3.501)--(7.857,3.501);
\draw[gp path] (1.504,3.624)--(1.594,3.624);
\draw[gp path] (7.947,3.624)--(7.857,3.624);
\draw[gp path] (1.504,3.711)--(1.594,3.711);
\draw[gp path] (7.947,3.711)--(7.857,3.711);
\draw[gp path] (1.504,3.779)--(1.594,3.779);
\draw[gp path] (7.947,3.779)--(7.857,3.779);
\draw[gp path] (1.504,3.834)--(1.594,3.834);
\draw[gp path] (7.947,3.834)--(7.857,3.834);
\draw[gp path] (1.504,3.881)--(1.594,3.881);
\draw[gp path] (7.947,3.881)--(7.857,3.881);
\draw[gp path] (1.504,3.922)--(1.594,3.922);
\draw[gp path] (7.947,3.922)--(7.857,3.922);
\draw[gp path] (1.504,3.957)--(1.594,3.957);
\draw[gp path] (7.947,3.957)--(7.857,3.957);
\gpcolor{color=gp lt color axes}
\gpsetlinetype{gp lt axes}
\gpsetdashtype{gp dt axes}
\gpsetlinewidth{0.50}
\draw[gp path] (1.504,3.989)--(7.947,3.989);
\gpcolor{color=gp lt color border}
\gpsetlinetype{gp lt border}
\gpsetdashtype{gp dt solid}
\gpsetlinewidth{1.00}
\draw[gp path] (1.504,3.989)--(1.684,3.989);
\draw[gp path] (7.947,3.989)--(7.767,3.989);
\node[gp node right] at (1.320,3.989) {$10^{-1}$};
\draw[gp path] (1.504,4.200)--(1.594,4.200);
\draw[gp path] (7.947,4.200)--(7.857,4.200);
\draw[gp path] (1.504,4.323)--(1.594,4.323);
\draw[gp path] (7.947,4.323)--(7.857,4.323);
\draw[gp path] (1.504,4.410)--(1.594,4.410);
\draw[gp path] (7.947,4.410)--(7.857,4.410);
\draw[gp path] (1.504,4.478)--(1.594,4.478);
\draw[gp path] (7.947,4.478)--(7.857,4.478);
\draw[gp path] (1.504,4.533)--(1.594,4.533);
\draw[gp path] (7.947,4.533)--(7.857,4.533);
\draw[gp path] (1.504,4.580)--(1.594,4.580);
\draw[gp path] (7.947,4.580)--(7.857,4.580);
\draw[gp path] (1.504,4.620)--(1.594,4.620);
\draw[gp path] (7.947,4.620)--(7.857,4.620);
\draw[gp path] (1.504,4.656)--(1.594,4.656);
\draw[gp path] (7.947,4.656)--(7.857,4.656);
\gpcolor{color=gp lt color axes}
\gpsetlinetype{gp lt axes}
\gpsetdashtype{gp dt axes}
\gpsetlinewidth{0.50}
\draw[gp path] (1.504,4.688)--(7.947,4.688);
\gpcolor{color=gp lt color border}
\gpsetlinetype{gp lt border}
\gpsetdashtype{gp dt solid}
\gpsetlinewidth{1.00}
\draw[gp path] (1.504,4.688)--(1.684,4.688);
\draw[gp path] (7.947,4.688)--(7.767,4.688);
\node[gp node right] at (1.320,4.688) {$10^{0}$};
\gpcolor{color=gp lt color axes}
\gpsetlinetype{gp lt axes}
\gpsetdashtype{gp dt axes}
\gpsetlinewidth{0.50}
\draw[gp path] (1.504,0.985)--(1.504,4.691);
\gpcolor{color=gp lt color border}
\gpsetlinetype{gp lt border}
\gpsetdashtype{gp dt solid}
\gpsetlinewidth{1.00}
\draw[gp path] (1.504,0.985)--(1.504,1.165);
\draw[gp path] (1.504,4.691)--(1.504,4.511);
\node[gp node center] at (1.504,0.677) {$0$};
\gpcolor{color=gp lt color axes}
\gpsetlinetype{gp lt axes}
\gpsetdashtype{gp dt axes}
\gpsetlinewidth{0.50}
\draw[gp path] (2.220,0.985)--(2.220,4.691);
\gpcolor{color=gp lt color border}
\gpsetlinetype{gp lt border}
\gpsetdashtype{gp dt solid}
\gpsetlinewidth{1.00}
\draw[gp path] (2.220,0.985)--(2.220,1.165);
\draw[gp path] (2.220,4.691)--(2.220,4.511);
\node[gp node center] at (2.220,0.677) {$20$};
\gpcolor{color=gp lt color axes}
\gpsetlinetype{gp lt axes}
\gpsetdashtype{gp dt axes}
\gpsetlinewidth{0.50}
\draw[gp path] (2.936,0.985)--(2.936,4.691);
\gpcolor{color=gp lt color border}
\gpsetlinetype{gp lt border}
\gpsetdashtype{gp dt solid}
\gpsetlinewidth{1.00}
\draw[gp path] (2.936,0.985)--(2.936,1.165);
\draw[gp path] (2.936,4.691)--(2.936,4.511);
\node[gp node center] at (2.936,0.677) {$40$};
\gpcolor{color=gp lt color axes}
\gpsetlinetype{gp lt axes}
\gpsetdashtype{gp dt axes}
\gpsetlinewidth{0.50}
\draw[gp path] (3.652,0.985)--(3.652,4.691);
\gpcolor{color=gp lt color border}
\gpsetlinetype{gp lt border}
\gpsetdashtype{gp dt solid}
\gpsetlinewidth{1.00}
\draw[gp path] (3.652,0.985)--(3.652,1.165);
\draw[gp path] (3.652,4.691)--(3.652,4.511);
\node[gp node center] at (3.652,0.677) {$60$};
\gpcolor{color=gp lt color axes}
\gpsetlinetype{gp lt axes}
\gpsetdashtype{gp dt axes}
\gpsetlinewidth{0.50}
\draw[gp path] (4.368,0.985)--(4.368,4.691);
\gpcolor{color=gp lt color border}
\gpsetlinetype{gp lt border}
\gpsetdashtype{gp dt solid}
\gpsetlinewidth{1.00}
\draw[gp path] (4.368,0.985)--(4.368,1.165);
\draw[gp path] (4.368,4.691)--(4.368,4.511);
\node[gp node center] at (4.368,0.677) {$80$};
\gpcolor{color=gp lt color axes}
\gpsetlinetype{gp lt axes}
\gpsetdashtype{gp dt axes}
\gpsetlinewidth{0.50}
\draw[gp path] (5.083,0.985)--(5.083,4.691);
\gpcolor{color=gp lt color border}
\gpsetlinetype{gp lt border}
\gpsetdashtype{gp dt solid}
\gpsetlinewidth{1.00}
\draw[gp path] (5.083,0.985)--(5.083,1.165);
\draw[gp path] (5.083,4.691)--(5.083,4.511);
\node[gp node center] at (5.083,0.677) {$100$};
\gpcolor{color=gp lt color axes}
\gpsetlinetype{gp lt axes}
\gpsetdashtype{gp dt axes}
\gpsetlinewidth{0.50}
\draw[gp path] (5.799,0.985)--(5.799,4.691);
\gpcolor{color=gp lt color border}
\gpsetlinetype{gp lt border}
\gpsetdashtype{gp dt solid}
\gpsetlinewidth{1.00}
\draw[gp path] (5.799,0.985)--(5.799,1.165);
\draw[gp path] (5.799,4.691)--(5.799,4.511);
\node[gp node center] at (5.799,0.677) {$120$};
\gpcolor{color=gp lt color axes}
\gpsetlinetype{gp lt axes}
\gpsetdashtype{gp dt axes}
\gpsetlinewidth{0.50}
\draw[gp path] (6.515,0.985)--(6.515,4.691);
\gpcolor{color=gp lt color border}
\gpsetlinetype{gp lt border}
\gpsetdashtype{gp dt solid}
\gpsetlinewidth{1.00}
\draw[gp path] (6.515,0.985)--(6.515,1.165);
\draw[gp path] (6.515,4.691)--(6.515,4.511);
\node[gp node center] at (6.515,0.677) {$140$};
\gpcolor{color=gp lt color axes}
\gpsetlinetype{gp lt axes}
\gpsetdashtype{gp dt axes}
\gpsetlinewidth{0.50}
\draw[gp path] (7.231,0.985)--(7.231,4.691);
\gpcolor{color=gp lt color border}
\gpsetlinetype{gp lt border}
\gpsetdashtype{gp dt solid}
\gpsetlinewidth{1.00}
\draw[gp path] (7.231,0.985)--(7.231,1.165);
\draw[gp path] (7.231,4.691)--(7.231,4.511);
\node[gp node center] at (7.231,0.677) {$160$};
\gpcolor{color=gp lt color axes}
\gpsetlinetype{gp lt axes}
\gpsetdashtype{gp dt axes}
\gpsetlinewidth{0.50}
\draw[gp path] (7.947,0.985)--(7.947,4.691);
\gpcolor{color=gp lt color border}
\gpsetlinetype{gp lt border}
\gpsetdashtype{gp dt solid}
\gpsetlinewidth{1.00}
\draw[gp path] (7.947,0.985)--(7.947,1.165);
\draw[gp path] (7.947,4.691)--(7.947,4.511);
\node[gp node center] at (7.947,0.677) {$180$};
\draw[gp path] (1.504,4.691)--(1.504,0.985)--(7.947,0.985)--(7.947,4.691)--cycle;
\node[gp node center,rotate=-270] at (0.292,2.838) {$\| \nabla J \|$};
\node[gp node center] at (4.725,0.215) {optim. iteration};
\gpcolor{rgb color={0.580,0.000,0.827}}
\gpsetlinewidth{2.00}
\draw[gp path] (1.504,4.569)--(1.540,4.525)--(1.576,4.673)--(1.611,4.631)--(1.647,4.486)%
  --(1.683,4.457)--(1.719,4.386)--(1.755,4.499)--(1.790,4.659)--(1.826,4.590)--(1.862,4.573)%
  --(1.898,4.591)--(1.934,4.522)--(1.969,4.368)--(2.005,4.423)--(2.041,4.393)--(2.077,4.265)%
  --(2.113,4.168)--(2.148,4.165)--(2.184,4.095)--(2.220,3.845)--(2.256,3.776)--(2.291,3.874)%
  --(2.327,3.348)--(2.363,3.127)--(2.399,2.976)--(2.435,2.611)--(2.470,2.167)--(2.506,1.984)%
  --(2.542,1.711)--(2.578,1.397)--(2.614,1.511)--(2.649,1.062);
\gpcolor{rgb color={0.000,0.620,0.451}}
\draw[gp path] (1.504,4.556)--(1.540,4.205)--(1.576,4.135)--(1.611,3.504)--(1.647,3.389)%
  --(1.683,3.183)--(1.719,3.231)--(1.755,3.332)--(1.790,3.110)--(1.826,2.976)--(1.862,2.523)%
  --(1.898,2.568)--(1.934,2.454)--(1.969,2.497)--(2.005,2.675)--(2.041,2.815)--(2.077,2.894)%
  --(2.113,3.025)--(2.148,2.947)--(2.184,3.178)--(2.220,3.433)--(2.256,3.233)--(2.291,3.138)%
  --(2.327,3.391)--(2.363,3.552)--(2.399,3.450)--(2.435,3.380)--(2.470,3.310)--(2.506,3.464)%
  --(2.542,3.505)--(2.578,3.335)--(2.614,3.268)--(2.649,3.313)--(2.685,3.373)--(2.721,3.492)%
  --(2.757,3.594)--(2.793,3.354)--(2.828,3.236)--(2.864,3.333)--(2.900,3.535)--(2.936,3.676)%
  --(2.972,3.722)--(3.007,3.450)--(3.043,3.373)--(3.079,3.437)--(3.115,3.318)--(3.151,3.197)%
  --(3.186,3.698)--(3.222,3.489)--(3.258,3.452)--(3.294,3.659)--(3.330,3.371)--(3.365,3.622)%
  --(3.401,3.452)--(3.437,3.482)--(3.473,3.472)--(3.508,3.333)--(3.544,3.321)--(3.580,3.243)%
  --(3.616,3.292)--(3.652,3.652)--(3.687,3.565)--(3.723,3.157)--(3.759,3.117)--(3.795,3.221)%
  --(3.831,3.286)--(3.866,3.132)--(3.902,3.222)--(3.938,3.459)--(3.974,3.205)--(4.010,3.151)%
  --(4.045,3.166)--(4.081,3.167)--(4.117,3.007)--(4.153,3.165)--(4.189,3.254)--(4.224,3.152)%
  --(4.260,3.033)--(4.296,3.045)--(4.332,2.961)--(4.368,2.889)--(4.403,3.023)--(4.439,3.205)%
  --(4.475,2.847)--(4.511,2.866)--(4.547,2.879)--(4.582,2.868)--(4.618,3.133)--(4.654,2.972)%
  --(4.690,2.928)--(4.726,2.978)--(4.761,3.009)--(4.797,2.843)--(4.833,2.891)--(4.869,3.029)%
  --(4.904,2.820)--(4.940,2.716)--(4.976,2.830)--(5.012,2.887)--(5.048,2.591)--(5.083,2.553)%
  --(5.119,2.695)--(5.155,2.771)--(5.191,2.525)--(5.227,2.409)--(5.262,2.581)--(5.298,2.586)%
  --(5.334,2.534)--(5.370,2.225)--(5.406,2.216)--(5.441,2.260)--(5.477,2.435)--(5.513,2.056)%
  --(5.549,2.216)--(5.585,2.284)--(5.620,2.314)--(5.656,1.993)--(5.692,1.823)--(5.728,1.928)%
  --(5.764,1.945)--(5.799,1.942)--(5.835,1.654)--(5.871,1.555)--(5.907,1.671)--(5.943,1.827)%
  --(5.978,1.854)--(6.014,1.732)--(6.050,1.377)--(6.086,1.399)--(6.121,1.508)--(6.157,1.598)%
  --(6.193,1.588)--(6.229,1.377)--(6.265,1.364)--(6.300,1.492)--(6.336,1.569)--(6.372,1.643)%
  --(6.408,1.451)--(6.444,1.403)--(6.479,1.475)--(6.515,1.320)--(6.551,1.246)--(6.587,1.240)%
  --(6.623,1.260)--(6.658,1.465)--(6.694,1.358)--(6.730,1.023);
\gpcolor{rgb color={0.902,0.624,0.000}}
\draw[gp path] (1.504,4.361)--(1.540,4.245)--(1.576,4.167)--(1.611,3.915)--(1.647,3.869)%
  --(1.683,4.092)--(1.719,3.904)--(1.755,3.737)--(1.790,3.525)--(1.826,3.467)--(1.862,3.541)%
  --(1.898,3.386)--(1.934,3.308)--(1.969,3.225)--(2.005,3.405)--(2.041,3.633)--(2.077,3.789)%
  --(2.113,3.826)--(2.148,3.813)--(2.184,4.050)--(2.220,3.805)--(2.256,3.688)--(2.291,3.825)%
  --(2.327,3.750)--(2.363,3.774)--(2.399,3.607)--(2.435,3.400)--(2.470,3.499)--(2.506,3.561)%
  --(2.542,3.374)--(2.578,3.600)--(2.614,3.312)--(2.649,3.640)--(2.685,3.483)--(2.721,3.531)%
  --(2.757,3.595)--(2.793,3.773)--(2.828,3.518)--(2.864,3.717)--(2.900,3.475)--(2.936,3.456)%
  --(2.972,3.496)--(3.007,3.426)--(3.043,3.223)--(3.079,3.176)--(3.115,3.323)--(3.151,3.426)%
  --(3.186,3.270)--(3.222,3.161)--(3.258,3.389)--(3.294,3.275)--(3.330,3.276)--(3.365,3.317)%
  --(3.401,3.381)--(3.437,3.130)--(3.473,3.052)--(3.508,2.846)--(3.544,3.095)--(3.580,3.088)%
  --(3.616,2.671)--(3.652,2.599)--(3.687,2.745)--(3.723,2.694)--(3.759,2.350)--(3.795,2.170)%
  --(3.831,2.115)--(3.866,1.956)--(3.902,1.681)--(3.938,1.519)--(3.974,1.391)--(4.010,1.154);
\gpcolor{rgb color={0.580,0.000,0.827}}
\gpsetdashtype{gp dt 4}
\draw[gp path] (1.504,4.569)--(1.540,4.527)--(1.576,4.679)--(1.611,4.618)--(1.647,4.500)%
  --(1.683,4.464)--(1.719,4.386)--(1.755,4.496)--(1.790,4.612)--(1.826,4.552)--(1.860,4.691);
\draw[gp path] (1.862,4.691)--(1.898,4.447)--(1.934,4.418)--(1.969,4.473)--(2.005,4.558)%
  --(2.041,4.411)--(2.077,4.203)--(2.113,4.165)--(2.148,4.290)--(2.184,3.979)--(2.220,3.969)%
  --(2.256,3.895)--(2.291,3.902)--(2.327,3.664)--(2.363,3.500)--(2.399,3.271)--(2.435,2.953)%
  --(2.470,2.844)--(2.506,2.858)--(2.542,2.903)--(2.578,3.071)--(2.614,3.165)--(2.649,3.190)%
  --(2.685,3.245)--(2.721,3.158)--(2.757,3.273)--(2.793,3.436)--(2.828,3.342)--(2.864,3.357)%
  --(2.900,3.259)--(2.936,3.471)--(2.972,3.194)--(3.007,3.310)--(3.043,3.315)--(3.079,3.414)%
  --(3.115,3.423)--(3.151,3.440)--(3.186,3.456)--(3.222,3.469)--(3.258,3.484)--(3.294,3.484)%
  --(3.330,3.519)--(3.365,3.396)--(3.401,3.460)--(3.437,3.277)--(3.473,3.390)--(3.508,3.295)%
  --(3.544,3.331)--(3.580,3.295)--(3.616,3.327)--(3.652,3.357)--(3.687,3.389)--(3.723,3.412)%
  --(3.759,3.455)--(3.795,3.407)--(3.831,3.347)--(3.866,3.304)--(3.902,3.213)--(3.938,3.147)%
  --(3.974,3.001)--(4.010,3.329)--(4.045,3.210)--(4.081,3.237)--(4.117,3.298)--(4.153,3.371)%
  --(4.189,3.478)--(4.224,3.323)--(4.260,3.466)--(4.296,3.306)--(4.332,3.458)--(4.368,3.533)%
  --(4.403,3.680)--(4.439,3.540)--(4.475,3.576)--(4.511,3.427)--(4.547,3.435)--(4.582,3.238)%
  --(4.618,3.331)--(4.654,3.310)--(4.690,3.564)--(4.726,3.430)--(4.761,3.447)--(4.797,3.226)%
  --(4.833,3.282)--(4.869,3.492)--(4.904,3.382)--(4.940,3.335)--(4.976,3.583)--(5.012,3.447)%
  --(5.048,3.447)--(5.083,3.485)--(5.119,3.315)--(5.155,3.508)--(5.191,3.349)--(5.227,3.274)%
  --(5.262,3.303)--(5.298,3.112)--(5.334,3.264)--(5.370,2.969)--(5.406,3.294)--(5.441,3.317)%
  --(5.477,3.439)--(5.513,3.544)--(5.549,3.507)--(5.585,3.542)--(5.620,3.299)--(5.656,3.413)%
  --(5.692,3.372)--(5.728,3.380)--(5.764,3.158)--(5.799,3.281)--(5.835,3.093)--(5.871,3.219)%
  --(5.907,3.370)--(5.943,3.168)--(5.978,3.239)--(6.014,3.036)--(6.050,3.397)--(6.086,3.118)%
  --(6.121,3.460)--(6.157,3.153)--(6.193,3.326)--(6.229,3.259)--(6.265,3.349)--(6.300,3.389)%
  --(6.336,3.343)--(6.372,3.410)--(6.408,3.297)--(6.444,3.347)--(6.479,3.234)--(6.515,3.315)%
  --(6.551,3.135)--(6.587,3.197)--(6.623,3.139)--(6.658,3.415)--(6.694,3.200)--(6.730,3.372)%
  --(6.766,3.273)--(6.802,3.182)--(6.837,3.358)--(6.873,3.209)--(6.909,3.236)--(6.945,3.251)%
  --(6.981,3.257)--(7.016,3.266)--(7.052,3.281)--(7.088,3.131)--(7.124,3.176)--(7.160,3.281)%
  --(7.195,3.219)--(7.231,3.253)--(7.267,3.091)--(7.303,2.938)--(7.338,3.357)--(7.374,3.172)%
  --(7.410,3.361)--(7.446,3.426)--(7.482,3.228)--(7.517,3.325)--(7.553,3.090)--(7.589,3.212)%
  --(7.625,3.387)--(7.661,3.233)--(7.696,3.311)--(7.732,3.171)--(7.768,3.256)--(7.804,3.074)%
  --(7.840,3.329)--(7.875,3.174)--(7.911,3.139)--(7.947,3.254);
\gpcolor{rgb color={0.000,0.620,0.451}}
\draw[gp path] (1.504,4.556)--(1.540,4.602)--(1.576,4.365)--(1.611,4.224)--(1.647,4.043)%
  --(1.683,3.708)--(1.719,3.609)--(1.755,3.674)--(1.790,3.798)--(1.826,3.813)--(1.862,3.816)%
  --(1.898,3.519)--(1.934,3.296)--(1.969,3.312)--(2.005,3.354)--(2.041,3.530)--(2.077,3.683)%
  --(2.113,3.546)--(2.148,3.530)--(2.184,3.494)--(2.220,3.644)--(2.256,3.537)--(2.291,3.586)%
  --(2.327,3.703)--(2.363,3.321)--(2.399,3.251)--(2.435,3.488)--(2.470,3.530)--(2.506,3.263)%
  --(2.542,3.202)--(2.578,3.265)--(2.614,3.356)--(2.649,3.460)--(2.685,3.634)--(2.721,3.524)%
  --(2.757,3.620)--(2.793,3.334)--(2.828,3.256)--(2.864,3.345)--(2.900,3.419)--(2.936,3.638)%
  --(2.972,3.461)--(3.007,3.276)--(3.043,3.318)--(3.079,3.328)--(3.115,3.333)--(3.151,3.300)%
  --(3.186,3.316)--(3.222,3.134)--(3.258,3.232)--(3.294,3.230)--(3.330,3.433)--(3.365,3.152)%
  --(3.401,3.159)--(3.437,3.217)--(3.473,3.506)--(3.508,3.105)--(3.544,3.205)--(3.580,3.141)%
  --(3.616,3.166)--(3.652,3.053)--(3.687,3.223)--(3.723,3.243)--(3.759,3.380)--(3.795,3.195)%
  --(3.831,2.935)--(3.866,3.127)--(3.902,3.123)--(3.938,3.144)--(3.974,3.075)--(4.010,3.225)%
  --(4.045,3.125)--(4.081,3.080)--(4.117,3.269)--(4.153,3.219)--(4.189,3.191)--(4.224,3.049)%
  --(4.260,3.095)--(4.296,3.009)--(4.332,3.187)--(4.368,2.957)--(4.403,3.341)--(4.439,2.979)%
  --(4.475,2.981)--(4.511,3.040)--(4.547,3.140)--(4.582,2.892)--(4.618,2.822)--(4.654,2.836)%
  --(4.690,3.239)--(4.726,2.985)--(4.761,2.985)--(4.797,2.846)--(4.833,2.909)--(4.869,2.921)%
  --(4.904,2.953)--(4.940,3.065)--(4.976,3.250)--(5.012,3.142)--(5.048,2.787)--(5.083,2.850)%
  --(5.119,2.886)--(5.155,2.914)--(5.191,3.101)--(5.227,2.971)--(5.262,2.805)--(5.298,2.805)%
  --(5.334,2.770)--(5.370,2.863)--(5.406,2.846)--(5.441,2.979)--(5.477,3.002)--(5.513,2.870)%
  --(5.549,2.745)--(5.585,2.635)--(5.620,2.784)--(5.656,3.107)--(5.692,2.983)--(5.728,2.932)%
  --(5.764,2.703)--(5.799,2.751)--(5.835,2.734)--(5.871,2.955)--(5.907,3.080)--(5.943,3.248)%
  --(5.978,3.066)--(6.014,2.850)--(6.050,2.867)--(6.086,2.827)--(6.121,3.129)--(6.157,3.109)%
  --(6.193,3.011)--(6.229,3.059)--(6.265,2.788)--(6.300,2.823)--(6.336,2.887)--(6.372,3.007)%
  --(6.408,2.819)--(6.444,2.745)--(6.479,2.950)--(6.515,2.947)--(6.551,2.735)--(6.587,2.945)%
  --(6.623,3.119)--(6.658,3.002)--(6.694,3.097)--(6.730,2.841)--(6.766,2.777)--(6.802,2.958)%
  --(6.837,2.969)--(6.873,3.052)--(6.909,3.198)--(6.945,2.898)--(6.981,2.841)--(7.016,3.083)%
  --(7.052,2.769)--(7.088,2.953)--(7.124,2.753)--(7.160,3.042)--(7.195,2.820)--(7.231,2.745)%
  --(7.267,3.251)--(7.303,3.052)--(7.338,2.948)--(7.374,3.156)--(7.410,2.921)--(7.446,3.113)%
  --(7.482,2.800)--(7.517,2.805)--(7.553,2.947)--(7.589,2.964)--(7.625,3.247)--(7.661,2.976)%
  --(7.696,2.871)--(7.732,2.760)--(7.768,2.981)--(7.804,2.889)--(7.840,2.998)--(7.875,3.047)%
  --(7.911,3.016)--(7.947,2.828);
\gpcolor{rgb color={0.902,0.624,0.000}}
\draw[gp path] (1.504,4.361)--(1.540,4.339)--(1.576,4.313)--(1.611,4.316)--(1.647,4.103)%
  --(1.683,3.991)--(1.719,3.874)--(1.755,3.780)--(1.790,3.706)--(1.826,3.855)--(1.862,3.607)%
  --(1.898,3.611)--(1.934,3.552)--(1.969,3.557)--(2.005,3.476)--(2.041,3.547)--(2.077,3.739)%
  --(2.113,3.574)--(2.148,3.445)--(2.184,3.527)--(2.220,3.468)--(2.256,3.393)--(2.291,3.407)%
  --(2.327,3.139)--(2.363,3.307)--(2.399,3.249)--(2.435,3.237)--(2.470,3.214)--(2.506,3.165)%
  --(2.542,3.114)--(2.578,3.158)--(2.614,3.247)--(2.649,3.122)--(2.685,2.930)--(2.721,3.201)%
  --(2.757,2.982)--(2.793,3.031)--(2.828,3.243)--(2.864,2.989)--(2.900,2.942)--(2.936,3.246)%
  --(2.972,2.986)--(3.007,2.802)--(3.043,3.060)--(3.079,2.923)--(3.115,3.217)--(3.151,2.998)%
  --(3.186,3.120)--(3.222,2.858)--(3.258,2.987)--(3.294,2.900)--(3.330,3.032)--(3.365,2.846)%
  --(3.401,2.870)--(3.437,2.616)--(3.473,2.859)--(3.508,2.936)--(3.544,2.770)--(3.580,2.828)%
  --(3.616,2.910)--(3.652,2.764)--(3.687,2.776)--(3.723,2.772)--(3.759,2.822)--(3.795,2.852)%
  --(3.831,2.647)--(3.866,2.666)--(3.902,2.754)--(3.938,2.703)--(3.974,2.961)--(4.010,2.894)%
  --(4.045,2.684)--(4.081,2.822)--(4.117,2.697)--(4.153,2.607)--(4.189,2.794)--(4.224,2.688)%
  --(4.260,2.962)--(4.296,2.797)--(4.332,2.921)--(4.368,2.775)--(4.403,2.849)--(4.439,2.884)%
  --(4.475,2.786)--(4.511,2.560)--(4.547,2.898)--(4.582,2.534)--(4.618,2.752)--(4.654,2.876)%
  --(4.690,2.624)--(4.726,2.742)--(4.761,2.557)--(4.797,2.697)--(4.833,2.499)--(4.869,2.610)%
  --(4.904,2.675)--(4.940,2.853)--(4.976,2.513)--(5.012,2.630)--(5.048,2.700)--(5.083,2.797)%
  --(5.119,2.623)--(5.155,2.827)--(5.191,2.632)--(5.227,2.847)--(5.262,2.739)--(5.298,2.860)%
  --(5.334,2.609)--(5.370,2.713)--(5.406,2.598)--(5.441,2.768)--(5.477,2.803)--(5.513,2.557)%
  --(5.549,2.764)--(5.585,2.698)--(5.620,2.574)--(5.656,2.838)--(5.692,2.588)--(5.728,2.901)%
  --(5.764,2.622)--(5.799,2.889)--(5.835,2.656)--(5.871,2.838)--(5.907,2.908)--(5.943,2.786)%
  --(5.978,2.876)--(6.014,2.484)--(6.050,2.696)--(6.086,2.621)--(6.121,2.876)--(6.157,2.626)%
  --(6.193,2.615)--(6.229,2.506)--(6.265,2.710)--(6.300,2.669)--(6.336,2.490)--(6.372,2.694)%
  --(6.408,2.828)--(6.444,2.752)--(6.479,2.563)--(6.515,2.604)--(6.551,2.589)--(6.587,2.560)%
  --(6.623,2.585)--(6.658,2.638)--(6.694,2.675)--(6.730,2.710)--(6.766,2.722)--(6.802,2.751)%
  --(6.837,2.824)--(6.873,2.607)--(6.909,2.701)--(6.945,2.544)--(6.981,2.643)--(7.016,2.636)%
  --(7.052,2.470)--(7.088,2.573)--(7.124,2.468)--(7.160,2.788)--(7.195,2.412)--(7.231,2.488)%
  --(7.267,2.487)--(7.303,2.632)--(7.338,2.889)--(7.374,2.592)--(7.410,2.730)--(7.446,2.830)%
  --(7.482,2.646)--(7.517,2.896)--(7.553,2.658)--(7.589,2.520)--(7.625,2.696)--(7.661,2.706)%
  --(7.696,2.617)--(7.732,2.422)--(7.768,2.689)--(7.804,2.462)--(7.840,2.854)--(7.875,2.524)%
  --(7.911,2.549)--(7.947,2.695);
\gpcolor{color=gp lt color border}
\gpsetdashtype{gp dt solid}
\gpsetlinewidth{1.00}
\draw[gp path] (1.504,4.691)--(1.504,0.985)--(7.947,0.985)--(7.947,4.691)--cycle;
\gpdefrectangularnode{gp plot 1}{\pgfpoint{1.504cm}{0.985cm}}{\pgfpoint{7.947cm}{4.691cm}}
\end{tikzpicture}

%% file: figures/cnot_gate/Hessian_eig_1to15.tex
\begin{tikzpicture}[gnuplot]
\tikzset{every node/.append style={font={\footnotesize}}}
\path (0.000,0.000) rectangle (8.500,6.800);
\gpcolor{color=gp lt color axes}
\gpsetlinetype{gp lt axes}
\gpsetdashtype{gp dt axes}
\gpsetlinewidth{0.50}
\draw[gp path] (1.504,0.985)--(7.947,0.985);
\gpcolor{color=gp lt color border}
\gpsetlinetype{gp lt border}
\gpsetdashtype{gp dt solid}
\gpsetlinewidth{1.00}
\draw[gp path] (1.504,0.985)--(1.684,0.985);
\draw[gp path] (7.947,0.985)--(7.767,0.985);
\node[gp node right] at (1.320,0.985) {$10^{-1}$};
\draw[gp path] (1.504,1.214)--(1.594,1.214);
\draw[gp path] (7.947,1.214)--(7.857,1.214);
\draw[gp path] (1.504,1.348)--(1.594,1.348);
\draw[gp path] (7.947,1.348)--(7.857,1.348);
\draw[gp path] (1.504,1.443)--(1.594,1.443);
\draw[gp path] (7.947,1.443)--(7.857,1.443);
\draw[gp path] (1.504,1.517)--(1.594,1.517);
\draw[gp path] (7.947,1.517)--(7.857,1.517);
\draw[gp path] (1.504,1.577)--(1.594,1.577);
\draw[gp path] (7.947,1.577)--(7.857,1.577);
\draw[gp path] (1.504,1.628)--(1.594,1.628);
\draw[gp path] (7.947,1.628)--(7.857,1.628);
\draw[gp path] (1.504,1.672)--(1.594,1.672);
\draw[gp path] (7.947,1.672)--(7.857,1.672);
\draw[gp path] (1.504,1.711)--(1.594,1.711);
\draw[gp path] (7.947,1.711)--(7.857,1.711);
\gpcolor{color=gp lt color axes}
\gpsetlinetype{gp lt axes}
\gpsetdashtype{gp dt axes}
\gpsetlinewidth{0.50}
\draw[gp path] (1.504,1.746)--(7.947,1.746);
\gpcolor{color=gp lt color border}
\gpsetlinetype{gp lt border}
\gpsetdashtype{gp dt solid}
\gpsetlinewidth{1.00}
\draw[gp path] (1.504,1.746)--(1.684,1.746);
\draw[gp path] (7.947,1.746)--(7.767,1.746);
\node[gp node right] at (1.320,1.746) {$10^{0}$};
\draw[gp path] (1.504,1.974)--(1.594,1.974);
\draw[gp path] (7.947,1.974)--(7.857,1.974);
\draw[gp path] (1.504,2.108)--(1.594,2.108);
\draw[gp path] (7.947,2.108)--(7.857,2.108);
\draw[gp path] (1.504,2.203)--(1.594,2.203);
\draw[gp path] (7.947,2.203)--(7.857,2.203);
\draw[gp path] (1.504,2.277)--(1.594,2.277);
\draw[gp path] (7.947,2.277)--(7.857,2.277);
\draw[gp path] (1.504,2.337)--(1.594,2.337);
\draw[gp path] (7.947,2.337)--(7.857,2.337);
\draw[gp path] (1.504,2.388)--(1.594,2.388);
\draw[gp path] (7.947,2.388)--(7.857,2.388);
\draw[gp path] (1.504,2.432)--(1.594,2.432);
\draw[gp path] (7.947,2.432)--(7.857,2.432);
\draw[gp path] (1.504,2.471)--(1.594,2.471);
\draw[gp path] (7.947,2.471)--(7.857,2.471);
\gpcolor{color=gp lt color axes}
\gpsetlinetype{gp lt axes}
\gpsetdashtype{gp dt axes}
\gpsetlinewidth{0.50}
\draw[gp path] (1.504,2.506)--(7.947,2.506);
\gpcolor{color=gp lt color border}
\gpsetlinetype{gp lt border}
\gpsetdashtype{gp dt solid}
\gpsetlinewidth{1.00}
\draw[gp path] (1.504,2.506)--(1.684,2.506);
\draw[gp path] (7.947,2.506)--(7.767,2.506);
\node[gp node right] at (1.320,2.506) {$10^{1}$};
\draw[gp path] (1.504,2.735)--(1.594,2.735);
\draw[gp path] (7.947,2.735)--(7.857,2.735);
\draw[gp path] (1.504,2.869)--(1.594,2.869);
\draw[gp path] (7.947,2.869)--(7.857,2.869);
\draw[gp path] (1.504,2.964)--(1.594,2.964);
\draw[gp path] (7.947,2.964)--(7.857,2.964);
\draw[gp path] (1.504,3.038)--(1.594,3.038);
\draw[gp path] (7.947,3.038)--(7.857,3.038);
\draw[gp path] (1.504,3.098)--(1.594,3.098);
\draw[gp path] (7.947,3.098)--(7.857,3.098);
\draw[gp path] (1.504,3.149)--(1.594,3.149);
\draw[gp path] (7.947,3.149)--(7.857,3.149);
\draw[gp path] (1.504,3.193)--(1.594,3.193);
\draw[gp path] (7.947,3.193)--(7.857,3.193);
\draw[gp path] (1.504,3.232)--(1.594,3.232);
\draw[gp path] (7.947,3.232)--(7.857,3.232);
\gpcolor{color=gp lt color axes}
\gpsetlinetype{gp lt axes}
\gpsetdashtype{gp dt axes}
\gpsetlinewidth{0.50}
\draw[gp path] (1.504,3.267)--(7.947,3.267);
\gpcolor{color=gp lt color border}
\gpsetlinetype{gp lt border}
\gpsetdashtype{gp dt solid}
\gpsetlinewidth{1.00}
\draw[gp path] (1.504,3.267)--(1.684,3.267);
\draw[gp path] (7.947,3.267)--(7.767,3.267);
\node[gp node right] at (1.320,3.267) {$10^{2}$};
\draw[gp path] (1.504,3.495)--(1.594,3.495);
\draw[gp path] (7.947,3.495)--(7.857,3.495);
\draw[gp path] (1.504,3.629)--(1.594,3.629);
\draw[gp path] (7.947,3.629)--(7.857,3.629);
\draw[gp path] (1.504,3.724)--(1.594,3.724);
\draw[gp path] (7.947,3.724)--(7.857,3.724);
\draw[gp path] (1.504,3.798)--(1.594,3.798);
\draw[gp path] (7.947,3.798)--(7.857,3.798);
\draw[gp path] (1.504,3.858)--(1.594,3.858);
\draw[gp path] (7.947,3.858)--(7.857,3.858);
\draw[gp path] (1.504,3.909)--(1.594,3.909);
\draw[gp path] (7.947,3.909)--(7.857,3.909);
\draw[gp path] (1.504,3.953)--(1.594,3.953);
\draw[gp path] (7.947,3.953)--(7.857,3.953);
\draw[gp path] (1.504,3.992)--(1.594,3.992);
\draw[gp path] (7.947,3.992)--(7.857,3.992);
\gpcolor{color=gp lt color axes}
\gpsetlinetype{gp lt axes}
\gpsetdashtype{gp dt axes}
\gpsetlinewidth{0.50}
\draw[gp path] (1.504,4.027)--(7.947,4.027);
\gpcolor{color=gp lt color border}
\gpsetlinetype{gp lt border}
\gpsetdashtype{gp dt solid}
\gpsetlinewidth{1.00}
\draw[gp path] (1.504,4.027)--(1.684,4.027);
\draw[gp path] (7.947,4.027)--(7.767,4.027);
\node[gp node right] at (1.320,4.027) {$10^{3}$};
\gpcolor{color=gp lt color axes}
\gpsetlinetype{gp lt axes}
\gpsetdashtype{gp dt axes}
\gpsetlinewidth{0.50}
\draw[gp path] (1.504,0.985)--(1.504,4.027);
\gpcolor{color=gp lt color border}
\gpsetlinetype{gp lt border}
\gpsetdashtype{gp dt solid}
\gpsetlinewidth{1.00}
\draw[gp path] (1.504,0.985)--(1.504,1.165);
\draw[gp path] (1.504,4.027)--(1.504,3.847);
\node[gp node center] at (1.504,0.677) {$0$};
\gpcolor{color=gp lt color axes}
\gpsetlinetype{gp lt axes}
\gpsetdashtype{gp dt axes}
\gpsetlinewidth{0.50}
\draw[gp path] (2.424,0.985)--(2.424,4.027);
\gpcolor{color=gp lt color border}
\gpsetlinetype{gp lt border}
\gpsetdashtype{gp dt solid}
\gpsetlinewidth{1.00}
\draw[gp path] (2.424,0.985)--(2.424,1.165);
\draw[gp path] (2.424,4.027)--(2.424,3.847);
\node[gp node center] at (2.424,0.677) {$2$};
\gpcolor{color=gp lt color axes}
\gpsetlinetype{gp lt axes}
\gpsetdashtype{gp dt axes}
\gpsetlinewidth{0.50}
\draw[gp path] (3.345,0.985)--(3.345,4.027);
\gpcolor{color=gp lt color border}
\gpsetlinetype{gp lt border}
\gpsetdashtype{gp dt solid}
\gpsetlinewidth{1.00}
\draw[gp path] (3.345,0.985)--(3.345,1.165);
\draw[gp path] (3.345,4.027)--(3.345,3.847);
\node[gp node center] at (3.345,0.677) {$4$};
\gpcolor{color=gp lt color axes}
\gpsetlinetype{gp lt axes}
\gpsetdashtype{gp dt axes}
\gpsetlinewidth{0.50}
\draw[gp path] (4.265,0.985)--(4.265,4.027);
\gpcolor{color=gp lt color border}
\gpsetlinetype{gp lt border}
\gpsetdashtype{gp dt solid}
\gpsetlinewidth{1.00}
\draw[gp path] (4.265,0.985)--(4.265,1.165);
\draw[gp path] (4.265,4.027)--(4.265,3.847);
\node[gp node center] at (4.265,0.677) {$6$};
\gpcolor{color=gp lt color axes}
\gpsetlinetype{gp lt axes}
\gpsetdashtype{gp dt axes}
\gpsetlinewidth{0.50}
\draw[gp path] (5.186,0.985)--(5.186,4.027);
\gpcolor{color=gp lt color border}
\gpsetlinetype{gp lt border}
\gpsetdashtype{gp dt solid}
\gpsetlinewidth{1.00}
\draw[gp path] (5.186,0.985)--(5.186,1.165);
\draw[gp path] (5.186,4.027)--(5.186,3.847);
\node[gp node center] at (5.186,0.677) {$8$};
\gpcolor{color=gp lt color axes}
\gpsetlinetype{gp lt axes}
\gpsetdashtype{gp dt axes}
\gpsetlinewidth{0.50}
\draw[gp path] (6.106,0.985)--(6.106,4.027);
\gpcolor{color=gp lt color border}
\gpsetlinetype{gp lt border}
\gpsetdashtype{gp dt solid}
\gpsetlinewidth{1.00}
\draw[gp path] (6.106,0.985)--(6.106,1.165);
\draw[gp path] (6.106,4.027)--(6.106,3.847);
\node[gp node center] at (6.106,0.677) {$10$};
\gpcolor{color=gp lt color axes}
\gpsetlinetype{gp lt axes}
\gpsetdashtype{gp dt axes}
\gpsetlinewidth{0.50}
\draw[gp path] (7.027,0.985)--(7.027,4.027);
\gpcolor{color=gp lt color border}
\gpsetlinetype{gp lt border}
\gpsetdashtype{gp dt solid}
\gpsetlinewidth{1.00}
\draw[gp path] (7.027,0.985)--(7.027,1.165);
\draw[gp path] (7.027,4.027)--(7.027,3.847);
\node[gp node center] at (7.027,0.677) {$12$};
\gpcolor{color=gp lt color axes}
\gpsetlinetype{gp lt axes}
\gpsetdashtype{gp dt axes}
\gpsetlinewidth{0.50}
\draw[gp path] (7.947,0.985)--(7.947,4.027);
\gpcolor{color=gp lt color border}
\gpsetlinetype{gp lt border}
\gpsetdashtype{gp dt solid}
\gpsetlinewidth{1.00}
\draw[gp path] (7.947,0.985)--(7.947,1.165);
\draw[gp path] (7.947,4.027)--(7.947,3.847);
\node[gp node center] at (7.947,0.677) {$14$};
\draw[gp path] (1.504,4.027)--(1.504,0.985)--(7.947,0.985)--(7.947,4.027)--cycle;
\node[gp node center,rotate=-270] at (0.292,2.506) {eigenvalue};
\node[gp node center] at (4.725,0.215) {index};
\node[gp node right] at (6.291,6.466) {basis, $J_{Frob}$};
\gpcolor{rgb color={0.580,0.000,0.827}}
\gpsetpointsize{8.00}
\gp3point{gp mark 6}{}{(1.504,3.646)}
\gp3point{gp mark 6}{}{(1.964,3.633)}
\gp3point{gp mark 6}{}{(2.424,3.609)}
\gp3point{gp mark 6}{}{(2.885,3.581)}
\gp3point{gp mark 6}{}{(3.345,3.500)}
\gp3point{gp mark 6}{}{(3.805,3.493)}
\gp3point{gp mark 6}{}{(4.265,3.454)}
\gp3point{gp mark 6}{}{(4.726,3.397)}
\gp3point{gp mark 6}{}{(5.186,3.381)}
\gp3point{gp mark 6}{}{(5.646,3.356)}
\gp3point{gp mark 6}{}{(6.106,3.333)}
\gp3point{gp mark 6}{}{(6.566,3.265)}
\gp3point{gp mark 6}{}{(7.027,3.235)}
\gp3point{gp mark 6}{}{(7.487,3.153)}
\gp3point{gp mark 6}{}{(7.947,3.087)}
\gp3point{gp mark 6}{}{(6.933,6.466)}
\gpcolor{color=gp lt color border}
\node[gp node right] at (6.291,6.158) {3 states, $J_{Frob}$};
\gpcolor{rgb color={0.000,0.620,0.451}}
\gp3point{gp mark 6}{}{(1.504,3.489)}
\gp3point{gp mark 6}{}{(1.964,3.478)}
\gp3point{gp mark 6}{}{(2.424,3.448)}
\gp3point{gp mark 6}{}{(2.885,3.407)}
\gp3point{gp mark 6}{}{(3.345,3.338)}
\gp3point{gp mark 6}{}{(3.805,3.252)}
\gp3point{gp mark 6}{}{(4.265,2.430)}
\gp3point{gp mark 6}{}{(4.726,2.408)}
\gp3point{gp mark 6}{}{(5.186,2.258)}
\gp3point{gp mark 6}{}{(5.646,2.219)}
\gp3point{gp mark 6}{}{(6.106,1.999)}
\gp3point{gp mark 6}{}{(6.566,1.939)}
\gp3point{gp mark 6}{}{(7.027,1.905)}
\gp3point{gp mark 6}{}{(7.487,1.600)}
\gp3point{gp mark 6}{}{(7.947,1.396)}
\gp3point{gp mark 6}{}{(6.933,6.158)}
\gpcolor{color=gp lt color border}
\node[gp node right] at (6.291,5.850) {3 states weighted, $J_{Frob}$};
\gpcolor{rgb color={0.902,0.624,0.000}}
\gp3point{gp mark 6}{}{(1.504,3.357)}
\gp3point{gp mark 6}{}{(1.964,3.324)}
\gp3point{gp mark 6}{}{(2.424,3.126)}
\gp3point{gp mark 6}{}{(2.885,3.097)}
\gp3point{gp mark 6}{}{(3.345,3.044)}
\gp3point{gp mark 6}{}{(3.805,3.024)}
\gp3point{gp mark 6}{}{(4.265,2.934)}
\gp3point{gp mark 6}{}{(4.726,2.834)}
\gp3point{gp mark 6}{}{(5.186,2.765)}
\gp3point{gp mark 6}{}{(5.646,2.704)}
\gp3point{gp mark 6}{}{(6.106,2.665)}
\gp3point{gp mark 6}{}{(6.566,2.583)}
\gp3point{gp mark 6}{}{(7.027,2.302)}
\gp3point{gp mark 6}{}{(7.487,2.271)}
\gp3point{gp mark 6}{}{(7.947,1.935)}
\gp3point{gp mark 6}{}{(6.933,5.850)}
\gpcolor{color=gp lt color border}
\node[gp node right] at (6.291,5.542) {basis, $J_{Tr}$};
\gpcolor{rgb color={0.580,0.000,0.827}}
\gp3point{gp mark 2}{}{(1.504,3.645)}
\gp3point{gp mark 2}{}{(1.964,3.631)}
\gp3point{gp mark 2}{}{(2.424,3.612)}
\gp3point{gp mark 2}{}{(2.885,3.546)}
\gp3point{gp mark 2}{}{(3.345,3.524)}
\gp3point{gp mark 2}{}{(3.805,3.483)}
\gp3point{gp mark 2}{}{(4.265,3.465)}
\gp3point{gp mark 2}{}{(4.726,3.421)}
\gp3point{gp mark 2}{}{(5.186,3.390)}
\gp3point{gp mark 2}{}{(5.646,3.386)}
\gp3point{gp mark 2}{}{(6.106,3.317)}
\gp3point{gp mark 2}{}{(6.566,3.248)}
\gp3point{gp mark 2}{}{(7.027,3.176)}
\gp3point{gp mark 2}{}{(7.487,3.160)}
\gp3point{gp mark 2}{}{(7.947,3.064)}
\gp3point{gp mark 2}{}{(6.933,5.542)}
\gpcolor{color=gp lt color border}
\node[gp node right] at (6.291,5.234) {3 states, $J_{Tr}$};
\gpcolor{rgb color={0.000,0.620,0.451}}
\gp3point{gp mark 2}{}{(1.504,3.483)}
\gp3point{gp mark 2}{}{(1.964,3.457)}
\gp3point{gp mark 2}{}{(2.424,3.449)}
\gp3point{gp mark 2}{}{(2.885,3.348)}
\gp3point{gp mark 2}{}{(3.345,3.318)}
\gp3point{gp mark 2}{}{(3.805,3.253)}
\gp3point{gp mark 2}{}{(4.265,2.447)}
\gp3point{gp mark 2}{}{(4.726,2.433)}
\gp3point{gp mark 2}{}{(5.186,2.250)}
\gp3point{gp mark 2}{}{(5.646,2.177)}
\gp3point{gp mark 2}{}{(6.106,2.055)}
\gp3point{gp mark 2}{}{(6.566,1.984)}
\gp3point{gp mark 2}{}{(7.027,1.819)}
\gp3point{gp mark 2}{}{(7.487,1.695)}
\gp3point{gp mark 2}{}{(7.947,1.510)}
\gp3point{gp mark 2}{}{(6.933,5.234)}
\gpcolor{color=gp lt color border}
\node[gp node right] at (6.291,4.926) {3 states weighted, $J_{Tr}$};
\gpcolor{rgb color={0.902,0.624,0.000}}
\gp3point{gp mark 2}{}{(1.504,3.360)}
\gp3point{gp mark 2}{}{(1.964,3.345)}
\gp3point{gp mark 2}{}{(2.424,3.132)}
\gp3point{gp mark 2}{}{(2.885,3.105)}
\gp3point{gp mark 2}{}{(3.345,3.088)}
\gp3point{gp mark 2}{}{(3.805,3.046)}
\gp3point{gp mark 2}{}{(4.265,2.904)}
\gp3point{gp mark 2}{}{(4.726,2.870)}
\gp3point{gp mark 2}{}{(5.186,2.745)}
\gp3point{gp mark 2}{}{(5.646,2.727)}
\gp3point{gp mark 2}{}{(6.106,2.632)}
\gp3point{gp mark 2}{}{(6.566,2.583)}
\gp3point{gp mark 2}{}{(7.027,2.350)}
\gp3point{gp mark 2}{}{(7.487,2.308)}
\gp3point{gp mark 2}{}{(7.947,1.959)}
\gp3point{gp mark 2}{}{(6.933,4.926)}
\gpcolor{color=gp lt color border}
\draw[gp path] (1.504,4.027)--(1.504,0.985)--(7.947,0.985)--(7.947,4.027)--cycle;
\node[gp node center] at (4.725,4.489) {first 15 eigenvalues};
\gpdefrectangularnode{gp plot 1}{\pgfpoint{1.504cm}{0.985cm}}{\pgfpoint{7.947cm}{4.027cm}}
\end{tikzpicture}

%% file: figures/cnot_gate/Hessian_eig_16to200.tex
\begin{tikzpicture}[gnuplot]
\tikzset{every node/.append style={font={\footnotesize}}}
\path (0.000,0.000) rectangle (8.500,5.000);
\gpcolor{color=gp lt color axes}
\gpsetlinetype{gp lt axes}
\gpsetdashtype{gp dt axes}
\gpsetlinewidth{0.50}
\draw[gp path] (1.688,1.178)--(7.947,1.178);
\gpcolor{color=gp lt color border}
\gpsetlinetype{gp lt border}
\gpsetdashtype{gp dt solid}
\gpsetlinewidth{1.00}
\draw[gp path] (1.688,1.178)--(1.868,1.178);
\draw[gp path] (7.947,1.178)--(7.767,1.178);
\node[gp node right] at (1.504,1.178) {-0.04};
\gpcolor{color=gp lt color axes}
\gpsetlinetype{gp lt axes}
\gpsetdashtype{gp dt axes}
\gpsetlinewidth{0.50}
\draw[gp path] (1.688,1.564)--(7.947,1.564);
\gpcolor{color=gp lt color border}
\gpsetlinetype{gp lt border}
\gpsetdashtype{gp dt solid}
\gpsetlinewidth{1.00}
\draw[gp path] (1.688,1.564)--(1.868,1.564);
\draw[gp path] (7.947,1.564)--(7.767,1.564);
\node[gp node right] at (1.504,1.564) {-0.02};
\gpcolor{color=gp lt color axes}
\gpsetlinetype{gp lt axes}
\gpsetdashtype{gp dt axes}
\gpsetlinewidth{0.50}
\draw[gp path] (1.688,1.951)--(7.947,1.951);
\gpcolor{color=gp lt color border}
\gpsetlinetype{gp lt border}
\gpsetdashtype{gp dt solid}
\gpsetlinewidth{1.00}
\draw[gp path] (1.688,1.951)--(1.868,1.951);
\draw[gp path] (7.947,1.951)--(7.767,1.951);
\node[gp node right] at (1.504,1.951) {0};
\gpcolor{color=gp lt color axes}
\gpsetlinetype{gp lt axes}
\gpsetdashtype{gp dt axes}
\gpsetlinewidth{0.50}
\draw[gp path] (1.688,2.337)--(7.947,2.337);
\gpcolor{color=gp lt color border}
\gpsetlinetype{gp lt border}
\gpsetdashtype{gp dt solid}
\gpsetlinewidth{1.00}
\draw[gp path] (1.688,2.337)--(1.868,2.337);
\draw[gp path] (7.947,2.337)--(7.767,2.337);
\node[gp node right] at (1.504,2.337) {0.02};
\gpcolor{color=gp lt color axes}
\gpsetlinetype{gp lt axes}
\gpsetdashtype{gp dt axes}
\gpsetlinewidth{0.50}
\draw[gp path] (1.688,2.723)--(7.947,2.723);
\gpcolor{color=gp lt color border}
\gpsetlinetype{gp lt border}
\gpsetdashtype{gp dt solid}
\gpsetlinewidth{1.00}
\draw[gp path] (1.688,2.723)--(1.868,2.723);
\draw[gp path] (7.947,2.723)--(7.767,2.723);
\node[gp node right] at (1.504,2.723) {0.04};
\gpcolor{color=gp lt color axes}
\gpsetlinetype{gp lt axes}
\gpsetdashtype{gp dt axes}
\gpsetlinewidth{0.50}
\draw[gp path] (1.688,3.109)--(7.947,3.109);
\gpcolor{color=gp lt color border}
\gpsetlinetype{gp lt border}
\gpsetdashtype{gp dt solid}
\gpsetlinewidth{1.00}
\draw[gp path] (1.688,3.109)--(1.868,3.109);
\draw[gp path] (7.947,3.109)--(7.767,3.109);
\node[gp node right] at (1.504,3.109) {0.06};
\gpcolor{color=gp lt color axes}
\gpsetlinetype{gp lt axes}
\gpsetdashtype{gp dt axes}
\gpsetlinewidth{0.50}
\draw[gp path] (1.688,3.496)--(7.947,3.496);
\gpcolor{color=gp lt color border}
\gpsetlinetype{gp lt border}
\gpsetdashtype{gp dt solid}
\gpsetlinewidth{1.00}
\draw[gp path] (1.688,3.496)--(1.868,3.496);
\draw[gp path] (7.947,3.496)--(7.767,3.496);
\node[gp node right] at (1.504,3.496) {0.08};
\gpcolor{color=gp lt color axes}
\gpsetlinetype{gp lt axes}
\gpsetdashtype{gp dt axes}
\gpsetlinewidth{0.50}
\draw[gp path] (1.688,3.882)--(7.947,3.882);
\gpcolor{color=gp lt color border}
\gpsetlinetype{gp lt border}
\gpsetdashtype{gp dt solid}
\gpsetlinewidth{1.00}
\draw[gp path] (1.688,3.882)--(1.868,3.882);
\draw[gp path] (7.947,3.882)--(7.767,3.882);
\node[gp node right] at (1.504,3.882) {0.1};
\gpcolor{color=gp lt color axes}
\gpsetlinetype{gp lt axes}
\gpsetdashtype{gp dt axes}
\gpsetlinewidth{0.50}
\draw[gp path] (1.688,0.985)--(1.688,4.075);
\gpcolor{color=gp lt color border}
\gpsetlinetype{gp lt border}
\gpsetdashtype{gp dt solid}
\gpsetlinewidth{1.00}
\draw[gp path] (1.688,0.985)--(1.688,1.165);
\draw[gp path] (1.688,4.075)--(1.688,3.895);
\node[gp node center] at (1.688,0.677) {$0$};
\gpcolor{color=gp lt color axes}
\gpsetlinetype{gp lt axes}
\gpsetdashtype{gp dt axes}
\gpsetlinewidth{0.50}
\draw[gp path] (3.253,0.985)--(3.253,4.075);
\gpcolor{color=gp lt color border}
\gpsetlinetype{gp lt border}
\gpsetdashtype{gp dt solid}
\gpsetlinewidth{1.00}
\draw[gp path] (3.253,0.985)--(3.253,1.165);
\draw[gp path] (3.253,4.075)--(3.253,3.895);
\node[gp node center] at (3.253,0.677) {$50$};
\gpcolor{color=gp lt color axes}
\gpsetlinetype{gp lt axes}
\gpsetdashtype{gp dt axes}
\gpsetlinewidth{0.50}
\draw[gp path] (4.818,0.985)--(4.818,4.075);
\gpcolor{color=gp lt color border}
\gpsetlinetype{gp lt border}
\gpsetdashtype{gp dt solid}
\gpsetlinewidth{1.00}
\draw[gp path] (4.818,0.985)--(4.818,1.165);
\draw[gp path] (4.818,4.075)--(4.818,3.895);
\node[gp node center] at (4.818,0.677) {$100$};
\gpcolor{color=gp lt color axes}
\gpsetlinetype{gp lt axes}
\gpsetdashtype{gp dt axes}
\gpsetlinewidth{0.50}
\draw[gp path] (6.382,0.985)--(6.382,4.075);
\gpcolor{color=gp lt color border}
\gpsetlinetype{gp lt border}
\gpsetdashtype{gp dt solid}
\gpsetlinewidth{1.00}
\draw[gp path] (6.382,0.985)--(6.382,1.165);
\draw[gp path] (6.382,4.075)--(6.382,3.895);
\node[gp node center] at (6.382,0.677) {$150$};
\gpcolor{color=gp lt color axes}
\gpsetlinetype{gp lt axes}
\gpsetdashtype{gp dt axes}
\gpsetlinewidth{0.50}
\draw[gp path] (7.947,0.985)--(7.947,4.075);
\gpcolor{color=gp lt color border}
\gpsetlinetype{gp lt border}
\gpsetdashtype{gp dt solid}
\gpsetlinewidth{1.00}
\draw[gp path] (7.947,0.985)--(7.947,1.165);
\draw[gp path] (7.947,4.075)--(7.947,3.895);
\node[gp node center] at (7.947,0.677) {$200$};
\draw[gp path] (1.688,4.075)--(1.688,0.985)--(7.947,0.985)--(7.947,4.075)--cycle;
\node[gp node center,rotate=-270] at (0.292,2.530) {eigenvalue};
\node[gp node center] at (4.817,0.215) {index};
\gpcolor{rgb color={0.580,0.000,0.827}}
\gpsetpointsize{8.00}
\gp3point{gp mark 6}{}{(2.157,1.952)}
\gp3point{gp mark 6}{}{(2.189,1.952)}
\gp3point{gp mark 6}{}{(2.220,1.950)}
\gp3point{gp mark 6}{}{(2.251,1.952)}
\gp3point{gp mark 6}{}{(2.283,1.950)}
\gp3point{gp mark 6}{}{(2.314,1.951)}
\gp3point{gp mark 6}{}{(2.345,1.951)}
\gp3point{gp mark 6}{}{(2.376,1.950)}
\gp3point{gp mark 6}{}{(2.408,1.951)}
\gp3point{gp mark 6}{}{(2.439,1.951)}
\gp3point{gp mark 6}{}{(2.470,1.950)}
\gp3point{gp mark 6}{}{(2.502,1.951)}
\gp3point{gp mark 6}{}{(2.533,1.951)}
\gp3point{gp mark 6}{}{(2.564,1.951)}
\gp3point{gp mark 6}{}{(2.596,1.951)}
\gp3point{gp mark 6}{}{(2.627,1.950)}
\gp3point{gp mark 6}{}{(2.658,1.950)}
\gp3point{gp mark 6}{}{(2.689,1.951)}
\gp3point{gp mark 6}{}{(2.721,1.951)}
\gp3point{gp mark 6}{}{(2.752,1.951)}
\gp3point{gp mark 6}{}{(2.783,1.950)}
\gp3point{gp mark 6}{}{(2.815,1.950)}
\gp3point{gp mark 6}{}{(2.846,1.950)}
\gp3point{gp mark 6}{}{(2.877,1.951)}
\gp3point{gp mark 6}{}{(2.909,1.951)}
\gp3point{gp mark 6}{}{(2.940,1.951)}
\gp3point{gp mark 6}{}{(2.971,1.950)}
\gp3point{gp mark 6}{}{(3.002,1.950)}
\gp3point{gp mark 6}{}{(3.034,1.951)}
\gp3point{gp mark 6}{}{(3.065,1.951)}
\gp3point{gp mark 6}{}{(3.096,1.950)}
\gp3point{gp mark 6}{}{(3.128,1.951)}
\gp3point{gp mark 6}{}{(3.159,1.951)}
\gp3point{gp mark 6}{}{(3.190,1.950)}
\gp3point{gp mark 6}{}{(3.221,1.950)}
\gp3point{gp mark 6}{}{(3.253,1.951)}
\gp3point{gp mark 6}{}{(3.284,1.950)}
\gp3point{gp mark 6}{}{(3.315,1.950)}
\gp3point{gp mark 6}{}{(3.347,1.950)}
\gp3point{gp mark 6}{}{(3.378,1.951)}
\gp3point{gp mark 6}{}{(3.409,1.950)}
\gp3point{gp mark 6}{}{(3.441,1.951)}
\gp3point{gp mark 6}{}{(3.472,1.950)}
\gp3point{gp mark 6}{}{(3.503,1.950)}
\gp3point{gp mark 6}{}{(3.534,1.951)}
\gp3point{gp mark 6}{}{(3.566,1.950)}
\gp3point{gp mark 6}{}{(3.597,1.951)}
\gp3point{gp mark 6}{}{(3.628,1.950)}
\gp3point{gp mark 6}{}{(3.660,1.951)}
\gp3point{gp mark 6}{}{(3.691,1.951)}
\gp3point{gp mark 6}{}{(3.722,1.951)}
\gp3point{gp mark 6}{}{(3.753,1.951)}
\gp3point{gp mark 6}{}{(3.785,1.951)}
\gp3point{gp mark 6}{}{(3.816,1.951)}
\gp3point{gp mark 6}{}{(3.847,1.951)}
\gp3point{gp mark 6}{}{(3.879,1.951)}
\gp3point{gp mark 6}{}{(3.910,1.951)}
\gp3point{gp mark 6}{}{(3.941,1.951)}
\gp3point{gp mark 6}{}{(3.973,1.951)}
\gp3point{gp mark 6}{}{(4.004,1.951)}
\gp3point{gp mark 6}{}{(4.035,1.951)}
\gp3point{gp mark 6}{}{(4.066,1.951)}
\gp3point{gp mark 6}{}{(4.098,1.951)}
\gp3point{gp mark 6}{}{(4.129,1.951)}
\gp3point{gp mark 6}{}{(4.160,1.951)}
\gp3point{gp mark 6}{}{(4.192,1.951)}
\gp3point{gp mark 6}{}{(4.223,1.951)}
\gp3point{gp mark 6}{}{(4.254,1.951)}
\gp3point{gp mark 6}{}{(4.285,1.951)}
\gp3point{gp mark 6}{}{(4.317,1.951)}
\gp3point{gp mark 6}{}{(4.348,1.951)}
\gp3point{gp mark 6}{}{(4.379,1.951)}
\gp3point{gp mark 6}{}{(4.411,1.951)}
\gp3point{gp mark 6}{}{(4.442,1.951)}
\gp3point{gp mark 6}{}{(4.473,1.951)}
\gp3point{gp mark 6}{}{(4.505,1.951)}
\gp3point{gp mark 6}{}{(4.536,1.951)}
\gp3point{gp mark 6}{}{(4.567,1.951)}
\gp3point{gp mark 6}{}{(4.598,1.951)}
\gp3point{gp mark 6}{}{(4.630,1.951)}
\gp3point{gp mark 6}{}{(4.661,1.951)}
\gp3point{gp mark 6}{}{(4.692,1.951)}
\gp3point{gp mark 6}{}{(4.724,1.951)}
\gp3point{gp mark 6}{}{(4.755,1.951)}
\gp3point{gp mark 6}{}{(4.786,1.951)}
\gp3point{gp mark 6}{}{(4.818,1.951)}
\gp3point{gp mark 6}{}{(4.849,1.951)}
\gp3point{gp mark 6}{}{(4.880,1.951)}
\gp3point{gp mark 6}{}{(4.911,1.951)}
\gp3point{gp mark 6}{}{(4.943,1.951)}
\gp3point{gp mark 6}{}{(4.974,1.951)}
\gp3point{gp mark 6}{}{(5.005,1.951)}
\gp3point{gp mark 6}{}{(5.037,1.951)}
\gp3point{gp mark 6}{}{(5.068,1.951)}
\gp3point{gp mark 6}{}{(5.099,1.951)}
\gp3point{gp mark 6}{}{(5.130,1.951)}
\gp3point{gp mark 6}{}{(5.162,1.951)}
\gp3point{gp mark 6}{}{(5.193,1.951)}
\gp3point{gp mark 6}{}{(5.224,1.951)}
\gp3point{gp mark 6}{}{(5.256,1.951)}
\gp3point{gp mark 6}{}{(5.287,1.951)}
\gp3point{gp mark 6}{}{(5.318,1.951)}
\gp3point{gp mark 6}{}{(5.350,1.951)}
\gp3point{gp mark 6}{}{(5.381,1.951)}
\gp3point{gp mark 6}{}{(5.412,1.951)}
\gp3point{gp mark 6}{}{(5.443,1.951)}
\gp3point{gp mark 6}{}{(5.475,1.951)}
\gp3point{gp mark 6}{}{(5.506,1.951)}
\gp3point{gp mark 6}{}{(5.537,1.951)}
\gp3point{gp mark 6}{}{(5.569,1.951)}
\gp3point{gp mark 6}{}{(5.600,1.951)}
\gp3point{gp mark 6}{}{(5.631,1.951)}
\gp3point{gp mark 6}{}{(5.662,1.951)}
\gp3point{gp mark 6}{}{(5.694,1.951)}
\gp3point{gp mark 6}{}{(5.725,1.951)}
\gp3point{gp mark 6}{}{(5.756,1.951)}
\gp3point{gp mark 6}{}{(5.788,1.951)}
\gp3point{gp mark 6}{}{(5.819,1.951)}
\gp3point{gp mark 6}{}{(5.850,1.951)}
\gp3point{gp mark 6}{}{(5.882,1.951)}
\gp3point{gp mark 6}{}{(5.913,1.951)}
\gp3point{gp mark 6}{}{(5.944,1.951)}
\gp3point{gp mark 6}{}{(5.975,1.951)}
\gp3point{gp mark 6}{}{(6.007,1.951)}
\gp3point{gp mark 6}{}{(6.038,1.951)}
\gp3point{gp mark 6}{}{(6.069,1.951)}
\gp3point{gp mark 6}{}{(6.101,1.951)}
\gp3point{gp mark 6}{}{(6.132,1.951)}
\gp3point{gp mark 6}{}{(6.163,1.951)}
\gp3point{gp mark 6}{}{(6.194,1.951)}
\gp3point{gp mark 6}{}{(6.226,1.951)}
\gp3point{gp mark 6}{}{(6.257,1.951)}
\gp3point{gp mark 6}{}{(6.288,1.951)}
\gp3point{gp mark 6}{}{(6.320,1.951)}
\gp3point{gp mark 6}{}{(6.351,1.951)}
\gp3point{gp mark 6}{}{(6.382,1.951)}
\gp3point{gp mark 6}{}{(6.414,1.951)}
\gp3point{gp mark 6}{}{(6.445,1.951)}
\gp3point{gp mark 6}{}{(6.476,1.951)}
\gp3point{gp mark 6}{}{(6.507,1.951)}
\gp3point{gp mark 6}{}{(6.539,1.951)}
\gp3point{gp mark 6}{}{(6.570,1.951)}
\gp3point{gp mark 6}{}{(6.601,1.951)}
\gp3point{gp mark 6}{}{(6.633,1.951)}
\gp3point{gp mark 6}{}{(6.664,1.951)}
\gp3point{gp mark 6}{}{(6.695,1.951)}
\gp3point{gp mark 6}{}{(6.726,1.951)}
\gp3point{gp mark 6}{}{(6.758,1.951)}
\gp3point{gp mark 6}{}{(6.789,1.951)}
\gp3point{gp mark 6}{}{(6.820,1.951)}
\gp3point{gp mark 6}{}{(6.852,1.951)}
\gp3point{gp mark 6}{}{(6.883,1.951)}
\gp3point{gp mark 6}{}{(6.914,1.951)}
\gp3point{gp mark 6}{}{(6.946,1.951)}
\gp3point{gp mark 6}{}{(6.977,1.951)}
\gp3point{gp mark 6}{}{(7.008,1.951)}
\gp3point{gp mark 6}{}{(7.039,1.951)}
\gp3point{gp mark 6}{}{(7.071,1.951)}
\gp3point{gp mark 6}{}{(7.102,1.951)}
\gp3point{gp mark 6}{}{(7.133,1.951)}
\gp3point{gp mark 6}{}{(7.165,1.951)}
\gp3point{gp mark 6}{}{(7.196,1.951)}
\gp3point{gp mark 6}{}{(7.227,1.951)}
\gp3point{gp mark 6}{}{(7.259,1.951)}
\gp3point{gp mark 6}{}{(7.290,1.951)}
\gp3point{gp mark 6}{}{(7.321,1.951)}
\gp3point{gp mark 6}{}{(7.352,1.951)}
\gp3point{gp mark 6}{}{(7.384,1.951)}
\gp3point{gp mark 6}{}{(7.415,1.951)}
\gp3point{gp mark 6}{}{(7.446,1.951)}
\gp3point{gp mark 6}{}{(7.478,1.951)}
\gp3point{gp mark 6}{}{(7.509,1.951)}
\gp3point{gp mark 6}{}{(7.540,1.951)}
\gp3point{gp mark 6}{}{(7.571,1.951)}
\gp3point{gp mark 6}{}{(7.603,1.951)}
\gp3point{gp mark 6}{}{(7.634,1.951)}
\gp3point{gp mark 6}{}{(7.665,1.951)}
\gp3point{gp mark 6}{}{(7.697,1.951)}
\gp3point{gp mark 6}{}{(7.728,1.951)}
\gp3point{gp mark 6}{}{(7.759,1.951)}
\gp3point{gp mark 6}{}{(7.791,1.951)}
\gp3point{gp mark 6}{}{(7.822,1.951)}
\gp3point{gp mark 6}{}{(7.853,1.951)}
\gp3point{gp mark 6}{}{(7.884,1.951)}
\gp3point{gp mark 6}{}{(7.916,1.951)}
\gp3point{gp mark 6}{}{(7.947,1.951)}
\gpcolor{rgb color={0.000,0.620,0.451}}
\gp3point{gp mark 6}{}{(2.157,1.946)}
\gp3point{gp mark 6}{}{(2.189,1.948)}
\gp3point{gp mark 6}{}{(2.220,1.948)}
\gp3point{gp mark 6}{}{(2.251,1.953)}
\gp3point{gp mark 6}{}{(2.283,1.948)}
\gp3point{gp mark 6}{}{(2.314,1.953)}
\gp3point{gp mark 6}{}{(2.345,1.949)}
\gp3point{gp mark 6}{}{(2.376,1.952)}
\gp3point{gp mark 6}{}{(2.408,1.949)}
\gp3point{gp mark 6}{}{(2.439,1.952)}
\gp3point{gp mark 6}{}{(2.470,1.949)}
\gp3point{gp mark 6}{}{(2.502,1.952)}
\gp3point{gp mark 6}{}{(2.533,1.949)}
\gp3point{gp mark 6}{}{(2.564,1.952)}
\gp3point{gp mark 6}{}{(2.596,1.949)}
\gp3point{gp mark 6}{}{(2.627,1.952)}
\gp3point{gp mark 6}{}{(2.658,1.950)}
\gp3point{gp mark 6}{}{(2.689,1.952)}
\gp3point{gp mark 6}{}{(2.721,1.950)}
\gp3point{gp mark 6}{}{(2.752,1.952)}
\gp3point{gp mark 6}{}{(2.783,1.952)}
\gp3point{gp mark 6}{}{(2.815,1.952)}
\gp3point{gp mark 6}{}{(2.846,1.950)}
\gp3point{gp mark 6}{}{(2.877,1.950)}
\gp3point{gp mark 6}{}{(2.909,1.951)}
\gp3point{gp mark 6}{}{(2.940,1.951)}
\gp3point{gp mark 6}{}{(2.971,1.950)}
\gp3point{gp mark 6}{}{(3.002,1.951)}
\gp3point{gp mark 6}{}{(3.034,1.950)}
\gp3point{gp mark 6}{}{(3.065,1.951)}
\gp3point{gp mark 6}{}{(3.096,1.950)}
\gp3point{gp mark 6}{}{(3.128,1.951)}
\gp3point{gp mark 6}{}{(3.159,1.951)}
\gp3point{gp mark 6}{}{(3.190,1.950)}
\gp3point{gp mark 6}{}{(3.221,1.951)}
\gp3point{gp mark 6}{}{(3.253,1.950)}
\gp3point{gp mark 6}{}{(3.284,1.951)}
\gp3point{gp mark 6}{}{(3.315,1.951)}
\gp3point{gp mark 6}{}{(3.347,1.951)}
\gp3point{gp mark 6}{}{(3.378,1.950)}
\gp3point{gp mark 6}{}{(3.409,1.951)}
\gp3point{gp mark 6}{}{(3.441,1.950)}
\gp3point{gp mark 6}{}{(3.472,1.950)}
\gp3point{gp mark 6}{}{(3.503,1.951)}
\gp3point{gp mark 6}{}{(3.534,1.950)}
\gp3point{gp mark 6}{}{(3.566,1.951)}
\gp3point{gp mark 6}{}{(3.597,1.950)}
\gp3point{gp mark 6}{}{(3.628,1.950)}
\gp3point{gp mark 6}{}{(3.660,1.951)}
\gp3point{gp mark 6}{}{(3.691,1.951)}
\gp3point{gp mark 6}{}{(3.722,1.951)}
\gp3point{gp mark 6}{}{(3.753,1.950)}
\gp3point{gp mark 6}{}{(3.785,1.951)}
\gp3point{gp mark 6}{}{(3.816,1.950)}
\gp3point{gp mark 6}{}{(3.847,1.951)}
\gp3point{gp mark 6}{}{(3.879,1.950)}
\gp3point{gp mark 6}{}{(3.910,1.951)}
\gp3point{gp mark 6}{}{(3.941,1.950)}
\gp3point{gp mark 6}{}{(3.973,1.951)}
\gp3point{gp mark 6}{}{(4.004,1.950)}
\gp3point{gp mark 6}{}{(4.035,1.951)}
\gp3point{gp mark 6}{}{(4.066,1.950)}
\gp3point{gp mark 6}{}{(4.098,1.951)}
\gp3point{gp mark 6}{}{(4.129,1.950)}
\gp3point{gp mark 6}{}{(4.160,1.951)}
\gp3point{gp mark 6}{}{(4.192,1.950)}
\gp3point{gp mark 6}{}{(4.223,1.951)}
\gp3point{gp mark 6}{}{(4.254,1.951)}
\gp3point{gp mark 6}{}{(4.285,1.951)}
\gp3point{gp mark 6}{}{(4.317,1.950)}
\gp3point{gp mark 6}{}{(4.348,1.951)}
\gp3point{gp mark 6}{}{(4.379,1.950)}
\gp3point{gp mark 6}{}{(4.411,1.951)}
\gp3point{gp mark 6}{}{(4.442,1.950)}
\gp3point{gp mark 6}{}{(4.473,1.951)}
\gp3point{gp mark 6}{}{(4.505,1.951)}
\gp3point{gp mark 6}{}{(4.536,1.950)}
\gp3point{gp mark 6}{}{(4.567,1.950)}
\gp3point{gp mark 6}{}{(4.598,1.950)}
\gp3point{gp mark 6}{}{(4.630,1.951)}
\gp3point{gp mark 6}{}{(4.661,1.951)}
\gp3point{gp mark 6}{}{(4.692,1.951)}
\gp3point{gp mark 6}{}{(4.724,1.950)}
\gp3point{gp mark 6}{}{(4.755,1.950)}
\gp3point{gp mark 6}{}{(4.786,1.951)}
\gp3point{gp mark 6}{}{(4.818,1.950)}
\gp3point{gp mark 6}{}{(4.849,1.951)}
\gp3point{gp mark 6}{}{(4.880,1.951)}
\gp3point{gp mark 6}{}{(4.911,1.950)}
\gp3point{gp mark 6}{}{(4.943,1.951)}
\gp3point{gp mark 6}{}{(4.974,1.950)}
\gp3point{gp mark 6}{}{(5.005,1.951)}
\gp3point{gp mark 6}{}{(5.037,1.951)}
\gp3point{gp mark 6}{}{(5.068,1.950)}
\gp3point{gp mark 6}{}{(5.099,1.950)}
\gp3point{gp mark 6}{}{(5.130,1.951)}
\gp3point{gp mark 6}{}{(5.162,1.950)}
\gp3point{gp mark 6}{}{(5.193,1.951)}
\gp3point{gp mark 6}{}{(5.224,1.951)}
\gp3point{gp mark 6}{}{(5.256,1.951)}
\gp3point{gp mark 6}{}{(5.287,1.951)}
\gp3point{gp mark 6}{}{(5.318,1.951)}
\gp3point{gp mark 6}{}{(5.350,1.951)}
\gp3point{gp mark 6}{}{(5.381,1.951)}
\gp3point{gp mark 6}{}{(5.412,1.951)}
\gp3point{gp mark 6}{}{(5.443,1.951)}
\gp3point{gp mark 6}{}{(5.475,1.951)}
\gp3point{gp mark 6}{}{(5.506,1.951)}
\gp3point{gp mark 6}{}{(5.537,1.951)}
\gp3point{gp mark 6}{}{(5.569,1.951)}
\gp3point{gp mark 6}{}{(5.600,1.951)}
\gp3point{gp mark 6}{}{(5.631,1.951)}
\gp3point{gp mark 6}{}{(5.662,1.951)}
\gp3point{gp mark 6}{}{(5.694,1.951)}
\gp3point{gp mark 6}{}{(5.725,1.951)}
\gp3point{gp mark 6}{}{(5.756,1.951)}
\gp3point{gp mark 6}{}{(5.788,1.951)}
\gp3point{gp mark 6}{}{(5.819,1.951)}
\gp3point{gp mark 6}{}{(5.850,1.951)}
\gp3point{gp mark 6}{}{(5.882,1.951)}
\gp3point{gp mark 6}{}{(5.913,1.951)}
\gp3point{gp mark 6}{}{(5.944,1.951)}
\gp3point{gp mark 6}{}{(5.975,1.951)}
\gp3point{gp mark 6}{}{(6.007,1.951)}
\gp3point{gp mark 6}{}{(6.038,1.951)}
\gp3point{gp mark 6}{}{(6.069,1.951)}
\gp3point{gp mark 6}{}{(6.101,1.951)}
\gp3point{gp mark 6}{}{(6.132,1.951)}
\gp3point{gp mark 6}{}{(6.163,1.951)}
\gp3point{gp mark 6}{}{(6.194,1.951)}
\gp3point{gp mark 6}{}{(6.226,1.951)}
\gp3point{gp mark 6}{}{(6.257,1.951)}
\gp3point{gp mark 6}{}{(6.288,1.951)}
\gp3point{gp mark 6}{}{(6.320,1.951)}
\gp3point{gp mark 6}{}{(6.351,1.951)}
\gp3point{gp mark 6}{}{(6.382,1.951)}
\gp3point{gp mark 6}{}{(6.414,1.951)}
\gp3point{gp mark 6}{}{(6.445,1.951)}
\gp3point{gp mark 6}{}{(6.476,1.951)}
\gp3point{gp mark 6}{}{(6.507,1.951)}
\gp3point{gp mark 6}{}{(6.539,1.951)}
\gp3point{gp mark 6}{}{(6.570,1.951)}
\gp3point{gp mark 6}{}{(6.601,1.951)}
\gp3point{gp mark 6}{}{(6.633,1.951)}
\gp3point{gp mark 6}{}{(6.664,1.951)}
\gp3point{gp mark 6}{}{(6.695,1.951)}
\gp3point{gp mark 6}{}{(6.726,1.951)}
\gp3point{gp mark 6}{}{(6.758,1.951)}
\gp3point{gp mark 6}{}{(6.789,1.951)}
\gp3point{gp mark 6}{}{(6.820,1.951)}
\gp3point{gp mark 6}{}{(6.852,1.951)}
\gp3point{gp mark 6}{}{(6.883,1.951)}
\gp3point{gp mark 6}{}{(6.914,1.951)}
\gp3point{gp mark 6}{}{(6.946,1.951)}
\gp3point{gp mark 6}{}{(6.977,1.951)}
\gp3point{gp mark 6}{}{(7.008,1.951)}
\gp3point{gp mark 6}{}{(7.039,1.951)}
\gp3point{gp mark 6}{}{(7.071,1.951)}
\gp3point{gp mark 6}{}{(7.102,1.951)}
\gp3point{gp mark 6}{}{(7.133,1.951)}
\gp3point{gp mark 6}{}{(7.165,1.951)}
\gp3point{gp mark 6}{}{(7.196,1.951)}
\gp3point{gp mark 6}{}{(7.227,1.951)}
\gp3point{gp mark 6}{}{(7.259,1.951)}
\gp3point{gp mark 6}{}{(7.290,1.951)}
\gp3point{gp mark 6}{}{(7.321,1.951)}
\gp3point{gp mark 6}{}{(7.352,1.951)}
\gp3point{gp mark 6}{}{(7.384,1.951)}
\gp3point{gp mark 6}{}{(7.415,1.951)}
\gp3point{gp mark 6}{}{(7.446,1.951)}
\gp3point{gp mark 6}{}{(7.478,1.951)}
\gp3point{gp mark 6}{}{(7.509,1.951)}
\gp3point{gp mark 6}{}{(7.540,1.951)}
\gp3point{gp mark 6}{}{(7.571,1.951)}
\gp3point{gp mark 6}{}{(7.603,1.951)}
\gp3point{gp mark 6}{}{(7.634,1.951)}
\gp3point{gp mark 6}{}{(7.665,1.951)}
\gp3point{gp mark 6}{}{(7.697,1.951)}
\gp3point{gp mark 6}{}{(7.728,1.951)}
\gp3point{gp mark 6}{}{(7.759,1.951)}
\gp3point{gp mark 6}{}{(7.791,1.951)}
\gp3point{gp mark 6}{}{(7.822,1.951)}
\gp3point{gp mark 6}{}{(7.853,1.951)}
\gp3point{gp mark 6}{}{(7.884,1.951)}
\gp3point{gp mark 6}{}{(7.916,1.951)}
\gp3point{gp mark 6}{}{(7.947,1.951)}
\gpcolor{rgb color={0.902,0.624,0.000}}
\gp3point{gp mark 6}{}{(2.157,1.945)}
\gp3point{gp mark 6}{}{(2.189,1.946)}
\gp3point{gp mark 6}{}{(2.220,1.947)}
\gp3point{gp mark 6}{}{(2.251,1.954)}
\gp3point{gp mark 6}{}{(2.283,1.948)}
\gp3point{gp mark 6}{}{(2.314,1.948)}
\gp3point{gp mark 6}{}{(2.345,1.948)}
\gp3point{gp mark 6}{}{(2.376,1.953)}
\gp3point{gp mark 6}{}{(2.408,1.949)}
\gp3point{gp mark 6}{}{(2.439,1.953)}
\gp3point{gp mark 6}{}{(2.470,1.952)}
\gp3point{gp mark 6}{}{(2.502,1.949)}
\gp3point{gp mark 6}{}{(2.533,1.952)}
\gp3point{gp mark 6}{}{(2.564,1.949)}
\gp3point{gp mark 6}{}{(2.596,1.949)}
\gp3point{gp mark 6}{}{(2.627,1.952)}
\gp3point{gp mark 6}{}{(2.658,1.949)}
\gp3point{gp mark 6}{}{(2.689,1.952)}
\gp3point{gp mark 6}{}{(2.721,1.949)}
\gp3point{gp mark 6}{}{(2.752,1.952)}
\gp3point{gp mark 6}{}{(2.783,1.949)}
\gp3point{gp mark 6}{}{(2.815,1.949)}
\gp3point{gp mark 6}{}{(2.846,1.952)}
\gp3point{gp mark 6}{}{(2.877,1.950)}
\gp3point{gp mark 6}{}{(2.909,1.952)}
\gp3point{gp mark 6}{}{(2.940,1.952)}
\gp3point{gp mark 6}{}{(2.971,1.950)}
\gp3point{gp mark 6}{}{(3.002,1.950)}
\gp3point{gp mark 6}{}{(3.034,1.952)}
\gp3point{gp mark 6}{}{(3.065,1.950)}
\gp3point{gp mark 6}{}{(3.096,1.951)}
\gp3point{gp mark 6}{}{(3.128,1.950)}
\gp3point{gp mark 6}{}{(3.159,1.951)}
\gp3point{gp mark 6}{}{(3.190,1.951)}
\gp3point{gp mark 6}{}{(3.221,1.950)}
\gp3point{gp mark 6}{}{(3.253,1.950)}
\gp3point{gp mark 6}{}{(3.284,1.951)}
\gp3point{gp mark 6}{}{(3.315,1.951)}
\gp3point{gp mark 6}{}{(3.347,1.950)}
\gp3point{gp mark 6}{}{(3.378,1.951)}
\gp3point{gp mark 6}{}{(3.409,1.951)}
\gp3point{gp mark 6}{}{(3.441,1.950)}
\gp3point{gp mark 6}{}{(3.472,1.950)}
\gp3point{gp mark 6}{}{(3.503,1.951)}
\gp3point{gp mark 6}{}{(3.534,1.951)}
\gp3point{gp mark 6}{}{(3.566,1.950)}
\gp3point{gp mark 6}{}{(3.597,1.950)}
\gp3point{gp mark 6}{}{(3.628,1.951)}
\gp3point{gp mark 6}{}{(3.660,1.950)}
\gp3point{gp mark 6}{}{(3.691,1.951)}
\gp3point{gp mark 6}{}{(3.722,1.950)}
\gp3point{gp mark 6}{}{(3.753,1.950)}
\gp3point{gp mark 6}{}{(3.785,1.951)}
\gp3point{gp mark 6}{}{(3.816,1.950)}
\gp3point{gp mark 6}{}{(3.847,1.951)}
\gp3point{gp mark 6}{}{(3.879,1.950)}
\gp3point{gp mark 6}{}{(3.910,1.950)}
\gp3point{gp mark 6}{}{(3.941,1.951)}
\gp3point{gp mark 6}{}{(3.973,1.951)}
\gp3point{gp mark 6}{}{(4.004,1.950)}
\gp3point{gp mark 6}{}{(4.035,1.950)}
\gp3point{gp mark 6}{}{(4.066,1.951)}
\gp3point{gp mark 6}{}{(4.098,1.951)}
\gp3point{gp mark 6}{}{(4.129,1.950)}
\gp3point{gp mark 6}{}{(4.160,1.950)}
\gp3point{gp mark 6}{}{(4.192,1.951)}
\gp3point{gp mark 6}{}{(4.223,1.951)}
\gp3point{gp mark 6}{}{(4.254,1.950)}
\gp3point{gp mark 6}{}{(4.285,1.951)}
\gp3point{gp mark 6}{}{(4.317,1.950)}
\gp3point{gp mark 6}{}{(4.348,1.950)}
\gp3point{gp mark 6}{}{(4.379,1.951)}
\gp3point{gp mark 6}{}{(4.411,1.951)}
\gp3point{gp mark 6}{}{(4.442,1.950)}
\gp3point{gp mark 6}{}{(4.473,1.951)}
\gp3point{gp mark 6}{}{(4.505,1.950)}
\gp3point{gp mark 6}{}{(4.536,1.951)}
\gp3point{gp mark 6}{}{(4.567,1.950)}
\gp3point{gp mark 6}{}{(4.598,1.950)}
\gp3point{gp mark 6}{}{(4.630,1.951)}
\gp3point{gp mark 6}{}{(4.661,1.950)}
\gp3point{gp mark 6}{}{(4.692,1.951)}
\gp3point{gp mark 6}{}{(4.724,1.950)}
\gp3point{gp mark 6}{}{(4.755,1.951)}
\gp3point{gp mark 6}{}{(4.786,1.950)}
\gp3point{gp mark 6}{}{(4.818,1.951)}
\gp3point{gp mark 6}{}{(4.849,1.950)}
\gp3point{gp mark 6}{}{(4.880,1.951)}
\gp3point{gp mark 6}{}{(4.911,1.950)}
\gp3point{gp mark 6}{}{(4.943,1.951)}
\gp3point{gp mark 6}{}{(4.974,1.950)}
\gp3point{gp mark 6}{}{(5.005,1.951)}
\gp3point{gp mark 6}{}{(5.037,1.951)}
\gp3point{gp mark 6}{}{(5.068,1.951)}
\gp3point{gp mark 6}{}{(5.099,1.951)}
\gp3point{gp mark 6}{}{(5.130,1.951)}
\gp3point{gp mark 6}{}{(5.162,1.951)}
\gp3point{gp mark 6}{}{(5.193,1.951)}
\gp3point{gp mark 6}{}{(5.224,1.951)}
\gp3point{gp mark 6}{}{(5.256,1.951)}
\gp3point{gp mark 6}{}{(5.287,1.951)}
\gp3point{gp mark 6}{}{(5.318,1.951)}
\gp3point{gp mark 6}{}{(5.350,1.951)}
\gp3point{gp mark 6}{}{(5.381,1.951)}
\gp3point{gp mark 6}{}{(5.412,1.951)}
\gp3point{gp mark 6}{}{(5.443,1.951)}
\gp3point{gp mark 6}{}{(5.475,1.951)}
\gp3point{gp mark 6}{}{(5.506,1.951)}
\gp3point{gp mark 6}{}{(5.537,1.951)}
\gp3point{gp mark 6}{}{(5.569,1.951)}
\gp3point{gp mark 6}{}{(5.600,1.951)}
\gp3point{gp mark 6}{}{(5.631,1.951)}
\gp3point{gp mark 6}{}{(5.662,1.951)}
\gp3point{gp mark 6}{}{(5.694,1.951)}
\gp3point{gp mark 6}{}{(5.725,1.951)}
\gp3point{gp mark 6}{}{(5.756,1.951)}
\gp3point{gp mark 6}{}{(5.788,1.951)}
\gp3point{gp mark 6}{}{(5.819,1.951)}
\gp3point{gp mark 6}{}{(5.850,1.951)}
\gp3point{gp mark 6}{}{(5.882,1.951)}
\gp3point{gp mark 6}{}{(5.913,1.951)}
\gp3point{gp mark 6}{}{(5.944,1.951)}
\gp3point{gp mark 6}{}{(5.975,1.951)}
\gp3point{gp mark 6}{}{(6.007,1.951)}
\gp3point{gp mark 6}{}{(6.038,1.951)}
\gp3point{gp mark 6}{}{(6.069,1.951)}
\gp3point{gp mark 6}{}{(6.101,1.951)}
\gp3point{gp mark 6}{}{(6.132,1.951)}
\gp3point{gp mark 6}{}{(6.163,1.951)}
\gp3point{gp mark 6}{}{(6.194,1.951)}
\gp3point{gp mark 6}{}{(6.226,1.951)}
\gp3point{gp mark 6}{}{(6.257,1.951)}
\gp3point{gp mark 6}{}{(6.288,1.951)}
\gp3point{gp mark 6}{}{(6.320,1.951)}
\gp3point{gp mark 6}{}{(6.351,1.951)}
\gp3point{gp mark 6}{}{(6.382,1.951)}
\gp3point{gp mark 6}{}{(6.414,1.951)}
\gp3point{gp mark 6}{}{(6.445,1.951)}
\gp3point{gp mark 6}{}{(6.476,1.951)}
\gp3point{gp mark 6}{}{(6.507,1.951)}
\gp3point{gp mark 6}{}{(6.539,1.951)}
\gp3point{gp mark 6}{}{(6.570,1.951)}
\gp3point{gp mark 6}{}{(6.601,1.951)}
\gp3point{gp mark 6}{}{(6.633,1.951)}
\gp3point{gp mark 6}{}{(6.664,1.951)}
\gp3point{gp mark 6}{}{(6.695,1.951)}
\gp3point{gp mark 6}{}{(6.726,1.951)}
\gp3point{gp mark 6}{}{(6.758,1.951)}
\gp3point{gp mark 6}{}{(6.789,1.951)}
\gp3point{gp mark 6}{}{(6.820,1.951)}
\gp3point{gp mark 6}{}{(6.852,1.951)}
\gp3point{gp mark 6}{}{(6.883,1.951)}
\gp3point{gp mark 6}{}{(6.914,1.951)}
\gp3point{gp mark 6}{}{(6.946,1.951)}
\gp3point{gp mark 6}{}{(6.977,1.951)}
\gp3point{gp mark 6}{}{(7.008,1.951)}
\gp3point{gp mark 6}{}{(7.039,1.951)}
\gp3point{gp mark 6}{}{(7.071,1.951)}
\gp3point{gp mark 6}{}{(7.102,1.951)}
\gp3point{gp mark 6}{}{(7.133,1.951)}
\gp3point{gp mark 6}{}{(7.165,1.951)}
\gp3point{gp mark 6}{}{(7.196,1.951)}
\gp3point{gp mark 6}{}{(7.227,1.951)}
\gp3point{gp mark 6}{}{(7.259,1.951)}
\gp3point{gp mark 6}{}{(7.290,1.951)}
\gp3point{gp mark 6}{}{(7.321,1.951)}
\gp3point{gp mark 6}{}{(7.352,1.951)}
\gp3point{gp mark 6}{}{(7.384,1.951)}
\gp3point{gp mark 6}{}{(7.415,1.951)}
\gp3point{gp mark 6}{}{(7.446,1.951)}
\gp3point{gp mark 6}{}{(7.478,1.951)}
\gp3point{gp mark 6}{}{(7.509,1.951)}
\gp3point{gp mark 6}{}{(7.540,1.951)}
\gp3point{gp mark 6}{}{(7.571,1.951)}
\gp3point{gp mark 6}{}{(7.603,1.951)}
\gp3point{gp mark 6}{}{(7.634,1.951)}
\gp3point{gp mark 6}{}{(7.665,1.951)}
\gp3point{gp mark 6}{}{(7.697,1.951)}
\gp3point{gp mark 6}{}{(7.728,1.951)}
\gp3point{gp mark 6}{}{(7.759,1.951)}
\gp3point{gp mark 6}{}{(7.791,1.951)}
\gp3point{gp mark 6}{}{(7.822,1.951)}
\gp3point{gp mark 6}{}{(7.853,1.951)}
\gp3point{gp mark 6}{}{(7.884,1.951)}
\gp3point{gp mark 6}{}{(7.916,1.951)}
\gp3point{gp mark 6}{}{(7.947,1.951)}
\gpcolor{rgb color={0.580,0.000,0.827}}
\gp3point{gp mark 2}{}{(2.157,2.967)}
\gp3point{gp mark 2}{}{(2.189,2.891)}
\gp3point{gp mark 2}{}{(2.220,1.072)}
\gp3point{gp mark 2}{}{(2.251,2.774)}
\gp3point{gp mark 2}{}{(2.283,2.667)}
\gp3point{gp mark 2}{}{(2.314,1.248)}
\gp3point{gp mark 2}{}{(2.345,1.287)}
\gp3point{gp mark 2}{}{(2.376,2.587)}
\gp3point{gp mark 2}{}{(2.408,2.578)}
\gp3point{gp mark 2}{}{(2.439,2.530)}
\gp3point{gp mark 2}{}{(2.470,2.476)}
\gp3point{gp mark 2}{}{(2.502,1.452)}
\gp3point{gp mark 2}{}{(2.533,2.396)}
\gp3point{gp mark 2}{}{(2.564,1.507)}
\gp3point{gp mark 2}{}{(2.596,2.385)}
\gp3point{gp mark 2}{}{(2.627,2.359)}
\gp3point{gp mark 2}{}{(2.658,2.353)}
\gp3point{gp mark 2}{}{(2.689,1.554)}
\gp3point{gp mark 2}{}{(2.721,1.600)}
\gp3point{gp mark 2}{}{(2.752,2.284)}
\gp3point{gp mark 2}{}{(2.783,2.254)}
\gp3point{gp mark 2}{}{(2.815,1.648)}
\gp3point{gp mark 2}{}{(2.846,2.245)}
\gp3point{gp mark 2}{}{(2.877,1.668)}
\gp3point{gp mark 2}{}{(2.909,2.223)}
\gp3point{gp mark 2}{}{(2.940,2.209)}
\gp3point{gp mark 2}{}{(2.971,1.693)}
\gp3point{gp mark 2}{}{(3.002,2.199)}
\gp3point{gp mark 2}{}{(3.034,1.706)}
\gp3point{gp mark 2}{}{(3.065,2.190)}
\gp3point{gp mark 2}{}{(3.096,2.161)}
\gp3point{gp mark 2}{}{(3.128,1.742)}
\gp3point{gp mark 2}{}{(3.159,2.153)}
\gp3point{gp mark 2}{}{(3.190,1.754)}
\gp3point{gp mark 2}{}{(3.221,1.769)}
\gp3point{gp mark 2}{}{(3.253,2.124)}
\gp3point{gp mark 2}{}{(3.284,2.118)}
\gp3point{gp mark 2}{}{(3.315,1.786)}
\gp3point{gp mark 2}{}{(3.347,1.794)}
\gp3point{gp mark 2}{}{(3.378,2.107)}
\gp3point{gp mark 2}{}{(3.409,1.795)}
\gp3point{gp mark 2}{}{(3.441,1.807)}
\gp3point{gp mark 2}{}{(3.472,2.092)}
\gp3point{gp mark 2}{}{(3.503,2.091)}
\gp3point{gp mark 2}{}{(3.534,1.818)}
\gp3point{gp mark 2}{}{(3.566,2.083)}
\gp3point{gp mark 2}{}{(3.597,2.078)}
\gp3point{gp mark 2}{}{(3.628,1.825)}
\gp3point{gp mark 2}{}{(3.660,1.832)}
\gp3point{gp mark 2}{}{(3.691,1.837)}
\gp3point{gp mark 2}{}{(3.722,2.061)}
\gp3point{gp mark 2}{}{(3.753,1.844)}
\gp3point{gp mark 2}{}{(3.785,1.846)}
\gp3point{gp mark 2}{}{(3.816,2.053)}
\gp3point{gp mark 2}{}{(3.847,2.052)}
\gp3point{gp mark 2}{}{(3.879,2.044)}
\gp3point{gp mark 2}{}{(3.910,1.859)}
\gp3point{gp mark 2}{}{(3.941,1.860)}
\gp3point{gp mark 2}{}{(3.973,2.037)}
\gp3point{gp mark 2}{}{(4.004,2.032)}
\gp3point{gp mark 2}{}{(4.035,1.871)}
\gp3point{gp mark 2}{}{(4.066,2.030)}
\gp3point{gp mark 2}{}{(4.098,1.873)}
\gp3point{gp mark 2}{}{(4.129,1.874)}
\gp3point{gp mark 2}{}{(4.160,2.026)}
\gp3point{gp mark 2}{}{(4.192,1.880)}
\gp3point{gp mark 2}{}{(4.223,2.018)}
\gp3point{gp mark 2}{}{(4.254,1.885)}
\gp3point{gp mark 2}{}{(4.285,2.014)}
\gp3point{gp mark 2}{}{(4.317,1.888)}
\gp3point{gp mark 2}{}{(4.348,2.009)}
\gp3point{gp mark 2}{}{(4.379,1.893)}
\gp3point{gp mark 2}{}{(4.411,1.894)}
\gp3point{gp mark 2}{}{(4.442,2.008)}
\gp3point{gp mark 2}{}{(4.473,1.896)}
\gp3point{gp mark 2}{}{(4.505,2.003)}
\gp3point{gp mark 2}{}{(4.536,1.901)}
\gp3point{gp mark 2}{}{(4.567,2.000)}
\gp3point{gp mark 2}{}{(4.598,1.904)}
\gp3point{gp mark 2}{}{(4.630,1.997)}
\gp3point{gp mark 2}{}{(4.661,1.995)}
\gp3point{gp mark 2}{}{(4.692,1.907)}
\gp3point{gp mark 2}{}{(4.724,1.993)}
\gp3point{gp mark 2}{}{(4.755,1.909)}
\gp3point{gp mark 2}{}{(4.786,1.991)}
\gp3point{gp mark 2}{}{(4.818,1.910)}
\gp3point{gp mark 2}{}{(4.849,1.911)}
\gp3point{gp mark 2}{}{(4.880,1.989)}
\gp3point{gp mark 2}{}{(4.911,1.914)}
\gp3point{gp mark 2}{}{(4.943,1.915)}
\gp3point{gp mark 2}{}{(4.974,1.916)}
\gp3point{gp mark 2}{}{(5.005,1.985)}
\gp3point{gp mark 2}{}{(5.037,1.917)}
\gp3point{gp mark 2}{}{(5.068,1.983)}
\gp3point{gp mark 2}{}{(5.099,1.982)}
\gp3point{gp mark 2}{}{(5.130,1.919)}
\gp3point{gp mark 2}{}{(5.162,1.920)}
\gp3point{gp mark 2}{}{(5.193,1.981)}
\gp3point{gp mark 2}{}{(5.224,1.921)}
\gp3point{gp mark 2}{}{(5.256,1.980)}
\gp3point{gp mark 2}{}{(5.287,1.979)}
\gp3point{gp mark 2}{}{(5.318,1.923)}
\gp3point{gp mark 2}{}{(5.350,1.979)}
\gp3point{gp mark 2}{}{(5.381,1.923)}
\gp3point{gp mark 2}{}{(5.412,1.925)}
\gp3point{gp mark 2}{}{(5.443,1.976)}
\gp3point{gp mark 2}{}{(5.475,1.926)}
\gp3point{gp mark 2}{}{(5.506,1.927)}
\gp3point{gp mark 2}{}{(5.537,1.974)}
\gp3point{gp mark 2}{}{(5.569,1.973)}
\gp3point{gp mark 2}{}{(5.600,1.929)}
\gp3point{gp mark 2}{}{(5.631,1.929)}
\gp3point{gp mark 2}{}{(5.662,1.929)}
\gp3point{gp mark 2}{}{(5.694,1.971)}
\gp3point{gp mark 2}{}{(5.725,1.971)}
\gp3point{gp mark 2}{}{(5.756,1.931)}
\gp3point{gp mark 2}{}{(5.788,1.970)}
\gp3point{gp mark 2}{}{(5.819,1.932)}
\gp3point{gp mark 2}{}{(5.850,1.969)}
\gp3point{gp mark 2}{}{(5.882,1.932)}
\gp3point{gp mark 2}{}{(5.913,1.933)}
\gp3point{gp mark 2}{}{(5.944,1.968)}
\gp3point{gp mark 2}{}{(5.975,1.933)}
\gp3point{gp mark 2}{}{(6.007,1.968)}
\gp3point{gp mark 2}{}{(6.038,1.934)}
\gp3point{gp mark 2}{}{(6.069,1.935)}
\gp3point{gp mark 2}{}{(6.101,1.966)}
\gp3point{gp mark 2}{}{(6.132,1.966)}
\gp3point{gp mark 2}{}{(6.163,1.935)}
\gp3point{gp mark 2}{}{(6.194,1.936)}
\gp3point{gp mark 2}{}{(6.226,1.936)}
\gp3point{gp mark 2}{}{(6.257,1.965)}
\gp3point{gp mark 2}{}{(6.288,1.937)}
\gp3point{gp mark 2}{}{(6.320,1.964)}
\gp3point{gp mark 2}{}{(6.351,1.938)}
\gp3point{gp mark 2}{}{(6.382,1.963)}
\gp3point{gp mark 2}{}{(6.414,1.939)}
\gp3point{gp mark 2}{}{(6.445,1.939)}
\gp3point{gp mark 2}{}{(6.476,1.962)}
\gp3point{gp mark 2}{}{(6.507,1.962)}
\gp3point{gp mark 2}{}{(6.539,1.939)}
\gp3point{gp mark 2}{}{(6.570,1.940)}
\gp3point{gp mark 2}{}{(6.601,1.961)}
\gp3point{gp mark 2}{}{(6.633,1.940)}
\gp3point{gp mark 2}{}{(6.664,1.961)}
\gp3point{gp mark 2}{}{(6.695,1.940)}
\gp3point{gp mark 2}{}{(6.726,1.961)}
\gp3point{gp mark 2}{}{(6.758,1.941)}
\gp3point{gp mark 2}{}{(6.789,1.941)}
\gp3point{gp mark 2}{}{(6.820,1.960)}
\gp3point{gp mark 2}{}{(6.852,1.959)}
\gp3point{gp mark 2}{}{(6.883,1.942)}
\gp3point{gp mark 2}{}{(6.914,1.942)}
\gp3point{gp mark 2}{}{(6.946,1.959)}
\gp3point{gp mark 2}{}{(6.977,1.942)}
\gp3point{gp mark 2}{}{(7.008,1.959)}
\gp3point{gp mark 2}{}{(7.039,1.943)}
\gp3point{gp mark 2}{}{(7.071,1.958)}
\gp3point{gp mark 2}{}{(7.102,1.943)}
\gp3point{gp mark 2}{}{(7.133,1.943)}
\gp3point{gp mark 2}{}{(7.165,1.944)}
\gp3point{gp mark 2}{}{(7.196,1.958)}
\gp3point{gp mark 2}{}{(7.227,1.944)}
\gp3point{gp mark 2}{}{(7.259,1.957)}
\gp3point{gp mark 2}{}{(7.290,1.957)}
\gp3point{gp mark 2}{}{(7.321,1.944)}
\gp3point{gp mark 2}{}{(7.352,1.945)}
\gp3point{gp mark 2}{}{(7.384,1.945)}
\gp3point{gp mark 2}{}{(7.415,1.956)}
\gp3point{gp mark 2}{}{(7.446,1.945)}
\gp3point{gp mark 2}{}{(7.478,1.956)}
\gp3point{gp mark 2}{}{(7.509,1.945)}
\gp3point{gp mark 2}{}{(7.540,1.946)}
\gp3point{gp mark 2}{}{(7.571,1.956)}
\gp3point{gp mark 2}{}{(7.603,1.956)}
\gp3point{gp mark 2}{}{(7.634,1.955)}
\gp3point{gp mark 2}{}{(7.665,1.946)}
\gp3point{gp mark 2}{}{(7.697,1.946)}
\gp3point{gp mark 2}{}{(7.728,1.955)}
\gp3point{gp mark 2}{}{(7.759,1.946)}
\gp3point{gp mark 2}{}{(7.791,1.955)}
\gp3point{gp mark 2}{}{(7.822,1.947)}
\gp3point{gp mark 2}{}{(7.853,1.955)}
\gp3point{gp mark 2}{}{(7.884,1.947)}
\gp3point{gp mark 2}{}{(7.916,1.947)}
\gp3point{gp mark 2}{}{(7.947,1.954)}
\gpcolor{rgb color={0.000,0.620,0.451}}
\gp3point{gp mark 2}{}{(2.157,3.869)}
\gp3point{gp mark 2}{}{(2.189,3.700)}
\gp3point{gp mark 2}{}{(2.220,3.320)}
\gp3point{gp mark 2}{}{(2.251,3.070)}
\gp3point{gp mark 2}{}{(2.283,3.004)}
\gp3point{gp mark 2}{}{(2.314,1.192)}
\gp3point{gp mark 2}{}{(2.345,2.703)}
\gp3point{gp mark 2}{}{(2.376,2.549)}
\gp3point{gp mark 2}{}{(2.408,1.374)}
\gp3point{gp mark 2}{}{(2.439,2.504)}
\gp3point{gp mark 2}{}{(2.470,2.447)}
\gp3point{gp mark 2}{}{(2.502,1.458)}
\gp3point{gp mark 2}{}{(2.533,1.480)}
\gp3point{gp mark 2}{}{(2.564,2.409)}
\gp3point{gp mark 2}{}{(2.596,1.499)}
\gp3point{gp mark 2}{}{(2.627,2.357)}
\gp3point{gp mark 2}{}{(2.658,1.551)}
\gp3point{gp mark 2}{}{(2.689,2.349)}
\gp3point{gp mark 2}{}{(2.721,2.316)}
\gp3point{gp mark 2}{}{(2.752,1.595)}
\gp3point{gp mark 2}{}{(2.783,1.604)}
\gp3point{gp mark 2}{}{(2.815,1.626)}
\gp3point{gp mark 2}{}{(2.846,1.662)}
\gp3point{gp mark 2}{}{(2.877,2.238)}
\gp3point{gp mark 2}{}{(2.909,1.687)}
\gp3point{gp mark 2}{}{(2.940,2.210)}
\gp3point{gp mark 2}{}{(2.971,1.693)}
\gp3point{gp mark 2}{}{(3.002,2.201)}
\gp3point{gp mark 2}{}{(3.034,1.703)}
\gp3point{gp mark 2}{}{(3.065,2.190)}
\gp3point{gp mark 2}{}{(3.096,2.176)}
\gp3point{gp mark 2}{}{(3.128,1.732)}
\gp3point{gp mark 2}{}{(3.159,1.746)}
\gp3point{gp mark 2}{}{(3.190,1.749)}
\gp3point{gp mark 2}{}{(3.221,1.753)}
\gp3point{gp mark 2}{}{(3.253,2.138)}
\gp3point{gp mark 2}{}{(3.284,2.135)}
\gp3point{gp mark 2}{}{(3.315,1.771)}
\gp3point{gp mark 2}{}{(3.347,2.124)}
\gp3point{gp mark 2}{}{(3.378,2.117)}
\gp3point{gp mark 2}{}{(3.409,1.786)}
\gp3point{gp mark 2}{}{(3.441,2.110)}
\gp3point{gp mark 2}{}{(3.472,1.796)}
\gp3point{gp mark 2}{}{(3.503,2.099)}
\gp3point{gp mark 2}{}{(3.534,1.803)}
\gp3point{gp mark 2}{}{(3.566,2.094)}
\gp3point{gp mark 2}{}{(3.597,1.809)}
\gp3point{gp mark 2}{}{(3.628,1.818)}
\gp3point{gp mark 2}{}{(3.660,2.083)}
\gp3point{gp mark 2}{}{(3.691,1.825)}
\gp3point{gp mark 2}{}{(3.722,1.828)}
\gp3point{gp mark 2}{}{(3.753,2.072)}
\gp3point{gp mark 2}{}{(3.785,1.837)}
\gp3point{gp mark 2}{}{(3.816,2.062)}
\gp3point{gp mark 2}{}{(3.847,1.842)}
\gp3point{gp mark 2}{}{(3.879,1.850)}
\gp3point{gp mark 2}{}{(3.910,2.051)}
\gp3point{gp mark 2}{}{(3.941,1.852)}
\gp3point{gp mark 2}{}{(3.973,2.047)}
\gp3point{gp mark 2}{}{(4.004,2.042)}
\gp3point{gp mark 2}{}{(4.035,1.861)}
\gp3point{gp mark 2}{}{(4.066,2.038)}
\gp3point{gp mark 2}{}{(4.098,2.035)}
\gp3point{gp mark 2}{}{(4.129,2.034)}
\gp3point{gp mark 2}{}{(4.160,1.868)}
\gp3point{gp mark 2}{}{(4.192,2.032)}
\gp3point{gp mark 2}{}{(4.223,1.871)}
\gp3point{gp mark 2}{}{(4.254,1.874)}
\gp3point{gp mark 2}{}{(4.285,2.025)}
\gp3point{gp mark 2}{}{(4.317,1.876)}
\gp3point{gp mark 2}{}{(4.348,1.880)}
\gp3point{gp mark 2}{}{(4.379,2.021)}
\gp3point{gp mark 2}{}{(4.411,1.885)}
\gp3point{gp mark 2}{}{(4.442,2.015)}
\gp3point{gp mark 2}{}{(4.473,2.013)}
\gp3point{gp mark 2}{}{(4.505,1.889)}
\gp3point{gp mark 2}{}{(4.536,2.009)}
\gp3point{gp mark 2}{}{(4.567,1.892)}
\gp3point{gp mark 2}{}{(4.598,2.007)}
\gp3point{gp mark 2}{}{(4.630,1.895)}
\gp3point{gp mark 2}{}{(4.661,2.005)}
\gp3point{gp mark 2}{}{(4.692,2.004)}
\gp3point{gp mark 2}{}{(4.724,1.897)}
\gp3point{gp mark 2}{}{(4.755,1.899)}
\gp3point{gp mark 2}{}{(4.786,2.001)}
\gp3point{gp mark 2}{}{(4.818,1.903)}
\gp3point{gp mark 2}{}{(4.849,1.903)}
\gp3point{gp mark 2}{}{(4.880,1.997)}
\gp3point{gp mark 2}{}{(4.911,1.997)}
\gp3point{gp mark 2}{}{(4.943,1.904)}
\gp3point{gp mark 2}{}{(4.974,1.908)}
\gp3point{gp mark 2}{}{(5.005,1.993)}
\gp3point{gp mark 2}{}{(5.037,1.993)}
\gp3point{gp mark 2}{}{(5.068,1.910)}
\gp3point{gp mark 2}{}{(5.099,1.911)}
\gp3point{gp mark 2}{}{(5.130,1.989)}
\gp3point{gp mark 2}{}{(5.162,1.913)}
\gp3point{gp mark 2}{}{(5.193,1.987)}
\gp3point{gp mark 2}{}{(5.224,1.914)}
\gp3point{gp mark 2}{}{(5.256,1.986)}
\gp3point{gp mark 2}{}{(5.287,1.986)}
\gp3point{gp mark 2}{}{(5.318,1.916)}
\gp3point{gp mark 2}{}{(5.350,1.985)}
\gp3point{gp mark 2}{}{(5.381,1.917)}
\gp3point{gp mark 2}{}{(5.412,1.917)}
\gp3point{gp mark 2}{}{(5.443,1.983)}
\gp3point{gp mark 2}{}{(5.475,1.920)}
\gp3point{gp mark 2}{}{(5.506,1.920)}
\gp3point{gp mark 2}{}{(5.537,1.980)}
\gp3point{gp mark 2}{}{(5.569,1.921)}
\gp3point{gp mark 2}{}{(5.600,1.979)}
\gp3point{gp mark 2}{}{(5.631,1.923)}
\gp3point{gp mark 2}{}{(5.662,1.977)}
\gp3point{gp mark 2}{}{(5.694,1.924)}
\gp3point{gp mark 2}{}{(5.725,1.925)}
\gp3point{gp mark 2}{}{(5.756,1.976)}
\gp3point{gp mark 2}{}{(5.788,1.975)}
\gp3point{gp mark 2}{}{(5.819,1.927)}
\gp3point{gp mark 2}{}{(5.850,1.974)}
\gp3point{gp mark 2}{}{(5.882,1.927)}
\gp3point{gp mark 2}{}{(5.913,1.973)}
\gp3point{gp mark 2}{}{(5.944,1.972)}
\gp3point{gp mark 2}{}{(5.975,1.929)}
\gp3point{gp mark 2}{}{(6.007,1.930)}
\gp3point{gp mark 2}{}{(6.038,1.930)}
\gp3point{gp mark 2}{}{(6.069,1.971)}
\gp3point{gp mark 2}{}{(6.101,1.970)}
\gp3point{gp mark 2}{}{(6.132,1.970)}
\gp3point{gp mark 2}{}{(6.163,1.932)}
\gp3point{gp mark 2}{}{(6.194,1.933)}
\gp3point{gp mark 2}{}{(6.226,1.933)}
\gp3point{gp mark 2}{}{(6.257,1.968)}
\gp3point{gp mark 2}{}{(6.288,1.968)}
\gp3point{gp mark 2}{}{(6.320,1.935)}
\gp3point{gp mark 2}{}{(6.351,1.966)}
\gp3point{gp mark 2}{}{(6.382,1.935)}
\gp3point{gp mark 2}{}{(6.414,1.966)}
\gp3point{gp mark 2}{}{(6.445,1.965)}
\gp3point{gp mark 2}{}{(6.476,1.965)}
\gp3point{gp mark 2}{}{(6.507,1.937)}
\gp3point{gp mark 2}{}{(6.539,1.964)}
\gp3point{gp mark 2}{}{(6.570,1.937)}
\gp3point{gp mark 2}{}{(6.601,1.937)}
\gp3point{gp mark 2}{}{(6.633,1.964)}
\gp3point{gp mark 2}{}{(6.664,1.963)}
\gp3point{gp mark 2}{}{(6.695,1.939)}
\gp3point{gp mark 2}{}{(6.726,1.939)}
\gp3point{gp mark 2}{}{(6.758,1.962)}
\gp3point{gp mark 2}{}{(6.789,1.939)}
\gp3point{gp mark 2}{}{(6.820,1.962)}
\gp3point{gp mark 2}{}{(6.852,1.961)}
\gp3point{gp mark 2}{}{(6.883,1.940)}
\gp3point{gp mark 2}{}{(6.914,1.961)}
\gp3point{gp mark 2}{}{(6.946,1.941)}
\gp3point{gp mark 2}{}{(6.977,1.961)}
\gp3point{gp mark 2}{}{(7.008,1.941)}
\gp3point{gp mark 2}{}{(7.039,1.960)}
\gp3point{gp mark 2}{}{(7.071,1.960)}
\gp3point{gp mark 2}{}{(7.102,1.960)}
\gp3point{gp mark 2}{}{(7.133,1.942)}
\gp3point{gp mark 2}{}{(7.165,1.942)}
\gp3point{gp mark 2}{}{(7.196,1.942)}
\gp3point{gp mark 2}{}{(7.227,1.959)}
\gp3point{gp mark 2}{}{(7.259,1.959)}
\gp3point{gp mark 2}{}{(7.290,1.958)}
\gp3point{gp mark 2}{}{(7.321,1.958)}
\gp3point{gp mark 2}{}{(7.352,1.943)}
\gp3point{gp mark 2}{}{(7.384,1.943)}
\gp3point{gp mark 2}{}{(7.415,1.958)}
\gp3point{gp mark 2}{}{(7.446,1.944)}
\gp3point{gp mark 2}{}{(7.478,1.957)}
\gp3point{gp mark 2}{}{(7.509,1.944)}
\gp3point{gp mark 2}{}{(7.540,1.957)}
\gp3point{gp mark 2}{}{(7.571,1.957)}
\gp3point{gp mark 2}{}{(7.603,1.945)}
\gp3point{gp mark 2}{}{(7.634,1.956)}
\gp3point{gp mark 2}{}{(7.665,1.945)}
\gp3point{gp mark 2}{}{(7.697,1.956)}
\gp3point{gp mark 2}{}{(7.728,1.945)}
\gp3point{gp mark 2}{}{(7.759,1.956)}
\gp3point{gp mark 2}{}{(7.791,1.956)}
\gp3point{gp mark 2}{}{(7.822,1.946)}
\gp3point{gp mark 2}{}{(7.853,1.946)}
\gp3point{gp mark 2}{}{(7.884,1.955)}
\gp3point{gp mark 2}{}{(7.916,1.946)}
\gp3point{gp mark 2}{}{(7.947,1.955)}
\gpcolor{rgb color={0.902,0.624,0.000}}
\gp3point{gp mark 2}{}{(2.157,3.557)}
\gp3point{gp mark 2}{}{(2.189,3.543)}
\gp3point{gp mark 2}{}{(2.220,2.831)}
\gp3point{gp mark 2}{}{(2.251,2.626)}
\gp3point{gp mark 2}{}{(2.283,2.539)}
\gp3point{gp mark 2}{}{(2.314,2.535)}
\gp3point{gp mark 2}{}{(2.345,2.505)}
\gp3point{gp mark 2}{}{(2.376,2.387)}
\gp3point{gp mark 2}{}{(2.408,1.528)}
\gp3point{gp mark 2}{}{(2.439,2.360)}
\gp3point{gp mark 2}{}{(2.470,2.301)}
\gp3point{gp mark 2}{}{(2.502,2.277)}
\gp3point{gp mark 2}{}{(2.533,1.635)}
\gp3point{gp mark 2}{}{(2.564,2.251)}
\gp3point{gp mark 2}{}{(2.596,1.654)}
\gp3point{gp mark 2}{}{(2.627,1.678)}
\gp3point{gp mark 2}{}{(2.658,2.218)}
\gp3point{gp mark 2}{}{(2.689,2.206)}
\gp3point{gp mark 2}{}{(2.721,1.699)}
\gp3point{gp mark 2}{}{(2.752,2.190)}
\gp3point{gp mark 2}{}{(2.783,1.722)}
\gp3point{gp mark 2}{}{(2.815,2.175)}
\gp3point{gp mark 2}{}{(2.846,2.166)}
\gp3point{gp mark 2}{}{(2.877,2.148)}
\gp3point{gp mark 2}{}{(2.909,1.761)}
\gp3point{gp mark 2}{}{(2.940,2.128)}
\gp3point{gp mark 2}{}{(2.971,1.778)}
\gp3point{gp mark 2}{}{(3.002,2.121)}
\gp3point{gp mark 2}{}{(3.034,1.783)}
\gp3point{gp mark 2}{}{(3.065,2.112)}
\gp3point{gp mark 2}{}{(3.096,1.799)}
\gp3point{gp mark 2}{}{(3.128,2.091)}
\gp3point{gp mark 2}{}{(3.159,1.816)}
\gp3point{gp mark 2}{}{(3.190,2.083)}
\gp3point{gp mark 2}{}{(3.221,1.824)}
\gp3point{gp mark 2}{}{(3.253,2.072)}
\gp3point{gp mark 2}{}{(3.284,2.066)}
\gp3point{gp mark 2}{}{(3.315,2.057)}
\gp3point{gp mark 2}{}{(3.347,1.846)}
\gp3point{gp mark 2}{}{(3.378,1.848)}
\gp3point{gp mark 2}{}{(3.409,2.053)}
\gp3point{gp mark 2}{}{(3.441,1.857)}
\gp3point{gp mark 2}{}{(3.472,2.044)}
\gp3point{gp mark 2}{}{(3.503,1.861)}
\gp3point{gp mark 2}{}{(3.534,2.039)}
\gp3point{gp mark 2}{}{(3.566,2.032)}
\gp3point{gp mark 2}{}{(3.597,1.870)}
\gp3point{gp mark 2}{}{(3.628,2.030)}
\gp3point{gp mark 2}{}{(3.660,1.874)}
\gp3point{gp mark 2}{}{(3.691,2.026)}
\gp3point{gp mark 2}{}{(3.722,2.025)}
\gp3point{gp mark 2}{}{(3.753,1.879)}
\gp3point{gp mark 2}{}{(3.785,2.017)}
\gp3point{gp mark 2}{}{(3.816,1.885)}
\gp3point{gp mark 2}{}{(3.847,2.015)}
\gp3point{gp mark 2}{}{(3.879,1.889)}
\gp3point{gp mark 2}{}{(3.910,2.012)}
\gp3point{gp mark 2}{}{(3.941,2.010)}
\gp3point{gp mark 2}{}{(3.973,2.007)}
\gp3point{gp mark 2}{}{(4.004,2.004)}
\gp3point{gp mark 2}{}{(4.035,2.001)}
\gp3point{gp mark 2}{}{(4.066,1.901)}
\gp3point{gp mark 2}{}{(4.098,1.902)}
\gp3point{gp mark 2}{}{(4.129,1.998)}
\gp3point{gp mark 2}{}{(4.160,1.905)}
\gp3point{gp mark 2}{}{(4.192,1.996)}
\gp3point{gp mark 2}{}{(4.223,1.994)}
\gp3point{gp mark 2}{}{(4.254,1.992)}
\gp3point{gp mark 2}{}{(4.285,1.910)}
\gp3point{gp mark 2}{}{(4.317,1.989)}
\gp3point{gp mark 2}{}{(4.348,1.988)}
\gp3point{gp mark 2}{}{(4.379,1.914)}
\gp3point{gp mark 2}{}{(4.411,1.985)}
\gp3point{gp mark 2}{}{(4.442,1.985)}
\gp3point{gp mark 2}{}{(4.473,1.918)}
\gp3point{gp mark 2}{}{(4.505,1.983)}
\gp3point{gp mark 2}{}{(4.536,1.918)}
\gp3point{gp mark 2}{}{(4.567,1.983)}
\gp3point{gp mark 2}{}{(4.598,1.981)}
\gp3point{gp mark 2}{}{(4.630,1.921)}
\gp3point{gp mark 2}{}{(4.661,1.979)}
\gp3point{gp mark 2}{}{(4.692,1.923)}
\gp3point{gp mark 2}{}{(4.724,1.979)}
\gp3point{gp mark 2}{}{(4.755,1.923)}
\gp3point{gp mark 2}{}{(4.786,1.977)}
\gp3point{gp mark 2}{}{(4.818,1.925)}
\gp3point{gp mark 2}{}{(4.849,1.975)}
\gp3point{gp mark 2}{}{(4.880,1.926)}
\gp3point{gp mark 2}{}{(4.911,1.975)}
\gp3point{gp mark 2}{}{(4.943,1.927)}
\gp3point{gp mark 2}{}{(4.974,1.974)}
\gp3point{gp mark 2}{}{(5.005,1.928)}
\gp3point{gp mark 2}{}{(5.037,1.973)}
\gp3point{gp mark 2}{}{(5.068,1.929)}
\gp3point{gp mark 2}{}{(5.099,1.972)}
\gp3point{gp mark 2}{}{(5.130,1.972)}
\gp3point{gp mark 2}{}{(5.162,1.971)}
\gp3point{gp mark 2}{}{(5.193,1.970)}
\gp3point{gp mark 2}{}{(5.224,1.931)}
\gp3point{gp mark 2}{}{(5.256,1.932)}
\gp3point{gp mark 2}{}{(5.287,1.969)}
\gp3point{gp mark 2}{}{(5.318,1.968)}
\gp3point{gp mark 2}{}{(5.350,1.934)}
\gp3point{gp mark 2}{}{(5.381,1.968)}
\gp3point{gp mark 2}{}{(5.412,1.967)}
\gp3point{gp mark 2}{}{(5.443,1.934)}
\gp3point{gp mark 2}{}{(5.475,1.934)}
\gp3point{gp mark 2}{}{(5.506,1.966)}
\gp3point{gp mark 2}{}{(5.537,1.966)}
\gp3point{gp mark 2}{}{(5.569,1.936)}
\gp3point{gp mark 2}{}{(5.600,1.965)}
\gp3point{gp mark 2}{}{(5.631,1.936)}
\gp3point{gp mark 2}{}{(5.662,1.964)}
\gp3point{gp mark 2}{}{(5.694,1.937)}
\gp3point{gp mark 2}{}{(5.725,1.964)}
\gp3point{gp mark 2}{}{(5.756,1.937)}
\gp3point{gp mark 2}{}{(5.788,1.963)}
\gp3point{gp mark 2}{}{(5.819,1.938)}
\gp3point{gp mark 2}{}{(5.850,1.963)}
\gp3point{gp mark 2}{}{(5.882,1.963)}
\gp3point{gp mark 2}{}{(5.913,1.939)}
\gp3point{gp mark 2}{}{(5.944,1.962)}
\gp3point{gp mark 2}{}{(5.975,1.940)}
\gp3point{gp mark 2}{}{(6.007,1.961)}
\gp3point{gp mark 2}{}{(6.038,1.961)}
\gp3point{gp mark 2}{}{(6.069,1.940)}
\gp3point{gp mark 2}{}{(6.101,1.941)}
\gp3point{gp mark 2}{}{(6.132,1.960)}
\gp3point{gp mark 2}{}{(6.163,1.941)}
\gp3point{gp mark 2}{}{(6.194,1.960)}
\gp3point{gp mark 2}{}{(6.226,1.960)}
\gp3point{gp mark 2}{}{(6.257,1.960)}
\gp3point{gp mark 2}{}{(6.288,1.942)}
\gp3point{gp mark 2}{}{(6.320,1.942)}
\gp3point{gp mark 2}{}{(6.351,1.959)}
\gp3point{gp mark 2}{}{(6.382,1.942)}
\gp3point{gp mark 2}{}{(6.414,1.959)}
\gp3point{gp mark 2}{}{(6.445,1.943)}
\gp3point{gp mark 2}{}{(6.476,1.943)}
\gp3point{gp mark 2}{}{(6.507,1.958)}
\gp3point{gp mark 2}{}{(6.539,1.943)}
\gp3point{gp mark 2}{}{(6.570,1.958)}
\gp3point{gp mark 2}{}{(6.601,1.958)}
\gp3point{gp mark 2}{}{(6.633,1.944)}
\gp3point{gp mark 2}{}{(6.664,1.958)}
\gp3point{gp mark 2}{}{(6.695,1.944)}
\gp3point{gp mark 2}{}{(6.726,1.957)}
\gp3point{gp mark 2}{}{(6.758,1.957)}
\gp3point{gp mark 2}{}{(6.789,1.944)}
\gp3point{gp mark 2}{}{(6.820,1.957)}
\gp3point{gp mark 2}{}{(6.852,1.945)}
\gp3point{gp mark 2}{}{(6.883,1.945)}
\gp3point{gp mark 2}{}{(6.914,1.956)}
\gp3point{gp mark 2}{}{(6.946,1.956)}
\gp3point{gp mark 2}{}{(6.977,1.945)}
\gp3point{gp mark 2}{}{(7.008,1.945)}
\gp3point{gp mark 2}{}{(7.039,1.956)}
\gp3point{gp mark 2}{}{(7.071,1.956)}
\gp3point{gp mark 2}{}{(7.102,1.946)}
\gp3point{gp mark 2}{}{(7.133,1.956)}
\gp3point{gp mark 2}{}{(7.165,1.946)}
\gp3point{gp mark 2}{}{(7.196,1.955)}
\gp3point{gp mark 2}{}{(7.227,1.946)}
\gp3point{gp mark 2}{}{(7.259,1.946)}
\gp3point{gp mark 2}{}{(7.290,1.955)}
\gp3point{gp mark 2}{}{(7.321,1.955)}
\gp3point{gp mark 2}{}{(7.352,1.946)}
\gp3point{gp mark 2}{}{(7.384,1.955)}
\gp3point{gp mark 2}{}{(7.415,1.947)}
\gp3point{gp mark 2}{}{(7.446,1.955)}
\gp3point{gp mark 2}{}{(7.478,1.947)}
\gp3point{gp mark 2}{}{(7.509,1.954)}
\gp3point{gp mark 2}{}{(7.540,1.954)}
\gp3point{gp mark 2}{}{(7.571,1.947)}
\gp3point{gp mark 2}{}{(7.603,1.954)}
\gp3point{gp mark 2}{}{(7.634,1.947)}
\gp3point{gp mark 2}{}{(7.665,1.947)}
\gp3point{gp mark 2}{}{(7.697,1.954)}
\gp3point{gp mark 2}{}{(7.728,1.954)}
\gp3point{gp mark 2}{}{(7.759,1.948)}
\gp3point{gp mark 2}{}{(7.791,1.954)}
\gp3point{gp mark 2}{}{(7.822,1.954)}
\gp3point{gp mark 2}{}{(7.853,1.948)}
\gp3point{gp mark 2}{}{(7.884,1.948)}
\gp3point{gp mark 2}{}{(7.916,1.953)}
\gp3point{gp mark 2}{}{(7.947,1.953)}
\gpcolor{color=gp lt color border}
\draw[gp path] (1.688,4.075)--(1.688,0.985)--(7.947,0.985)--(7.947,4.075)--cycle;
\node[gp node center] at (4.817,4.537) {eigenvalue 16 - 200};
\gpdefrectangularnode{gp plot 1}{\pgfpoint{1.688cm}{0.985cm}}{\pgfpoint{7.947cm}{4.075cm}}
\end{tikzpicture}

%% file: figures/petsc_scaling/strong_scaling.tex
\begin{tikzpicture}[gnuplot]
\tikzset{every node/.append style={font={\footnotesize}}}
\path (0.000,0.000) rectangle (8.500,6.000);
\gpcolor{color=gp lt color axes}
\gpsetlinetype{gp lt axes}
\gpsetdashtype{gp dt axes}
\gpsetlinewidth{0.50}
\draw[gp path] (1.504,1.481)--(7.947,1.481);
\gpcolor{color=gp lt color border}
\gpsetlinetype{gp lt border}
\gpsetdashtype{gp dt solid}
\gpsetlinewidth{1.00}
\draw[gp path] (1.504,1.481)--(1.684,1.481);
\draw[gp path] (7.947,1.481)--(7.767,1.481);
\node[gp node right] at (1.320,1.481) {$0.25$};
\gpcolor{color=gp lt color axes}
\gpsetlinetype{gp lt axes}
\gpsetdashtype{gp dt axes}
\gpsetlinewidth{0.50}
\draw[gp path] (1.504,2.231)--(7.947,2.231);
\gpcolor{color=gp lt color border}
\gpsetlinetype{gp lt border}
\gpsetdashtype{gp dt solid}
\gpsetlinewidth{1.00}
\draw[gp path] (1.504,2.231)--(1.684,2.231);
\draw[gp path] (7.947,2.231)--(7.767,2.231);
\node[gp node right] at (1.320,2.231) {$1$};
\gpcolor{color=gp lt color axes}
\gpsetlinetype{gp lt axes}
\gpsetdashtype{gp dt axes}
\gpsetlinewidth{0.50}
\draw[gp path] (1.504,2.980)--(7.947,2.980);
\gpcolor{color=gp lt color border}
\gpsetlinetype{gp lt border}
\gpsetdashtype{gp dt solid}
\gpsetlinewidth{1.00}
\draw[gp path] (1.504,2.980)--(1.684,2.980);
\draw[gp path] (7.947,2.980)--(7.767,2.980);
\node[gp node right] at (1.320,2.980) {$4$};
\gpcolor{color=gp lt color axes}
\gpsetlinetype{gp lt axes}
\gpsetdashtype{gp dt axes}
\gpsetlinewidth{0.50}
\draw[gp path] (1.504,3.730)--(5.191,3.730);
\draw[gp path] (7.763,3.730)--(7.947,3.730);
\gpcolor{color=gp lt color border}
\gpsetlinetype{gp lt border}
\gpsetdashtype{gp dt solid}
\gpsetlinewidth{1.00}
\draw[gp path] (1.504,3.730)--(1.684,3.730);
\draw[gp path] (7.947,3.730)--(7.767,3.730);
\node[gp node right] at (1.320,3.730) {$16$};
\gpcolor{color=gp lt color axes}
\gpsetlinetype{gp lt axes}
\gpsetdashtype{gp dt axes}
\gpsetlinewidth{0.50}
\draw[gp path] (1.504,4.480)--(5.191,4.480);
\draw[gp path] (7.763,4.480)--(7.947,4.480);
\gpcolor{color=gp lt color border}
\gpsetlinetype{gp lt border}
\gpsetdashtype{gp dt solid}
\gpsetlinewidth{1.00}
\draw[gp path] (1.504,4.480)--(1.684,4.480);
\draw[gp path] (7.947,4.480)--(7.767,4.480);
\node[gp node right] at (1.320,4.480) {$64$};
\gpcolor{color=gp lt color axes}
\gpsetlinetype{gp lt axes}
\gpsetdashtype{gp dt axes}
\gpsetlinewidth{0.50}
\draw[gp path] (1.504,5.230)--(5.191,5.230);
\draw[gp path] (7.763,5.230)--(7.947,5.230);
\gpcolor{color=gp lt color border}
\gpsetlinetype{gp lt border}
\gpsetdashtype{gp dt solid}
\gpsetlinewidth{1.00}
\draw[gp path] (1.504,5.230)--(1.684,5.230);
\draw[gp path] (7.947,5.230)--(7.767,5.230);
\node[gp node right] at (1.320,5.230) {$256$};
\gpcolor{color=gp lt color axes}
\gpsetlinetype{gp lt axes}
\gpsetdashtype{gp dt axes}
\gpsetlinewidth{0.50}
\draw[gp path] (1.504,0.985)--(1.504,5.691);
\gpcolor{color=gp lt color border}
\gpsetlinetype{gp lt border}
\gpsetdashtype{gp dt solid}
\gpsetlinewidth{1.00}
\draw[gp path] (1.504,0.985)--(1.504,1.165);
\draw[gp path] (1.504,5.691)--(1.504,5.511);
\node[gp node center] at (1.504,0.677) {$1$};
\gpcolor{color=gp lt color axes}
\gpsetlinetype{gp lt axes}
\gpsetdashtype{gp dt axes}
\gpsetlinewidth{0.50}
\draw[gp path] (2.495,0.985)--(2.495,5.691);
\gpcolor{color=gp lt color border}
\gpsetlinetype{gp lt border}
\gpsetdashtype{gp dt solid}
\gpsetlinewidth{1.00}
\draw[gp path] (2.495,0.985)--(2.495,1.165);
\draw[gp path] (2.495,5.691)--(2.495,5.511);
\node[gp node center] at (2.495,0.677) {$4$};
\gpcolor{color=gp lt color axes}
\gpsetlinetype{gp lt axes}
\gpsetdashtype{gp dt axes}
\gpsetlinewidth{0.50}
\draw[gp path] (3.486,0.985)--(3.486,5.691);
\gpcolor{color=gp lt color border}
\gpsetlinetype{gp lt border}
\gpsetdashtype{gp dt solid}
\gpsetlinewidth{1.00}
\draw[gp path] (3.486,0.985)--(3.486,1.165);
\draw[gp path] (3.486,5.691)--(3.486,5.511);
\node[gp node center] at (3.486,0.677) {$16$};
\gpcolor{color=gp lt color axes}
\gpsetlinetype{gp lt axes}
\gpsetdashtype{gp dt axes}
\gpsetlinewidth{0.50}
\draw[gp path] (4.478,0.985)--(4.478,5.691);
\gpcolor{color=gp lt color border}
\gpsetlinetype{gp lt border}
\gpsetdashtype{gp dt solid}
\gpsetlinewidth{1.00}
\draw[gp path] (4.478,0.985)--(4.478,1.165);
\draw[gp path] (4.478,5.691)--(4.478,5.511);
\node[gp node center] at (4.478,0.677) {$64$};
\gpcolor{color=gp lt color axes}
\gpsetlinetype{gp lt axes}
\gpsetdashtype{gp dt axes}
\gpsetlinewidth{0.50}
\draw[gp path] (5.469,0.985)--(5.469,3.355);
\draw[gp path] (5.469,5.511)--(5.469,5.691);
\gpcolor{color=gp lt color border}
\gpsetlinetype{gp lt border}
\gpsetdashtype{gp dt solid}
\gpsetlinewidth{1.00}
\draw[gp path] (5.469,0.985)--(5.469,1.165);
\draw[gp path] (5.469,5.691)--(5.469,5.511);
\node[gp node center] at (5.469,0.677) {$256$};
\gpcolor{color=gp lt color axes}
\gpsetlinetype{gp lt axes}
\gpsetdashtype{gp dt axes}
\gpsetlinewidth{0.50}
\draw[gp path] (6.460,0.985)--(6.460,3.355);
\draw[gp path] (6.460,5.511)--(6.460,5.691);
\gpcolor{color=gp lt color border}
\gpsetlinetype{gp lt border}
\gpsetdashtype{gp dt solid}
\gpsetlinewidth{1.00}
\draw[gp path] (6.460,0.985)--(6.460,1.165);
\draw[gp path] (6.460,5.691)--(6.460,5.511);
\node[gp node center] at (6.460,0.677) {$1024$};
\gpcolor{color=gp lt color axes}
\gpsetlinetype{gp lt axes}
\gpsetdashtype{gp dt axes}
\gpsetlinewidth{0.50}
\draw[gp path] (7.451,0.985)--(7.451,3.355);
\draw[gp path] (7.451,5.511)--(7.451,5.691);
\gpcolor{color=gp lt color border}
\gpsetlinetype{gp lt border}
\gpsetdashtype{gp dt solid}
\gpsetlinewidth{1.00}
\draw[gp path] (7.451,0.985)--(7.451,1.165);
\draw[gp path] (7.451,5.691)--(7.451,5.511);
\node[gp node center] at (7.451,0.677) {$4096$};
\draw[gp path] (1.504,5.691)--(1.504,0.985)--(7.947,0.985)--(7.947,5.691)--cycle;
\gpcolor{rgb color={0.580,0.000,0.827}}
\node[gp node left] at (2.400,2.203) {Qutip};
\gpcolor{rgb color={0.000,0.620,0.451}}
\node[gp node left] at (2.400,3.927) {Qutip};
\gpcolor{rgb color={0.337,0.706,0.914}}
\node[gp node left] at (2.400,5.592) {Qutip};
\gpcolor{color=gp lt color border}
\node[gp node center,rotate=-270] at (0.292,3.338) {Time (sec) for $N_T=10$ time steps};
\node[gp node center] at (4.725,0.215) {Number of cores};
\node[gp node right] at (6.479,5.357) {$SU(2^8)$};
\gpcolor{rgb color={0.580,0.000,0.827}}
\gpsetdashtype{gp dt 4}
\gpsetlinewidth{2.00}
\draw[gp path] (6.663,5.357)--(7.579,5.357);
\draw[gp path] (1.504,2.267)--(2.000,2.168)--(2.495,1.897)--(2.991,1.647)--(3.486,1.412)%
  --(3.982,1.303);
\gpsetpointsize{4.00}
\gp3point{gp mark 7}{}{(1.504,2.267)}
\gp3point{gp mark 7}{}{(2.000,2.168)}
\gp3point{gp mark 7}{}{(2.495,1.897)}
\gp3point{gp mark 7}{}{(2.991,1.647)}
\gp3point{gp mark 7}{}{(3.486,1.412)}
\gp3point{gp mark 7}{}{(3.982,1.303)}
\gp3point{gp mark 7}{}{(7.121,5.357)}
\gpcolor{color=gp lt color border}
\node[gp node right] at (6.479,5.049) {$SU(16^2)$};
\gpcolor{rgb color={0.580,0.000,0.827}}
\draw[gp path] (6.663,5.049)--(7.579,5.049);
\draw[gp path] (1.504,2.174)--(2.000,1.989)--(2.495,1.693)--(2.991,1.502)--(3.486,1.332)%
  --(3.982,1.239);
\gp3point{gp mark 8}{}{(1.504,2.174)}
\gp3point{gp mark 8}{}{(2.000,1.989)}
\gp3point{gp mark 8}{}{(2.495,1.693)}
\gp3point{gp mark 8}{}{(2.991,1.502)}
\gp3point{gp mark 8}{}{(3.486,1.332)}
\gp3point{gp mark 8}{}{(3.982,1.239)}
\gp3point{gp mark 8}{}{(7.121,5.049)}
\gpcolor{color=gp lt color border}
\node[gp node right] at (6.479,4.741) {$SU(2^{10})$};
\gpcolor{rgb color={0.000,0.620,0.451}}
\draw[gp path] (6.663,4.741)--(7.579,4.741);
\draw[gp path] (1.504,3.861)--(2.000,3.752)--(2.495,3.393)--(2.991,3.061)--(3.486,2.764)%
  --(3.982,2.501)--(4.478,2.203)--(4.973,1.998)--(5.469,1.845)--(5.965,1.799);
\gp3point{gp mark 7}{}{(1.504,3.861)}
\gp3point{gp mark 7}{}{(2.000,3.752)}
\gp3point{gp mark 7}{}{(2.495,3.393)}
\gp3point{gp mark 7}{}{(2.991,3.061)}
\gp3point{gp mark 7}{}{(3.486,2.764)}
\gp3point{gp mark 7}{}{(3.982,2.501)}
\gp3point{gp mark 7}{}{(4.478,2.203)}
\gp3point{gp mark 7}{}{(4.973,1.998)}
\gp3point{gp mark 7}{}{(5.469,1.845)}
\gp3point{gp mark 7}{}{(5.965,1.799)}
\gp3point{gp mark 7}{}{(7.121,4.741)}
\gpcolor{color=gp lt color border}
\node[gp node right] at (6.479,4.433) {$SU(32^2)$};
\gpcolor{rgb color={0.000,0.620,0.451}}
\draw[gp path] (6.663,4.433)--(7.579,4.433);
\draw[gp path] (1.504,3.685)--(2.000,3.429)--(2.495,3.162)--(2.991,2.805)--(3.486,2.543)%
  --(3.982,2.334)--(4.478,2.155)--(4.973,1.822)--(5.469,1.561)--(5.965,1.597);
\gp3point{gp mark 8}{}{(1.504,3.685)}
\gp3point{gp mark 8}{}{(2.000,3.429)}
\gp3point{gp mark 8}{}{(2.495,3.162)}
\gp3point{gp mark 8}{}{(2.991,2.805)}
\gp3point{gp mark 8}{}{(3.486,2.543)}
\gp3point{gp mark 8}{}{(3.982,2.334)}
\gp3point{gp mark 8}{}{(4.478,2.155)}
\gp3point{gp mark 8}{}{(4.973,1.822)}
\gp3point{gp mark 8}{}{(5.469,1.561)}
\gp3point{gp mark 8}{}{(5.965,1.597)}
\gp3point{gp mark 8}{}{(7.121,4.433)}
\gpcolor{color=gp lt color border}
\node[gp node right] at (6.479,4.125) {$SU(2^{12})$};
\gpcolor{rgb color={0.337,0.706,0.914}}
\draw[gp path] (6.663,4.125)--(7.579,4.125);
\draw[gp path] (1.504,5.510)--(2.000,5.430)--(2.495,5.063)--(2.991,4.715)--(3.486,4.417)%
  --(3.982,4.236)--(4.478,3.884)--(4.973,3.487)--(5.469,3.108)--(5.965,2.803)--(6.460,2.676)%
  --(6.956,2.518)--(7.451,2.377)--(7.947,2.635);
\gp3point{gp mark 7}{}{(1.504,5.510)}
\gp3point{gp mark 7}{}{(2.000,5.430)}
\gp3point{gp mark 7}{}{(2.495,5.063)}
\gp3point{gp mark 7}{}{(2.991,4.715)}
\gp3point{gp mark 7}{}{(3.486,4.417)}
\gp3point{gp mark 7}{}{(3.982,4.236)}
\gp3point{gp mark 7}{}{(4.478,3.884)}
\gp3point{gp mark 7}{}{(4.973,3.487)}
\gp3point{gp mark 7}{}{(5.469,3.108)}
\gp3point{gp mark 7}{}{(5.965,2.803)}
\gp3point{gp mark 7}{}{(6.460,2.676)}
\gp3point{gp mark 7}{}{(6.956,2.518)}
\gp3point{gp mark 7}{}{(7.451,2.377)}
\gp3point{gp mark 7}{}{(7.947,2.635)}
\gp3point{gp mark 7}{}{(7.121,4.125)}
\gpcolor{color=gp lt color border}
\node[gp node right] at (6.479,3.817) {$SU(64^2)$};
\gpcolor{rgb color={0.337,0.706,0.914}}
\draw[gp path] (6.663,3.817)--(7.579,3.817);
\draw[gp path] (1.504,5.203)--(2.000,5.011)--(2.495,4.708)--(2.991,4.351)--(3.486,4.075)%
  --(3.982,3.863)--(4.478,3.518)--(4.973,3.151)--(5.469,2.807)--(5.965,2.485)--(6.460,2.385)%
  --(6.956,2.393)--(7.451,2.311)--(7.947,2.803);
\gp3point{gp mark 8}{}{(1.504,5.203)}
\gp3point{gp mark 8}{}{(2.000,5.011)}
\gp3point{gp mark 8}{}{(2.495,4.708)}
\gp3point{gp mark 8}{}{(2.991,4.351)}
\gp3point{gp mark 8}{}{(3.486,4.075)}
\gp3point{gp mark 8}{}{(3.982,3.863)}
\gp3point{gp mark 8}{}{(4.478,3.518)}
\gp3point{gp mark 8}{}{(4.973,3.151)}
\gp3point{gp mark 8}{}{(5.469,2.807)}
\gp3point{gp mark 8}{}{(5.965,2.485)}
\gp3point{gp mark 8}{}{(6.460,2.385)}
\gp3point{gp mark 8}{}{(6.956,2.393)}
\gp3point{gp mark 8}{}{(7.451,2.311)}
\gp3point{gp mark 8}{}{(7.947,2.803)}
\gp3point{gp mark 8}{}{(7.121,3.817)}
\gpcolor{color=gp lt color border}
\node[gp node right] at (6.479,3.509) {perfect};
\gpcolor{rgb color={0.000,0.000,0.000}}
\gpsetdashtype{gp dt 2}
\gpsetlinewidth{1.00}
\draw[gp path] (6.663,3.509)--(7.579,3.509);
\draw[gp path] (1.504,5.316)--(1.569,5.267)--(1.634,5.218)--(1.699,5.168)--(1.764,5.119)%
  --(1.829,5.070)--(1.894,5.021)--(1.960,4.971)--(2.025,4.922)--(2.090,4.873)--(2.155,4.824)%
  --(2.220,4.774)--(2.285,4.725)--(2.350,4.676)--(2.415,4.627)--(2.480,4.577)--(2.545,4.528)%
  --(2.610,4.479)--(2.675,4.430)--(2.741,4.381)--(2.806,4.331)--(2.871,4.282)--(2.936,4.233)%
  --(3.001,4.184)--(3.066,4.134)--(3.131,4.085)--(3.196,4.036)--(3.261,3.987)--(3.326,3.937)%
  --(3.391,3.888)--(3.456,3.839)--(3.522,3.790)--(3.587,3.740)--(3.652,3.691)--(3.717,3.642)%
  --(3.782,3.593)--(3.847,3.544)--(3.912,3.494)--(3.977,3.445)--(4.042,3.396)--(4.107,3.347)%
  --(4.172,3.297)--(4.237,3.248)--(4.302,3.199)--(4.368,3.150)--(4.433,3.100)--(4.498,3.051)%
  --(4.563,3.002)--(4.628,2.953)--(4.693,2.903)--(4.758,2.854)--(4.823,2.805)--(4.888,2.756)%
  --(4.953,2.706)--(5.018,2.657)--(5.083,2.608)--(5.149,2.559)--(5.214,2.510)--(5.279,2.460)%
  --(5.344,2.411)--(5.409,2.362)--(5.474,2.313)--(5.539,2.263)--(5.604,2.214)--(5.669,2.165)%
  --(5.734,2.116)--(5.799,2.066)--(5.864,2.017)--(5.929,1.968)--(5.995,1.919)--(6.060,1.869)%
  --(6.125,1.820)--(6.190,1.771)--(6.255,1.722)--(6.320,1.673)--(6.385,1.623)--(6.450,1.574)%
  --(6.515,1.525)--(6.580,1.476)--(6.645,1.426)--(6.710,1.377)--(6.776,1.328)--(6.841,1.279)%
  --(6.906,1.229)--(6.971,1.180)--(7.036,1.131)--(7.101,1.082)--(7.166,1.032)--(7.228,0.985);
\gpcolor{rgb color={0.580,0.000,0.827}}
\gpsetdashtype{gp dt 4}
\gpsetlinewidth{2.00}
\draw[gp path] (1.504,2.075)--(2.000,2.075)--(2.495,2.075)--(2.991,2.075)--(3.486,2.075);
\gpcolor{rgb color={0.000,0.620,0.451}}
\draw[gp path] (1.504,3.757)--(2.000,3.757)--(2.495,3.757)--(2.991,3.757)--(3.486,3.757);
\gpcolor{rgb color={0.337,0.706,0.914}}
\draw[gp path] (1.504,5.440)--(2.000,5.440)--(2.495,5.440)--(2.991,5.440)--(3.486,5.440);
\gpcolor{color=gp lt color border}
\gpsetdashtype{gp dt solid}
\gpsetlinewidth{1.00}
\draw[gp path] (1.504,5.691)--(1.504,0.985)--(7.947,0.985)--(7.947,5.691)--cycle;
\gpdefrectangularnode{gp plot 1}{\pgfpoint{1.504cm}{0.985cm}}{\pgfpoint{7.947cm}{5.691cm}}
\end{tikzpicture}

%% file: figures/petsc_scaling/speedup.tex
\begin{tikzpicture}[gnuplot]
\tikzset{every node/.append style={font={\footnotesize}}}
\path (0.000,0.000) rectangle (8.500,6.000);
\gpcolor{color=gp lt color axes}
\gpsetlinetype{gp lt axes}
\gpsetdashtype{gp dt axes}
\gpsetlinewidth{0.50}
\draw[gp path] (1.320,0.985)--(7.947,0.985);
\gpcolor{color=gp lt color border}
\gpsetlinetype{gp lt border}
\gpsetdashtype{gp dt solid}
\gpsetlinewidth{1.00}
\draw[gp path] (1.320,0.985)--(1.500,0.985);
\draw[gp path] (7.947,0.985)--(7.767,0.985);
\node[gp node right] at (1.136,0.985) {$1$};
\gpcolor{color=gp lt color axes}
\gpsetlinetype{gp lt axes}
\gpsetdashtype{gp dt axes}
\gpsetlinewidth{0.50}
\draw[gp path] (1.320,1.457)--(7.947,1.457);
\gpcolor{color=gp lt color border}
\gpsetlinetype{gp lt border}
\gpsetdashtype{gp dt solid}
\gpsetlinewidth{1.00}
\draw[gp path] (1.320,1.457)--(1.500,1.457);
\draw[gp path] (7.947,1.457)--(7.767,1.457);
\node[gp node right] at (1.136,1.457) {$2$};
\gpcolor{color=gp lt color axes}
\gpsetlinetype{gp lt axes}
\gpsetdashtype{gp dt axes}
\gpsetlinewidth{0.50}
\draw[gp path] (1.320,1.929)--(7.947,1.929);
\gpcolor{color=gp lt color border}
\gpsetlinetype{gp lt border}
\gpsetdashtype{gp dt solid}
\gpsetlinewidth{1.00}
\draw[gp path] (1.320,1.929)--(1.500,1.929);
\draw[gp path] (7.947,1.929)--(7.767,1.929);
\node[gp node right] at (1.136,1.929) {$4$};
\gpcolor{color=gp lt color axes}
\gpsetlinetype{gp lt axes}
\gpsetdashtype{gp dt axes}
\gpsetlinewidth{0.50}
\draw[gp path] (1.320,2.402)--(7.947,2.402);
\gpcolor{color=gp lt color border}
\gpsetlinetype{gp lt border}
\gpsetdashtype{gp dt solid}
\gpsetlinewidth{1.00}
\draw[gp path] (1.320,2.402)--(1.500,2.402);
\draw[gp path] (7.947,2.402)--(7.767,2.402);
\node[gp node right] at (1.136,2.402) {$8$};
\gpcolor{color=gp lt color axes}
\gpsetlinetype{gp lt axes}
\gpsetdashtype{gp dt axes}
\gpsetlinewidth{0.50}
\draw[gp path] (1.320,2.874)--(7.947,2.874);
\gpcolor{color=gp lt color border}
\gpsetlinetype{gp lt border}
\gpsetdashtype{gp dt solid}
\gpsetlinewidth{1.00}
\draw[gp path] (1.320,2.874)--(1.500,2.874);
\draw[gp path] (7.947,2.874)--(7.767,2.874);
\node[gp node right] at (1.136,2.874) {$16$};
\gpcolor{color=gp lt color axes}
\gpsetlinetype{gp lt axes}
\gpsetdashtype{gp dt axes}
\gpsetlinewidth{0.50}
\draw[gp path] (1.320,3.346)--(7.947,3.346);
\gpcolor{color=gp lt color border}
\gpsetlinetype{gp lt border}
\gpsetdashtype{gp dt solid}
\gpsetlinewidth{1.00}
\draw[gp path] (1.320,3.346)--(1.500,3.346);
\draw[gp path] (7.947,3.346)--(7.767,3.346);
\node[gp node right] at (1.136,3.346) {$32$};
\gpcolor{color=gp lt color axes}
\gpsetlinetype{gp lt axes}
\gpsetdashtype{gp dt axes}
\gpsetlinewidth{0.50}
\draw[gp path] (1.320,3.818)--(1.504,3.818);
\draw[gp path] (4.076,3.818)--(7.947,3.818);
\gpcolor{color=gp lt color border}
\gpsetlinetype{gp lt border}
\gpsetdashtype{gp dt solid}
\gpsetlinewidth{1.00}
\draw[gp path] (1.320,3.818)--(1.500,3.818);
\draw[gp path] (7.947,3.818)--(7.767,3.818);
\node[gp node right] at (1.136,3.818) {$64$};
\gpcolor{color=gp lt color axes}
\gpsetlinetype{gp lt axes}
\gpsetdashtype{gp dt axes}
\gpsetlinewidth{0.50}
\draw[gp path] (1.320,4.291)--(1.504,4.291);
\draw[gp path] (4.076,4.291)--(7.947,4.291);
\gpcolor{color=gp lt color border}
\gpsetlinetype{gp lt border}
\gpsetdashtype{gp dt solid}
\gpsetlinewidth{1.00}
\draw[gp path] (1.320,4.291)--(1.500,4.291);
\draw[gp path] (7.947,4.291)--(7.767,4.291);
\node[gp node right] at (1.136,4.291) {$128$};
\gpcolor{color=gp lt color axes}
\gpsetlinetype{gp lt axes}
\gpsetdashtype{gp dt axes}
\gpsetlinewidth{0.50}
\draw[gp path] (1.320,4.763)--(1.504,4.763);
\draw[gp path] (4.076,4.763)--(7.947,4.763);
\gpcolor{color=gp lt color border}
\gpsetlinetype{gp lt border}
\gpsetdashtype{gp dt solid}
\gpsetlinewidth{1.00}
\draw[gp path] (1.320,4.763)--(1.500,4.763);
\draw[gp path] (7.947,4.763)--(7.767,4.763);
\node[gp node right] at (1.136,4.763) {$256$};
\gpcolor{color=gp lt color axes}
\gpsetlinetype{gp lt axes}
\gpsetdashtype{gp dt axes}
\gpsetlinewidth{0.50}
\draw[gp path] (1.320,5.235)--(1.504,5.235);
\draw[gp path] (4.076,5.235)--(7.947,5.235);
\gpcolor{color=gp lt color border}
\gpsetlinetype{gp lt border}
\gpsetdashtype{gp dt solid}
\gpsetlinewidth{1.00}
\draw[gp path] (1.320,5.235)--(1.500,5.235);
\draw[gp path] (7.947,5.235)--(7.767,5.235);
\node[gp node right] at (1.136,5.235) {$512$};
\gpcolor{color=gp lt color axes}
\gpsetlinetype{gp lt axes}
\gpsetdashtype{gp dt axes}
\gpsetlinewidth{0.50}
\draw[gp path] (1.320,0.985)--(1.320,5.691);
\gpcolor{color=gp lt color border}
\gpsetlinetype{gp lt border}
\gpsetdashtype{gp dt solid}
\gpsetlinewidth{1.00}
\draw[gp path] (1.320,0.985)--(1.320,1.165);
\draw[gp path] (1.320,5.691)--(1.320,5.511);
\node[gp node center] at (1.320,0.677) {$1$};
\gpcolor{color=gp lt color axes}
\gpsetlinetype{gp lt axes}
\gpsetdashtype{gp dt axes}
\gpsetlinewidth{0.50}
\draw[gp path] (2.340,0.985)--(2.340,3.663);
\draw[gp path] (2.340,5.511)--(2.340,5.691);
\gpcolor{color=gp lt color border}
\gpsetlinetype{gp lt border}
\gpsetdashtype{gp dt solid}
\gpsetlinewidth{1.00}
\draw[gp path] (2.340,0.985)--(2.340,1.165);
\draw[gp path] (2.340,5.691)--(2.340,5.511);
\node[gp node center] at (2.340,0.677) {$4$};
\gpcolor{color=gp lt color axes}
\gpsetlinetype{gp lt axes}
\gpsetdashtype{gp dt axes}
\gpsetlinewidth{0.50}
\draw[gp path] (3.359,0.985)--(3.359,3.663);
\draw[gp path] (3.359,5.511)--(3.359,5.691);
\gpcolor{color=gp lt color border}
\gpsetlinetype{gp lt border}
\gpsetdashtype{gp dt solid}
\gpsetlinewidth{1.00}
\draw[gp path] (3.359,0.985)--(3.359,1.165);
\draw[gp path] (3.359,5.691)--(3.359,5.511);
\node[gp node center] at (3.359,0.677) {$16$};
\gpcolor{color=gp lt color axes}
\gpsetlinetype{gp lt axes}
\gpsetdashtype{gp dt axes}
\gpsetlinewidth{0.50}
\draw[gp path] (4.379,0.985)--(4.379,5.691);
\gpcolor{color=gp lt color border}
\gpsetlinetype{gp lt border}
\gpsetdashtype{gp dt solid}
\gpsetlinewidth{1.00}
\draw[gp path] (4.379,0.985)--(4.379,1.165);
\draw[gp path] (4.379,5.691)--(4.379,5.511);
\node[gp node center] at (4.379,0.677) {$64$};
\gpcolor{color=gp lt color axes}
\gpsetlinetype{gp lt axes}
\gpsetdashtype{gp dt axes}
\gpsetlinewidth{0.50}
\draw[gp path] (5.398,0.985)--(5.398,5.691);
\gpcolor{color=gp lt color border}
\gpsetlinetype{gp lt border}
\gpsetdashtype{gp dt solid}
\gpsetlinewidth{1.00}
\draw[gp path] (5.398,0.985)--(5.398,1.165);
\draw[gp path] (5.398,5.691)--(5.398,5.511);
\node[gp node center] at (5.398,0.677) {$256$};
\gpcolor{color=gp lt color axes}
\gpsetlinetype{gp lt axes}
\gpsetdashtype{gp dt axes}
\gpsetlinewidth{0.50}
\draw[gp path] (6.418,0.985)--(6.418,5.691);
\gpcolor{color=gp lt color border}
\gpsetlinetype{gp lt border}
\gpsetdashtype{gp dt solid}
\gpsetlinewidth{1.00}
\draw[gp path] (6.418,0.985)--(6.418,1.165);
\draw[gp path] (6.418,5.691)--(6.418,5.511);
\node[gp node center] at (6.418,0.677) {$1024$};
\gpcolor{color=gp lt color axes}
\gpsetlinetype{gp lt axes}
\gpsetdashtype{gp dt axes}
\gpsetlinewidth{0.50}
\draw[gp path] (7.437,0.985)--(7.437,5.691);
\gpcolor{color=gp lt color border}
\gpsetlinetype{gp lt border}
\gpsetdashtype{gp dt solid}
\gpsetlinewidth{1.00}
\draw[gp path] (7.437,0.985)--(7.437,1.165);
\draw[gp path] (7.437,5.691)--(7.437,5.511);
\node[gp node center] at (7.437,0.677) {$4096$};
\draw[gp path] (1.320,5.691)--(1.320,0.985)--(7.947,0.985)--(7.947,5.691)--cycle;
\node[gp node center,rotate=-270] at (0.292,3.338) {Speedup};
\node[gp node center] at (4.633,0.215) {Number of cores};
\node[gp node right] at (2.792,5.357) {$SU(2^8)$};
\gpcolor{rgb color={0.580,0.000,0.827}}
\gpsetdashtype{gp dt 4}
\gpsetlinewidth{2.00}
\draw[gp path] (2.976,5.357)--(3.892,5.357);
\draw[gp path] (1.320,0.985)--(1.830,1.110)--(2.340,1.451)--(2.849,1.766)--(3.359,2.063)%
  --(3.869,2.199);
\gpsetpointsize{4.00}
\gp3point{gp mark 7}{}{(1.320,0.985)}
\gp3point{gp mark 7}{}{(1.830,1.110)}
\gp3point{gp mark 7}{}{(2.340,1.451)}
\gp3point{gp mark 7}{}{(2.849,1.766)}
\gp3point{gp mark 7}{}{(3.359,2.063)}
\gp3point{gp mark 7}{}{(3.869,2.199)}
\gp3point{gp mark 7}{}{(3.434,5.357)}
\gpcolor{color=gp lt color border}
\node[gp node right] at (2.792,5.049) {$SU(16^2)$};
\gpcolor{rgb color={0.580,0.000,0.827}}
\draw[gp path] (2.976,5.049)--(3.892,5.049);
\draw[gp path] (1.320,0.985)--(1.830,1.217)--(2.340,1.591)--(2.849,1.831)--(3.359,2.045)%
  --(3.869,2.162);
\gp3point{gp mark 8}{}{(1.320,0.985)}
\gp3point{gp mark 8}{}{(1.830,1.217)}
\gp3point{gp mark 8}{}{(2.340,1.591)}
\gp3point{gp mark 8}{}{(2.849,1.831)}
\gp3point{gp mark 8}{}{(3.359,2.045)}
\gp3point{gp mark 8}{}{(3.869,2.162)}
\gp3point{gp mark 8}{}{(3.434,5.049)}
\gpcolor{color=gp lt color border}
\node[gp node right] at (2.792,4.741) {$SU(2^{10})$};
\gpcolor{rgb color={0.000,0.620,0.451}}
\draw[gp path] (2.976,4.741)--(3.892,4.741);
\draw[gp path] (1.320,0.985)--(1.830,1.122)--(2.340,1.575)--(2.849,1.993)--(3.359,2.367)%
  --(3.869,2.697)--(4.379,3.073)--(4.888,3.332)--(5.398,3.524)--(5.908,3.582);
\gp3point{gp mark 7}{}{(1.320,0.985)}
\gp3point{gp mark 7}{}{(1.830,1.122)}
\gp3point{gp mark 7}{}{(2.340,1.575)}
\gp3point{gp mark 7}{}{(2.849,1.993)}
\gp3point{gp mark 7}{}{(3.359,2.367)}
\gp3point{gp mark 7}{}{(3.869,2.697)}
\gp3point{gp mark 7}{}{(4.379,3.073)}
\gp3point{gp mark 7}{}{(4.888,3.332)}
\gp3point{gp mark 7}{}{(5.398,3.524)}
\gp3point{gp mark 7}{}{(5.908,3.582)}
\gp3point{gp mark 7}{}{(3.434,4.741)}
\gpcolor{color=gp lt color border}
\node[gp node right] at (2.792,4.433) {$SU(32^2)$};
\gpcolor{rgb color={0.000,0.620,0.451}}
\draw[gp path] (2.976,4.433)--(3.892,4.433);
\draw[gp path] (1.320,0.985)--(1.830,1.306)--(2.340,1.644)--(2.849,2.093)--(3.359,2.423)%
  --(3.869,2.686)--(4.379,2.911)--(4.888,3.331)--(5.398,3.659)--(5.908,3.614);
\gp3point{gp mark 8}{}{(1.320,0.985)}
\gp3point{gp mark 8}{}{(1.830,1.306)}
\gp3point{gp mark 8}{}{(2.340,1.644)}
\gp3point{gp mark 8}{}{(2.849,2.093)}
\gp3point{gp mark 8}{}{(3.359,2.423)}
\gp3point{gp mark 8}{}{(3.869,2.686)}
\gp3point{gp mark 8}{}{(4.379,2.911)}
\gp3point{gp mark 8}{}{(4.888,3.331)}
\gp3point{gp mark 8}{}{(5.398,3.659)}
\gp3point{gp mark 8}{}{(5.908,3.614)}
\gp3point{gp mark 8}{}{(3.434,4.433)}
\gpcolor{color=gp lt color border}
\node[gp node right] at (2.792,4.125) {$SU(2^{12})$};
\gpcolor{rgb color={0.337,0.706,0.914}}
\draw[gp path] (2.976,4.125)--(3.892,4.125);
\draw[gp path] (1.320,0.985)--(1.830,1.086)--(2.340,1.548)--(2.849,1.986)--(3.359,2.362)%
  --(3.869,2.590)--(4.379,3.033)--(4.888,3.533)--(5.398,4.011)--(5.908,4.395)--(6.418,4.554)%
  --(6.927,4.754)--(7.437,4.931)--(7.947,4.607);
\gp3point{gp mark 7}{}{(1.320,0.985)}
\gp3point{gp mark 7}{}{(1.830,1.086)}
\gp3point{gp mark 7}{}{(2.340,1.548)}
\gp3point{gp mark 7}{}{(2.849,1.986)}
\gp3point{gp mark 7}{}{(3.359,2.362)}
\gp3point{gp mark 7}{}{(3.869,2.590)}
\gp3point{gp mark 7}{}{(4.379,3.033)}
\gp3point{gp mark 7}{}{(4.888,3.533)}
\gp3point{gp mark 7}{}{(5.398,4.011)}
\gp3point{gp mark 7}{}{(5.908,4.395)}
\gp3point{gp mark 7}{}{(6.418,4.554)}
\gp3point{gp mark 7}{}{(6.927,4.754)}
\gp3point{gp mark 7}{}{(7.437,4.931)}
\gp3point{gp mark 7}{}{(7.947,4.607)}
\gp3point{gp mark 7}{}{(3.434,4.125)}
\gpcolor{color=gp lt color border}
\node[gp node right] at (2.792,3.817) {$SU(64^2)$};
\gpcolor{rgb color={0.337,0.706,0.914}}
\draw[gp path] (2.976,3.817)--(3.892,3.817);
\draw[gp path] (1.320,0.985)--(1.830,1.226)--(2.340,1.608)--(2.849,2.058)--(3.359,2.404)%
  --(3.869,2.672)--(4.379,3.107)--(4.888,3.569)--(5.398,4.002)--(5.908,4.408)--(6.418,4.534)%
  --(6.927,4.523)--(7.437,4.627)--(7.947,4.007);
\gp3point{gp mark 8}{}{(1.320,0.985)}
\gp3point{gp mark 8}{}{(1.830,1.226)}
\gp3point{gp mark 8}{}{(2.340,1.608)}
\gp3point{gp mark 8}{}{(2.849,2.058)}
\gp3point{gp mark 8}{}{(3.359,2.404)}
\gp3point{gp mark 8}{}{(3.869,2.672)}
\gp3point{gp mark 8}{}{(4.379,3.107)}
\gp3point{gp mark 8}{}{(4.888,3.569)}
\gp3point{gp mark 8}{}{(5.398,4.002)}
\gp3point{gp mark 8}{}{(5.908,4.408)}
\gp3point{gp mark 8}{}{(6.418,4.534)}
\gp3point{gp mark 8}{}{(6.927,4.523)}
\gp3point{gp mark 8}{}{(7.437,4.627)}
\gp3point{gp mark 8}{}{(7.947,4.007)}
\gp3point{gp mark 8}{}{(3.434,3.817)}
\gpcolor{rgb color={0.000,0.000,0.000}}
\gpsetdashtype{gp dt 2}
\gpsetlinewidth{1.00}
\draw[gp path] (1.320,0.985)--(1.387,1.047)--(1.454,1.109)--(1.521,1.171)--(1.588,1.233)%
  --(1.655,1.295)--(1.722,1.357)--(1.789,1.419)--(1.856,1.481)--(1.922,1.543)--(1.989,1.605)%
  --(2.056,1.667)--(2.123,1.729)--(2.190,1.791)--(2.257,1.853)--(2.324,1.915)--(2.391,1.977)%
  --(2.458,2.039)--(2.525,2.101)--(2.592,2.163)--(2.659,2.225)--(2.726,2.287)--(2.793,2.349)%
  --(2.860,2.411)--(2.927,2.473)--(2.993,2.535)--(3.060,2.597)--(3.127,2.659)--(3.194,2.721)%
  --(3.261,2.783)--(3.328,2.845)--(3.395,2.907)--(3.462,2.969)--(3.529,3.031)--(3.596,3.093)%
  --(3.663,3.155)--(3.730,3.217)--(3.797,3.279)--(3.864,3.341)--(3.931,3.403)--(3.998,3.465)%
  --(4.065,3.527)--(4.131,3.589)--(4.198,3.651)--(4.265,3.713)--(4.332,3.775)--(4.399,3.837)%
  --(4.466,3.899)--(4.533,3.961)--(4.600,4.023)--(4.667,4.085)--(4.734,4.147)--(4.801,4.209)%
  --(4.868,4.271)--(4.935,4.333)--(5.002,4.395)--(5.069,4.457)--(5.136,4.519)--(5.202,4.581)%
  --(5.269,4.643)--(5.336,4.705)--(5.403,4.767)--(5.470,4.830)--(5.537,4.892)--(5.604,4.954)%
  --(5.671,5.016)--(5.738,5.078)--(5.805,5.140)--(5.872,5.202)--(5.939,5.264)--(6.006,5.326)%
  --(6.073,5.388)--(6.140,5.450)--(6.207,5.512)--(6.274,5.574)--(6.340,5.636)--(6.399,5.691);
\gpcolor{color=gp lt color border}
\gpsetdashtype{gp dt solid}
\draw[gp path] (1.320,5.691)--(1.320,0.985)--(7.947,0.985)--(7.947,5.691)--cycle;
\gpdefrectangularnode{gp plot 1}{\pgfpoint{1.320cm}{0.985cm}}{\pgfpoint{7.947cm}{5.691cm}}
\end{tikzpicture}

%% file: figures/petsc_scaling/weakscaling.tex
\begin{tikzpicture}[gnuplot]
\tikzset{every node/.append style={font={\footnotesize}}}
\path (0.000,0.000) rectangle (8.500,6.000);
\gpcolor{color=gp lt color axes}
\gpsetlinetype{gp lt axes}
\gpsetdashtype{gp dt axes}
\gpsetlinewidth{0.50}
\draw[gp path] (2.240,1.199)--(7.947,1.199);
\gpcolor{color=gp lt color border}
\gpsetlinetype{gp lt border}
\gpsetdashtype{gp dt solid}
\gpsetlinewidth{1.00}
\draw[gp path] (2.240,1.199)--(2.420,1.199);
\draw[gp path] (7.947,1.199)--(7.767,1.199);
\node[gp node right] at (2.056,1.199) {$0.015625$};
\gpcolor{color=gp lt color axes}
\gpsetlinetype{gp lt axes}
\gpsetdashtype{gp dt axes}
\gpsetlinewidth{0.50}
\draw[gp path] (2.240,1.627)--(7.947,1.627);
\gpcolor{color=gp lt color border}
\gpsetlinetype{gp lt border}
\gpsetdashtype{gp dt solid}
\gpsetlinewidth{1.00}
\draw[gp path] (2.240,1.627)--(2.420,1.627);
\draw[gp path] (7.947,1.627)--(7.767,1.627);
\node[gp node right] at (2.056,1.627) {$0.0625$};
\gpcolor{color=gp lt color axes}
\gpsetlinetype{gp lt axes}
\gpsetdashtype{gp dt axes}
\gpsetlinewidth{0.50}
\draw[gp path] (2.240,2.055)--(7.947,2.055);
\gpcolor{color=gp lt color border}
\gpsetlinetype{gp lt border}
\gpsetdashtype{gp dt solid}
\gpsetlinewidth{1.00}
\draw[gp path] (2.240,2.055)--(2.420,2.055);
\draw[gp path] (7.947,2.055)--(7.767,2.055);
\node[gp node right] at (2.056,2.055) {$0.25$};
\gpcolor{color=gp lt color axes}
\gpsetlinetype{gp lt axes}
\gpsetdashtype{gp dt axes}
\gpsetlinewidth{0.50}
\draw[gp path] (2.240,2.482)--(7.947,2.482);
\gpcolor{color=gp lt color border}
\gpsetlinetype{gp lt border}
\gpsetdashtype{gp dt solid}
\gpsetlinewidth{1.00}
\draw[gp path] (2.240,2.482)--(2.420,2.482);
\draw[gp path] (7.947,2.482)--(7.767,2.482);
\node[gp node right] at (2.056,2.482) {$1$};
\gpcolor{color=gp lt color axes}
\gpsetlinetype{gp lt axes}
\gpsetdashtype{gp dt axes}
\gpsetlinewidth{0.50}
\draw[gp path] (2.240,2.910)--(7.947,2.910);
\gpcolor{color=gp lt color border}
\gpsetlinetype{gp lt border}
\gpsetdashtype{gp dt solid}
\gpsetlinewidth{1.00}
\draw[gp path] (2.240,2.910)--(2.420,2.910);
\draw[gp path] (7.947,2.910)--(7.767,2.910);
\node[gp node right] at (2.056,2.910) {$4$};
\gpcolor{color=gp lt color axes}
\gpsetlinetype{gp lt axes}
\gpsetdashtype{gp dt axes}
\gpsetlinewidth{0.50}
\draw[gp path] (2.240,3.338)--(7.947,3.338);
\gpcolor{color=gp lt color border}
\gpsetlinetype{gp lt border}
\gpsetdashtype{gp dt solid}
\gpsetlinewidth{1.00}
\draw[gp path] (2.240,3.338)--(2.420,3.338);
\draw[gp path] (7.947,3.338)--(7.767,3.338);
\node[gp node right] at (2.056,3.338) {$16$};
\gpcolor{color=gp lt color axes}
\gpsetlinetype{gp lt axes}
\gpsetdashtype{gp dt axes}
\gpsetlinewidth{0.50}
\draw[gp path] (2.240,3.766)--(7.947,3.766);
\gpcolor{color=gp lt color border}
\gpsetlinetype{gp lt border}
\gpsetdashtype{gp dt solid}
\gpsetlinewidth{1.00}
\draw[gp path] (2.240,3.766)--(2.420,3.766);
\draw[gp path] (7.947,3.766)--(7.767,3.766);
\node[gp node right] at (2.056,3.766) {$64$};
\gpcolor{color=gp lt color axes}
\gpsetlinetype{gp lt axes}
\gpsetdashtype{gp dt axes}
\gpsetlinewidth{0.50}
\draw[gp path] (2.240,4.194)--(7.947,4.194);
\gpcolor{color=gp lt color border}
\gpsetlinetype{gp lt border}
\gpsetdashtype{gp dt solid}
\gpsetlinewidth{1.00}
\draw[gp path] (2.240,4.194)--(2.420,4.194);
\draw[gp path] (7.947,4.194)--(7.767,4.194);
\node[gp node right] at (2.056,4.194) {$256$};
\gpcolor{color=gp lt color axes}
\gpsetlinetype{gp lt axes}
\gpsetdashtype{gp dt axes}
\gpsetlinewidth{0.50}
\draw[gp path] (2.240,4.621)--(7.947,4.621);
\gpcolor{color=gp lt color border}
\gpsetlinetype{gp lt border}
\gpsetdashtype{gp dt solid}
\gpsetlinewidth{1.00}
\draw[gp path] (2.240,4.621)--(2.420,4.621);
\draw[gp path] (7.947,4.621)--(7.767,4.621);
\node[gp node right] at (2.056,4.621) {$1024$};
\gpcolor{color=gp lt color axes}
\gpsetlinetype{gp lt axes}
\gpsetdashtype{gp dt axes}
\gpsetlinewidth{0.50}
\draw[gp path] (2.240,5.049)--(2.424,5.049);
\draw[gp path] (5.180,5.049)--(7.947,5.049);
\gpcolor{color=gp lt color border}
\gpsetlinetype{gp lt border}
\gpsetdashtype{gp dt solid}
\gpsetlinewidth{1.00}
\draw[gp path] (2.240,5.049)--(2.420,5.049);
\draw[gp path] (7.947,5.049)--(7.767,5.049);
\node[gp node right] at (2.056,5.049) {$4096$};
\gpcolor{color=gp lt color axes}
\gpsetlinetype{gp lt axes}
\gpsetdashtype{gp dt axes}
\gpsetlinewidth{0.50}
\draw[gp path] (2.240,5.477)--(2.424,5.477);
\draw[gp path] (5.180,5.477)--(7.947,5.477);
\gpcolor{color=gp lt color border}
\gpsetlinetype{gp lt border}
\gpsetdashtype{gp dt solid}
\gpsetlinewidth{1.00}
\draw[gp path] (2.240,5.477)--(2.420,5.477);
\draw[gp path] (7.947,5.477)--(7.767,5.477);
\node[gp node right] at (2.056,5.477) {$16384$};
\gpcolor{color=gp lt color axes}
\gpsetlinetype{gp lt axes}
\gpsetdashtype{gp dt axes}
\gpsetlinewidth{0.50}
\draw[gp path] (2.514,0.985)--(2.514,4.895);
\draw[gp path] (2.514,5.511)--(2.514,5.691);
\gpcolor{color=gp lt color border}
\gpsetlinetype{gp lt border}
\gpsetdashtype{gp dt solid}
\gpsetlinewidth{1.00}
\draw[gp path] (2.514,0.985)--(2.514,1.165);
\draw[gp path] (2.514,5.691)--(2.514,5.511);
\node[gp node center] at (2.514,0.677) {$3$};
\gpcolor{color=gp lt color axes}
\gpsetlinetype{gp lt axes}
\gpsetdashtype{gp dt axes}
\gpsetlinewidth{0.50}
\draw[gp path] (3.063,0.985)--(3.063,4.895);
\draw[gp path] (3.063,5.511)--(3.063,5.691);
\gpcolor{color=gp lt color border}
\gpsetlinetype{gp lt border}
\gpsetdashtype{gp dt solid}
\gpsetlinewidth{1.00}
\draw[gp path] (3.063,0.985)--(3.063,1.165);
\draw[gp path] (3.063,5.691)--(3.063,5.511);
\node[gp node center] at (3.063,0.677) {$4$};
\gpcolor{color=gp lt color axes}
\gpsetlinetype{gp lt axes}
\gpsetdashtype{gp dt axes}
\gpsetlinewidth{0.50}
\draw[gp path] (3.612,0.985)--(3.612,4.895);
\draw[gp path] (3.612,5.511)--(3.612,5.691);
\gpcolor{color=gp lt color border}
\gpsetlinetype{gp lt border}
\gpsetdashtype{gp dt solid}
\gpsetlinewidth{1.00}
\draw[gp path] (3.612,0.985)--(3.612,1.165);
\draw[gp path] (3.612,5.691)--(3.612,5.511);
\node[gp node center] at (3.612,0.677) {$5$};
\gpcolor{color=gp lt color axes}
\gpsetlinetype{gp lt axes}
\gpsetdashtype{gp dt axes}
\gpsetlinewidth{0.50}
\draw[gp path] (4.161,0.985)--(4.161,4.895);
\draw[gp path] (4.161,5.511)--(4.161,5.691);
\gpcolor{color=gp lt color border}
\gpsetlinetype{gp lt border}
\gpsetdashtype{gp dt solid}
\gpsetlinewidth{1.00}
\draw[gp path] (4.161,0.985)--(4.161,1.165);
\draw[gp path] (4.161,5.691)--(4.161,5.511);
\node[gp node center] at (4.161,0.677) {$6$};
\gpcolor{color=gp lt color axes}
\gpsetlinetype{gp lt axes}
\gpsetdashtype{gp dt axes}
\gpsetlinewidth{0.50}
\draw[gp path] (4.709,0.985)--(4.709,4.895);
\draw[gp path] (4.709,5.511)--(4.709,5.691);
\gpcolor{color=gp lt color border}
\gpsetlinetype{gp lt border}
\gpsetdashtype{gp dt solid}
\gpsetlinewidth{1.00}
\draw[gp path] (4.709,0.985)--(4.709,1.165);
\draw[gp path] (4.709,5.691)--(4.709,5.511);
\node[gp node center] at (4.709,0.677) {$7$};
\gpcolor{color=gp lt color axes}
\gpsetlinetype{gp lt axes}
\gpsetdashtype{gp dt axes}
\gpsetlinewidth{0.50}
\draw[gp path] (5.258,0.985)--(5.258,5.691);
\gpcolor{color=gp lt color border}
\gpsetlinetype{gp lt border}
\gpsetdashtype{gp dt solid}
\gpsetlinewidth{1.00}
\draw[gp path] (5.258,0.985)--(5.258,1.165);
\draw[gp path] (5.258,5.691)--(5.258,5.511);
\node[gp node center] at (5.258,0.677) {$8$};
\gpcolor{color=gp lt color axes}
\gpsetlinetype{gp lt axes}
\gpsetdashtype{gp dt axes}
\gpsetlinewidth{0.50}
\draw[gp path] (5.807,0.985)--(5.807,5.691);
\gpcolor{color=gp lt color border}
\gpsetlinetype{gp lt border}
\gpsetdashtype{gp dt solid}
\gpsetlinewidth{1.00}
\draw[gp path] (5.807,0.985)--(5.807,1.165);
\draw[gp path] (5.807,5.691)--(5.807,5.511);
\node[gp node center] at (5.807,0.677) {$9$};
\gpcolor{color=gp lt color axes}
\gpsetlinetype{gp lt axes}
\gpsetdashtype{gp dt axes}
\gpsetlinewidth{0.50}
\draw[gp path] (6.356,0.985)--(6.356,5.691);
\gpcolor{color=gp lt color border}
\gpsetlinetype{gp lt border}
\gpsetdashtype{gp dt solid}
\gpsetlinewidth{1.00}
\draw[gp path] (6.356,0.985)--(6.356,1.165);
\draw[gp path] (6.356,5.691)--(6.356,5.511);
\node[gp node center] at (6.356,0.677) {$10$};
\gpcolor{color=gp lt color axes}
\gpsetlinetype{gp lt axes}
\gpsetdashtype{gp dt axes}
\gpsetlinewidth{0.50}
\draw[gp path] (6.904,0.985)--(6.904,5.691);
\gpcolor{color=gp lt color border}
\gpsetlinetype{gp lt border}
\gpsetdashtype{gp dt solid}
\gpsetlinewidth{1.00}
\draw[gp path] (6.904,0.985)--(6.904,1.165);
\draw[gp path] (6.904,5.691)--(6.904,5.511);
\node[gp node center] at (6.904,0.677) {$11$};
\gpcolor{color=gp lt color axes}
\gpsetlinetype{gp lt axes}
\gpsetdashtype{gp dt axes}
\gpsetlinewidth{0.50}
\draw[gp path] (7.453,0.985)--(7.453,5.691);
\gpcolor{color=gp lt color border}
\gpsetlinetype{gp lt border}
\gpsetdashtype{gp dt solid}
\gpsetlinewidth{1.00}
\draw[gp path] (7.453,0.985)--(7.453,1.165);
\draw[gp path] (7.453,5.691)--(7.453,5.511);
\node[gp node center] at (7.453,0.677) {$12$};
\draw[gp path] (2.240,5.691)--(2.240,0.985)--(7.947,0.985)--(7.947,5.691)--cycle;
\node[gp node left] at (3.392,1.400) {1 core};
\node[gp node left] at (3.941,1.801) {2 cores};
\node[gp node left] at (4.490,2.111) {4 cores};
\node[gp node left] at (5.039,2.298) {16 cores};
\node[gp node left] at (5.587,2.497) {64 cores};
\node[gp node left] at (6.136,2.696) {256 cores};
\node[gp node left] at (6.685,2.906) {1024 cores};
\node[gp node left] at (7.234,3.476) {4096 cores};
\node[gp node center,rotate=-270] at (0.292,3.338) {Time (min) for $\nabla J$ with T=1ns};
\node[gp node center] at (5.093,0.215) {Number of qubits};
\node[gp node right] at (3.896,5.357) {serial};
\gpcolor{rgb color={0.580,0.000,0.827}}
\gpsetdashtype{gp dt 4}
\gpsetlinewidth{2.00}
\draw[gp path] (4.080,5.357)--(4.996,5.357);
\draw[gp path] (2.514,0.999)--(3.063,1.400)--(3.612,1.627)--(4.161,2.114)--(4.709,2.633)%
  --(5.258,3.210)--(5.807,3.748)--(6.356,4.452)--(6.904,5.061)--(7.453,5.587);
\gpsetpointsize{4.00}
\gp3point{gp mark 7}{}{(2.514,0.999)}
\gp3point{gp mark 7}{}{(3.063,1.400)}
\gp3point{gp mark 7}{}{(3.612,1.627)}
\gp3point{gp mark 7}{}{(4.161,2.114)}
\gp3point{gp mark 7}{}{(4.709,2.633)}
\gp3point{gp mark 7}{}{(5.258,3.210)}
\gp3point{gp mark 7}{}{(5.807,3.748)}
\gp3point{gp mark 7}{}{(6.356,4.452)}
\gp3point{gp mark 7}{}{(6.904,5.061)}
\gp3point{gp mark 7}{}{(7.453,5.587)}
\gp3point{gp mark 7}{}{(4.538,5.357)}
\gpcolor{color=gp lt color border}
\node[gp node right] at (3.896,5.049) {parallel};
\gpcolor{rgb color={0.000,0.620,0.451}}
\draw[gp path] (4.080,5.049)--(4.996,5.049);
\draw[gp path] (2.514,0.999)--(3.063,1.400)--(3.612,1.627)--(4.161,2.009)--(4.709,2.364)%
  --(5.258,2.556)--(5.807,2.765)--(6.356,2.920)--(6.904,3.081)--(7.453,3.263);
\gp3point{gp mark 7}{}{(2.514,0.999)}
\gp3point{gp mark 7}{}{(3.063,1.400)}
\gp3point{gp mark 7}{}{(3.612,1.627)}
\gp3point{gp mark 7}{}{(4.161,2.009)}
\gp3point{gp mark 7}{}{(4.709,2.364)}
\gp3point{gp mark 7}{}{(5.258,2.556)}
\gp3point{gp mark 7}{}{(5.807,2.765)}
\gp3point{gp mark 7}{}{(6.356,2.920)}
\gp3point{gp mark 7}{}{(6.904,3.081)}
\gp3point{gp mark 7}{}{(7.453,3.263)}
\gp3point{gp mark 7}{}{(4.538,5.049)}
\gpcolor{color=gp lt color border}
\gpsetdashtype{gp dt solid}
\gpsetlinewidth{1.00}
\draw[gp path] (2.240,5.691)--(2.240,0.985)--(7.947,0.985)--(7.947,5.691)--cycle;
\gpdefrectangularnode{gp plot 1}{\pgfpoint{2.240cm}{0.985cm}}{\pgfpoint{7.947cm}{5.691cm}}
\end{tikzpicture}

%% file: figures/swap03/optim_history.tex
\begin{tikzpicture}[gnuplot]
\tikzset{every node/.append style={font={\footnotesize}}}
\path (0.000,0.000) rectangle (8.500,5.000);
\gpcolor{color=gp lt color axes}
\gpsetlinetype{gp lt axes}
\gpsetdashtype{gp dt axes}
\gpsetlinewidth{0.50}
\draw[gp path] (1.196,0.985)--(7.947,0.985);
\gpcolor{color=gp lt color border}
\gpsetlinetype{gp lt border}
\gpsetdashtype{gp dt solid}
\gpsetlinewidth{1.00}
\draw[gp path] (1.196,0.985)--(1.376,0.985);
\draw[gp path] (7.947,0.985)--(7.767,0.985);
\node[gp node right] at (1.012,0.985) {$10^{-6}$};
\draw[gp path] (1.196,1.144)--(1.286,1.144);
\draw[gp path] (7.947,1.144)--(7.857,1.144);
\draw[gp path] (1.196,1.238)--(1.286,1.238);
\draw[gp path] (7.947,1.238)--(7.857,1.238);
\draw[gp path] (1.196,1.304)--(1.286,1.304);
\draw[gp path] (7.947,1.304)--(7.857,1.304);
\draw[gp path] (1.196,1.355)--(1.286,1.355);
\draw[gp path] (7.947,1.355)--(7.857,1.355);
\draw[gp path] (1.196,1.397)--(1.286,1.397);
\draw[gp path] (7.947,1.397)--(7.857,1.397);
\draw[gp path] (1.196,1.432)--(1.286,1.432);
\draw[gp path] (7.947,1.432)--(7.857,1.432);
\draw[gp path] (1.196,1.463)--(1.286,1.463);
\draw[gp path] (7.947,1.463)--(7.857,1.463);
\draw[gp path] (1.196,1.490)--(1.286,1.490);
\draw[gp path] (7.947,1.490)--(7.857,1.490);
\gpcolor{color=gp lt color axes}
\gpsetlinetype{gp lt axes}
\gpsetdashtype{gp dt axes}
\gpsetlinewidth{0.50}
\draw[gp path] (1.196,1.514)--(7.947,1.514);
\gpcolor{color=gp lt color border}
\gpsetlinetype{gp lt border}
\gpsetdashtype{gp dt solid}
\gpsetlinewidth{1.00}
\draw[gp path] (1.196,1.514)--(1.376,1.514);
\draw[gp path] (7.947,1.514)--(7.767,1.514);
\node[gp node right] at (1.012,1.514) {$10^{-5}$};
\draw[gp path] (1.196,1.674)--(1.286,1.674);
\draw[gp path] (7.947,1.674)--(7.857,1.674);
\draw[gp path] (1.196,1.767)--(1.286,1.767);
\draw[gp path] (7.947,1.767)--(7.857,1.767);
\draw[gp path] (1.196,1.833)--(1.286,1.833);
\draw[gp path] (7.947,1.833)--(7.857,1.833);
\draw[gp path] (1.196,1.884)--(1.286,1.884);
\draw[gp path] (7.947,1.884)--(7.857,1.884);
\draw[gp path] (1.196,1.926)--(1.286,1.926);
\draw[gp path] (7.947,1.926)--(7.857,1.926);
\draw[gp path] (1.196,1.962)--(1.286,1.962);
\draw[gp path] (7.947,1.962)--(7.857,1.962);
\draw[gp path] (1.196,1.993)--(1.286,1.993);
\draw[gp path] (7.947,1.993)--(7.857,1.993);
\draw[gp path] (1.196,2.020)--(1.286,2.020);
\draw[gp path] (7.947,2.020)--(7.857,2.020);
\gpcolor{color=gp lt color axes}
\gpsetlinetype{gp lt axes}
\gpsetdashtype{gp dt axes}
\gpsetlinewidth{0.50}
\draw[gp path] (1.196,2.044)--(7.947,2.044);
\gpcolor{color=gp lt color border}
\gpsetlinetype{gp lt border}
\gpsetdashtype{gp dt solid}
\gpsetlinewidth{1.00}
\draw[gp path] (1.196,2.044)--(1.376,2.044);
\draw[gp path] (7.947,2.044)--(7.767,2.044);
\node[gp node right] at (1.012,2.044) {$10^{-4}$};
\draw[gp path] (1.196,2.203)--(1.286,2.203);
\draw[gp path] (7.947,2.203)--(7.857,2.203);
\draw[gp path] (1.196,2.296)--(1.286,2.296);
\draw[gp path] (7.947,2.296)--(7.857,2.296);
\draw[gp path] (1.196,2.363)--(1.286,2.363);
\draw[gp path] (7.947,2.363)--(7.857,2.363);
\draw[gp path] (1.196,2.414)--(1.286,2.414);
\draw[gp path] (7.947,2.414)--(7.857,2.414);
\draw[gp path] (1.196,2.456)--(1.286,2.456);
\draw[gp path] (7.947,2.456)--(7.857,2.456);
\draw[gp path] (1.196,2.491)--(1.286,2.491);
\draw[gp path] (7.947,2.491)--(7.857,2.491);
\draw[gp path] (1.196,2.522)--(1.286,2.522);
\draw[gp path] (7.947,2.522)--(7.857,2.522);
\draw[gp path] (1.196,2.549)--(1.286,2.549);
\draw[gp path] (7.947,2.549)--(7.857,2.549);
\gpcolor{color=gp lt color axes}
\gpsetlinetype{gp lt axes}
\gpsetdashtype{gp dt axes}
\gpsetlinewidth{0.50}
\draw[gp path] (1.196,2.573)--(7.947,2.573);
\gpcolor{color=gp lt color border}
\gpsetlinetype{gp lt border}
\gpsetdashtype{gp dt solid}
\gpsetlinewidth{1.00}
\draw[gp path] (1.196,2.573)--(1.376,2.573);
\draw[gp path] (7.947,2.573)--(7.767,2.573);
\node[gp node right] at (1.012,2.573) {$10^{-3}$};
\draw[gp path] (1.196,2.733)--(1.286,2.733);
\draw[gp path] (7.947,2.733)--(7.857,2.733);
\draw[gp path] (1.196,2.826)--(1.286,2.826);
\draw[gp path] (7.947,2.826)--(7.857,2.826);
\draw[gp path] (1.196,2.892)--(1.286,2.892);
\draw[gp path] (7.947,2.892)--(7.857,2.892);
\draw[gp path] (1.196,2.943)--(1.286,2.943);
\draw[gp path] (7.947,2.943)--(7.857,2.943);
\draw[gp path] (1.196,2.985)--(1.286,2.985);
\draw[gp path] (7.947,2.985)--(7.857,2.985);
\draw[gp path] (1.196,3.021)--(1.286,3.021);
\draw[gp path] (7.947,3.021)--(7.857,3.021);
\draw[gp path] (1.196,3.051)--(1.286,3.051);
\draw[gp path] (7.947,3.051)--(7.857,3.051);
\draw[gp path] (1.196,3.078)--(1.286,3.078);
\draw[gp path] (7.947,3.078)--(7.857,3.078);
\gpcolor{color=gp lt color axes}
\gpsetlinetype{gp lt axes}
\gpsetdashtype{gp dt axes}
\gpsetlinewidth{0.50}
\draw[gp path] (1.196,3.103)--(7.947,3.103);
\gpcolor{color=gp lt color border}
\gpsetlinetype{gp lt border}
\gpsetdashtype{gp dt solid}
\gpsetlinewidth{1.00}
\draw[gp path] (1.196,3.103)--(1.376,3.103);
\draw[gp path] (7.947,3.103)--(7.767,3.103);
\node[gp node right] at (1.012,3.103) {$10^{-2}$};
\draw[gp path] (1.196,3.262)--(1.286,3.262);
\draw[gp path] (7.947,3.262)--(7.857,3.262);
\draw[gp path] (1.196,3.355)--(1.286,3.355);
\draw[gp path] (7.947,3.355)--(7.857,3.355);
\draw[gp path] (1.196,3.421)--(1.286,3.421);
\draw[gp path] (7.947,3.421)--(7.857,3.421);
\draw[gp path] (1.196,3.473)--(1.286,3.473);
\draw[gp path] (7.947,3.473)--(7.857,3.473);
\draw[gp path] (1.196,3.515)--(1.286,3.515);
\draw[gp path] (7.947,3.515)--(7.857,3.515);
\draw[gp path] (1.196,3.550)--(1.286,3.550);
\draw[gp path] (7.947,3.550)--(7.857,3.550);
\draw[gp path] (1.196,3.581)--(1.286,3.581);
\draw[gp path] (7.947,3.581)--(7.857,3.581);
\draw[gp path] (1.196,3.608)--(1.286,3.608);
\draw[gp path] (7.947,3.608)--(7.857,3.608);
\gpcolor{color=gp lt color axes}
\gpsetlinetype{gp lt axes}
\gpsetdashtype{gp dt axes}
\gpsetlinewidth{0.50}
\draw[gp path] (1.196,3.632)--(7.947,3.632);
\gpcolor{color=gp lt color border}
\gpsetlinetype{gp lt border}
\gpsetdashtype{gp dt solid}
\gpsetlinewidth{1.00}
\draw[gp path] (1.196,3.632)--(1.376,3.632);
\draw[gp path] (7.947,3.632)--(7.767,3.632);
\node[gp node right] at (1.012,3.632) {$10^{-1}$};
\draw[gp path] (1.196,3.792)--(1.286,3.792);
\draw[gp path] (7.947,3.792)--(7.857,3.792);
\draw[gp path] (1.196,3.885)--(1.286,3.885);
\draw[gp path] (7.947,3.885)--(7.857,3.885);
\draw[gp path] (1.196,3.951)--(1.286,3.951);
\draw[gp path] (7.947,3.951)--(7.857,3.951);
\draw[gp path] (1.196,4.002)--(1.286,4.002);
\draw[gp path] (7.947,4.002)--(7.857,4.002);
\draw[gp path] (1.196,4.044)--(1.286,4.044);
\draw[gp path] (7.947,4.044)--(7.857,4.044);
\draw[gp path] (1.196,4.080)--(1.286,4.080);
\draw[gp path] (7.947,4.080)--(7.857,4.080);
\draw[gp path] (1.196,4.110)--(1.286,4.110);
\draw[gp path] (7.947,4.110)--(7.857,4.110);
\draw[gp path] (1.196,4.137)--(1.286,4.137);
\draw[gp path] (7.947,4.137)--(7.857,4.137);
\gpcolor{color=gp lt color axes}
\gpsetlinetype{gp lt axes}
\gpsetdashtype{gp dt axes}
\gpsetlinewidth{0.50}
\draw[gp path] (1.196,4.162)--(5.007,4.162);
\draw[gp path] (7.763,4.162)--(7.947,4.162);
\gpcolor{color=gp lt color border}
\gpsetlinetype{gp lt border}
\gpsetdashtype{gp dt solid}
\gpsetlinewidth{1.00}
\draw[gp path] (1.196,4.162)--(1.376,4.162);
\draw[gp path] (7.947,4.162)--(7.767,4.162);
\node[gp node right] at (1.012,4.162) {$10^{0}$};
\draw[gp path] (1.196,4.321)--(1.286,4.321);
\draw[gp path] (7.947,4.321)--(7.857,4.321);
\draw[gp path] (1.196,4.414)--(1.286,4.414);
\draw[gp path] (7.947,4.414)--(7.857,4.414);
\draw[gp path] (1.196,4.480)--(1.286,4.480);
\draw[gp path] (7.947,4.480)--(7.857,4.480);
\draw[gp path] (1.196,4.532)--(1.286,4.532);
\draw[gp path] (7.947,4.532)--(7.857,4.532);
\draw[gp path] (1.196,4.574)--(1.286,4.574);
\draw[gp path] (7.947,4.574)--(7.857,4.574);
\draw[gp path] (1.196,4.609)--(1.286,4.609);
\draw[gp path] (7.947,4.609)--(7.857,4.609);
\draw[gp path] (1.196,4.640)--(1.286,4.640);
\draw[gp path] (7.947,4.640)--(7.857,4.640);
\draw[gp path] (1.196,4.667)--(1.286,4.667);
\draw[gp path] (7.947,4.667)--(7.857,4.667);
\gpcolor{color=gp lt color axes}
\gpsetlinetype{gp lt axes}
\gpsetdashtype{gp dt axes}
\gpsetlinewidth{0.50}
\draw[gp path] (1.196,4.691)--(7.947,4.691);
\gpcolor{color=gp lt color border}
\gpsetlinetype{gp lt border}
\gpsetdashtype{gp dt solid}
\gpsetlinewidth{1.00}
\draw[gp path] (1.196,4.691)--(1.376,4.691);
\draw[gp path] (7.947,4.691)--(7.767,4.691);
\node[gp node right] at (1.012,4.691) {$10^{1}$};
\gpcolor{color=gp lt color axes}
\gpsetlinetype{gp lt axes}
\gpsetdashtype{gp dt axes}
\gpsetlinewidth{0.50}
\draw[gp path] (1.196,0.985)--(1.196,4.691);
\gpcolor{color=gp lt color border}
\gpsetlinetype{gp lt border}
\gpsetdashtype{gp dt solid}
\gpsetlinewidth{1.00}
\draw[gp path] (1.196,0.985)--(1.196,1.165);
\draw[gp path] (1.196,4.691)--(1.196,4.511);
\node[gp node center] at (1.196,0.677) {$0$};
\gpcolor{color=gp lt color axes}
\gpsetlinetype{gp lt axes}
\gpsetdashtype{gp dt axes}
\gpsetlinewidth{0.50}
\draw[gp path] (2.546,0.985)--(2.546,4.691);
\gpcolor{color=gp lt color border}
\gpsetlinetype{gp lt border}
\gpsetdashtype{gp dt solid}
\gpsetlinewidth{1.00}
\draw[gp path] (2.546,0.985)--(2.546,1.165);
\draw[gp path] (2.546,4.691)--(2.546,4.511);
\node[gp node center] at (2.546,0.677) {$200$};
\gpcolor{color=gp lt color axes}
\gpsetlinetype{gp lt axes}
\gpsetdashtype{gp dt axes}
\gpsetlinewidth{0.50}
\draw[gp path] (3.896,0.985)--(3.896,4.691);
\gpcolor{color=gp lt color border}
\gpsetlinetype{gp lt border}
\gpsetdashtype{gp dt solid}
\gpsetlinewidth{1.00}
\draw[gp path] (3.896,0.985)--(3.896,1.165);
\draw[gp path] (3.896,4.691)--(3.896,4.511);
\node[gp node center] at (3.896,0.677) {$400$};
\gpcolor{color=gp lt color axes}
\gpsetlinetype{gp lt axes}
\gpsetdashtype{gp dt axes}
\gpsetlinewidth{0.50}
\draw[gp path] (5.247,0.985)--(5.247,3.895);
\draw[gp path] (5.247,4.511)--(5.247,4.691);
\gpcolor{color=gp lt color border}
\gpsetlinetype{gp lt border}
\gpsetdashtype{gp dt solid}
\gpsetlinewidth{1.00}
\draw[gp path] (5.247,0.985)--(5.247,1.165);
\draw[gp path] (5.247,4.691)--(5.247,4.511);
\node[gp node center] at (5.247,0.677) {$600$};
\gpcolor{color=gp lt color axes}
\gpsetlinetype{gp lt axes}
\gpsetdashtype{gp dt axes}
\gpsetlinewidth{0.50}
\draw[gp path] (6.597,0.985)--(6.597,3.895);
\draw[gp path] (6.597,4.511)--(6.597,4.691);
\gpcolor{color=gp lt color border}
\gpsetlinetype{gp lt border}
\gpsetdashtype{gp dt solid}
\gpsetlinewidth{1.00}
\draw[gp path] (6.597,0.985)--(6.597,1.165);
\draw[gp path] (6.597,4.691)--(6.597,4.511);
\node[gp node center] at (6.597,0.677) {$800$};
\gpcolor{color=gp lt color axes}
\gpsetlinetype{gp lt axes}
\gpsetdashtype{gp dt axes}
\gpsetlinewidth{0.50}
\draw[gp path] (7.947,0.985)--(7.947,4.691);
\gpcolor{color=gp lt color border}
\gpsetlinetype{gp lt border}
\gpsetdashtype{gp dt solid}
\gpsetlinewidth{1.00}
\draw[gp path] (7.947,0.985)--(7.947,1.165);
\draw[gp path] (7.947,4.691)--(7.947,4.511);
\node[gp node center] at (7.947,0.677) {$1000$};
\draw[gp path] (1.196,4.691)--(1.196,0.985)--(7.947,0.985)--(7.947,4.691)--cycle;
\node[gp node center] at (4.571,0.215) {iteration};
\node[gp node right] at (6.479,4.357) {$J$};
\gpcolor{rgb color={0.580,0.000,0.827}}
\draw[gp path] (6.663,4.357)--(7.579,4.357);
\draw[gp path] (1.196,3.777)--(1.203,3.774)--(1.210,3.774)--(1.216,3.767)--(1.223,3.752)%
  --(1.230,3.658)--(1.237,3.628)--(1.243,3.517)--(1.250,3.461)--(1.257,3.402)--(1.264,3.393)%
  --(1.270,3.388)--(1.277,3.376)--(1.284,3.358)--(1.291,3.344)--(1.297,3.319)--(1.304,3.310)%
  --(1.311,3.304)--(1.318,3.297)--(1.324,3.290)--(1.331,3.274)--(1.338,3.266)--(1.345,3.260)%
  --(1.351,3.253)--(1.358,3.246)--(1.365,3.232)--(1.372,3.224)--(1.378,3.214)--(1.385,3.209)%
  --(1.392,3.202)--(1.399,3.188)--(1.405,3.182)--(1.412,3.171)--(1.419,3.167)--(1.426,3.163)%
  --(1.432,3.153)--(1.439,3.148)--(1.446,3.132)--(1.453,3.124)--(1.459,3.119)--(1.466,3.113)%
  --(1.473,3.103)--(1.480,3.088)--(1.486,3.085)--(1.493,3.078)--(1.500,3.074)--(1.507,3.066)%
  --(1.513,3.054)--(1.520,3.048)--(1.527,3.039)--(1.534,3.032)--(1.540,3.026)--(1.547,3.017)%
  --(1.554,3.011)--(1.561,3.001)--(1.567,2.997)--(1.574,2.991)--(1.581,2.984)--(1.588,2.978)%
  --(1.594,2.974)--(1.601,2.971)--(1.608,2.966)--(1.615,2.956)--(1.621,2.953)--(1.628,2.947)%
  --(1.635,2.944)--(1.642,2.939)--(1.648,2.932)--(1.655,2.926)--(1.662,2.923)--(1.669,2.918)%
  --(1.675,2.913)--(1.682,2.908)--(1.689,2.903)--(1.696,2.898)--(1.702,2.891)--(1.709,2.886)%
  --(1.716,2.880)--(1.723,2.877)--(1.729,2.874)--(1.736,2.868)--(1.743,2.863)--(1.750,2.858)%
  --(1.756,2.854)--(1.763,2.851)--(1.770,2.843)--(1.777,2.836)--(1.783,2.831)--(1.790,2.827)%
  --(1.797,2.823)--(1.804,2.816)--(1.810,2.808)--(1.817,2.805)--(1.824,2.801)--(1.831,2.798)%
  --(1.837,2.793)--(1.844,2.789)--(1.851,2.784)--(1.858,2.782)--(1.864,2.779)--(1.871,2.776)%
  --(1.878,2.774)--(1.885,2.771)--(1.891,2.769)--(1.898,2.766)--(1.905,2.762)--(1.912,2.760)%
  --(1.918,2.757)--(1.925,2.755)--(1.932,2.752)--(1.939,2.749)--(1.945,2.745)--(1.952,2.743)%
  --(1.959,2.741)--(1.966,2.737)--(1.972,2.735)--(1.979,2.733)--(1.986,2.731)--(1.993,2.728)%
  --(1.999,2.722)--(2.006,2.718)--(2.013,2.714)--(2.020,2.711)--(2.026,2.707)--(2.033,2.706)%
  --(2.040,2.700)--(2.047,2.699)--(2.053,2.696)--(2.060,2.691)--(2.067,2.689)--(2.074,2.685)%
  --(2.080,2.683)--(2.087,2.680)--(2.094,2.678)--(2.101,2.676)--(2.107,2.674)--(2.114,2.672)%
  --(2.121,2.669)--(2.128,2.667)--(2.134,2.665)--(2.141,2.663)--(2.148,2.661)--(2.155,2.658)%
  --(2.161,2.655)--(2.168,2.653)--(2.175,2.651)--(2.182,2.648)--(2.188,2.644)--(2.195,2.641)%
  --(2.202,2.640)--(2.209,2.639)--(2.215,2.636)--(2.222,2.634)--(2.229,2.633)--(2.236,2.631)%
  --(2.242,2.629)--(2.249,2.626)--(2.256,2.624)--(2.263,2.622)--(2.269,2.620)--(2.276,2.618)%
  --(2.283,2.615)--(2.290,2.613)--(2.296,2.611)--(2.303,2.610)--(2.310,2.606)--(2.317,2.604)%
  --(2.323,2.603)--(2.330,2.601)--(2.337,2.599)--(2.344,2.596)--(2.350,2.594)--(2.357,2.592)%
  --(2.364,2.591)--(2.371,2.588)--(2.377,2.584)--(2.384,2.581)--(2.391,2.579)--(2.398,2.577)%
  --(2.404,2.574)--(2.411,2.573)--(2.418,2.569)--(2.425,2.568)--(2.431,2.566)--(2.438,2.563)%
  --(2.445,2.560)--(2.452,2.558)--(2.458,2.557)--(2.465,2.555)--(2.472,2.552)--(2.479,2.550)%
  --(2.485,2.548)--(2.492,2.545)--(2.499,2.543)--(2.506,2.540)--(2.512,2.536)--(2.519,2.535)%
  --(2.526,2.533)--(2.533,2.529)--(2.539,2.526)--(2.546,2.523)--(2.553,2.521)--(2.560,2.520)%
  --(2.566,2.517)--(2.573,2.514)--(2.580,2.513)--(2.587,2.511)--(2.593,2.508)--(2.600,2.505)%
  --(2.607,2.501)--(2.614,2.499)--(2.620,2.498)--(2.627,2.496)--(2.634,2.493)--(2.641,2.490)%
  --(2.647,2.487)--(2.654,2.485)--(2.661,2.482)--(2.668,2.481)--(2.674,2.479)--(2.681,2.476)%
  --(2.688,2.474)--(2.695,2.470)--(2.701,2.467)--(2.708,2.464)--(2.715,2.459)--(2.722,2.454)%
  --(2.728,2.451)--(2.735,2.449)--(2.742,2.447)--(2.749,2.445)--(2.755,2.443)--(2.762,2.440)%
  --(2.769,2.437)--(2.776,2.435)--(2.782,2.434)--(2.789,2.432)--(2.796,2.430)--(2.803,2.426)%
  --(2.809,2.424)--(2.816,2.421)--(2.823,2.419)--(2.830,2.417)--(2.836,2.416)--(2.843,2.414)%
  --(2.850,2.411)--(2.857,2.409)--(2.863,2.407)--(2.870,2.405)--(2.877,2.403)--(2.884,2.401)%
  --(2.891,2.397)--(2.897,2.394)--(2.904,2.391)--(2.911,2.388)--(2.918,2.386)--(2.924,2.383)%
  --(2.931,2.381)--(2.938,2.378)--(2.945,2.375)--(2.951,2.374)--(2.958,2.372)--(2.965,2.370)%
  --(2.972,2.367)--(2.978,2.364)--(2.985,2.361)--(2.992,2.358)--(2.999,2.353)--(3.005,2.352)%
  --(3.012,2.351)--(3.019,2.349)--(3.026,2.346)--(3.032,2.343)--(3.039,2.340)--(3.046,2.338)%
  --(3.053,2.336)--(3.059,2.334)--(3.066,2.330)--(3.073,2.328)--(3.080,2.325)--(3.086,2.323)%
  --(3.093,2.319)--(3.100,2.317)--(3.107,2.314)--(3.113,2.309)--(3.120,2.307)--(3.127,2.305)%
  --(3.134,2.303)--(3.140,2.300)--(3.147,2.298)--(3.154,2.295)--(3.161,2.294)--(3.167,2.291)%
  --(3.174,2.287)--(3.181,2.285)--(3.188,2.282)--(3.194,2.279)--(3.201,2.276)--(3.208,2.275)%
  --(3.215,2.271)--(3.221,2.270)--(3.228,2.269)--(3.235,2.266)--(3.242,2.264)--(3.248,2.261)%
  --(3.255,2.259)--(3.262,2.255)--(3.269,2.253)--(3.275,2.251)--(3.282,2.249)--(3.289,2.246)%
  --(3.296,2.242)--(3.302,2.238)--(3.309,2.235)--(3.316,2.231)--(3.323,2.228)--(3.329,2.225)%
  --(3.336,2.222)--(3.343,2.219)--(3.350,2.217)--(3.356,2.215)--(3.363,2.214)--(3.370,2.210)%
  --(3.377,2.206)--(3.383,2.205)--(3.390,2.201)--(3.397,2.200)--(3.404,2.198)--(3.410,2.194)%
  --(3.417,2.192)--(3.424,2.190)--(3.431,2.188)--(3.437,2.186)--(3.444,2.185)--(3.451,2.182)%
  --(3.458,2.177)--(3.464,2.176)--(3.471,2.172)--(3.478,2.171)--(3.485,2.168)--(3.491,2.165)%
  --(3.498,2.162)--(3.505,2.159)--(3.512,2.156)--(3.518,2.154)--(3.525,2.151)--(3.532,2.148)%
  --(3.539,2.146)--(3.545,2.144)--(3.552,2.141)--(3.559,2.138)--(3.566,2.134)--(3.572,2.132)%
  --(3.579,2.127)--(3.586,2.125)--(3.593,2.121)--(3.599,2.117)--(3.606,2.115)--(3.613,2.112)%
  --(3.620,2.108)--(3.626,2.103)--(3.633,2.100)--(3.640,2.097)--(3.647,2.095)--(3.653,2.091)%
  --(3.660,2.090)--(3.667,2.088)--(3.674,2.084)--(3.680,2.080)--(3.687,2.079)--(3.694,2.075)%
  --(3.701,2.073)--(3.707,2.071)--(3.714,2.068)--(3.721,2.067)--(3.728,2.063)--(3.734,2.061)%
  --(3.741,2.059)--(3.748,2.058)--(3.755,2.055)--(3.761,2.052)--(3.768,2.051)--(3.775,2.047)%
  --(3.782,2.044)--(3.788,2.039)--(3.795,2.037)--(3.802,2.034)--(3.809,2.030)--(3.815,2.027)%
  --(3.822,2.023)--(3.829,2.020)--(3.836,2.016)--(3.842,2.014)--(3.849,2.007)--(3.856,2.005)%
  --(3.863,2.003)--(3.869,2.002)--(3.876,1.996)--(3.883,1.994)--(3.890,1.992)--(3.896,1.990)%
  --(3.903,1.986)--(3.910,1.983)--(3.917,1.980)--(3.923,1.978)--(3.930,1.974)--(3.937,1.973)%
  --(3.944,1.970)--(3.950,1.969)--(3.957,1.966)--(3.964,1.963)--(3.971,1.960)--(3.977,1.959)%
  --(3.984,1.957)--(3.991,1.955)--(3.998,1.953)--(4.004,1.951)--(4.011,1.950)--(4.018,1.947)%
  --(4.025,1.945)--(4.031,1.944)--(4.038,1.942)--(4.045,1.939)--(4.052,1.938)--(4.058,1.937)%
  --(4.065,1.936)--(4.072,1.934)--(4.079,1.933)--(4.085,1.932)--(4.092,1.930)--(4.099,1.929)%
  --(4.106,1.927)--(4.112,1.925)--(4.119,1.923)--(4.126,1.921)--(4.133,1.920)--(4.139,1.918)%
  --(4.146,1.915)--(4.153,1.912)--(4.160,1.910)--(4.166,1.909)--(4.173,1.908)--(4.180,1.906)%
  --(4.187,1.904)--(4.193,1.902)--(4.200,1.900)--(4.207,1.898)--(4.214,1.896)--(4.220,1.894)%
  --(4.227,1.892)--(4.234,1.889)--(4.241,1.888)--(4.247,1.886)--(4.254,1.884)--(4.261,1.882)%
  --(4.268,1.880)--(4.274,1.878)--(4.281,1.875)--(4.288,1.873)--(4.295,1.872)--(4.301,1.869)%
  --(4.308,1.868)--(4.315,1.865)--(4.322,1.862)--(4.328,1.860)--(4.335,1.856)--(4.342,1.855)%
  --(4.349,1.852)--(4.355,1.851)--(4.362,1.850)--(4.369,1.848)--(4.376,1.845)--(4.382,1.842)%
  --(4.389,1.840)--(4.396,1.838)--(4.403,1.836)--(4.409,1.834)--(4.416,1.832)--(4.423,1.829)%
  --(4.430,1.826)--(4.436,1.825)--(4.443,1.823)--(4.450,1.822)--(4.457,1.819)--(4.463,1.818)%
  --(4.470,1.815)--(4.477,1.812)--(4.484,1.810)--(4.490,1.809)--(4.497,1.806)--(4.504,1.803)%
  --(4.511,1.801)--(4.517,1.800)--(4.524,1.796)--(4.531,1.794)--(4.538,1.792)--(4.544,1.790)%
  --(4.551,1.788)--(4.558,1.786)--(4.565,1.785)--(4.572,1.784)--(4.578,1.784)--(4.585,1.783)%
  --(4.592,1.783)--(4.599,1.780)--(4.605,1.780)--(4.612,1.779)--(4.619,1.777)--(4.626,1.776)%
  --(4.632,1.774)--(4.639,1.774)--(4.646,1.773)--(4.653,1.773)--(4.659,1.772)--(4.666,1.771)%
  --(4.673,1.770)--(4.680,1.770)--(4.686,1.769)--(4.693,1.768)--(4.700,1.767)--(4.707,1.767)%
  --(4.713,1.766)--(4.720,1.765)--(4.727,1.764)--(4.734,1.762)--(4.740,1.761)--(4.747,1.760)%
  --(4.754,1.759)--(4.761,1.758)--(4.767,1.757)--(4.774,1.756)--(4.781,1.755)--(4.788,1.753)%
  --(4.794,1.752)--(4.801,1.750)--(4.808,1.750)--(4.815,1.749)--(4.821,1.747)--(4.828,1.747)%
  --(4.835,1.745)--(4.842,1.745)--(4.848,1.744)--(4.855,1.743)--(4.862,1.741)--(4.869,1.739)%
  --(4.875,1.739)--(4.882,1.738)--(4.889,1.737)--(4.896,1.736)--(4.902,1.735)--(4.909,1.734)%
  --(4.916,1.733)--(4.923,1.732)--(4.929,1.732)--(4.936,1.731)--(4.943,1.730)--(4.950,1.729)%
  --(4.956,1.728)--(4.963,1.727)--(4.970,1.725)--(4.977,1.724)--(4.983,1.722)--(4.990,1.720)%
  --(4.997,1.719)--(5.004,1.718)--(5.010,1.717)--(5.017,1.715)--(5.024,1.713)--(5.031,1.712)%
  --(5.037,1.711)--(5.044,1.710)--(5.051,1.709)--(5.058,1.708)--(5.064,1.707)--(5.071,1.706)%
  --(5.078,1.705)--(5.085,1.704)--(5.091,1.703)--(5.098,1.702)--(5.105,1.701)--(5.112,1.699)%
  --(5.118,1.698)--(5.125,1.697)--(5.132,1.696)--(5.139,1.694)--(5.145,1.692)--(5.152,1.692)%
  --(5.159,1.690)--(5.166,1.689)--(5.172,1.688)--(5.179,1.688)--(5.186,1.686)--(5.193,1.685)%
  --(5.199,1.683)--(5.206,1.681)--(5.213,1.681)--(5.220,1.679)--(5.226,1.678)--(5.233,1.676)%
  --(5.240,1.674)--(5.247,1.672)--(5.253,1.669)--(5.260,1.667)--(5.267,1.665)--(5.274,1.664)%
  --(5.280,1.662)--(5.287,1.659)--(5.294,1.658)--(5.301,1.655)--(5.307,1.653)--(5.314,1.651)%
  --(5.321,1.648)--(5.328,1.646)--(5.334,1.644)--(5.341,1.642)--(5.348,1.639)--(5.355,1.637)%
  --(5.361,1.634)--(5.368,1.632)--(5.375,1.630)--(5.382,1.626)--(5.388,1.624)--(5.395,1.622)%
  --(5.402,1.621)--(5.409,1.619)--(5.415,1.617)--(5.422,1.614)--(5.429,1.612)--(5.436,1.609)%
  --(5.442,1.606)--(5.449,1.605)--(5.456,1.602)--(5.463,1.601)--(5.469,1.598)--(5.476,1.594)%
  --(5.483,1.593)--(5.490,1.591)--(5.496,1.589)--(5.503,1.587)--(5.510,1.586)--(5.517,1.583)%
  --(5.523,1.582)--(5.530,1.580)--(5.537,1.578)--(5.544,1.575)--(5.550,1.573)--(5.557,1.571)%
  --(5.564,1.569)--(5.571,1.568)--(5.577,1.564)--(5.584,1.564)--(5.591,1.561)--(5.598,1.560)%
  --(5.604,1.557)--(5.611,1.556)--(5.618,1.555)--(5.625,1.553)--(5.631,1.551)--(5.638,1.549)%
  --(5.645,1.546)--(5.652,1.544)--(5.658,1.542)--(5.665,1.541)--(5.672,1.538)--(5.679,1.534)%
  --(5.685,1.532)--(5.692,1.530)--(5.699,1.529)--(5.706,1.527)--(5.712,1.526)--(5.719,1.524)%
  --(5.726,1.522)--(5.733,1.520)--(5.739,1.517)--(5.746,1.515)--(5.753,1.514)--(5.760,1.512)%
  --(5.766,1.509)--(5.773,1.506)--(5.780,1.504)--(5.787,1.502)--(5.793,1.500)--(5.800,1.498)%
  --(5.807,1.496)--(5.814,1.495)--(5.820,1.492)--(5.827,1.490)--(5.834,1.488)--(5.841,1.484)%
  --(5.847,1.483)--(5.854,1.481)--(5.861,1.479)--(5.868,1.477)--(5.874,1.476)--(5.881,1.474)%
  --(5.888,1.473)--(5.895,1.472)--(5.901,1.470)--(5.908,1.468)--(5.915,1.467)--(5.922,1.464)%
  --(5.928,1.463)--(5.935,1.462)--(5.942,1.462)--(5.949,1.460)--(5.955,1.458)--(5.962,1.456)%
  --(5.969,1.454)--(5.976,1.451)--(5.982,1.450)--(5.989,1.449)--(5.996,1.448)--(6.003,1.447)%
  --(6.009,1.445)--(6.016,1.443)--(6.023,1.442)--(6.030,1.439)--(6.036,1.437)--(6.043,1.435)%
  --(6.050,1.434)--(6.057,1.433)--(6.063,1.432)--(6.070,1.430)--(6.077,1.430)--(6.084,1.429)%
  --(6.090,1.427)--(6.097,1.426)--(6.104,1.424)--(6.111,1.422)--(6.117,1.421)--(6.124,1.420)%
  --(6.131,1.419)--(6.138,1.418)--(6.144,1.416)--(6.151,1.415)--(6.158,1.414)--(6.165,1.413)%
  --(6.171,1.411)--(6.178,1.410)--(6.185,1.408)--(6.192,1.407)--(6.198,1.405)--(6.205,1.404)%
  --(6.212,1.403)--(6.219,1.402)--(6.225,1.400)--(6.232,1.399)--(6.239,1.398)--(6.246,1.397)%
  --(6.252,1.396)--(6.259,1.394)--(6.266,1.393)--(6.273,1.392)--(6.280,1.390)--(6.286,1.388)%
  --(6.293,1.387)--(6.300,1.386)--(6.307,1.384)--(6.313,1.382)--(6.320,1.381)--(6.327,1.379)%
  --(6.334,1.378)--(6.340,1.376)--(6.347,1.374)--(6.354,1.373)--(6.361,1.371)--(6.367,1.370)%
  --(6.374,1.368)--(6.381,1.367)--(6.388,1.365)--(6.394,1.363)--(6.401,1.361)--(6.408,1.359)%
  --(6.415,1.357)--(6.421,1.355)--(6.428,1.353)--(6.435,1.351)--(6.442,1.350)--(6.448,1.349)%
  --(6.455,1.347)--(6.462,1.346)--(6.469,1.345)--(6.475,1.344)--(6.482,1.343)--(6.489,1.342)%
  --(6.496,1.340)--(6.502,1.339)--(6.509,1.338)--(6.516,1.337)--(6.523,1.335)--(6.529,1.334)%
  --(6.536,1.332)--(6.543,1.329)--(6.550,1.328)--(6.556,1.323)--(6.563,1.322)--(6.570,1.320)%
  --(6.577,1.318)--(6.583,1.315)--(6.590,1.312)--(6.597,1.311)--(6.604,1.310)--(6.610,1.308)%
  --(6.617,1.307)--(6.624,1.306)--(6.631,1.304)--(6.637,1.303)--(6.644,1.302)--(6.651,1.300)%
  --(6.658,1.299)--(6.664,1.296)--(6.671,1.295)--(6.678,1.293)--(6.685,1.292)--(6.691,1.291)%
  --(6.698,1.289)--(6.705,1.287)--(6.712,1.285)--(6.718,1.284)--(6.725,1.283)--(6.732,1.282)%
  --(6.739,1.279)--(6.745,1.278)--(6.752,1.277)--(6.759,1.275)--(6.766,1.274)--(6.772,1.273)%
  --(6.779,1.272)--(6.786,1.270)--(6.793,1.269)--(6.799,1.267)--(6.806,1.266)--(6.813,1.264)%
  --(6.820,1.262)--(6.826,1.261)--(6.833,1.260)--(6.840,1.256)--(6.847,1.255)--(6.853,1.254)%
  --(6.860,1.253)--(6.867,1.251)--(6.874,1.249)--(6.880,1.248)--(6.887,1.246)--(6.894,1.245)%
  --(6.901,1.244)--(6.907,1.243)--(6.914,1.241)--(6.921,1.240)--(6.928,1.239)--(6.934,1.238)%
  --(6.941,1.237)--(6.948,1.236)--(6.955,1.236)--(6.961,1.235)--(6.968,1.233)--(6.975,1.232)%
  --(6.982,1.230)--(6.988,1.229)--(6.995,1.228)--(7.002,1.227)--(7.009,1.227)--(7.015,1.226)%
  --(7.022,1.225)--(7.029,1.224)--(7.036,1.222)--(7.042,1.221)--(7.049,1.220)--(7.056,1.220)%
  --(7.063,1.219)--(7.069,1.218)--(7.076,1.217)--(7.083,1.217)--(7.090,1.216)--(7.096,1.216)%
  --(7.103,1.215)--(7.110,1.214)--(7.117,1.213)--(7.123,1.212)--(7.130,1.211)--(7.137,1.210)%
  --(7.144,1.209)--(7.150,1.208)--(7.157,1.207)--(7.164,1.207)--(7.171,1.206)--(7.177,1.205)%
  --(7.184,1.205)--(7.191,1.204)--(7.198,1.204)--(7.204,1.203)--(7.211,1.203)--(7.218,1.203)%
  --(7.225,1.202)--(7.231,1.201)--(7.238,1.201)--(7.245,1.200)--(7.252,1.200)--(7.258,1.199)%
  --(7.265,1.198)--(7.272,1.198)--(7.279,1.197)--(7.285,1.197)--(7.292,1.196)--(7.299,1.196)%
  --(7.306,1.195)--(7.312,1.195)--(7.319,1.194)--(7.326,1.194)--(7.333,1.193)--(7.339,1.192)%
  --(7.346,1.192)--(7.353,1.191)--(7.360,1.190)--(7.366,1.190)--(7.373,1.189)--(7.380,1.188)%
  --(7.387,1.187)--(7.393,1.187)--(7.400,1.187)--(7.407,1.186)--(7.414,1.185)--(7.420,1.184)%
  --(7.427,1.184)--(7.434,1.183)--(7.441,1.182)--(7.447,1.181)--(7.454,1.180)--(7.461,1.180)%
  --(7.468,1.179)--(7.474,1.178)--(7.481,1.177)--(7.488,1.177)--(7.495,1.176)--(7.501,1.175)%
  --(7.508,1.175)--(7.515,1.174)--(7.522,1.173)--(7.528,1.173)--(7.535,1.172)--(7.542,1.171)%
  --(7.549,1.171)--(7.555,1.170)--(7.562,1.170)--(7.569,1.169)--(7.576,1.169)--(7.582,1.168)%
  --(7.589,1.168)--(7.596,1.168)--(7.603,1.167)--(7.609,1.167)--(7.616,1.166)--(7.623,1.166)%
  --(7.630,1.165)--(7.636,1.165)--(7.643,1.164)--(7.650,1.164)--(7.657,1.164)--(7.663,1.163)%
  --(7.670,1.163)--(7.677,1.162)--(7.684,1.162)--(7.690,1.161)--(7.697,1.161)--(7.704,1.161)%
  --(7.711,1.161)--(7.717,1.160)--(7.724,1.160)--(7.731,1.160)--(7.738,1.159)--(7.744,1.159)%
  --(7.751,1.158)--(7.758,1.158)--(7.765,1.158)--(7.771,1.157)--(7.778,1.157)--(7.785,1.157)%
  --(7.792,1.156)--(7.798,1.156)--(7.805,1.155)--(7.812,1.155)--(7.819,1.155)--(7.825,1.154)%
  --(7.832,1.154)--(7.839,1.154)--(7.846,1.153)--(7.852,1.153)--(7.859,1.152)--(7.866,1.152)%
  --(7.873,1.152)--(7.879,1.151)--(7.886,1.150)--(7.893,1.149)--(7.900,1.149)--(7.906,1.148)%
  --(7.913,1.148)--(7.920,1.148)--(7.927,1.147)--(7.933,1.147)--(7.940,1.146)--(7.947,1.146);
\gpcolor{color=gp lt color border}
\node[gp node right] at (6.479,4.049) {$\|\nabla J \|$};
\gpcolor{rgb color={0.000,0.620,0.451}}
\draw[gp path] (6.663,4.049)--(7.579,4.049);
\draw[gp path] (1.196,3.262)--(1.203,4.045)--(1.210,4.046)--(1.216,4.077)--(1.223,4.217)%
  --(1.230,4.312)--(1.237,4.300)--(1.243,4.193)--(1.250,4.169)--(1.257,3.955)--(1.264,3.788)%
  --(1.270,3.754)--(1.277,3.825)--(1.284,3.922)--(1.291,4.003)--(1.297,3.827)--(1.304,3.750)%
  --(1.311,3.762)--(1.318,3.761)--(1.324,3.770)--(1.331,3.863)--(1.338,3.757)--(1.345,3.691)%
  --(1.351,3.695)--(1.358,3.722)--(1.365,3.801)--(1.372,3.864)--(1.378,3.679)--(1.385,3.639)%
  --(1.392,3.713)--(1.399,3.790)--(1.405,3.860)--(1.412,3.666)--(1.419,3.605)--(1.426,3.627)%
  --(1.432,3.699)--(1.439,3.890)--(1.446,3.686)--(1.453,3.576)--(1.459,3.592)--(1.466,3.738)%
  --(1.473,3.678)--(1.480,3.641)--(1.486,3.768)--(1.493,3.583)--(1.500,3.576)--(1.507,3.678)%
  --(1.513,3.698)--(1.520,3.746)--(1.527,3.600)--(1.534,3.575)--(1.540,3.627)--(1.547,3.626)%
  --(1.554,3.727)--(1.561,3.558)--(1.567,3.510)--(1.574,3.573)--(1.581,3.641)--(1.588,3.590)%
  --(1.594,3.464)--(1.601,3.530)--(1.608,3.576)--(1.615,3.655)--(1.621,3.660)--(1.628,3.443)%
  --(1.635,3.428)--(1.642,3.494)--(1.648,3.643)--(1.655,3.542)--(1.662,3.437)--(1.669,3.454)%
  --(1.675,3.496)--(1.682,3.550)--(1.689,3.424)--(1.696,3.492)--(1.702,3.553)--(1.709,3.603)%
  --(1.716,3.441)--(1.723,3.380)--(1.729,3.430)--(1.736,3.513)--(1.743,3.554)--(1.750,3.431)%
  --(1.756,3.423)--(1.763,3.469)--(1.770,3.588)--(1.777,3.527)--(1.783,3.406)--(1.790,3.401)%
  --(1.797,3.475)--(1.804,3.473)--(1.810,3.454)--(1.817,3.511)--(1.824,3.376)--(1.831,3.362)%
  --(1.837,3.436)--(1.844,3.545)--(1.851,3.384)--(1.858,3.320)--(1.864,3.367)--(1.871,3.390)%
  --(1.878,3.490)--(1.885,3.288)--(1.891,3.296)--(1.898,3.379)--(1.905,3.412)--(1.912,3.434)%
  --(1.918,3.305)--(1.925,3.311)--(1.932,3.369)--(1.939,3.498)--(1.945,3.316)--(1.952,3.282)%
  --(1.959,3.338)--(1.966,3.393)--(1.972,3.424)--(1.979,3.268)--(1.986,3.298)--(1.993,3.375)%
  --(1.999,3.448)--(2.006,3.461)--(2.013,3.315)--(2.020,3.283)--(2.026,3.319)--(2.033,3.549)%
  --(2.040,3.293)--(2.047,3.247)--(2.053,3.331)--(2.060,3.367)--(2.067,3.426)--(2.074,3.258)%
  --(2.080,3.205)--(2.087,3.282)--(2.094,3.335)--(2.101,3.267)--(2.107,3.220)--(2.114,3.260)%
  --(2.121,3.322)--(2.128,3.316)--(2.134,3.225)--(2.141,3.226)--(2.148,3.281)--(2.155,3.447)%
  --(2.161,3.291)--(2.168,3.214)--(2.175,3.257)--(2.182,3.340)--(2.188,3.395)--(2.195,3.264)%
  --(2.202,3.191)--(2.209,3.210)--(2.215,3.287)--(2.222,3.292)--(2.229,3.180)--(2.236,3.213)%
  --(2.242,3.270)--(2.249,3.295)--(2.256,3.321)--(2.263,3.193)--(2.269,3.196)--(2.276,3.268)%
  --(2.283,3.280)--(2.290,3.267)--(2.296,3.221)--(2.303,3.197)--(2.310,3.232)--(2.317,3.306)%
  --(2.323,3.182)--(2.330,3.163)--(2.337,3.248)--(2.344,3.281)--(2.350,3.308)--(2.357,3.186)%
  --(2.364,3.196)--(2.371,3.252)--(2.377,3.375)--(2.384,3.274)--(2.391,3.170)--(2.398,3.224)%
  --(2.404,3.270)--(2.411,3.426)--(2.418,3.172)--(2.425,3.138)--(2.431,3.228)--(2.438,3.318)%
  --(2.445,3.291)--(2.452,3.154)--(2.458,3.166)--(2.465,3.222)--(2.472,3.321)--(2.479,3.285)%
  --(2.485,3.169)--(2.492,3.204)--(2.499,3.253)--(2.506,3.415)--(2.512,3.190)--(2.519,3.138)%
  --(2.526,3.193)--(2.533,3.286)--(2.539,3.378)--(2.546,3.200)--(2.553,3.126)--(2.560,3.153)%
  --(2.566,3.350)--(2.573,3.185)--(2.580,3.127)--(2.587,3.199)--(2.593,3.280)--(2.600,3.249)%
  --(2.607,3.224)--(2.614,3.255)--(2.620,3.127)--(2.627,3.147)--(2.634,3.207)--(2.641,3.360)%
  --(2.647,3.175)--(2.654,3.130)--(2.661,3.162)--(2.668,3.197)--(2.674,3.119)--(2.681,3.160)%
  --(2.688,3.262)--(2.695,3.202)--(2.701,3.162)--(2.708,3.250)--(2.715,3.261)--(2.722,3.274)%
  --(2.728,3.254)--(2.735,3.106)--(2.742,3.104)--(2.749,3.179)--(2.755,3.307)--(2.762,3.147)%
  --(2.769,3.091)--(2.776,3.118)--(2.782,3.207)--(2.789,3.107)--(2.796,3.079)--(2.803,3.173)%
  --(2.809,3.328)--(2.816,3.130)--(2.823,3.066)--(2.830,3.070)--(2.836,3.185)--(2.843,3.079)%
  --(2.850,3.109)--(2.857,3.156)--(2.863,3.246)--(2.870,3.083)--(2.877,3.102)--(2.884,3.170)%
  --(2.891,3.225)--(2.897,3.202)--(2.904,3.057)--(2.911,3.146)--(2.918,3.133)--(2.924,3.197)%
  --(2.931,3.095)--(2.938,3.116)--(2.945,3.140)--(2.951,3.239)--(2.958,3.077)--(2.965,3.081)%
  --(2.972,3.147)--(2.978,3.227)--(2.985,3.130)--(2.992,3.086)--(2.999,3.135)--(3.005,3.106)%
  --(3.012,3.031)--(3.019,3.056)--(3.026,3.149)--(3.032,3.209)--(3.039,3.072)--(3.046,3.061)%
  --(3.053,3.092)--(3.059,3.298)--(3.066,3.076)--(3.073,3.065)--(3.080,3.125)--(3.086,3.315)%
  --(3.093,3.096)--(3.100,3.068)--(3.107,3.152)--(3.113,3.178)--(3.120,3.168)--(3.127,3.037)%
  --(3.134,3.080)--(3.140,3.139)--(3.147,3.218)--(3.154,3.038)--(3.161,3.011)--(3.167,3.107)%
  --(3.174,3.127)--(3.181,3.182)--(3.188,3.044)--(3.194,3.073)--(3.201,3.120)--(3.208,3.313)%
  --(3.215,3.031)--(3.221,2.981)--(3.228,3.062)--(3.235,3.100)--(3.242,3.174)--(3.248,3.050)%
  --(3.255,3.020)--(3.262,3.118)--(3.269,3.087)--(3.275,3.026)--(3.282,3.005)--(3.289,3.078)%
  --(3.296,3.177)--(3.302,3.116)--(3.309,3.054)--(3.316,3.062)--(3.323,3.176)--(3.329,3.034)%
  --(3.336,3.054)--(3.343,3.102)--(3.350,3.161)--(3.356,2.966)--(3.363,2.971)--(3.370,3.059)%
  --(3.377,3.145)--(3.383,3.232)--(3.390,2.982)--(3.397,2.954)--(3.404,3.028)--(3.410,3.106)%
  --(3.417,3.104)--(3.424,2.943)--(3.431,2.988)--(3.437,3.020)--(3.444,2.999)--(3.451,2.998)%
  --(3.458,3.094)--(3.464,3.209)--(3.471,2.946)--(3.478,2.906)--(3.485,3.023)--(3.491,3.131)%
  --(3.498,3.030)--(3.505,2.983)--(3.512,3.056)--(3.518,3.074)--(3.525,2.962)--(3.532,3.010)%
  --(3.539,3.029)--(3.545,3.097)--(3.552,2.961)--(3.559,3.030)--(3.566,3.065)--(3.572,3.240)%
  --(3.579,2.989)--(3.586,2.970)--(3.593,3.038)--(3.599,3.081)--(3.606,3.155)--(3.613,3.003)%
  --(3.620,3.026)--(3.626,3.069)--(3.633,3.110)--(3.640,2.985)--(3.647,2.954)--(3.653,3.007)%
  --(3.660,3.020)--(3.667,2.949)--(3.674,3.025)--(3.680,3.080)--(3.687,3.209)--(3.694,2.958)%
  --(3.701,2.881)--(3.707,2.969)--(3.714,3.007)--(3.721,3.181)--(3.728,2.935)--(3.734,2.876)%
  --(3.741,2.928)--(3.748,3.169)--(3.755,2.961)--(3.761,2.864)--(3.768,2.914)--(3.775,2.996)%
  --(3.782,3.131)--(3.788,2.931)--(3.795,2.913)--(3.802,2.994)--(3.809,3.062)--(3.815,2.940)%
  --(3.822,2.915)--(3.829,3.020)--(3.836,2.969)--(3.842,3.203)--(3.849,2.922)--(3.856,2.863)%
  --(3.863,2.877)--(3.869,3.203)--(3.876,2.968)--(3.883,2.831)--(3.890,2.849)--(3.896,2.941)%
  --(3.903,2.971)--(3.910,2.916)--(3.917,2.935)--(3.923,3.102)--(3.930,2.839)--(3.937,2.860)%
  --(3.944,2.883)--(3.950,3.053)--(3.957,2.914)--(3.964,2.867)--(3.971,2.891)--(3.977,2.969)%
  --(3.984,2.834)--(3.991,2.815)--(3.998,2.900)--(4.004,2.966)--(4.011,2.833)--(4.018,2.847)%
  --(4.025,2.867)--(4.031,2.964)--(4.038,2.823)--(4.045,2.784)--(4.052,2.922)--(4.058,2.767)%
  --(4.065,2.761)--(4.072,2.854)--(4.079,3.014)--(4.085,2.792)--(4.092,2.775)--(4.099,2.843)%
  --(4.106,2.973)--(4.112,2.822)--(4.119,2.795)--(4.126,2.804)--(4.133,2.927)--(4.139,2.862)%
  --(4.146,2.851)--(4.153,2.956)--(4.160,2.825)--(4.166,2.771)--(4.173,2.852)--(4.180,2.856)%
  --(4.187,2.818)--(4.193,2.970)--(4.200,2.828)--(4.207,2.811)--(4.214,2.884)--(4.220,2.980)%
  --(4.227,2.817)--(4.234,2.797)--(4.241,2.843)--(4.247,2.806)--(4.254,2.809)--(4.261,2.866)%
  --(4.268,2.947)--(4.274,2.825)--(4.281,2.849)--(4.288,2.863)--(4.295,3.062)--(4.301,2.787)%
  --(4.308,2.768)--(4.315,2.859)--(4.322,2.923)--(4.328,3.060)--(4.335,2.803)--(4.342,2.807)%
  --(4.349,2.816)--(4.355,2.958)--(4.362,2.784)--(4.369,2.834)--(4.376,2.892)--(4.382,2.931)%
  --(4.389,2.988)--(4.396,2.792)--(4.403,2.818)--(4.409,2.834)--(4.416,2.883)--(4.423,2.771)%
  --(4.430,2.827)--(4.436,2.941)--(4.443,2.763)--(4.450,2.773)--(4.457,2.842)--(4.463,3.020)%
  --(4.470,2.809)--(4.477,2.749)--(4.484,2.775)--(4.490,2.972)--(4.497,2.794)--(4.504,2.766)%
  --(4.511,2.803)--(4.517,3.050)--(4.524,2.803)--(4.531,2.746)--(4.538,2.802)--(4.544,2.787)%
  --(4.551,2.862)--(4.558,2.692)--(4.565,2.693)--(4.572,2.829)--(4.578,2.829)--(4.585,2.732)%
  --(4.592,2.702)--(4.599,2.710)--(4.605,2.748)--(4.612,2.705)--(4.619,2.752)--(4.626,2.831)%
  --(4.632,2.752)--(4.639,2.675)--(4.646,2.620)--(4.653,2.638)--(4.659,2.756)--(4.666,2.786)%
  --(4.673,2.619)--(4.680,2.634)--(4.686,2.681)--(4.693,2.883)--(4.700,2.707)--(4.707,2.615)%
  --(4.713,2.725)--(4.720,2.791)--(4.727,2.988)--(4.734,2.752)--(4.740,2.654)--(4.747,2.687)%
  --(4.754,2.754)--(4.761,2.919)--(4.767,2.755)--(4.774,2.677)--(4.781,2.749)--(4.788,2.805)%
  --(4.794,2.878)--(4.801,2.716)--(4.808,2.638)--(4.815,2.716)--(4.821,2.776)--(4.828,2.934)%
  --(4.835,2.682)--(4.842,2.617)--(4.848,2.703)--(4.855,2.793)--(4.862,2.933)--(4.869,2.756)%
  --(4.875,2.644)--(4.882,2.673)--(4.889,2.706)--(4.896,2.940)--(4.902,2.657)--(4.909,2.571)%
  --(4.916,2.666)--(4.923,2.739)--(4.929,2.900)--(4.936,2.672)--(4.943,2.636)--(4.950,2.721)%
  --(4.956,2.777)--(4.963,2.945)--(4.970,2.728)--(4.977,2.633)--(4.983,2.757)--(4.990,2.803)%
  --(4.997,2.858)--(5.004,2.721)--(5.010,2.647)--(5.017,2.753)--(5.024,2.830)--(5.031,2.803)%
  --(5.037,2.632)--(5.044,2.645)--(5.051,2.708)--(5.058,2.886)--(5.064,2.632)--(5.071,2.616)%
  --(5.078,2.656)--(5.085,2.895)--(5.091,2.668)--(5.098,2.639)--(5.105,2.759)--(5.112,2.781)%
  --(5.118,2.870)--(5.125,2.664)--(5.132,2.672)--(5.139,2.746)--(5.145,2.757)--(5.152,2.821)%
  --(5.159,2.617)--(5.166,2.654)--(5.172,2.691)--(5.179,2.834)--(5.186,2.661)--(5.193,2.703)%
  --(5.199,2.786)--(5.206,2.815)--(5.213,2.885)--(5.220,2.637)--(5.226,2.640)--(5.233,2.758)%
  --(5.240,2.786)--(5.247,2.762)--(5.253,2.746)--(5.260,2.748)--(5.267,2.802)--(5.274,2.687)%
  --(5.280,2.736)--(5.287,2.819)--(5.294,2.947)--(5.301,2.719)--(5.307,2.673)--(5.314,2.779)%
  --(5.321,2.827)--(5.328,2.848)--(5.334,2.692)--(5.341,2.741)--(5.348,2.815)--(5.355,2.931)%
  --(5.361,2.743)--(5.368,2.696)--(5.375,2.778)--(5.382,2.830)--(5.388,2.837)--(5.395,2.662)%
  --(5.402,2.682)--(5.409,2.739)--(5.415,2.920)--(5.422,2.725)--(5.429,2.687)--(5.436,2.778)%
  --(5.442,2.814)--(5.449,2.921)--(5.456,2.663)--(5.463,2.647)--(5.469,2.761)--(5.476,2.819)%
  --(5.483,2.839)--(5.490,2.690)--(5.496,2.637)--(5.503,2.696)--(5.510,2.850)--(5.517,2.615)%
  --(5.523,2.611)--(5.530,2.732)--(5.537,2.815)--(5.544,2.750)--(5.550,2.627)--(5.557,2.727)%
  --(5.564,2.742)--(5.571,2.925)--(5.577,2.633)--(5.584,2.591)--(5.591,2.685)--(5.598,2.871)%
  --(5.604,2.640)--(5.611,2.598)--(5.618,2.660)--(5.625,2.778)--(5.631,2.583)--(5.638,2.638)%
  --(5.645,2.756)--(5.652,2.767)--(5.658,2.635)--(5.665,2.666)--(5.672,2.763)--(5.679,2.795)%
  --(5.685,2.777)--(5.692,2.603)--(5.699,2.613)--(5.706,2.647)--(5.712,2.779)--(5.719,2.593)%
  --(5.726,2.669)--(5.733,2.734)--(5.739,2.820)--(5.746,2.763)--(5.753,2.613)--(5.760,2.649)%
  --(5.766,2.727)--(5.773,2.732)--(5.780,2.777)--(5.787,2.630)--(5.793,2.624)--(5.800,2.654)%
  --(5.807,2.675)--(5.814,2.601)--(5.820,2.696)--(5.827,2.704)--(5.834,2.636)--(5.841,2.666)%
  --(5.847,2.818)--(5.854,2.655)--(5.861,2.550)--(5.868,2.591)--(5.874,2.778)--(5.881,2.555)%
  --(5.888,2.563)--(5.895,2.669)--(5.901,2.682)--(5.908,2.622)--(5.915,2.569)--(5.922,2.670)%
  --(5.928,2.730)--(5.935,2.501)--(5.942,2.509)--(5.949,2.625)--(5.955,2.680)--(5.962,2.690)%
  --(5.969,2.608)--(5.976,2.646)--(5.982,2.686)--(5.989,2.471)--(5.996,2.528)--(6.003,2.639)%
  --(6.009,2.714)--(6.016,2.592)--(6.023,2.556)--(6.030,2.653)--(6.036,2.778)--(6.043,2.593)%
  --(6.050,2.524)--(6.057,2.561)--(6.063,2.733)--(6.070,2.530)--(6.077,2.458)--(6.084,2.509)%
  --(6.090,2.610)--(6.097,2.600)--(6.104,2.533)--(6.111,2.563)--(6.117,2.719)--(6.124,2.490)%
  --(6.131,2.464)--(6.138,2.546)--(6.144,2.707)--(6.151,2.551)--(6.158,2.486)--(6.165,2.540)%
  --(6.171,2.632)--(6.178,2.549)--(6.185,2.534)--(6.192,2.591)--(6.198,2.565)--(6.205,2.609)%
  --(6.212,2.499)--(6.219,2.517)--(6.225,2.629)--(6.232,2.537)--(6.239,2.496)--(6.246,2.574)%
  --(6.252,2.655)--(6.259,2.497)--(6.266,2.518)--(6.273,2.602)--(6.280,2.700)--(6.286,2.618)%
  --(6.293,2.508)--(6.300,2.546)--(6.307,2.585)--(6.313,2.707)--(6.320,2.541)--(6.327,2.539)%
  --(6.334,2.602)--(6.340,2.737)--(6.347,2.544)--(6.354,2.502)--(6.361,2.583)--(6.367,2.674)%
  --(6.374,2.609)--(6.381,2.498)--(6.388,2.569)--(6.394,2.627)--(6.401,2.692)--(6.408,2.606)%
  --(6.415,2.539)--(6.421,2.577)--(6.428,2.623)--(6.435,2.591)--(6.442,2.512)--(6.448,2.519)%
  --(6.455,2.532)--(6.462,2.687)--(6.469,2.466)--(6.475,2.473)--(6.482,2.559)--(6.489,2.697)%
  --(6.496,2.525)--(6.502,2.470)--(6.509,2.510)--(6.516,2.753)--(6.523,2.473)--(6.529,2.461)%
  --(6.536,2.618)--(6.543,2.688)--(6.550,2.827)--(6.556,2.603)--(6.563,2.471)--(6.570,2.569)%
  --(6.577,2.640)--(6.583,2.782)--(6.590,2.575)--(6.597,2.452)--(6.604,2.514)--(6.610,2.542)%
  --(6.617,2.580)--(6.624,2.451)--(6.631,2.522)--(6.637,2.540)--(6.644,2.607)--(6.651,2.473)%
  --(6.658,2.494)--(6.664,2.603)--(6.671,2.679)--(6.678,2.458)--(6.685,2.435)--(6.691,2.506)%
  --(6.698,2.629)--(6.705,2.640)--(6.712,2.448)--(6.718,2.488)--(6.725,2.522)--(6.732,2.783)%
  --(6.739,2.486)--(6.745,2.425)--(6.752,2.500)--(6.759,2.528)--(6.766,2.645)--(6.772,2.434)%
  --(6.779,2.445)--(6.786,2.518)--(6.793,2.597)--(6.799,2.525)--(6.806,2.440)--(6.813,2.524)%
  --(6.820,2.518)--(6.826,2.510)--(6.833,2.523)--(6.840,2.599)--(6.847,2.501)--(6.853,2.430)%
  --(6.860,2.480)--(6.867,2.517)--(6.874,2.568)--(6.880,2.471)--(6.887,2.502)--(6.894,2.418)%
  --(6.901,2.391)--(6.907,2.557)--(6.914,2.526)--(6.921,2.452)--(6.928,2.518)--(6.934,2.530)%
  --(6.941,2.401)--(6.948,2.331)--(6.955,2.389)--(6.961,2.684)--(6.968,2.454)--(6.975,2.404)%
  --(6.982,2.453)--(6.988,2.506)--(6.995,2.597)--(7.002,2.387)--(7.009,2.366)--(7.015,2.410)%
  --(7.022,2.511)--(7.029,2.425)--(7.036,2.423)--(7.042,2.454)--(7.049,2.359)--(7.056,2.379)%
  --(7.063,2.493)--(7.069,2.528)--(7.076,2.333)--(7.083,2.335)--(7.090,2.382)--(7.096,2.562)%
  --(7.103,2.335)--(7.110,2.348)--(7.117,2.399)--(7.123,2.436)--(7.130,2.578)--(7.137,2.392)%
  --(7.144,2.377)--(7.150,2.434)--(7.157,2.364)--(7.164,2.363)--(7.171,2.369)--(7.177,2.349)%
  --(7.184,2.309)--(7.191,2.363)--(7.198,2.549)--(7.204,2.301)--(7.211,2.236)--(7.218,2.271)%
  --(7.225,2.447)--(7.231,2.331)--(7.238,2.309)--(7.245,2.474)--(7.252,2.309)--(7.258,2.342)%
  --(7.265,2.372)--(7.272,2.450)--(7.279,2.358)--(7.285,2.298)--(7.292,2.411)--(7.299,2.266)%
  --(7.306,2.240)--(7.312,2.398)--(7.319,2.382)--(7.326,2.334)--(7.333,2.415)--(7.339,2.466)%
  --(7.346,2.347)--(7.353,2.247)--(7.360,2.327)--(7.366,2.557)--(7.373,2.327)--(7.380,2.306)%
  --(7.387,2.333)--(7.393,2.461)--(7.400,2.356)--(7.407,2.311)--(7.414,2.406)--(7.420,2.375)%
  --(7.427,2.322)--(7.434,2.387)--(7.441,2.477)--(7.447,2.360)--(7.454,2.286)--(7.461,2.309)%
  --(7.468,2.350)--(7.474,2.552)--(7.481,2.343)--(7.488,2.301)--(7.495,2.323)--(7.501,2.472)%
  --(7.508,2.329)--(7.515,2.261)--(7.522,2.397)--(7.528,2.307)--(7.535,2.315)--(7.542,2.468)%
  --(7.549,2.403)--(7.555,2.283)--(7.562,2.276)--(7.569,2.366)--(7.576,2.269)--(7.582,2.288)%
  --(7.589,2.489)--(7.596,2.263)--(7.603,2.211)--(7.609,2.309)--(7.616,2.341)--(7.623,2.451)%
  --(7.630,2.300)--(7.636,2.268)--(7.643,2.431)--(7.650,2.265)--(7.657,2.231)--(7.663,2.301)%
  --(7.670,2.472)--(7.677,2.290)--(7.684,2.274)--(7.690,2.293)--(7.697,2.480)--(7.704,2.209)%
  --(7.711,2.205)--(7.717,2.258)--(7.724,2.297)--(7.731,2.440)--(7.738,2.256)--(7.744,2.266)%
  --(7.751,2.301)--(7.758,2.296)--(7.765,2.212)--(7.771,2.314)--(7.778,2.294)--(7.785,2.267)%
  --(7.792,2.291)--(7.798,2.317)--(7.805,2.275)--(7.812,2.242)--(7.819,2.280)--(7.825,2.352)%
  --(7.832,2.250)--(7.839,2.246)--(7.846,2.397)--(7.852,2.276)--(7.859,2.285)--(7.866,2.329)%
  --(7.873,2.280)--(7.879,2.309)--(7.886,2.434)--(7.893,2.351)--(7.900,2.302)--(7.906,2.298)%
  --(7.913,2.226)--(7.920,2.225)--(7.927,2.452)--(7.933,2.285)--(7.940,2.237)--(7.947,2.288);
\gpcolor{color=gp lt color border}
\draw[gp path] (1.196,4.691)--(1.196,0.985)--(7.947,0.985)--(7.947,4.691)--cycle;
\gpdefrectangularnode{gp plot 1}{\pgfpoint{1.196cm}{0.985cm}}{\pgfpoint{7.947cm}{4.691cm}}
\end{tikzpicture}